%% file: LSQT_main.tex
\documentclass[a4paper,5p,times]{elsarticle}
\usepackage{lipsum}
\usepackage{etoolbox}
\usepackage{url}
\usepackage[numbers]{natbib}
\setcitestyle{square,comma,numbers,sort&compress}

\usepackage{dsfont}
\usepackage{amsmath}
\usepackage{amssymb}
\usepackage{upgreek}
\usepackage{bm}
\usepackage{physics}

\usepackage{algorithm}
\usepackage{algpseudocode}

\def\tsc#1{\csdef{#1}{\textsc{\lowercase{#1}}\xspace}}
\tsc{WGM}
\tsc{QE}
\tsc{EP}
\tsc{PMS}
\tsc{BEC}
\tsc{DE}
\begin{document}
\let\WriteBookmarks\relax
\def\floatpagepagefraction{1}
\def\textpagefraction{.001}

%Title
\title {Linear Scaling Quantum Transport Methodologies}

\author[jinzhou,aalto]{Zheyong Fan\corref{cor2}}%[orcid=0000-0002-2253-8210]
\author[barcelona]{Jose Hugo Garcia\corref{cor2}}%[ orcid=0000-0002-5752-4759] 

\author[barcelona]{Aron W. Cummings\corref{cor2}}%[ orcid=0000-0003-2307-497X]
\author[mexico]{Jose Eduardo Barrios-Vargas}%[ orcid=0000-0002-6880-8941]
\author[dresden]{Michel Panhans}%[ ]
\author[aalto]{Ari Harju}%[ orcid=0000-0002-2233-2896]
\author[dresden]{Frank Ortmann}%[ orcid=0000-0002-5884-5749]
\author[barcelona,icrea]{Stephan Roche \corref{cor1}}%[ orcid=0000-0003-0323-4665]
\ead{stephan.roche@icn2.cat}

\address[jinzhou]{School of Mathematics and Physics, Bohai University, Jinzhou, China}
\address[aalto]{QTF Centre of Excellence, Department of Applied Physics, Aalto University, FI-00076 Aalto, Finland}
\address[barcelona]{Catalan Institute of Nanoscience and Nanotechnology (ICN2), CSIC and The Barcelona Institute of Science and Technology, Campus UAB, Bellaterra, 08193 Barcelona, Spain}
\address[mexico]{Departamento de F\'isica y Química Te\'orica, Facultad de Qu\'imica, UNAM, Mexico City 04510, Mexico}
\address[dresden]{Center for Advancing Electronics Dresden Technische Universit\"{a}t Dresden 01062 Dresden, Germany}
\address[icrea]{ICREA – Instituci\'{o} Catalana de Recerca i Estudis Avan\c cats, 08010 Barcelona, Spain}

\cortext[cor1]{Corresponding author}
\cortext[cor2]{Equal contribution}

\begin{abstract}
In recent years, predictive computational modeling has become a cornerstone for the study of fundamental electronic, optical, and thermal properties in complex forms of condensed matter, including Dirac and topological materials. The simulation of quantum transport in realistic models calls for the development of linear scaling, or order-$N$, numerical methods, which then become enabling tools for guiding experimental research and for supporting the interpretation of measurements. In this review, we describe and compare different order-$N$ computational methods that have been developed during the past twenty years, and which have been used extensively to explore quantum transport phenomena in disordered media. We place particular focus on the zero-frequency electrical conductivities derived within the Kubo-Greenwood and Kubo-Streda formalisms, and illustrate the capabilities of these methods to tackle the quasi-ballistic, diffusive, and localization regimes of quantum transport in the noninteracting limit. The fundamental issue of computational cost versus accuracy of various proposed numerical schemes is addressed in depth. We then illustrate the usefulness of these methods with various examples of transport in disordered materials, such as polycrystalline and defected graphene models, 3D metals and Dirac semimetals, carbon nanotubes, and organic semiconductors. Finally, we extend the review to the study of spin dynamics and topological transport, for which efficient approaches for calculating charge, spin, and valley Hall conductivities are described.
\end{abstract}


\maketitle

\tableofcontents

\input{secI_introduction}

\input{secII_kubo_formulas}

\input{secIII_numerical_techniques}

\input{secIV_landauer}

\input{secV_numerical_examples}

\input{secVI_applications}

\input{secVII_hall_spins}

\section{Summary and conclusions\label{section:summary}}

This paper has reviewed the development of linear-scaling numerical methods applied to quantum transport based on the Kubo-Greenwood and Kubo-Streda formalisms. These methods provide insight into the transport physics in the ballistic, diffusive, and localized regimes, as well as in topological regimes such as the quantum Hall effect.

The fundamental issue of computational cost versus numerical accuracy of various proposed numerical schemes has been addressed in detail, illustrating the capabilities and limitations of each. The usefulness of the time-propagation methods has been shown for the calculation of the dissipative conductivity, since it allows one to track the conduction regime in which the quantum conductivity is computed. This is actually critical for disordered systems since the onset of (anti)localization effects will reveal the (up)downscaling behavior of the conductivity. Meanwhile, the implementations based on KPM-type of polynomial expansions become much more practical in the presence of topological gaps when compared with time-propagation methods, and allow for a faster convergence of the results. Finally, we have illustrated the applicability of such approaches to spin and valley Hall conductivities as well as to the time evolution of spin densities, while some references to the efforts to improve the scaling behavior of computational approaches for the Landauer--B\"uttiker conductance were also outlined.

Today, linear-scaling quantum transport methodologies stand as unique computational methodologies to explore many emerging and complex quantum transport phenomena in modern condensed matter physics, including disordered topological materials such as topological Anderson insulators \cite{GrothPRL2009,JianPRL2009,ZhangPRB2012}, three-dimensional models of Dirac semimetals \cite{YoungPRL2012,KobayashiPRL2014,PixleyPRL2015,Louvet_2018}, and topological insulators  \cite{HasanRMP2010,FuPRL2007,Soriano2012,KobayashiPRL2014, Liao2015,Chiu2016,Araki2019}, which all display nontrivial transport features difficult to fully tackle with perturbative approaches and simplified effective models.

We hope that interested readers will harness such enabling tools to investigate unexplored quantum transport phenomena in complex matter, and that the clarification of the capabilities of such methods, as well as their dissemination through various dedicated open sources, will also promote their use in machine learning strategies \cite{Schleder2019}, therefore taking part in the global efforts to bring materials simulation to its highest level of predictability.

\section*{Acknowledgements}
We are indebted to D.\ Mayou, F.\ Triozon, S.\ Latil, A.\ Lherbier, Y.M.\ Niquet, N.\ Leconte, D.\ Van Tuan,  C. \ Benoit,  M.\ Settnes, R.\ Mart{\'i}nez-Gordillo, T.\ Rappoport, W.\ Li, H.\ Sevin\c{c}li, J.P.\ Gaspard, F.\ Ducastelle, Ph.\ Lambin, R.\ Saito, L.\ Covaci, G.\ Cuniberti, M.\ Brandbyge, D.\ Soriano, S.R.\ Power, Th.\ Louvet, J.\ Jiang, T. \ Fujiwara, A.\ Ferreira, J.C.\ Charlier and A.\ Pekka-Jauho for fruitful discussions, comments, and inspiring collaborations. ZF was supported by National Natural Science Foundation of China under Grant No.\ 11974059. FO would like to thank the Deutsche Forschungsgemeinschaft for financial support (project OR-349/1). ZF and AH acknowledge the computational resources provided by Aalto Science-IT project and Finland's IT Center for Science (CSC). SR, AWC and JHG acknowledge PRACE and the Barcelona Supercomputing Center (Project No.\ 2015133194), and the European Union Seventh Framework Programme under Grant Agreement No.\ 785219 Graphene Flagship. ICN2 is supported by the Severo Ochoa program from Spanish MINECO (grant no.\ SEV-2017-0706) and funded by the CERCA Programme / Generalitat de Catalunya.

\bibliographystyle{elsarticle-num-names}\biboptions{sort&compress}
\bibliography{refs}

\end{document}

%% file: secI_introduction.tex
\section{Introduction}
\label{section:introduction}

The study and development of new materials and devices often involves three complementary avenues of exploration -- experiments, theory, and numerical simulations. Theory provides a framework to explain or predict material or device behavior, while numerical simulations are often needed to apply these theories to the complex situations that are encountered in experiments. In electronic devices, understanding the flow of electrons in response to an electric field is of central importance, and it is also a fundamental issue in condensed matter physics. The performance of devices in many applications in electronics, thermoelectrics, spintronics, optoelectronics, and photovoltaics is intricately connected to the material's electrical conductivity or charge carrier mobility. To this end, there is a need for simulation tools that can accurately describe electronic transport in complex materials and devices, and do so in an efficient manner in order to reach the length scales typically seen in experiments.

To study electronic transport in disordered materials and devices on experimental length scales, one often needs to consider large systems consisting of many millions or billions of atoms. To simulate such systems accurately and efficiently, two basic ingredients are needed. The first is a realistic description of the structure and electronic properties of the material of interest. This can be achieved by using {\it ab initio} electronic structure methods such as density functional theory (DFT) \cite{hohenberg1964pr, kohn1965pr, kohn1999rmp, jones2015rmp}. DFT has proven to be highly successful for describing the electronic, optical, and vibrational properties of a large number of materials. However, the computational cost of {\it ab initio} methods severely limits the size of the systems that can be studied. This limitation can be overcome with quasiparticle-based real-space tight-binding (TB) models, for which the Hamiltonian describing the electronic properties of the system becomes highly sparse, allowing for efficient numerical simulation. Therefore, using {\it ab initio} methods as a basis for the construction of appropriate TB models is currently the most successful approach for describing electronic and transport properties of large-area, spatially complex disordered or nanostructured materials. 

The second ingredient needed to study large disordered systems is an efficient numerical method for simulating electronic transport. Ideally, to reach experimental length scales this method should be linear-scaling, i.e., its computational cost should be directly proportional to the number of atoms $N$. Such methods are also called order-$N$ or $\mathcal{O}(N)$ methods. The development of stable $\mathcal{O}(N)$ algorithms for the calculation of spectral and transport quantities was initiated by the seminal works of Roger Haydock \cite{haydock1980ssp, haydock1975jpc,haydock1972jpc}, who first derived a real-space approach to compute spectral functions in disordered materials using the so-called recursion and continued fraction expansion technique. This was followed by the introduction of improved techniques for the computation of density of states, correlation functions, and transport coefficients in disordered materials. Techniques employing orthogonal polynomials (such as Chebyshev expansion)  \cite{ezer1984jcp, leforestier1991jcp, petitfor1985book} and the kernel polynomial method (KPM), have shown superior performance \cite{weisse2006rmp} and are experiencing growing popularity for studying the dynamics of quantum systems \cite{fehske2009pla, JunPRB2007}. Consequently, they find a considerable range of applications in chemistry and physics, including the fields of disordered systems, electron-phonon interactions, quantum spin systems, and strongly correlated quantum systems \cite{viswanath1994book, BoehnkePRB2011, Ganahl2014}.

A fully quantum treatment of charge transport is undoubtedly a great asset and marks a significant advance over simpler classical models. However, under the proper conditions electronic transport can be treated classically, with electrons in a material behaving as point particles that are scattered by various sources, such as lattice defects, impurities, or phonons, which can have qualitative similarities to the regime of Variable-Range Hopping for disordered systems \cite{2020PhysRep}.
 In this semiclassical picture, the electrical conductivity is proportional to the momentum relaxation time, $\sigma \propto \tau_\text{p}$, which is the average time it takes for scattering processes to randomize an electron's direction of motion. The conductivity is also independent of the system size in this regime. Each scattering process can be treated fully quantum mechanically, but between scattering events the electrons behave as point particles. This semiclassical picture is appropriate in the limit $k_\text{F} l_\text{e} \gg 1$, where $l_\text{e}$ is the average distance between scattering events, $k_\text{F} = 2\pi / \lambda_\text{F}$, and $\lambda_\text{F}$ is the Fermi wavelength of the electrons.

The limits of this semiclassical description are met when the wave nature of electrons starts to play a role, i.e., when $\lambda_\text{F}$ becomes comparable to $l_\text{e}$. In this regime, quantum interference effects can become important. In particular, constructive interference in closed scattering loops can lead to the coherent localization of electrons, resulting in a decrease of the conductivity compared to its semiclassical value. The calculation of quantum corrections to the semiclassical conductivity was pioneered in 1979 by Abrahams \textit{et al.}\ \cite{abrahams1979prl}, who developed a scaling theory of localization in which the zero-temperature conductivity of a disordered material depends universally on its length scale $L$, and transitions smoothly from a logarithmic or slower decay to an exponential decay with increasing $L$. At the same time, the leading quantum corrections to the semiclassical conductivity were shown to be driven by coherent backscattering of electrons from momenta $k$ to $-k$ \cite{Gorkov1979, Altshuler1980}. This phenomenon of weak localization (WL) has now been studied in many different materials and has been the topic of extensive reviews \cite{lee1985rmp, rammer1986rmp, Belitz94, Montambaux}.

An important consequence of localization is that the conductivity becomes dependent on the system size. Thus, in low-dimensional materials and devices where localization effects are more pronounced, an accurate treatment of the impact of quantum effects on electronic transport is crucial. Over the past several decades, this has been revealed by the study of quantum interference in low-dimensional semiconductor systems, including quantum wells, superlattices, and nanowires, as well as in a wide variety of organic systems \cite{rurali2010rmp, dasgupta2014am, charlier2007rmp, laird2015rmp}. More recently, the growing interest in low-dimensional materials such as carbon nanotubes, graphene, and transition metal dichalcogenides \cite{castro2009rmp, das2011rmp, ferrari2015nanoscale, mucciolo2010jpcm, geim2013nature, novoselov2016science}, among many others, also highlights the need for efficient ways to calculate the electrical conductivity while fully accounting for quantum effects.

There are several common approaches for simulating electronic transport, including the Boltzmann transport equation, the Landauer-B\"{u}ttiker formalism, and the Kubo formula. The Boltzmann transport equation describes the dynamics of the electron distribution function, and is traditionally applied to the semiclassical regime of transport described above. However, with appropriate extensions it can also describe quantum interference effects, for example through the use of Wigner distribution functions \cite{Bordone1999, Nedjalkov2004} or by introducing nonlocal terms into the collision integral \cite{Hershfield1986}. The Landauer-B\"{u}ttiker formalism expresses the electrical conductance in terms of transmission probabilities, which are calculated from the full quantum mechanical scattering matrix, and thus naturally captures the impact of quantum effects on electron transport. Traditionally the Landauer-B\"{u}ttiker formalism has been applied to devices with two or more electrical contacts \cite{landauer1957ibm, landauer1970pm, buttiker1985prb}, but with the proper choice of self energies it can also handle bulk systems \cite{StoneIBM1988, Baranger89, Nikolic2001}.

In this review, we focus on efficient numerical calculations of the Kubo and Kubo-Bastin formulas for the electrical and Hall conductivities. In its most general form, the Kubo formula describes the linear response of a system to a time-dependent perturbation. It does so by connecting the linear response with time-dependent correlation functions in the absence of the perturbation \cite{Pines89, Doniach74}. In the case of electrical conductivity, it allows one to calculate the charge current response to an electric field through the current-current correlation function in the absence of the electric field. We note that by making a connection between (i) the response of a system to an external perturbation, and (ii) its response to spontaneous system fluctuations, the Kubo formula is a manifestation of the fluctuation-dissipation theorem \cite{kubo1985book, kubo1965rpp}. It is also known as the Kubo-Greenwood approach for non-interacting electrons.

The calculation of the Kubo conductivity using $\mathcal{O}(N)$ techniques was pioneered by Thouless and Kirkpatrick in their study of a one-dimensional linear chain \cite{thouless1981jpc}. A subsequent attempt to perform real-space calculations of the Kubo formula was made in higher-dimensional models \cite{Bose1993}, but an important step forward was accomplished by Mayou and Khanna who extended recursion methods to compute the frequency-dependent conductivity \cite{mayou1988epl, mayou1995jpi}. Roche and Mayou further combined real-space $\mathcal{O}(N)$ recursion techniques with time-propagation methods to evaluate the Kubo conductivity in its zero-frequency version \cite{roche1997prl, roche1999prb}. One of the main advantages of such approaches is the ability to identify different regimes of quantum transport -- ballistic, diffusive, and localized -- by following the time-dependent spatial spreading of quantum wavepackets. Similar types of methodology, as well as other algorithms using the KPM technique, have extended the capability of these methods to the study of other quantities such as the Hall conductivity \cite{ortmann2013prl, garcia2015prl, ortmann2015prb}, spin dynamics \cite{vantuan2014np, Cummings2017prl, vierimaa2017prb}, and lattice thermal conductivity in disordered systems \cite{li2010prb, li2011prb, sevincli2011prb}.

In recent years, the predictive power of such methods has been demonstrated in a large variety of realistic models of disordered graphene and two-dimensional materials \cite{lherbier2008prl_chemical, WehlingPRL2010, yuan2010prb, radchenko2013prb, laissardiere2013prl, GargiuloPRL2014, zhao2015prb, ferreira2015prl}, multilayer graphene \cite{yuan2010prb_b, Missaoui_2018}, organic semiconductors \cite{ortmann2011prb, Ishii2015, Ishii2017, FratiniNatMat2017, Ishii2018} and conducting polymers \cite{Ihnatsenka2015, Adjizian2016, TonneleJPM2019}, quasicrystals and aperiodic systems \cite{roche1997prl, roche1997jmp, DELAISSARDIERE201470,TramblyPRL2006}, silicon nanowires \cite{markussen2006prb, persson2008nl}, carbon nanotubes \cite{latil2004prl,LatilNL2005,ishii2010prl} and three-dimensional models of topological insulators \cite{Soriano2012, wehling2014aip, cresti2016rnc}. Charge, spin, and Hall transport coefficients have been numerically computed in different transport regimes, including the quasi-ballistic, diffusive, weak localization, weak antilocalization (WAL), and strong (Anderson) localization regimes, providing in-depth quantitative analysis directly comparable with experimental data. Today these approaches have become a cornerstone for the simulation of quantum transport in complex situations that are out of reach of analytical treatments and perturbative methods, especially in the presence of weak magnetic fields and for experimentally relevant disordered systems containing many millions of atoms.

This review covers more than twenty years of research dedicated to the numerical implementations of the Kubo formula for the electrical conductivity. Its purpose is to provide a comprehensive description of the most efficient linear-scaling algorithms for studying electronic transport in complex forms of disordered materials. The review is organized as follows. In Sec.\ \ref{section:kubo_formulas} we derive a few forms of the single-particle Kubo formula, emphasizing those that are based on the velocity autocorrelation function and the mean-square displacement. Section \ref{section:numerical_techniques} discusses the numerical implementations that enable linear-scaling calculations of quantum transport using these formulas. Section \ref{section:lb} describes how the linear-scaling techniques described in this review can be applied to the Landauer-B{\"u}ttiker formalism. In Sec.\ \ref{section:numerical_examples} we provide explicit examples of how these methods may be used to describe electrical transport in the ballistic, diffusive, and localized regimes, and highlight similarities and differences between different approaches with respect to accuracy and computational cost. Section \ref{section:applications} summarizes and illustrates a variety of applications of this methodology to charge transport in disordered graphene, 3D metals and Dirac semimetals, carbon nanotubes, and organic semiconductors. Section \ref{section:hall_spin} presents further extensions of this method to calculations of the Hall conductivity and to spin dynamics. Finally, a summary and general conclusions are given in Sec.\ \ref{section:summary}. This review is intended to communicate essential knowledge about physics and algorithms on equal footing, and we hope it will serve as a valuable resource for future developers and users of such methodologies, which can be applied to the large variety of materials of current interest for fundamental science and advanced technologies.

%% file: secII_kubo_formulas.tex
\section{Quantum linear response theory and Kubo formulas }
\label{section:kubo_formulas}

A fundamental method for extracting information about the intrinsic properties of a system is to measure its response to an external perturbation. A perturbation can be, e.g., an electric field or a temperature gradient, and the response can be an electric current or a heat flux. In general, the response of a system can be very complex, but for perturbations that are small enough, one intuitively expects that the response will be proportional to the perturbation. This is the fundamental assumption of linear response theory, and is the starting point of the work of Ryogo Kubo \cite{kubo1957jpsj}, who showed under general conditions that if the perturbation is applied sufficiently slowly such that the system always remains close to its equilibrium, one can express the response of the system in terms of its equilibrium properties. This result, currently known as the Kubo formula, is one of the pillars of modern quantum transport theory, and serves as the starting point for the different linear scaling quantum transport (LSQT) methodologies discussed in this review.

Although the Kubo formula can be used to extract transport coefficients in the presence of many-body interactions \cite{Dugaev2005, Bohr2006, Langer1962}, we will focus only on its applications to non-interacting disordered systems. This choice is dictated by the fact that the LSQT methodologies have been developed and optimized for these systems, and their full capabilities are thus only attained when used within this context. On the other hand, the treatment of disordered systems is necessary due to the unavoidable presence of defects and disorder in real materials. In this section, we will first outline a derivation of the general Kubo formula based on quantum linear response theory along the same lines as G.\ D.\ Mahan \cite{mahan2000book}. Then we will proceed to derive a non-interacting Kubo formula that allows for defining the single-particle density matrix, which is useful for obtaining the expectation values of physical observables in systems out of equilibrium. After the general derivations, we will focus on the specific case of electrical conductivity and derive different but equivalent representations of the non-interacting Kubo formula, which will serve as the starting points for the different LSQT methodologies. Next we will discuss the meaning of Green's function regularization and its effect on the dissipative conductivity, as well as its relation to the different transport regimes. Such a discussion is crucial for understanding the numerical simulations that will appear in later sections. Finally, we will define time- and length- dependent forms of the electrical conductivity, which are needed to identify the ballistic, diffusive, and localized regimes of transport, and to make a proper comparison to experiments.

\subsection{Quantum linear response theory and the many-body Kubo formula} 
\label{ManyBody}

The Kubo formula is derived under four fundamental assumptions \cite{ventra2008book, mahan2000book}:
\begin{enumerate}
    \item The system is at thermal equilibrium before the application of the perturbation.
    \item The response is linear with respect to the perturbation strength.
    \item The perturbation is turned on adiabatically.
    \item The system is closed (although not isolated) and evolves under unitary Hamiltonian dynamics.
\end{enumerate}
Although these assumptions can be relaxed, for example, by employing the Keldysh formalism \cite{Rammer2007}, they are sufficient for the systems we are dealing with in this review and allow for a general and rigorous derivation of the Kubo formula. 

A general way to describe the state of a quantum system is by specifying its density matrix $\hat{\rho}$, which can then be used to compute the expectation value $\langle\hat{A}(t)\rangle$ of a general physical quantity, described by the quantum mechanical operator $\hat{A}$ at time $t$ as \cite{sakurai_napolitano_2017}
\begin{equation}
\langle\hat{A}(t)\rangle = \text{Tr} \left[ \hat{A}\, \hat{\rho}(t) \right], 
\label{equation:expectation-of-A}
\end{equation}
where $\text{Tr}[...]$ denotes the trace over a complete basis set. Therefore, in order to determine the evolution of a certain observable, one must first determine the evolution of the density matrix $\hat{\rho}(t)$ after the perturbation is turned on. The evolution of the density matrix can be described by the von Neumann equation, also called the quantum Liouville equation,
\begin{equation}
i \hbar \frac{d \hat{\rho}(t)}{dt} = \left[ \hat{H}_{\rm tot}(t), \hat{\rho}(t) \right],
\label{equation:von-Neumann}
\end{equation}
where $\hat{H}_{\rm tot}(t)$ is the total Hamiltonian of the system, generally time dependent, and the brackets represent a commutator.

Assumption 1 states that at some long time in the past, the system is in an equilibrium state described by $\hat{\rho}_{\rm eq}$, which is obtained from the unperturbed Hamiltonian $\hat{H}$. After some time, a perturbation $\hat{H}'$ is then switched on adiabatically from $t = -\infty$ to the present time $t$ \cite{ventra2008book,mahan2000book}. As long as this process is sufficiently slow, the exact time dependence of the perturbation is not important and one can choose an arbitrary function to describe it. Here we assume an exponential increase of the perturbation with a rate of $1/\tau_{\phi}$ and add the perturbation to the equilibrium Hamiltonian to define the total Hamiltonian of the system,
\begin{equation}
\hat{H}_{\rm tot}(t) = \hat{H} + \lim_{\tau_{\phi} \to \infty} e^{t/\tau_{\phi}} \hat{H}'.
\label{equation:total-hamiltonian}
\end{equation}
The limit $\tau_{\phi} \rightarrow \infty$ has to be taken in order to force an adiabatic evolution in agreement with assumption 3. At this point, it is convenient to point out the importance of the order of the limits; in order to agree with the assumption that the perturbation vanishes at $t=-\infty$ -- $\lim_{t\rightarrow -\infty} \hat{H}_{\rm tot}(t)= \hat{H}$ -- one must make sure that $\tau_\phi \rightarrow \infty$ at a slower rate than $t \rightarrow -\infty$. 

In this review we are interested in the zero-frequency electrical response, and therefore we will focus on the case where the perturbation is a static electric field. Due to gauge invariance, there is no unique way to introduce the electric field into the Hamiltonian. We choose to express the static electric field $\bm{E}_0(\bm{r})$ in terms of a scalar potential $\phi(\bm{r}) = -\bm{r}\cdot\bm{E}_0$, and write the perturbation as
\begin{equation}
    \hat{H}' = \int d^3 \bm{r} \rho(\bm{r}) \phi(\bm{r}), \label{equation:electric_field}
\end{equation}
where $\rho(\bm{r})$ is the charge density and the time dependence of the perturbation is assumed to be only embodied in the exponential function in Eq.\ (\ref{equation:total-hamiltonian}). This choice of gauge also implies that we are neglecting any induced magnetic field due to the change of the electric field, a condition which is justified by assumptions 2 and 3 above.

Solving the von Neumann equation in Eq.\ (\ref{equation:von-Neumann}) for the Hamiltonian in Eq.\ (\ref{equation:total-hamiltonian}) is generally a challenging task. However, based on the second and third assumptions, we can assume that the density matrix is just slightly out of equilibrium and make an ansatz that it follows a similar time evolution as the perturbation,
\begin{equation}
\hat{\rho}(t) = \hat{\rho}_{\rm eq} + \lim_{\tau_{\phi} \to \infty} e^{ t /\tau_{\phi}} \Delta \hat{\rho}(t)\label{equation:nonequilibrium-density-matrix}.
\end{equation}
Here, $\Delta \hat{\rho}(t)$ is a small deviation of the density matrix from its equilibrium value, and is assumed to vanish in the limit $t\to -\infty$ and to be linearly proportional to $\bm{E}_0(\bm{r})$, in the same way as the perturbation. By invoking Maxwell's equations, substituting Eq.\ (\ref{equation:total-hamiltonian}) and Eq.\ (\ref{equation:nonequilibrium-density-matrix}) into Eq.\ (\ref{equation:von-Neumann}), and dropping terms that are nonlinear in  $\bm{E}_0(\bm{r})$ (assumption 2), one can obtain an expression for $\Delta \hat{\rho}(t)$. We then substitute this into Eq.\ (\ref{equation:expectation-of-A}) to obtain an expression for the nonequilibrium expectation value of operator $\hat{A}$ under the action of the static external electric field $\bm{E}_0(\bm{r})$ \cite{mahan2000book, ventra2008book},
\begin{align}
\langle \hat{A} \rangle &= \lim_{\tau_{\phi} \to \infty}\Omega \int_{0}^{\infty}  dt e^{- t/\tau_{\phi}}
\int_0^{\beta} d\lambda \nonumber \\
&\times \text{Tr} \left[\hat{\rho}_{\rm eq} \hat{A}(0)\, \hat{\bm{J}}(t +i\hbar\lambda )\right]\cdot \bm{E}_0(\bm{r})
\label{equation:expectation-operator},
\end{align}
where $\Omega$ is the volume of the system and $\beta=1/k_{\rm B}T$ is the inverse thermal energy. The time-dependent current density operator is defined in the interaction picture as  
\begin{equation}
\hat{\bm{J}}(t)=\hat{U}^{\dagger}(t)\hat{\bm{J}}\hat{U}(t),
\end{equation}
where
\begin{equation}
\hat{U}(t)=e^{-i\hat{H}t/\hbar}
\end{equation}
is the time evolution operator associated with the unperturbed Hamiltonian $\hat{H}$.

Equation (\ref{equation:expectation-operator}) is the direct-current (DC) Kubo formula for electrical response. This formula was first derived by Kubo for computing the dissipative electrical conductivity \cite{kubo1957jpsj}, for which $\hat{A}$ is chosen as the current density $\hat{J}_{\alpha}$ in the same direction $\alpha$ as the electric field. Equation (\ref{equation:expectation-operator}) then becomes
\begin{align}
\langle \hat{J_\alpha} \rangle &= E_0(\bm{r}) \left\{\lim_{\tau_{\phi} \to \infty}\Omega \int_{0}^{\infty}  dt e^{- t/\tau_{\phi} }
\int_0^{\beta} d\lambda \right.\nonumber \\
&\times \text{Tr} \left.\left[\hat{\rho}_{\rm eq}  \hat{J_\alpha}(0) \hat{J_\alpha}(t +i\hbar\lambda )\right] \right\},
\label{equation:current-current}
\end{align}
where $E_0(\bm{r})$ is the magnitude of $\bm{E}_0(\bm{r})$. According to Ohm's law, the expression in the braces is just the DC electrical conductivity $\sigma$. The Kubo formula can also be interpreted as a manifestation of the fluctuation-dissipation theorem \cite{kubo1985book,kubo1965rpp}, which states that the response of a system to a small external perturbation is equivalent to its spontaneous fluctuations at equilibrium. In this case, the response of the electrical current to an electric field is equivalent to the spontaneous fluctuations of the equilibrium current, captured by current-current correlation function $[\hat{\rho}_{\rm eq}  \hat{J_\alpha}(0) \hat{J_\alpha}(t +i\hbar\lambda )]$.

\subsection{Kubo formulas for noninteracting electrons \label{section:kubo-noninteracting}}

In many situations, many-body effects driven by electron-electron interactions remain weaker than the effects of disorder. Therefore, it would be overkill and often impractical to use the general many-body Kubo formula. The noninteracting problem of $N$ particles is equivalent to solving a single-particle problem and occupying the single-particle states with $N$ particles with correct statistics. In the noninteracting approximation, all the many-body operators can be conveniently represented in second quantization notation \cite{mahan2000book} using a complete set of orthonormal eigenvectors $\{\ket{n}\}$ of the single-particle Hamiltonian $\hat{H}$, $\hat{H}\ket{n}=E_n\ket{n}$. In this notation, any operator can be expressed as
\begin{equation}
\hat{A} = \sum_{m,n} c_m^\dagger c_n \bra{m} \hat{A} \ket{n},
\label{equation:J_nu_0}
\end{equation}
where $c_m^\dagger$ and $c_n$ are the creation and annihilation operators of an electron in the single-particle eigenstates $\ket{m}$ and $\ket{n}$ respectively, and $\bra{m} \hat{A} \ket{n}$ is the matrix element of the single-particle operator. The time-dependent current density operator in Eq.\ (\ref{equation:current-current}) can then be expressed in second quantization notation as
\begin{equation}
\hat{J}_{\alpha}(t+i\hbar\lambda) = \sum_{p,q} c_p^\dagger c_q \bra{p} \hat{J}_{\alpha} \ket{q} e^{i(E_p-E_q)(t+i\hbar\lambda)/\hbar},
\label{equation:J_mu_t}
\end{equation}
where the exponential comes from the time evolution operator. 
Inserting Eqs.\ (\ref{equation:J_nu_0}) and (\ref{equation:J_mu_t}) into Eq.\ (\ref{equation:expectation-operator}), making use of the identity \cite{allen2006book}
\begin{align}
\text{Tr}[\hat{\rho}_{\rm eq}c_m^\dagger c_n c^\dagger_p c_q]
&= \delta_{mq}\delta_{np}f(E_m)[1-f(E_n) ] \nonumber \\
&+ \delta_{mn}\delta_{pq}f(E_m)f(E_p),
\end{align}
and performing the integration over $\lambda$ and $t$, one obtains the single-particle Kubo formula 
\begin{align}
\langle \hat{A(\mu)} \rangle
&= i \hbar\Omega \lim_{\tau_{\phi} \to \infty} \sum_{m,n} \frac{f(E_m-\mu)-f(E_n-\mu)}{(E_n-E_m)(E_n-E_m+i\hbar/\tau_{\phi} )} \nonumber \\
&\times \bra{m} \hat{A}\ket{n}\bra{n}\,(\hat{\bm{J}}\cdot\bm{E}_0)\,\ket{m},
\label{equation:sigma_mu_nu_without_t}
\end{align}
where
\begin{equation}
f(E_m-\mu)=\frac{1}{e^{\beta(E_m-\mu)}+1}\label{equation:fermi_dirac}
\end{equation}
is the Fermi-Dirac distribution function, with $\mu$ being the chemical potential or Fermi level.

Equation (\ref{equation:sigma_mu_nu_without_t}) expresses the conductivity in terms of the eigenvalues and eigenvectors of the Hamiltonian, which in general are difficult to obtain. Therefore, it is desirable to find an expression that depends solely on $\hat{H}$ instead. This can be done by using the definition of the trace as well as the following identities ($F$ is a general function):
\begin{align}
    \int dE' \,F(E')\delta(E'-E_n) &=F(E_n), \nonumber\\
    \sum_n F(E_n) \ket{n}\bra{n} &= F(\hat{H}). \label{equation:spectral_identities}
\end{align}
After some algebra, we can rewrite Eq.\ (\ref{equation:sigma_mu_nu_without_t}) as
{\small \begin{align}
&\langle \hat{A}(\mu) \rangle= \lim_{\tau_{\phi}\rightarrow \infty}-i \hbar\Omega  \int dE' f(E'-\mu)\times \nonumber \\
&\text{Tr}\Big[ \delta(E'-\hat{H})\hat{A}\frac{1}{(\hat{H}-E')(\hat{H}-E'-i\hbar/\tau_{\phi})}(\hat{\bm{J}}\cdot\bm{E}_0)- \text{h.c.}\Big], \label{equation:kubo_temp}
\end{align}
}
where h.c.\ stands for the Hermitian conjugate of the preceding operator within the trace, and $\delta(E-\hat{H})$ is the projector onto the eigenstates of the Hamiltonian with energy $E$. Furthermore, one can express Eq.\ (\ref{equation:kubo_temp}) in terms of Green's functions, which will then allow its evaluation using Green's function techniques. To this end, we first define the retarded ($G^{+}$) and advanced ($G^{-}$) regularized Green's functions, 
\begin{equation}
    G^\pm(E;\tau_{\phi} , \tau )= \mp i \int_{0}^{ \tau } \frac{dt}{\hbar }\,e^{\pm i(E-\hat{H}\pm i \hbar/ \tau_{\phi})t/\hbar }, \label{equation:reg_green}
\end{equation}
where $\tau>0$ and $\tau_{\phi}>0$ are regularization parameters which can be interpreted as a finite evolution time and a finite quasi-particle lifetime, respectively. These functions can be used to obtain the exact Green's functions through the limits
\begin{align}
    G^\pm(E) &= \lim_{\tau_{\phi} \rightarrow \infty} \lim_{\tau\rightarrow \infty} G^\pm(E;\tau_{\phi} , \tau) \nonumber \\
    &= \lim_{\tau_{\phi} \rightarrow \infty} \frac{1}{E-\hat{H}\pm i\hbar/\tau_{\phi}},
\end{align}
where the order of the limits should be respected. Finally, using the identity
\begin{equation}
\lim_{h\rightarrow 0} \,   \frac{1}{x(x+h)}=-\lim_{h\rightarrow 0} \,\frac{d}{dx} \left(\frac{1}{x+h}\right),
\end{equation}
one can replace the energy factor in Eq.\ (\ref{equation:kubo_temp}) by the derivative of the retarded Green's function, leading to the Kubo-Bastin formula \cite{bastinjphyschem1971}
\begin{align}
\langle \hat{A}(\mu) \rangle
=~&i \hbar\Omega  \int dE' f(E'-\mu)\nonumber\times \\
&\text{Tr}\Big[  \delta(E'-\hat{H})\hat{A}\frac{dG^+(E')}{dE'}(\hat{\bm{J}}\cdot\bm{E}_0) - \text{h.c.}\Big]. \label{equation:sigma_mu_nu_Bastin}
\end{align}
The Kubo-Bastin formula is as general as the original Kubo formula in Eq.\ (\ref{equation:expectation-operator}), but within the noninteracting approximation. It can be used to compute the electrical response of \emph{any} operator, with the advantage that one only needs to know the Hamiltonian and the Green's functions of the system. The LSQT methods are primarily based on methodologies for effectively computing the Green functions and the energy projection operator, as we will show in the following sections.

An important point here concerns the roles of $\tau_\phi$ and $\tau$. As can be seen from Eqs.\ (\ref{equation:kubo_temp}) and (\ref{equation:reg_green}), $\tau_\phi$ can be interpreted as a finite quasi-particle lifetime, arising from interactions between the system and the external electric field that provide a means of exchanging energy inelastically at a rate of $1/\tau_{\phi}$. Adiabatic evolution ensures that this rate is vanishingly small, but it is always nonzero in practical numerical calculations. A nonzero $1/\tau_{\phi}$ has an important effect on the transport coefficients, and the interplay between $\tau_\phi$ and the evolution time $\tau$ can affect the convergence of the results in numerical simulations \cite{Schleede2010, thouless1981jpc}, as we will discuss in greater detail below. 

\subsection{The dissipative conductivity}\label{diss_cond}

In this section we will focus on the dissipative conductivity, which is one of the main probes of the electronic properties of materials. This section serves to prepare the reader for the numerical methods to be introduced in Sec.\ \ref{section:numerical_techniques}. Therefore, we will clearly define the main quantities needed for calculating the dissipative conductivity. An important quantity here is the current density which, in the single-particle approximation, is proportional to the velocity operator $\hat{\bm{V}}$, $\hat{\bm{J}}=q\hat{\bm{V}}/\Omega$, with $q$ being the charge of a single carrier ($q = -e$ for electrons). The velocity operator can be  calculated using the Heisenberg equation of motion,
\begin{equation}
    \hat{\bm{V}} = \frac{i}{\hbar}[\hat{H},\hat{\bm{R}}], \label{equation:x_and_v}
\end{equation}
where $\hat{\bm{R}}\equiv(\hat{X},\hat{Y},\hat{Z})$ is the position operator. Equation (\ref{equation:x_and_v}) is of limited use when working in momentum space, but for disordered systems without translational invariance, it is more convenient to use a real-space basis set, $\{\ket{\bm{R}_i}\}$, where $\ket{\bm{R}_i}$ is a state centered at site $i$ of the system. Such a basis set can be, for example, formed by local atomic orbitals or Wannier functions. Using a real-space basis set, the matrix element of the velocity operator can be expressed in terms of the overlap integral $t_{ij}=\bra{\bm{R}_i}\hat{H}\ket{\bm{R}_j}$ as
\begin{equation}
    \bra{\bm{R}_i}\hat{\bm{V}}\ket{\bm{R}_j} = -\frac{i}{\hbar}t_{ij}(\bm{R}_i-\bm{R}_j),
\end{equation}
which is the expression used in all the LSQT methodologies discussed in this review. Additionally, from this point onward, we will denote the velocity pointing in the same direction of the electric field as $\hat{V}$, which defines the diagonal elements of the conductivity tensor. Finally, by inserting $\hat{\bm{J}}=q\hat{\bm{V}}/\Omega$ into the Kubo-Bastin formula (Eq.\ (\ref{equation:sigma_mu_nu_Bastin})), setting $\hat{A}=\hat{\bm{J}}$, and noting that $\langle \hat{\bm{J}} \rangle \equiv \sigma \bm{E}_0$, the DC conductivity $\sigma$ can be expressed in terms of the velocity operator as 
\begin{align}
\sigma(\mu)
=& \frac{i \hbar e^2}{\Omega}  \int dE' f(E'-\mu)\times \nonumber\\
&\text{Tr}\Big[  \delta(E'-\hat{H})\hat{V}\frac{d(G^+(E')-G^-(E'))}{dE'}\hat{V}\Big]. \label{equation:sigma_mu_nu_Greenwood}
\end{align}

\subsubsection{Kubo-Greenwood and Chester-Thellung formulas}

According to standard transport theories \cite{mahan2000book, rammer1986rmp}, the dissipative conductivity should depend only on the properties of the system around the Fermi level. However, the conductivity from the Kubo-Bastin formula seems to depend on the whole set of occupied states, as indicated by the presence of the Fermi-Dirac distribution function in the integral. It was first shown by Streda \cite{Streda1982PRC} that this is indeed not the case and that the contribution from the occupied states (called the topological or Fermi sea contribution) vanishes for the dissipative conductivity. To show this, one first needs to use the following identity relating the Green's functions to the Dirac delta function
\begin{equation}
G^+(E) - G^-(E) = -2\pi i \delta(E-\hat{H}),
\label{equation:imaginary_green} 
\end{equation}
which implies the adiabatic limit ($\tau,\tau_{\phi}\rightarrow \infty)$. Using this identity and integrating by parts, the Kubo-Bastin formula can be rewritten as 
\begin{align}
\sigma(\mu,T)
&=\frac{\pi \hbar e^2}{\Omega}  \int dE' \left[-\frac{\partial f(E'-\mu)}{\partial E'}\right]  \nonumber \\
&\times
\text{Tr}
  \Big[  \delta(E'-\hat{H})\hat{V}\delta(E'-\hat{H})\hat{V}\Big], 
\label{equation:sigma_KG_T}
\end{align}
where we have explicitly included the temperature dependence of $\sigma$ through the Fermi-Dirac distribution. The factor $-\partial f(E'-\mu)/\partial E'$ is known as the Fermi energy window, which selects the energies close to the chemical potential. At zero temperature, the chemical potential $\mu$ equals the Fermi energy $E$ and  this factor transforms into a Dirac delta function
\begin{equation}
    \lim_{T\rightarrow  0 } -\frac{\partial f(E'-\mu)}{\partial E'}= \delta(E'-E), \label{fdtodelta}
\end{equation}
which allows us to identify 
\begin{equation}
\sigma(E)
= \frac{\pi \hbar e^2}{\Omega}  \text{Tr}
  \Big[  \delta(E-\hat{H})\hat{V}\delta(E-\hat{H})\hat{V}\Big]
\label{equation:sigma_KG_T0}
\end{equation}
as the zero temperature conductivity. This is known as the Kubo-Greenwood formula \cite{greenwood1958pps}, and is at the core of different numerical \cite{roche1997prl, ferreira2015prl} and diagrammatic \cite{rammer1986rmp} methods for computing the conductivity. Additionally, this expression and Eq.\ (\ref{equation:sigma_KG_T}) demonstrate that the role of temperature is to smear the zero-temperature conductivity across the Fermi window. 

At this point, it is important to point out that the two seemingly equivalent Dirac delta functions in the Kubo Greenwood formula have different origins: one coming from the regularized Green's function, and the other as a true delta function arising from the identities of Eq.\ (\ref{equation:spectral_identities}). This can be seen explicitly in Eq.\ (\ref{equation:sigma_mu_nu_Greenwood}). Therefore, in order to use the Kubo Greenwood formula for dissipative conductivity in numerical calculations, a regularization of one of the Dirac delta functions should be performed. Typically, a Gaussian or Lorentzian regularization is considered \cite{thouless1981jpc, weisse2006rmp}. If one chooses a Lorentzian representation with width $\hbar/\tau_\phi$ (corresponding to the limit $\tau \rightarrow \infty$),
\begin{equation}
    \delta(E-\hat{H})=\frac{1}{\pi}\lim_{\tau_{\phi} \rightarrow \infty} 
    \frac{\hbar/\tau_\phi}{(E-\hat{H})^2 + (\hbar/\tau_\phi)^2},
\end{equation}
one can immediately identify the regularization parameter $\tau_{\phi}$ as a dephasing time, which can be attributed to the coupling of the system to some external inelastic source such as the electric field. Although such a source of dephasing can be thought of as the effect of inelastic scattering at nonzero temperature in a real system, it must originate from uncorrelated random events, which is not entirely realistic \cite{thouless1981jpc}. Nevertheless, under this regularization one can consider the limit $\tau_{\phi} \rightarrow \infty$ as a convergence to the adiabatic limit. 

An alternative regularization procedure can be done by using a finite time $\tau$ and a vanishingly small dephasing rate $1/\tau_{\phi}$. This process corresponds to time evolution toward the steady state. To show this, first replace one of the Dirac delta functions by its Fourier representation, 
\begin{equation}
\delta(E-\hat{H}) = \lim_{\tau\rightarrow \infty}\int_{-\tau}^{\tau} \frac{dt}{2\pi\hbar} \text{e}^{ i(E-\hat{H} ) t/\hbar }, 
\label{equation:dirac_delta_fourier}
\end{equation}
which is equivalent to taking the limits in Eq.\ (\ref{equation:reg_green}) in the order $\tau_{\phi} \rightarrow \infty$ and then $\tau \rightarrow \infty$. Then, by using the identity,
\begin{equation}
    -\frac{\partial f(E-\mu)}{\partial E} = \frac{\partial f(E-\mu)}{\partial \mu}
\end{equation}
and the identities in Eq.\ (\ref{equation:spectral_identities}), we can obtain the following expression from Eq.\ (\ref{equation:sigma_KG_T}):
\begin{gather}
\sigma(\mu,T) = 
\lim_{\tau\rightarrow \infty} \frac{ e^2}{2\Omega}\int_{0}^{\tau}dt 
\text{Tr}\left[  \frac{\partial f(\hat{H}-\mu)}{\partial \mu}\{\hat{V}(t),\hat{V}(0)\}\right],
\end{gather}
where the braces represent the anti-commutator, and $\hat{V}(t)\equiv {\rm e}^{i \hat{H}t/\hbar}\hat{V}{\rm e}^{-i \hat{H}t/\hbar}$ is the time-dependent velocity operator. This expression is known as the Chester-Thellung formula \cite{chester1959pps,chester1961pps}.

Next, one can define a single-particle density matrix as
\begin{equation}
\hat{\rho}_{\rm eq}(\mu,T) = \frac{1}{\Omega} \frac{1}{\rho(\mu,T)} \frac{\partial f(\hat{H}-\mu)}{\partial \mu}, 
\label{single-particle-density}
\end{equation}
where
\begin{equation}
 \rho(\mu,T) =  \frac{dn(\mu,T)}{d\mu} = \frac{1}{\Omega} {\rm Tr} \Big[ \frac{\partial f(\hat{H}-\mu)}{\partial \mu} \Big]
\end{equation}
is the density of states (DOS), and $n(\mu,T) = {\rm Tr}[ f(\hat{H}-\mu)]/\Omega$ is the charge density. This definition of the single-particle density matrix allows the conductivity to be expressed as
\begin{equation}
\sigma(\mu,T)
= \lim_{\tau\rightarrow \infty} e^2\rho(\mu) 
\int_{0}^{\tau}dt C_\text{vv}(\mu,t),
\label{equation:sigma_t_vac}
\end{equation}
where we have defined the quantity
\begin{align}
 C_\text{vv}(\mu,t) &\equiv \frac{1}{2}{\rm Tr} \left[ \hat{\rho}_{\rm eq} 
\{ \hat{V}(t),\hat{V}(0)\} \right] \nonumber \\
&= {\rm Re}\left({\rm Tr} \left[ \hat{\rho}_{\rm eq} 
 \hat{V}(t)\hat{V}(0) \right]\right),
\label{equation:vac_def}
\end{align}
which is the velocity autocorrelation (VAC) function. Comparison to Eq.\ (\ref{equation:current-current}) shows that this expression for conductivity offers a direct connection between the single-particle Kubo formula and the fluctuation-dissipation theorem. The DOS in Eq.\ (\ref{equation:sigma_t_vac}) appears as a consequence of multiple electrons taking part in the transport at the Fermi level.

Equation (\ref{single-particle-density}) indicates that the temperature dependence is embedded in $\hat{\rho}_{\rm eq}(\mu,T)$ and that its role is to smear the zero-temperature conductivity around the Fermi energy. Therefore, the temperature dependence can be included later provided that one knows the zero-temperature conductivity at all energies. Because of this, from this point forward we will focus on explaining how to obtain the conductivity in this limit, and unless otherwise specified we will refer to the zero-temperature conductivity simply as the conductivity. The zero-temperature limit is achieved simply by using Eq.\ (\ref{fdtodelta}), which allows the density matrix and the DOS to be expressed as 
\begin{align}
 \hat{\rho}_{\rm eq} (E) &= \frac{1}{\Omega} \frac{1}{\rho(E)} \delta(E-\hat{H}), \\
 \rho(E) &= \frac{1}{\Omega} {\rm Tr} 
 \left[\delta(E-\hat{H}) \right],
\label{equation:DOS}
\end{align}
while the rest of the formalism remains unchanged. 

\subsubsection{Relation between conductivity and diffusion}

Starting from the Chester-Thellung formula, one can easily obtain the Einstein relation, which relates the conductivity to the diffusion coefficient. To this end, we first define the mean-square displacement (MSD) as
\begin{equation}
\Delta X^2(E,t) \equiv  {\rm Tr} \left[ \hat{\rho}_{\rm eq}(E) (\hat{X}(t)-\hat{X}(0))^2 \right].
\end{equation}
Using the identity
\begin{equation}
\frac{d^2\Delta X^2(E,t) }{dt^2} =  
{\rm Tr} \left[ \hat{\rho}_{\rm eq}(E) \{ V(t),V(0)\} \right], 
\label{equation:identity_msd_vac}
\end{equation}
we have
\begin{equation}
\sigma(E) =e^2 \rho(E) \lim_{t \rightarrow \infty} \frac{1}{2} \frac{d\Delta X^2(E,t)}{dt}.
\label{equation:sigma_t_msd}
\end{equation}
If we define 
\begin{equation}
D(E)=\lim_{t \rightarrow \infty} D(E,t)=\lim_{t \rightarrow \infty} \frac{1}{2}\frac{d\Delta X^2(E,t)}{dt}, 
\label{charge_cond_timereg}
\end{equation}
we have $\sigma(E)=e^2\rho(E)D(E)$. The quantity $D(E)$ is called the diffusion coefficient and can be considered as one of the diagonal entries of the diffusion tensor \cite{Nakajima1958}. This result shows that choosing a regularization with time yields a formulation that is formally equivalent to semiclassical theory in the diffusive regime.

If instead of the time regularization we use the regularization provided by the dephasing time $\tau_\phi$, we obtain a slightly different expression for $D(E)$, which nevertheless provides the correct results for $\tau_\phi\rightarrow \infty$. Using the regularized retarded Green's function in Eq.\ (\ref{equation:reg_green}) to represent one of the Dirac delta functions in Eq.\ (\ref{equation:sigma_KG_T0}), in a similar way as in the derivation of the Chester-Thellung formula, yields a diffusion coefficient that depends on both the evolution time $\tau$ and the dephasing time $\tau_{\phi}$,
\begin{equation}
D(E;\tau_{\phi},\tau)
= \int_{0}^{\tau} dt e^{-t/\tau_{\phi}}
\frac{1}{2 }\frac{d^2\Delta X^2(E,t) }{dt^2},
\end{equation}
which when integrating by parts reduces to
\begin{align}
D(E;\tau_{\phi},\tau) 
&= \frac{1}{2} e^{-t/\tau_{\phi} }\left[\frac{d\Delta X^2(E,t) }{dt}  +\frac{\Delta X^2(E,t)}{\tau_{\phi}}  \right]_{t=0}^\tau \nonumber \\
&+ \frac{1}{2\tau_{\phi}^2}\int_{0}^{\tau}dt e^{-t/\tau_{\phi} }\Delta X^2(E,t).
\end{align}
If the limit $\tau_\phi\rightarrow \infty$ is taken first, we recover Eq.\ (\ref{charge_cond_timereg}). Otherwise, if the limit $\tau \rightarrow \infty$ is taken first, we have 
\begin{align}
D(E) &=\lim_{\tau_{\phi} \rightarrow \infty} D(E,\tau_{\phi}) \nonumber \\
&= \lim_{\tau_{\phi} \rightarrow \infty}  \frac{1}{2\tau_{\phi}^2}\int_{0}^{\infty}dt
e^{-t/\tau_{\phi} } \Delta X^2(E,t). 
\label{charge_cond_dephreg}
\end{align}
The last expression shows that when we incorporate a finite dephasing time, the diffusion coefficient becomes essentially an average of $\Delta X^2(E,t)/2\tau_{\phi}$ over the time scale defined by $\tau_{\phi}$. Equations (\ref{charge_cond_timereg}) and (\ref{charge_cond_dephreg}), although formally different, produce the same result when $\tau$ and $\tau_\phi$ are larger than the characteristic times of the system. This will be shown in the next section where we discuss the different transport regimes and the possible outcomes of these formulas.

\subsubsection{Transport regimes and length scales}
\label{section:transport_regimes}

So far we have derived different representations of the Kubo formula that can be used to obtain the conductivity at a stationary state, i.e., in the limit of infinite time. However, these Kubo formulas can also be used to determine the behavior of the system in different transport regimes that occur at finite time. This is one of the benefits of using time-dependent approaches for quantum transport. To show this, we start with a discussion of the different transport regimes and some relevant physical quantities.

Consider a perfect crystal material, which by definition is a periodic array of atoms. An electron in this environment will be subjected to a periodic potential due to the Coulomb field of the atoms. By virtue of Bloch's theorem, one can describe this system as a free electron gas whose components possess an effective mass accounting for the change in the group velocity due to a change in the crystal momentum. Therefore, under the action of a small external electric field, the electrons will move freely along the direction of the electric field at an average speed of the Fermi velocity $v_{\rm F}(E)$, leading to ballistic transport.

However, in disordered systems the electrons will be scattered by imperfections. After some time $\tau_{\rm p}(E)$, which is known as the momentum relaxation time, the system will have undergone many random scatterings that make it lose all memory about the initial conditions, leading to a steady state known as the diffusive regime. Finally, if the disorder is strong enough, a phenomenon known as Anderson localization will take place. In this situation, the electron's wave function is no longer extended. Instead, due to quantum interference effects the wave function becomes localized within a volume whose radius is usually defined as the localization length $\xi(E)$. These are the canonical transport regimes, and in the following we will see how to identify each of these with quantum transport simulations. The first thing to address is how to compute the characteristic parameters of each regime: the Fermi velocity, momentum relaxation time, and localization length. 

\begin{figure}[htb]
\begin{center}
\includegraphics[width=\columnwidth]{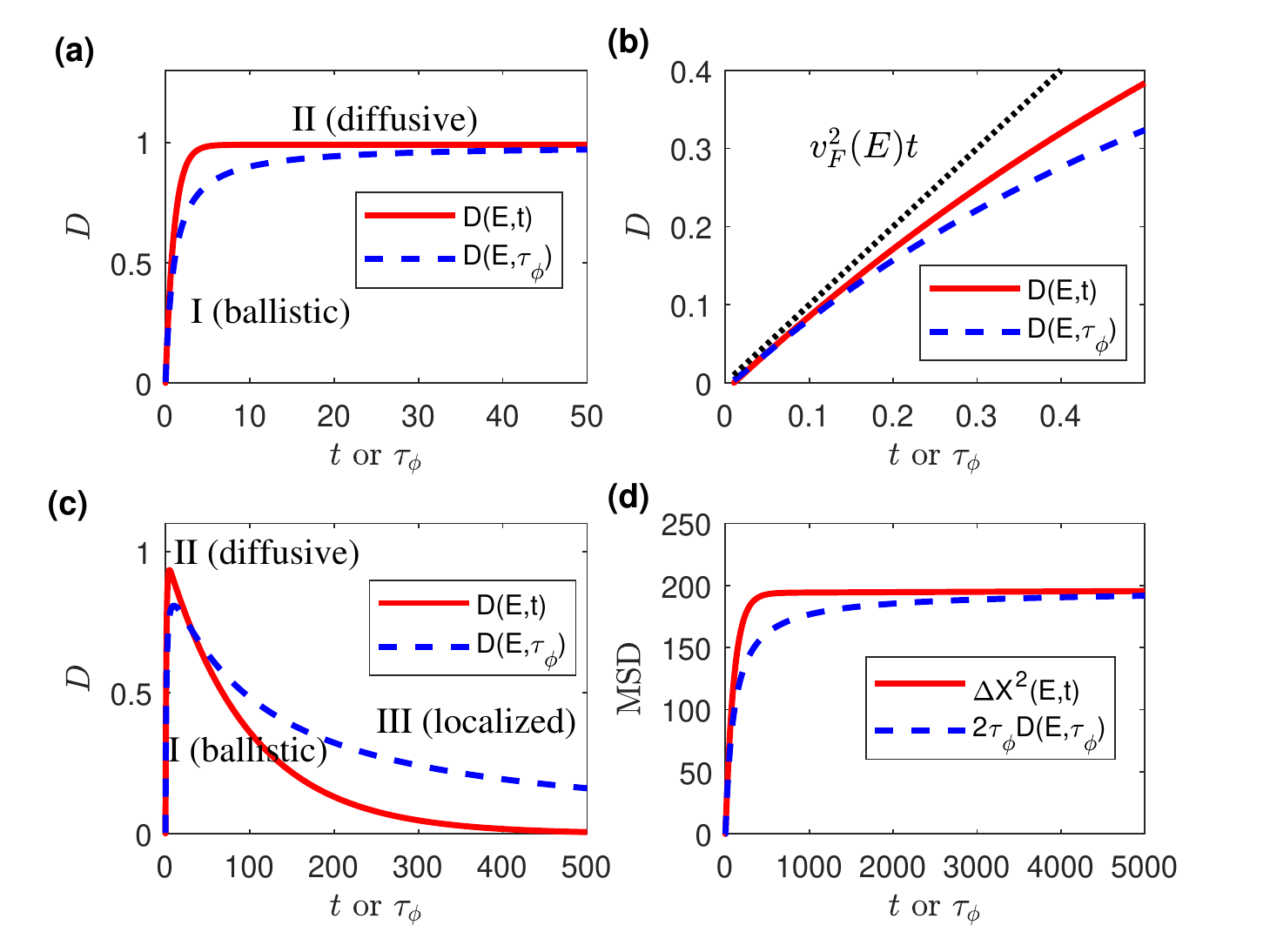}
\caption{(a) The diffusion coefficient $D(E,t)$ defined in Eq.\ (\ref{charge_cond_timereg}) as a function of the evolution time $t$ and the diffusion coefficient $D(E,\tau_{\phi})$ defined in Eq.\ (\ref{charge_cond_dephreg}) as a function of the dephasing time $\tau_{\phi}$. When there is no localization, the two diffusion coefficients converge to the same $D_{\rm sc}$. (b) A closer look at the ballistic limit in (a), where both diffusion coefficients have a slope of $v_{\rm F}^2(E)$ at short times. (c) Similar to (a), but for a system with strong localization. (d) The corresponding MSD as a function of $t$ and $2\tau_{\phi}D(E,\tau_{\phi})$ as a function of $\tau_{\phi}$. In the presence of strong localization they converge to the same value $\pi^2\xi^2(E)$, where $\xi(E)$ is the localization length.}
\label{figure:transport_regimes}
\end{center}
\end{figure}

For ballistic transport, the $\Delta X^2(E,t)$ grows quadratically as $v_{\rm F}^2(E) t^2$. Inserting this into Eqs.\ (\ref{charge_cond_timereg}) and (\ref{charge_cond_dephreg}) gives $D(E) = v_{\rm F}^2\tau$ and $D(E) = v_{\rm F}^2\tau_{\phi}$, respectively. This means that the conductivity diverges linearly with time, as expected for ballistic transport. Therefore, in the ballistic regime both regularization procedures give the same result, as illustrated in Fig.\ \ref{figure:transport_regimes}(b).

At longer time or length scales, the electrons will be scattered by imperfections and will lose the memory of their previous momenta after a time of order $\tau_{\rm p}(E)$. In this limit the conductivity becomes independent of length and time, which implies that it can be expressed in terms of a constant diffusion coefficient $D_{\rm sc}(E)$, which is commonly referred to as the semiclassical (sc) diffusion constant. This is a consequence of a linearly increasing MSD, which in the diffusive regime is proportional to the diffusion constant
\begin{equation}
    \Delta X^2(E,t) = 2 D_{\rm sc}(E) t.
\end{equation}
Inserting this into Eq.\ (\ref{charge_cond_timereg}) gives directly what we expect 
\begin{equation}
D(E) = D_{\rm sc}(E). 
\label{diffusive constant}
\end{equation}
As for Eq.\ (\ref{charge_cond_dephreg}), one can divide the integral into two contributions -- before reaching the momentum relaxation time $\tau_{\rm p}(E)$ the system behaves ballistically, while after reaching $\tau_{\rm p}(E)$ the system enters the diffusive regime. For simplicity, let us assume a sharp transition at $t=\tau_{\rm p}$ and write the diffusion coefficient as
\begin{equation}
D(E) 
= \lim_{\tau_{\phi} \rightarrow \infty}  \frac{1}{2\tau_{\phi}^2}\left(\int_{0}^{\tau_{\rm p}}+\int_{\tau_{\rm p}}^{\infty} \right) dt\, e^{-t/\tau_{\phi} }\Delta X^2(E,t).
\end{equation}
The first integral goes to zero as $\tau_\phi \rightarrow \infty$ and the second integral gives the same expression as Eq.\ (\ref{diffusive constant}). This result can also be derived by assuming an exponentially decaying VAC \cite{beenakker1991ssp},
\begin{equation}
C_\text{vv}(E,t) = \frac{1}{2} v_{\rm F}^2(E) e^{ -t/\tau_{\rm p}(E)},
\end{equation}
which leads to a connection between the diffusion constant and the momentum relaxation time,
\begin{equation}
\label{equation:diffusivity}
D_{\rm sc}(E) = \frac{1}{2} v_{\rm F}^2(E) \tau_{\rm p}(E).
\end{equation}
One can also define a mean free path
\begin{equation}
    l_\text{e}(E)=v_{\rm F}(E) \tau_{\rm p}(E)
\end{equation}
and write $D_{\rm sc}(E)=v_{\rm F}(E) l_\text{e}(E) / 2$. We thus see that the semiclassical electrical conductivity has the same form as that obtained from the Boltzmann equation within the relaxation time approximation \cite{ashcroft1976book}. The equivalence between the two regularizations of the Green's function in the diffusive limit is illustrated in Fig.\ \ref{figure:transport_regimes}(a).

Finally, in the weak and strong localization regimes the conductivity is expected to decay with increasing system length $L$ \cite{lee1985rmp}. The weak localization regime is characterized by a logarithmic decay of the conductivity, $\sigma(E,L)-\sigma_{\rm sc}(E) \propto -\ln (L/l_\text{e}(E))$, while the strong localization regime is associated with an exponential decay of the conductivity, $\sigma(E,L) \propto e^{-L/\xi(E)}$, where $\xi(E)$ is the localization length. 

In the strong localization regime, the MSD saturates to a constant value and Eq.\ (\ref{charge_cond_timereg}) predicts a zero diffusion coefficient and conductivity. The convergence towards this limit is usually assumed to be exponential with respect to the system length, and from this scaling the localization length can be obtained. One can also establish a relation between the localization length and the saturated value of the MSD \cite{triozon2000riken}, which was found quantitatively to be \cite{uppstu2014prb}
\begin{equation}
\xi(E)= \lim_{t \to \infty} \frac{\sqrt{\Delta X^2(E,t)}}{\pi}.
\label{equation:xi_from_MSD}
\end{equation}
This definition of localization length conforms with the standard definition in terms of the length scaling of conductance \cite{anderson1980} and is in accordance with the original definition of Anderson localization \cite{anderson1958pr}, namely, the absence of diffusion.

Meanwhile, Eq.\ (\ref{charge_cond_dephreg}) gives a different convergence behavior. If we define $\tau_{\xi}$ as the time after which strong localization dominates and impose the condition $\tau_{\xi}/\tau_{\phi}\ll 1$, we obtain a diffusion coefficient going to zero as the inverse dephasing time,
\begin{equation}
D(E) =\lim_{\tau_{\phi} \rightarrow \infty} D(E,\tau_{\phi}) = \lim_{\tau_{\phi} \rightarrow \infty}  \frac{\pi ^2\xi(E)^2}{2\tau_{\phi}}.\label{LocalizationLengthDef}
\end{equation}
This difference in convergence behavior is demonstrated in Fig.\ \ref{figure:transport_regimes}(c). Despite their different scaling, $2\tau_{\phi}D(E,\tau_{\phi})$ and the MSD converge to the same value $\pi ^2\xi^2(E)$ in the limit of infinite time, as illustrated in Fig.\ \ref{figure:transport_regimes}(d). Therefore, the two regularizations of the Green's function give the same results in the ballistic, diffusive, and localized regimes, when the appropriate limits are taken.

\subsubsection{Time- and length-dependent conductivity}
\label{section:scale_dependent_conductivity}

In Eqs.\ (\ref{equation:sigma_t_vac}) and (\ref{equation:sigma_t_msd}), we have defined expressions for the electronic conductivity at infinite time and sample size. However, following the discussion in Sec.\ \ref{section:transport_regimes} above, it is crucial to define a conductivity that is explicitly dependent on time or length. This allows one to numerically track the evolution of a system through ballistic, diffusive, and localized transport and identify the relevant quantities associated with each regime. Defining a length-dependent conductivity is also necessary for comparison to experiments, which are always limited to measuring transport properties at a fixed device size.

We therefore define time-dependent versions of the conductivity by cutting off Eqs.\ (\ref{equation:sigma_t_vac}) and (\ref{equation:sigma_t_msd}) at finite correlation time (and taking the zero-temperature limit),
\begin{gather}
\sigma(E,t) \equiv e^2 \rho(E) \int_{0}^{t}dt' C_\text{vv}(E,t'),
\label{equation:sigma_t_vac_t} \\
\text{and} \nonumber \\
\sigma(E,t) \equiv e^2 \rho(E) \frac{1}{2} \frac{d\Delta X^2(E,t)}{dt}.\label{equation:sigma_t_msd_t}
\end{gather}
The length-dependent conductivity is then defined as
\begin{gather}
\sigma(E,L) \equiv \sigma(E,t) \bigg|_{L = L(E,t)}
\label{equation:sigma_L} \\
\text{where} \nonumber \\
L(E,t) \equiv 2 \sqrt{\Delta X^2(E,t)}.
\label{equation:L}
\end{gather}
Here we have defined the system size $L$ as the propagation length associated with the MSD, and not the size of the simulation cell. In the time-dependent LSQT methods, one then considers a sufficiently large simulation cell and calculates the time-dependent conductivity $\sigma(E,t)$ and propagation length $L(E,t)$ to obtain the length-dependent conductivity $\sigma(E,L)$. To avoid finite size and edge effects, one should use periodic boundary conditions and ensure that $L$ remains smaller than the size of the simulation cell.

This definition of $L(E,t)$ can be justified by considering the ballistic regime of transport. In this regime, the conductivity scales linearly with time, $\sigma(E,t)=e^2\rho(E)v_{\rm F}^2(E)t$, and thus diverges. However, the conductance $g(E)$ is defined as 
\begin{equation}
\label{equation:sigma_to_g}
g(E) \equiv \frac{A}{L(E,t)}\sigma(E,t),
\end{equation}
where $A$ is the cross-sectional area through which the current flows. The conductance is a geometry-dependent quantity and is therefore finite for finite systems. Using the definition of length in Eq.\ (\ref{equation:L}), we have $L(E, t) = 2 v_{\rm F}(E) t$ in the ballistic regime and
\begin{equation}
g(E)=\frac{A}{2}e^2\rho(E)v_{\rm F}(E),
\label{equation:ballistic_conductance}
\end{equation}
which is independent of any length or time scale and is completely characterized by the DOS and Fermi velocity, consistent with the picture of ballistic transport. For a strictly one-dimensional system, the DOS is $\rho(E)=2/\pi\hbar v_{\rm F}(E)$ and we finally get $g(E)=2e^2/h$. This is the expected conductance quantum for ballistic transport, as has been measured in quantum point contacts \cite{vanWees1988prl,wharam1988jpc} and carbon nanotubes \cite{frank1998science}. The factor of two in Eq.\ (\ref{equation:L}) means that electrons propagate in two opposite directions. In early works \cite{roche2001prl, roche2001prb}, this factor of two was not included, but the conductivity was defined by substituting the derivative in Eq.\ (\ref{equation:sigma_t_msd_t}) with division by $t$, which exactly reduces the conductivity by half in the ballistic limit and results in the same ballistic conductance as in Eq.\ (\ref{equation:ballistic_conductance}).

%% file: secIII_numerical_techniques.tex
\section{Linear scaling numerical techniques }
\label{section:numerical_techniques}
In Sec.\ \ref{section:kubo_formulas}, we have presented three different representations of the Kubo formula for non-interacting electrons: the Kubo-Greenwood formula in Eq.\ (\ref{equation:sigma_KG_T0}), the VAC-based formula in Eq.\ (\ref{equation:sigma_t_vac_t}), and the MSD-based formula in Eq.\ (\ref{equation:sigma_t_msd_t}). The aim of this section is to review the various numerical techniques for efficiently evaluating these formulas. We will focus on dissipative transport in this section and discuss numerical evaluation of the Kubo-Bastin formula -- Eq.\ (\ref{equation:sigma_mu_nu_Bastin}) -- in Sec.\ \ref{section:hall_spin}, which can be used to compute other transport properties \cite{garcia2015prl}.

The major concern in numerical implementations is the scaling of the computational cost with respect to the Hamiltonian size $N$. A common feature of the above Kubo formulas is that the trace can be evaluated using any complete set of single-particle wave functions that obey periodic boundary conditions \cite{chester1959pps, chester1961pps}. An immediate choice would be to use the set of eigenvectors of the Hamiltonian, but this requires full diagonalization, which is usually prohibitive for large systems due to its $\mathcal{O}(N^3)$-scaling computational cost with respect to the system size. To enable the study of large systems (e.g., $N > 10^6$), a linear scaling, or $\mathcal{O}(N)$ algorithm is mandatory. To achieve linear scaling, we avoid using the Hamiltonian's eigenspace and instead work with a real-space tight-binding representation, where the basis functions are not eigenfunctions of the Hamiltonian but rather the electron orbitals around individual atoms. Because of this, the methods discussed in this review are usually referred to as real-space LSQT methods.

Before discussing the relevant numerical techniques for achieving linear scaling, we list the quantities to be calculated for each implementation. A prominent quantity is the DOS defined in Eq.\ (\ref{equation:DOS}), which contains information about the electronic structure of the system. In the Kubo-Greenwood representation, one directly evaluates the electrical conductivity as given in Eq.\ (\ref{equation:sigma_KG_T0}), but needs to represent one of the Dirac delta functions as a regularized Green's function. In the VAC representation of conductivity given in Eq.\ (\ref{equation:sigma_t_vac_t}), one first calculates the product of the DOS and the VAC, 
\begin{equation}
  \rho (E) C_\text{vv}(E, t) = \frac{1}{\Omega}
 {\rm Re}\left[ 
 \textmd{Tr}\left( \hat{U}(t)\hat{V} \delta(E - \hat{H}) \hat{U}^{\dagger}(t)\hat{V}
 \right)
 \right],
\end{equation}
and then performs a numerical time integration to obtain the time-dependent electrical conductivity $\sigma(E,t)$. In the MSD representation of Eq.\ (\ref{equation:sigma_t_msd_t}), one first calculates the product of the DOS and the MSD, 
\begin{equation}
 \rho (E) \Delta X^2(E, t) =
 \frac{1}{\Omega}
 \textmd{Tr}\left[
 \delta(E - \hat{H})(\hat{X}(t)-\hat{X})^2
 \right], \label{msd_ver1}
\end{equation}
and then performs a numerical time derivative to obtain the time-dependent electrical conductivity $\sigma(E,t)$. In periodic systems it is problematic to use the absolute position operator $\hat{X}$. Instead, one can use the identity $\hat{X}(t)-\hat{X} = \hat{U}^\dagger(t)[ \hat{X}, \hat{U}(t)]$ to change the above equation to an equivalent one  \cite{triozon2002prb,triozon2004prb},
\begin{equation}
 \rho (E) \Delta X^2(E, t) =
 \frac{1}{\Omega}
 \textmd{Tr}\left[
 [\hat{X}, \hat{U}(t)]^{\dagger}\delta(E - \hat{H})[\hat{X}, \hat{U}(t)]
 \right].
\end{equation}
After a polynomial expansion of the time evolution operator $\hat{U}(t)$ (as discussed below), the commutator $[\hat{X}, \hat{U}(t)]$ only depends on the velocity operator, which only depends on the difference between the positions of the orbitals and is well defined in periodic systems.  

There are some common features in these quantities: they are all represented as a trace and involve the quantum projection operator $\delta(E-\hat{H})$, and the time evolution operator $\hat{U}(t)$ appears in the VAC and MSD. Linear scaling techniques have been developed to evaluate these operators and we will discuss them in detail below.

\subsection{Evaluating the trace using a stochastic approach}
\label{section:trace}

Recall that the trace of an operator $\hat{A}$ is defined as
\begin{equation}
\textmd{Tr} [\hat{A}] = \sum_{n=1}^{N} \bra{n} \hat{A} \ket{n},
\end{equation}
where $\{\ket{n}\}_{n=1}^{N}$ is a complete basis set of the problem, which is taken as a real-space tight-binding basis set in this review. The operator $\hat{A}$ relevant to this review will be essentially polynomials formed by the Hamiltonian and other quantities such as the velocity operator. What is important here is that even if $\hat{A}$ is highly sparse, such that the operation $\hat{A} \ket{n}$ scales linearly, the total computation still has an $\mathcal{O}(N^2)$ scaling, which is better than $\mathcal{O}(N^3)$ for a non-sparse $\hat{A}$ but is still usually prohibitive. To achieve $\mathcal{O}(N)$ scaling, the trace must be approximated. A powerful method for approximating the trace of large matrices is to use random vectors, a stochastic approach that was developed along with the methods of calculating the spectral properties of large Hamiltonians \cite{skilling1989, silver1994ijmpc, drabold1993prl, silver1996jcp, weisse2006rmp}.

In this stochastic method, one approximates the trace by using $N_r$ random vectors $\{\ket{\phi_r}\}_{r=1}^{N_r}$,
\begin{equation}
\textmd{Tr}[\hat{A}] \approx \frac{1}{N_r} \sum_{r=1}^{N_r} \bra{\phi_r} \hat{A} \ket{\phi_r}.
\label{equation:trace_A}
\end{equation}
Each random vector $\ket{\phi_r}$ is constructed from $N$ random coefficients, 
\begin{equation}
\ket{\phi_r} = \sum_{n=1}^{N} \xi_{rn} \ket{n}.
\label{equation:rp_state}
\end{equation}
Here, $\xi_{rn} \in \mathbb{C}$ are independent identically distributed random variables which have zero mean and unit variance, which implies that each vector is normalized to $\sqrt{N}$. It has been shown that \cite{iitaka2004pre, weisse2006rmp} the statistical error for the trace is proportional to $1/\sqrt{N_rN}$, with the proportionality constant being related to the properties of the matrix $\hat{A}$. The statistical accuracy can be systematically improved by increasing $N_r$. In practice, for large $N$ a small $N_r$ on the order of unity is sufficient to achieve a high statistical accuracy.

For simplicity, we only use a single random vector $|\phi\rangle$ to present the subsequent formulas. In practice, one needs to check the convergence of the results with respect to $N_r$. Under the condition of sufficient average, in the following we use the ``$=$" sign instead of the ``$\approx$" sign as in Eq.\ (\ref{equation:trace_A}). Using this, we can express the quantities that need to be calculated as the following inner products:
\begin{equation}
\label{equation:trace_DOS}
 \rho(E) = \frac{1}{\Omega} \langle \phi | \delta(E - \hat{H}) |\phi \rangle;
\end{equation}
\begin{equation}
\label{equation:trace_KG}
\sigma(E) = \frac{\pi \hbar e^2}{\Omega} \langle\phi| \delta(E-\hat{H}) \hat{V} \delta(E-\hat{H}) \hat{V} |\phi \rangle;
\end{equation}
\begin{equation}
\label{equation:trace_VAC_modified}
 \rho (E) C_\text{vv}(E, t) =
 \frac{1}{\Omega} \textmd{Re} 
 \left[
 \langle\phi^{\rm vac}_\text{L}(t)|\delta(E - \hat{H})|\phi^{\rm vac}_\text{R}(t) \rangle
 \right];
\end{equation}
\begin{equation}
\label{equation:trace_MSD_modified}
 \rho (E) \Delta X^2(E, t) =
 \frac{1}{\Omega}
 \langle\phi^{\rm msd}_\text{L}(t)|\delta(E - \hat{H})|\phi^{\rm msd}_\text{R}(t) \rangle;
\end{equation}
where 
\begin{equation}
 |\phi^{\rm vac}_\text{L}(t)\rangle = \hat{V} \hat{U}(t)^{\dagger}|\phi\rangle; \quad
 |\phi^{\rm vac}_\text{R}(t)\rangle = \hat{U}(t)^{\dagger} \hat{V}|\phi\rangle;
 \label{equation:phi_vac}
\end{equation}
\begin{equation}
 |\phi^{\rm msd}_\text{L}(t)\rangle=|\phi^{\rm msd}_\text{R}(t)\rangle=[\hat{X},\hat{U}(t)]|\phi\rangle.
  \label{equation:phi_msd}
\end{equation}

The remaining task is to evaluate these inner products in a linear scaling way. We will discuss linear scaling numerical techniques related to the time evolution operator $\hat{U}(t)$ in Sec.\ \ref{section:time_evolution} and those related to the quantum projection operator $\delta(E-\hat{H})$ in Sec.\ \ref{section:delta}. Before doing these, we first review a crucial numerical technique, namely the Chebyshev polynomial expansion.

\subsection{Chebyshev polynomial expansion}
\label{subsection:ChebPol}

We have presented different theoretical frameworks that can be used to determine the conductivity. We saw that a numerical evaluation of the conductivity requires one to compute functions of the Hamiltonian matrix such as the time evolution operator and the quantum projection operator. We also discussed the need to choose an appropriate basis set so that the Hamiltonian can be represented as a sparse matrix. Finally, if we want to exploit this feature we need to find a way to avoid explicit evaluation of these quantities, because an arbitrary function of a sparse matrix is generally not a sparse matrix. The use of polynomial expansion provides a way to achieve the goal of linear scaling computation. Among various polynomials, the Chebyshev polynomials are usually the optimal choice \cite{boyd2001book}. 

The Chebyshev polynomials are a family of orthogonal polynomials which can be defined recursively. In this review we are using the Chebyshev polynomials of the first kind, $\{T_m(x)\}$, which are defined as $T_m(\cos(x))=\cos(mx)$ and have the recurrence relation
\begin{align}
T_0(x) &= 1; \quad
T_1(x) = x; \nonumber\\
T_m(x) &= 2 x\, T_{m-1}(x) \,-\, T_{m-2}(x) \quad (m\geq 2).
\end{align}
These polynomials form a complete basis for functions defined on the real axis within the interval $[-1,1]$. As such, they can be used to expand a function $f(x)$ defined within the same interval in a polynomial series
\begin{equation}
f(x) = \sum_{m=0}^\infty \bar{f}_m T_m(x),
\label{equation:expansion_1}
\end{equation}
where 
\begin{equation}
\bar{f}_m= (2-\delta_{m0})  \int_{-1}^{1} \frac{f(x)T_m(x)}{\pi\sqrt{1-x^2}} dx 
\label{chebint} 
\end{equation}
are the expansion coefficients and $\delta_{m0}$ is the Kronecker delta. 

In general, one deals with functions of the Hamiltonian $\hat{H}$, whose energy spectrum may exceed the interval $[-1,1]$ in a particular system. In order to use the Chebyshev polynomial expansion, one must first scale and shift $\hat{H}$ such that the modified energy spectrum is within $[-1,1]$. Specifically, this can be done by the linear transformation
\begin{equation}
\label{equation:scale}
\widetilde{H}=\frac{\hat{H}-\bar{E}}{\Delta E},
\end{equation}
where $\Delta E=(E_{\rm max}-E_{\rm min})/2$ and $\bar{E}=(E_{\rm max}+E_{\rm min})/2$, with $E_{\rm max}$ and $E_{\rm min}$ being the maximum and minimum eigenvalues of $\hat{H}$. To avoid numerical problems, one usually makes $\Delta E$ slightly larger than $(E_{\rm max}-E_{\rm min})/2$. Any function of $\widetilde{H}$ can thus be expanded in a manner similar to Eq.\ (\ref{equation:expansion_1}). To see this, we assume that $\widetilde{H}$ has the eigenvalues $\{\widetilde{E}_n\}$ and eigenvectors $\{\ket{n}\}$. For a general function $f(\widetilde{H})$, we have
\begin{align}
f(\widetilde{H}) 
=& \sum_{n} f(\widetilde{H}) \ket{n}\bra{n} =\sum_{n} f(\widetilde{E}_n) \ket{n}\bra{n} \nonumber \\
=&\sum_{n} \sum_{m=0}^\infty \overline{f}_m T_m(\widetilde{E}_n)\ket{n}\bra{n}  \nonumber \\
=&\sum_{n} \sum_{m=0}^\infty \overline{f}_m T_m(\widetilde{H})\ket{n}\bra{n}\nonumber \\
=&\sum_{m=0}^\infty \overline{f}_m T_m(\widetilde{H}). \label{spectral_expansion}
\end{align}

The inner products listed at the end of Sec.\ \ref{section:trace} are of the form $\bra{\phi_\text{L}} F(\widetilde{H}) \ket{\phi_\text{R}}$. These quantities can thus be evaluated iteratively by exploiting the recurrence relation of the Chebyshev polynomials and the whole computation reduces to a number of sparse matrix-vector multiplications, which scale linearly with the vector length $N$. In the next section, we discuss Chebyshev polynomial expansions of the time evolution operator and the regularized Green's function.

\subsection{The time evolution operator and the regularized Green's function \label{section:time_evolution}}

Both the VAC and MSD formalisms involve a time evolution operator $\hat{U}(t)$, and one of the Dirac delta functions $\delta(E-\hat{H})$ in the Kubo-Greenwood formula can be substituted by a regularized Green's function. In this subsection, we discuss the expansion of the time evolution operator and the regularized Green's function in terms of the Chebyshev polynomials.

In the VAC and MSD formalisms, in addition to applying the random vector approximation for the trace we need to evaluate the application of the time evolution operator to a vector, as can be seen from Eqs.\ (\ref{equation:phi_vac}) and (\ref{equation:phi_msd}). Because we need information at a discrete set of time points, we need to construct an iterative scheme for evaluating the time evolution. The strategy is to divide the total simulation time into a number of steps. From time $t$ to time $t+\Delta t$  (the time steps $\Delta t$ need not to be uniform), we have the following iterative relations for the vectors defined in Eqs.\ (\ref{equation:phi_vac}) and (\ref{equation:phi_msd}):
\begin{equation}
\label{equation:time_iterative_VAC_1}
\hat{V} \hat{U}^{\dagger}(t+\Delta t) |\phi\rangle = \hat{V} \hat{U}^{\dagger}(\Delta t) \hat{U}^{\dagger}(t) |\phi\rangle;
\end{equation}
\begin{equation}
\label{equation:time_iterative_VAC_2}
\hat{U}^{\dagger}(t+\Delta t) \hat{V} |\phi\rangle = \hat{U}^{\dagger}(\Delta t) \hat{U}^{\dagger}(t)\hat{V} |\phi\rangle;
\end{equation}
\begin{align}
\label{equation:time_iterative_MSD}
[\hat{X}, \hat{U}(t+\Delta t)] |\phi\rangle &=  \hat{U}(\Delta t) [\hat{X},\hat{U}(t)] |\phi\rangle \nonumber \\
&+[\hat{X},\hat{U}(\Delta t)] \hat{U}(t) |\phi\rangle.
\end{align}
Therefore, the task breaks down to evaluating the application of the operators $\hat{U}(\Delta t)$ and $[\hat{X},\hat{U}(\Delta t)]$ on some vectors.

\begin{figure}[htb]
\begin{center}
\includegraphics[width=\columnwidth]{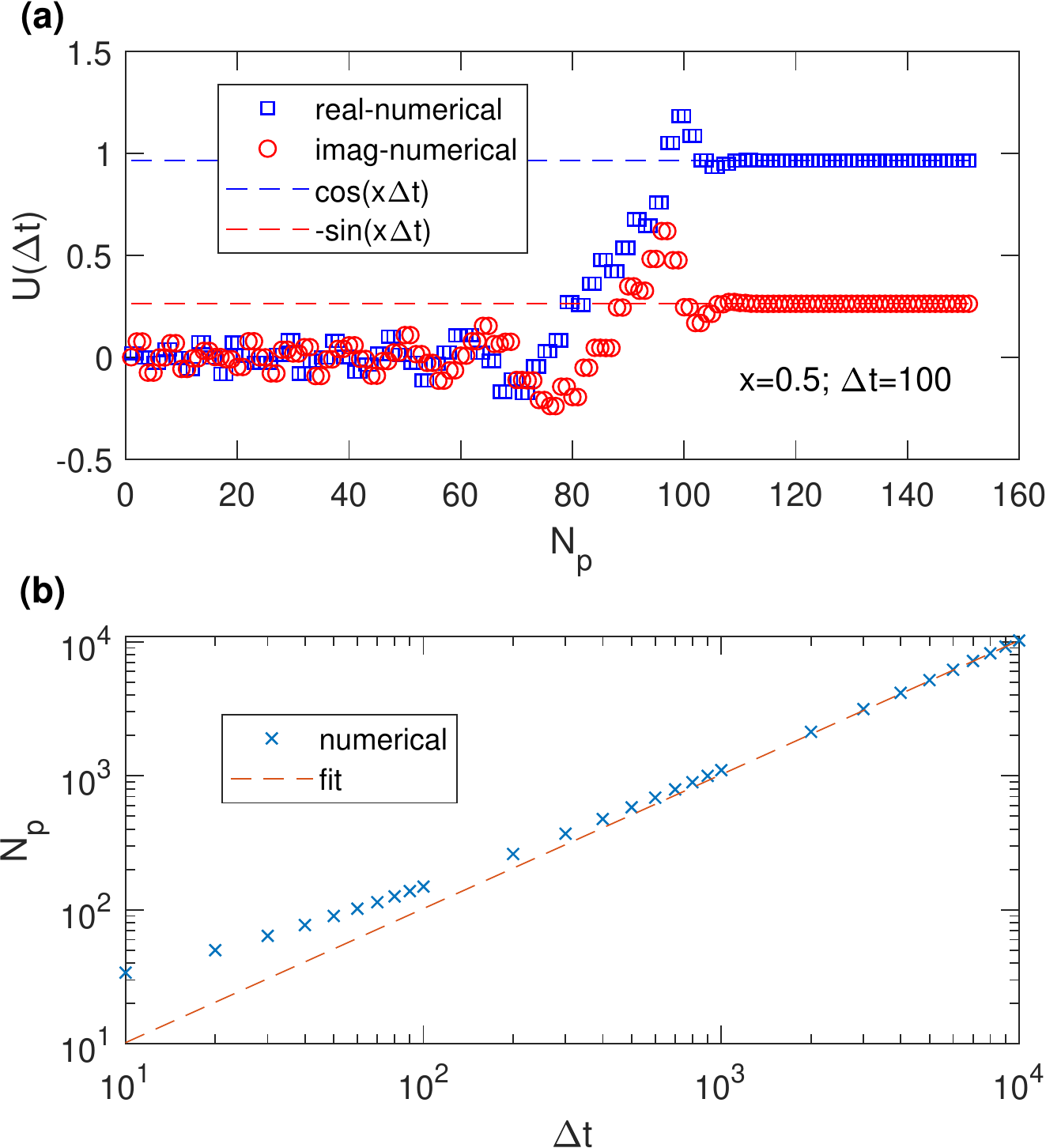}
\caption{Demonstration of the accuracy of the Chebyshev polynomial expansion of the time evolution operator using a simple function $U(\Delta t)=\exp(-ix\Delta t)$. (a) The real part $\cos(x\Delta t)$ and the imaginary part $-\sin(x\Delta t)$ of $U(\Delta t)$, calculated analytically (dashed lines) or by a numerical expansion (markers) similar to that in Eq.\ (\ref{equation:evolve}). Here, $x$ plays the role of the Hamiltonian in the time evolution operator and we choose $x=0.5$ and $\Delta t=100$. (b) The number of Chebyshev polynomials $N_\text{p}$ required for achieving a precision of $10^{-15}$ as a function of the time interval $\Delta t$. In the large $\Delta t$ limit, $N_\text{p} \propto \Delta t$, as indicated by the dashed line.}
\label{figure:test_evolution}
\end{center}
\end{figure}

The Chebyshev polynomial expansion is particularly efficient when the expanded function is regular and differentiable. One of its first uses was the expansion of the time evolution operator $\hat{U}(\Delta t)$. Solving the integral in Eq.\ (\ref{chebint}) for this operator leads to an expansion in the form of Eq.\ (\ref{spectral_expansion}) \cite{ezer1984jcp},
\begin{gather}
\label{equation:evolve}
 \hat{U}(\Delta t)
 \approx \sum_{m=0}^{N_\text{p}} \overline{U}_m(\Delta t) T_m( \widetilde{H}), \\
%\end{equation}
%\begin{equation}
\label{equation:evolve_coefficients}
\overline{U}_m(\Delta t) = (2-\delta_{m0})(-i)^m J_m\left( \omega_0 \Delta t \right),
\end{gather}
where $\omega_0=\Delta E/\hbar$ and $J_m(x)$ is the $m$th-order Bessel function of first kind.

The operator $[\hat{X}, \hat{U}(\Delta t)]$ can be similarly expanded in terms of the Chebyshev polynomials,
\begin{equation}
\label{equation:evolvex}
 [\hat{X}, \hat{U}(\Delta t)]
 \approx \sum_{m=0}^{N_\text{p}} \overline{U}_m(\Delta t)
[X, T_m(\widetilde{H})],
\end{equation}
where the commutator $[X, T_m(\widetilde{H})]$ can be calculated iteratively using the recurrence relation
\begin{align}
[\hat{X}, T_m(\widetilde{H})] &= 2[\hat{X}, \widetilde{H}] T_{m-1}(\widetilde{H}) \nonumber \\
&+ 2\widetilde{H} [\hat{X}, T_{m-1}(\widetilde{H})] \nonumber \\
&- [\hat{X}, T_{m-2}(\widetilde{H})].
\end{align}
Algorithms \ref{algorithm:evolve} and \ref{algorithm:evolvex} give explicit steps for evaluating $|\phi_{\rm out}\rangle=\hat{U}(\Delta t) |\phi_{\rm in}\rangle$ and $|\phi_{\rm out}\rangle=[\hat{X},\hat{U}(\Delta t)] |\phi_{\rm in}\rangle$. A demonstration of the accuracy of the Chebyshev expansion of the time evolution operator can be seen in Fig.\ \ref{figure:test_evolution}. Panel (a) shows how $\hat{U}(\Delta t)$ quickly converges to its expected value after a finite number of iteration steps. Panel (b) indicates the number of Chebyshev polynomials $N_\text{p}$ needed to achieve a precision of $10^{-15}$ for a given time step $\Delta t$. In the limit of large $\Delta t$, $N_\text{p} \propto \Delta t$, as indicated by the dashed line.

\begin{algorithm}[H] % must use [H]
\caption{Evaluating $|\phi_{\rm out}\rangle=\hat{U}(\Delta t) |\phi_{\rm in}\rangle$}
\label{algorithm:evolve}
\begin{algorithmic}[1]
\State $|\phi_{0}\rangle \gets |\phi_{\rm in}\rangle$
\State $|\phi_{1}\rangle \gets \widetilde{H} |\phi_{\rm in}\rangle$
\State $|\phi_{\rm out}\rangle \gets J_0\left( \omega_0 \Delta t \right) |\phi_{0}\rangle+2(-i)J_1\left( \omega_0 \Delta t \right) |\phi_{1}\rangle$
\State $m \gets 2$
\While {$\textmd{abs}\left[J_m\left( \omega_0 \Delta t\right)\right]>10^{-15}$}
   \State $|\phi_{2}\rangle \gets 2\widetilde{H} |\phi_{1}\rangle - |\phi_{0}\rangle$
   \State $|\phi_{\rm out}\rangle \gets |\phi_{\rm out}\rangle + 2(-i)^mJ_m\left( \omega_0 \Delta t \right) |\phi_{2}\rangle$
   \State $|\phi_{0}\rangle \gets |\phi_{1}\rangle$
   \State $|\phi_{1}\rangle \gets |\phi_{2}\rangle$
   \State $m \gets m+1$
\EndWhile
\end{algorithmic}
\end{algorithm}

\begin{algorithm}[H] % must use [H]
\caption{Evaluating $|\phi_{\rm out}\rangle=[\hat{X},\hat{U}(\Delta t)] |\phi_{\rm in}\rangle$}
\label{algorithm:evolvex}
\begin{algorithmic}[1]
\State $|\phi_{0}\rangle \gets |\phi_{\rm in}\rangle$
\State $|\phi_{0}^{x}\rangle \gets0$
\State $|\phi_{1}\rangle \gets \widetilde{H} |\phi_{0}\rangle$
\State $|\phi_{1}^x\rangle \gets [\hat{X},\widetilde{H}] |\phi_{0}\rangle$
\State $|\phi_{\rm out}\rangle \gets 2(- i)J_1\left( \omega_0 \Delta t\right) |\phi_{1}^x\rangle$
\State $m \gets 2$
\While {$\textmd{abs}\left[J_m\left( \omega_0 \Delta t\right)\right]>10^{-15}$}
   \State $|\phi_{2}\rangle \gets 2\widetilde{H} |\phi_{1}\rangle - |\phi_{0}\rangle$
      \State $|\phi_{2}^x\rangle \gets 2 [\hat{X},\widetilde{H}]  |\phi_{1}\rangle + 2  \widetilde{H}  |\phi_1^x\rangle- |\phi_{0}^x\rangle$
   \State $|\phi_{\rm out}\rangle \gets |\phi_{\rm out}\rangle + 2(- i)^mJ_m\left( \omega_0 \Delta t \right) |\phi_{2}^x\rangle$
   \State $|\phi_{0}\rangle \gets |\phi_{1}\rangle$
   \State $|\phi_{1}\rangle \gets |\phi_{2}\rangle$
   \State $|\phi_{0}^x\rangle \gets |\phi_{1}^x\rangle$
   \State $|\phi_{1}^x\rangle \gets |\phi_{2}^x\rangle$
   \State $m \gets m+1$
\EndWhile
\end{algorithmic}
\end{algorithm}

The Green's functions are spectral quantities, and as such, can also be evaluated using the Chebyshev polynomial expansion. However, special care should be taken as they have singularities which carry physical information of the system. In Sec.\ \ref{section:kubo_formulas} we defined a regularized version of the Green's functions in Eq.\ (\ref{equation:reg_green}), where we introduced an imaginary rate $1/\tau_{\phi}$ and a finite time $\tau$. These parameters broaden the singularities of the Green's functions and serve as a mathematical regularization that enables the approximation of the Green's functions to a given precision using a \emph{finite} Chebyshev polynomial expansion. In the limit $\tau \rightarrow \infty$, one can approximate the regularized retarded Green's function as
\begin{equation}
\label{equation:CPGF-1}
G^+(\tilde{E};\tau_\phi) = \frac{1}{\Delta E} \sum_m \bar{G}_m^+ T_m(\widetilde{H}). 
\end{equation}
Here, $\bar{G}_m^+$ are the Chebyshev coefficients defined in Eq.\ (\ref{chebint}), which can be evaluated using a Laplace transform of the     Bessel function \cite{gradshteyn_book} as
\begin{equation}
\label{equation:CPGF}
\bar{G}_m^+ = (2-\delta_{m0}) \frac{\left(z-i\sqrt{1-z^2}\right)^m}{i\sqrt{1-z^2}}, 
\end{equation}
where $z=( E+ i \hbar /\tau_\phi)/\Delta E$. This expansion has been used by Vijay {\it et al.}\ \cite{vijay2004jpca} in the context of spectral filters and by Braun and Schmitteckert \cite{braun2014prb} to determine the impurity Green's function of the interacting resonant level model.
Recently, it was applied by Ferreira and Mucciolo for the first time to quantum transport, where it was dubbed the Chebyshev-polynomial Green's function (CPGF) method \cite{ferreira2015prl}. One can also use analytic continuation of the logarithms to the complex plane to express Eq.\ (\ref{equation:CPGF}) in terms of an exponential, 
 \begin{equation}
\bar{G}_m^+= (2-\delta_{m0}) \frac{\exp[-i \, m \, \text{arccos}(z)]}{i\sqrt{1-z^2}}.
\label{KPMGreenFuncEta0}
\end{equation}
This shows that the CPGF method involves an analytic continuation of an expression previously obtained by other authors \cite{weisse2006rmp, Covaci2010PRL, garcia2015prl}, where the singularities have been smoothed. In the limit of infinite $\tau_\phi$, these coefficients do not decay but oscillate with increasing $m$, and it is the presence of a finite dephasing time which provides a damping of the coefficients and permits the convergence of the expansion.

In the next subsection, we discuss different approaches to deal with singular functions in the context of approximating the quantum projection operator $\delta(E-\hat{H})$, a common factor in all the representations of the dissipative conductivity.

\subsection{Evaluating the quantum projection operator \label{section:delta}}

We now discuss the evaluation of the quantum projection operator $\delta(E-\hat{H})$ involved in all the conductivity formulas. There are several linear scaling techniques for approximating this operator, including the Lanczos recursion method (LRM) \cite{haydock1972jpc, haydock1975jpc, haydock1980ssp, petitfor1985book, dagotto1994rmp}, the Fourier transform method (FTM) \cite{alben1975prb, feit1982jcp, hams2000pre}, the kernel polynomial method (KPM) \cite{silver1994ijmpc, wang1994prl, wang1994prb, silver1996jcp, weisse2006rmp}, and the maximum entropy method (MEM) \cite{skilling1989, drabold1993prl, silver1997pre}. We will only review the first three methods (LRM, FTM, and KPM), as the last one (MEM) has not been used in LSQT calculations. A comparison between the MEM and the KPM can be found in a previous review \cite{weisse2006rmp}. All of these methods have been used recently to compute transport properties in different systems \cite{yuan2010prb, yuan2010prb_b, ferreira2011prb, zhao2015prb, fan2014cpc, garcia2015prl, Cummings2017prl, weisse2006rmp}. Although we have a few different quantities to calculate, it suffices to discuss these methods in terms of the DOS as given in Eq.\ (\ref{equation:trace_DOS}). Generalizations to other quantities are straightforward.

\subsubsection{The Lanczos recursion method}\label{sec:lrm}

The LRM is based on the Lanczos algorithm for tridiagonalizing sparse Hermitian matrices \cite{lanczos1950jrnbs}. The Lanczos algorithm is usually used to obtain extremal eigenvalues and the corresponding eigenstates \cite{cullum1985}, but it can also be used to calculate spectral properties \cite{haydock1972jpc, haydock1975jpc, haydock1980ssp, petitfor1985book}.

The first step of the LRM is to project the Hamiltonian onto an orthogonal basis in a Krylov subspace, generating a tridiagonal matrix 
\begin{equation}
T = \left(
\begin{array}{ccccc}
a_1 & b_2 & 0 & \cdots & 0 \\
b_2 & a_2 & b_3 & \ddots & 0 \\
0 & \ddots & \ddots & \ddots & 0 \\
\vdots & \ddots & b_{M-1} & a_{M-1} & b_{M} \\
0 & \cdots & 0 & b_M & a_M
\end{array}
\right).
\end{equation}
The dimension $M$ of the tridiagonal matrix can be much smaller than the dimension $N$ of the original matrix. The matrix elements $\{a_n\}$ and $\{b_n\}$ are obtained from a Lanczos algorithm. There are multiple versions of the Lanczos algorithm and the most numerically stable one is given in Algorithm \ref{algorithm:lanczos} \cite{saad2003book}. The computational effort of the LRM is thus proportional to $NM$, which is $\mathcal{O}(N)$ when $M \ll N$.

\begin{algorithm}[H] % must use [H]
\caption{Lanczos algorithm \cite{saad2003book}}
\label{algorithm:lanczos}
\begin{algorithmic}[1]
\Require $|\phi_1\rangle=|\phi\rangle$ is the normalized random vector
\State $b_1 \gets 0$
\State $|\phi_0 \rangle \gets 0$
\For {$m$ = 1 to $M$}
   \State $|\psi_m\rangle \gets H |\phi_m\rangle - b_m |\phi_{m-1}\rangle$
   \State $a_m \gets \langle \psi_m | \phi_m\rangle$
   \State $|\psi_m\rangle \gets | \psi_m\rangle - a_m |\phi_m\rangle$
   \State $b_{m+1} \gets \sqrt{\langle \psi_m |\psi_m\rangle}$
   \State $|\phi_{m+1}\rangle \gets |\psi_m\rangle/b_{m+1}$
\EndFor
\end{algorithmic}
\end{algorithm}

The second step of the LRM is to calculate the first element of the retarded Green's function $G^{+}(E)=(E+i\eta-\hat{H})^{-1}$ in the Lanczos basis $\{|\phi_m\rangle\}$ using the continued fraction
\begin{equation}
\langle \phi| G^{+}(E) |\phi\rangle = \frac{1}{E+i\eta-a_1-\frac{b_2^2}{E+i\eta-a_2-\cdots}}.
\end{equation}
The DOS of Eq.\ (\ref{equation:trace_DOS}) can then be calculated using the relation between the quantum projection operator and the Green's function given in Eq.\ (\ref{equation:imaginary_green}). The computation time for the second step is proportional to $MN_\text{e}$, where $N_\text{e}$ is the number of energy points considered in the calculation. Usually, $N_\text{e} \ll N$, and the computation time for the second step is thus negligible compared to the first step. Because of this, the overall computational effort almost does not scale with respect to $N_\text{e}$. We can thus say that the algorithm is parallel in energy, which is a common feature for all the methods presented below.

An important issue is the energy resolution $\delta E$ achievable using a given number of recursion steps $M$. The energy resolution is actually set by the imaginary energy $i\eta$ in the Green's function, i.e., $\delta E=\eta$. One should therefore make sure that a sufficiently large $M$ is used to ensure converged results. However, it is well known that in its basic forms such as the one presented in Algorithm \ref{algorithm:lanczos}, the Lanczos algorithm can become numerically unstable when $M$ is large, due to the loss of orthogonality in the Lanczos basis vectors. The Lanczos basis vectors can be explicitly orthogonalized \cite{saad2003book}, but this will increase the computational complexity of the algorithm, making it less efficient than other methods. 

The Lanczos method has been also used for computing off-diagonal Green's functions, which are key quantities for computing correlation functions. A simple approach can be used when the Hamiltonian is symmetric; off-diagonal Green's function elements (such as $\langle \Psi_i| (z-\hat{H})^{-1}| \Psi_j \rangle$) can be derived from linear combinations of on-diagonal elements ($\langle \Psi_i\pm \Psi_j | (z-\hat{H})^{-1}| \Psi_i\pm \Psi_j\rangle$) \cite{petitfor1985book, viswanath1994book}. For non-symmetric matrices a procedure similar to the Lanczos method, based on the construction of a bi-orthogonal expansion, has been proposed \cite{Benoit_1994, Benoit_1995, Triozon2005}. The construction of the bi-orthogonal basis is calculated as
\begin{eqnarray}
|\psi_{n+1}\rangle&=&\hat{H}|\psi_{n}\rangle-a_{n+1}|\psi_{n}\rangle-b_{n}|\psi_{n-1}\rangle, \label{rec1} \\
\langle\phi_{n+1}|&=&\langle\phi_{n}|\hat{H}-\langle\phi_{n}|a_{n+1}
-\langle\phi_{n-1}|b_{n},
\label{rec2}
\end{eqnarray}
with the initial conditions $|\psi_{-1}\rangle \,{=}\, |\phi_{-1}\rangle \,{=}\, 0$, $|\psi_{0}\rangle \,{=}\, |\phi_{0}\rangle \,{=}\, |\psi \rangle$, and the bi-orthogonality condition $\langle \phi_{n}|\psi_{m}\rangle = 0 \ \text{if} \ n \neq m$. This last condition is equivalent to the following relations for $a_{n}$ and $b_{n}$:
\begin{eqnarray}
a_{n+1} & = & \frac{\langle\phi_{n}|
\hat{H}|{\psi}_{n}\rangle}{\langle\phi_{n}|{\psi}_{n}\rangle}, \label{cond1} \\
b_{n} & = & \frac{\langle\phi_{n-1}|\hat{H}|{\psi}_{n}\rangle}{\langle\phi_{n-1}|{\psi}_{n-1}\rangle} = \frac{\langle\phi_{n}|{\psi}_{n}\rangle}{\langle\phi_{n-1}|{\psi}_{n-1}\rangle}.
\label{cond2}
\end{eqnarray}
In the basis $\{|\psi_{n}\rangle \}$, $\hat{H}$ can be written as
\begin{align}
\hat{H} = \left(
\begin{array}{ccccc}
a_{1} & b_{1} & & & \\
1 & a_{2} & b_{2} & & \\
& 1 & a_{3} & b_{3} & \\
& & 1 & . & . \\
& & & . & . \\
\end{array}
\right) .
\label{eq:ham_tridiag}
\end{align}

The quantity $\langle \psi |G^\dagger(z=E\pm i0^{+})|\psi \rangle=\langle\phi_{0}|\frac{1}{z-\hat{H}}|\psi_{0}\rangle$ can then be computed by the continued fraction method. This quantity is equal to the first diagonal element of $(z-\hat{H})^{-1}$, where $\hat{H}$ is the tridiagonal matrix in Eq.\ (\ref{eq:ham_tridiag}). Let us call this matrix element $G_{0}(z)$ and define $G_{n}(z)$ to be the first diagonal element of the matrix $(z-\hat{H}_{n})^{-1}$, with $\hat{H}_{n}$ the matrix $\hat{H}$ without its $n$ first lines and columns,
\begin{align}
\hat{H}_{n} = \left(
\begin{array}{ccccc}
a_{n+1} & b_{n+1} & & & \\
1 & a_{n+2} & b_{n+2} & & \\
& 1 & a_{n+3} & b_{n+3} & \\
& & 1 & . & . \\
& & & . & . \\
\end{array}
\right) .
\label{eq:ham_n}
\end{align}
From standard linear algebra, it can be shown that
\begin{align}
\label{eq:G0_G1}
G_{0}(z) = \frac{1}{z-a_{1}-b_{1}G_{1}(z)},
\end{align}
and repeating this algorithm leads to a continued fraction expansion of $G_{0}(z)$,
\begin{align}
\label{eq:fraction}
G_{0}(z) =
\frac{1}{z-a_{1}-\frac{\displaystyle b_{1}}{\displaystyle z-a_{2}-\frac{\displaystyle b_{2}}{...}}}.
\end{align}
\noindent

One should note that in contrast to the standard Lanczos recursion, the coefficients $a_{n}$ and $b_{n}$ do not show any simple behavior for large $n$, but simple truncation of the continued fraction at sufficiently large $n$ was found to yield reasonably good convergence. This method was implemented for developing an $\mathcal{O}(N)$ approach of the Landauer-B\"{u}ttiker conductance formula \cite{Triozon2005} as further discussed in Sec.\ \ref{section:lb}.

\subsubsection{The Fourier transform method}

The FTM is very simple conceptually: it is based on the Fourier transform of the Dirac delta function as given in Eq.\ (\ref{equation:dirac_delta_fourier}). Ideally, the time integral is over the whole real axis, but in practice one can only reach a finite time with a finite time step $\Delta \tau$. Therefore, one should be satisfied with a truncated discrete Fourier transform,
\begin{equation}
\delta(E-\hat{H}) \approx \frac{\Delta \tau}{2\pi\hbar} \sum_{m=-M}^{M} e^{i(E-\hat{H})m\Delta \tau/\hbar},
\end{equation}
where $M\Delta\tau$ represents the upper limit of the time integral in Eq.\ (\ref{equation:dirac_delta_fourier}). A direct expansion in this way leads to Gibbs oscillations, and a window function is usually used to suppress them. A frequently used one is the Hann window,
\begin{equation}
\label{equation:w_m}
w_m = \frac{1}{2} \left[ 1 + \cos \left(  \frac{\pi m}{M+1}\right) \right].
\end{equation}
Using the discrete Fourier transform, we can write the DOS in Eq.\ (\ref{equation:trace_DOS}) as
\begin{equation}
\label{equation:rho_FTM}
\rho(E) \approx \frac{\Delta \tau}{2\pi\hbar\Omega} \sum_{m=-M}^{M} e^{iEm\Delta \tau/\hbar} w_m F_m,
\end{equation}
where
\begin{equation}
\label{equation:F_m}
F_m = \langle \phi| e^{-i\hat{H}m\Delta \tau/\hbar} |\phi\rangle = \langle \phi| \hat{U}(m\Delta \tau) |\phi\rangle
\end{equation}
is the $m$th Fourier moment.

Based on the formulas above, we can see that the FTM consists of the following two steps: (1) construct a set of Fourier moments $\{F_m\}$ as defined in Eq.\ (\ref{equation:F_m}), and (2) calculate physical properties such as the DOS from the Fourier moments through a discrete Fourier transform as given by Eq.\ (\ref{equation:rho_FTM}). Similar to the case of the LRM, the computation time for the second step is negligible compared to the first one and the algorithm is essentially parallel in energy.

As the Fourier moments are the expectation values of the time evolution operator, this method is also usually called the equation of motion method \cite{alben1975prb} or the time-dependent Schr\"{o}dinger equation  method \cite{feit1982jcp, hams2000pre}. Note that we have used $\Delta \tau$ here to distinguish it from the correlation time step $\Delta t$ in the VAC and MSD formalisms. Based on the Nyquist sampling theorem, which states that the sampling rate must be no less than the Nyquist rate $2f_{\rm max}$ to perfectly reconstruct a signal with spectrum between $0$ and $ f_{\rm max}$, the optimal value of $\Delta \tau$ can be determined to be $\Delta \tau = \pi\hbar/\Delta E$, giving $\omega_0\Delta \tau=\pi$. Using this $\Delta \tau$, the energy resolution is given by $\delta E \sim \Delta E/M$ \cite{feit1982jcp}.

\subsubsection{The kernel polynomial method}
\label{subsection:kpm}

In Sec.\ \ref{subsection:ChebPol} we introduced the Chebyshev polynomial expansion as a useful tool for approximating regular functions and discussed additionally the problem of expanding a singular function such as the Green's function using the CPGF method \cite{ferreira2015prl}. There, the singularity in the Green's function was regularized by introducing a small imaginary energy $i\eta$. There is another widely used approach to handle the singularity in the function to be expanded in terms of Chebyshev polynomials, which is called the kernel polynomial method (KPM) \cite{silver1994ijmpc, silver1996jcp, wang1994prl, wang1994prb, weisse2006rmp}. 

When the expansion in Eq.\ (\ref{equation:expansion_1}) is truncated to a finite order $M$, there will be Gibbs oscillations near the points where the expanded function $f(x)$ is not continuously differentiable. These can be damped by a convolution of the function with a kernel $K(x)$ \cite{silver1996jcp, weisse2006rmp}. The advantage of the Chebyshev expansion is that this convolution can be included by multiplying the Chebyshev coefficients with a damping factor $g_m$, transforming Eq.\ (\ref{equation:expansion_1}) into
\begin{equation}
f(x) \approx \sum_{m=0}^{M} \bar{f}_m g_m T_m(x)\label{equation:expansion_2}.
\end{equation}

To derive an expression for the DOS using the KPM, we start by exploiting the following scaling property of the delta function,
\begin{equation}
\delta(E-\hat{H})=\frac{1}{\Delta E} \delta(\widetilde{E}-\widetilde{H}),
\end{equation}
where $\widetilde{H}$ is defined in Eq.\ (\ref{equation:scale}) and $\widetilde{E}$ is defined similarly. Then, using the identity in Eq.\ (\ref{spectral_expansion}), it is straightforward to express the projection operator in the form of Eq.\ (\ref{equation:expansion_2}),
\begin{equation}
\delta(\widetilde{E}-\widetilde{H}) =  \sum_{m=0}^{M}\bar{\delta}_m(\widetilde{E}) g_m T_m(\widetilde{H}),
\end{equation}
where
\begin{equation}
  \bar{\delta}_m(\widetilde{E}) =  \frac{(2-\delta_{m0}) T_m(\widetilde{E})}{\pi\sqrt{1-\widetilde{E}^2}} 
\end{equation}
are the Chebyshev coefficients computed as defined in Eq.\ (\ref{chebint}). Finally, the DOS is obtained by computing the trace of the projection operator as defined in Eq.\ (\ref{equation:trace_DOS}),
\begin{equation}
\label{equation:rho_KPM}
\rho(\widetilde{E}) = \frac{1}{\Delta E \cdot \Omega}\sum_{m=0}^{M}  C^{\rm DOS}_m g_m \bar{\delta}_m(\widetilde{E}),
\end{equation}
where 
\begin{equation}
\label{equation:rho_moments}
C^{\rm DOS}_m=\langle\phi|T_m(\widetilde{H})|\phi\rangle
\end{equation}
are the Chebyshev moments for the DOS.

Up to this point we have shown that the DOS can be approximated using Chebyshev polynomials, but we have not specified any choice for the kernel, which will vary with the specific application. For the expansion of the quantum resolution operator, which is essentially a set of delta peaks, the Jackson kernel with the damping factor 
\begin{equation}
\label{equation:damping_Jackson}
 g^{\rm J}_m =\frac{ (M+1-m)\cos\left(\frac{\pi m}{M+1}  \right) +  \sin\left( \frac{\pi m}{M+1}  \right) \cot\left(\frac{\pi}{M+1} \right)}{M+1}
\end{equation}
has been found to be optimal, as it produces the smallest broadening for a given value of $M$. In particular, away from $\widetilde{E}=\pm1$ it approximates a Gaussian broadening with a width of $\delta E = \pi \cdot \Delta E / M$ \cite{silver1996jcp, weisse2006rmp}. If one considers the Green's function, the Lorentz kernel with the damping factor
\begin{equation}
\label{equation:damping_Lorentz}
 g^{\rm L}_m(\lambda) = \frac{\sinh[\lambda(1-m/M)]}{\sinh(\lambda)}
\end{equation} 
may offer a better choice ($\lambda$ is a free parameter which is usually chosen to be 3-5) due to the fact that it regularizes the imaginary part of the Green's function into a Lorentzian with a broadening given by $\delta E = \lambda \cdot \Delta E / M$ when away from $\widetilde{E}=\pm1$, which is closer to physical reality.

\begin{figure}[htb]
\begin{center}
\includegraphics[width=\columnwidth]{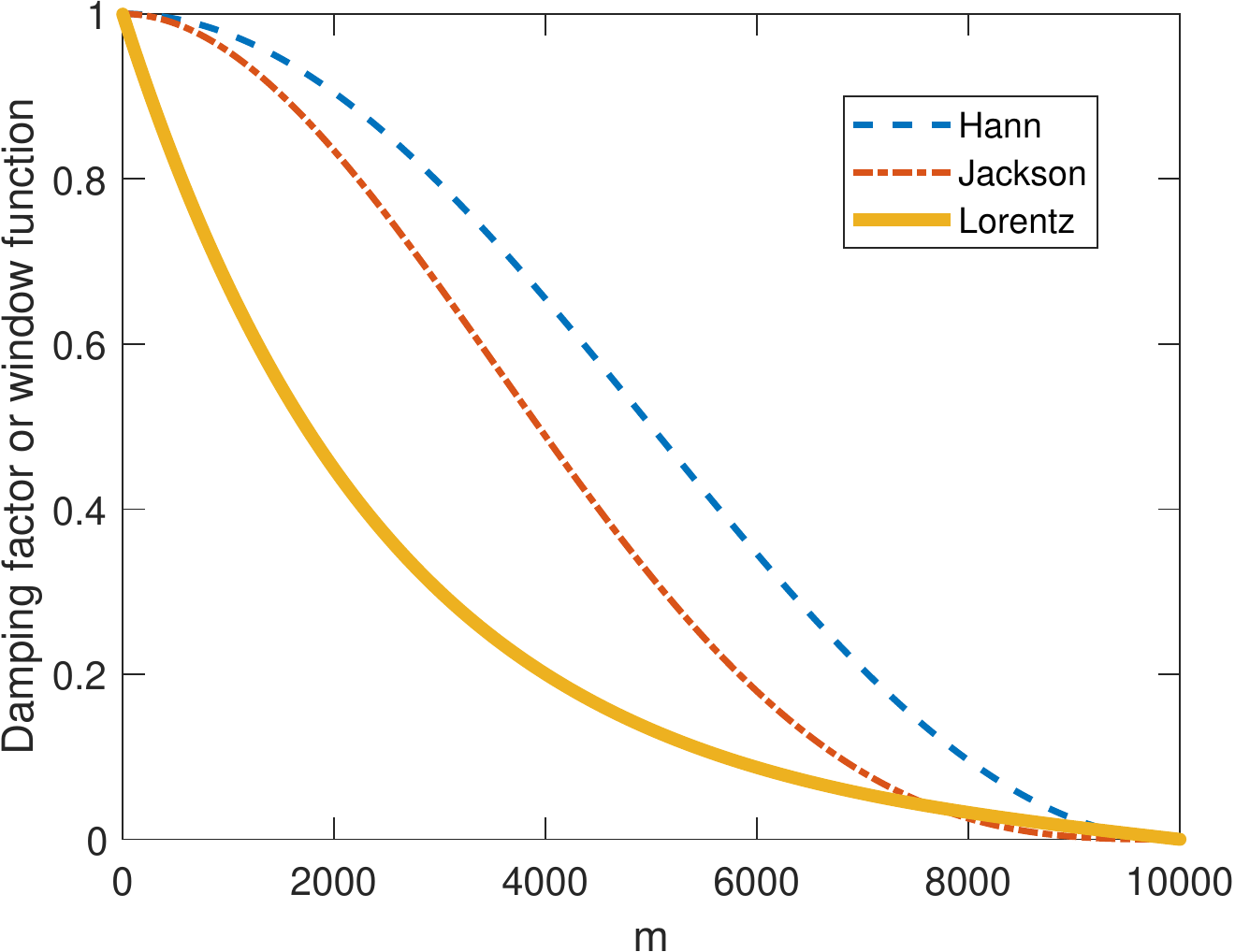}
\caption{Comparison between different damping factors and a window function, including the Jackson damping factor $g^{\rm J}_m$ defined in Eq.\ (\ref{equation:damping_Jackson}), the Lorentz damping factor $g^{\rm L}_m(\lambda=4)$ defined in Eq.\ (\ref{equation:damping_Lorentz}), and the Hann window function $w_m$ defined in Eq.\ (\ref{equation:w_m}).}
\label{figure:damping_1}
\end{center}
\end{figure}

\begin{figure}[htb]
\begin{center}
\includegraphics[width=\columnwidth]{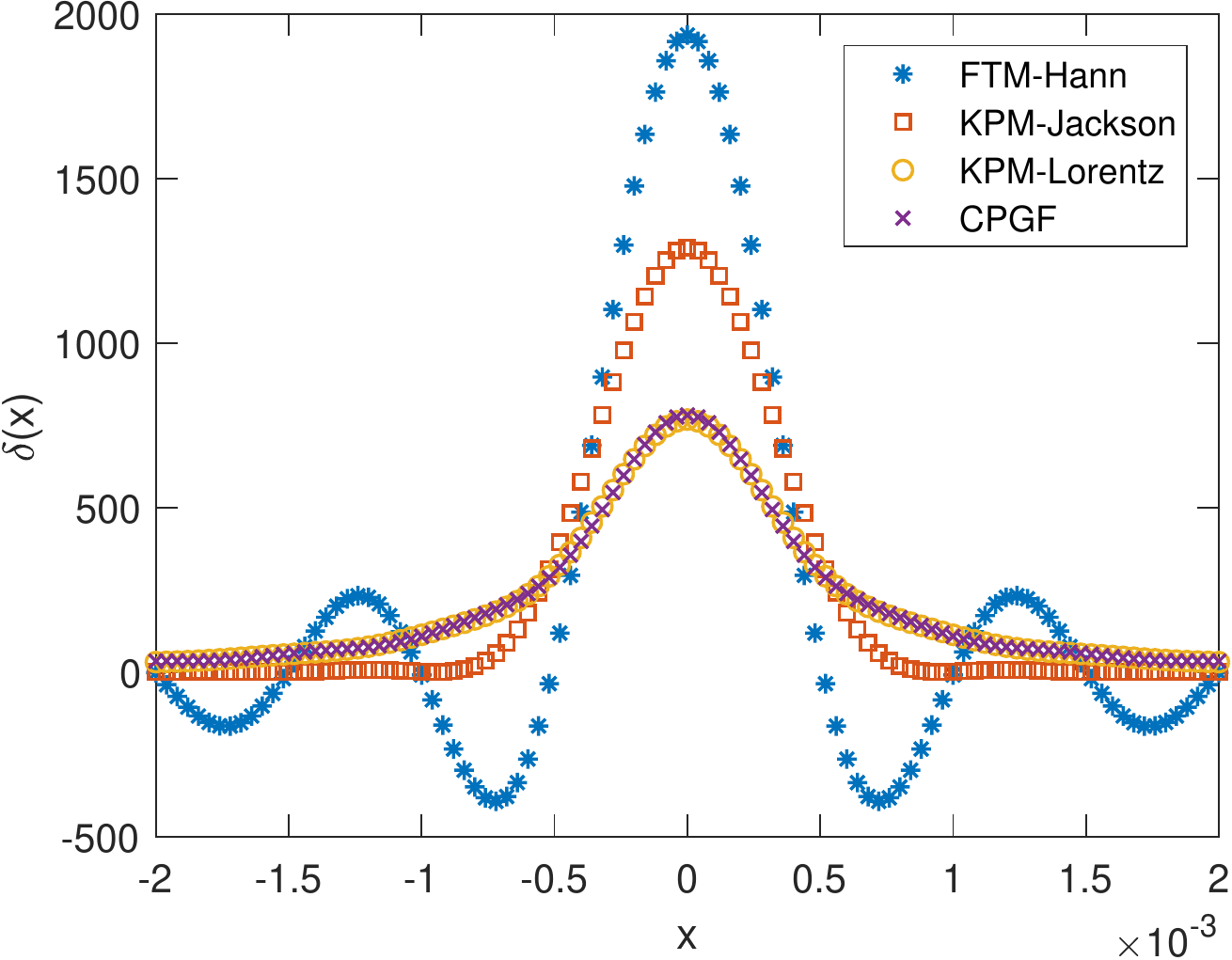}
\caption{Approximation of the Dirac delta function $\delta(x)$ using the KPM with the Jackson damping $g^{\rm J}_m$ defined in Eq.\ (\ref{equation:damping_Jackson}), the Lorentz damping $g^{\rm L}_m(\lambda=4)$ defined in Eq.\ (\ref{equation:damping_Lorentz}), the CPGF method, and the FTM with the Hann window $w_m$ defined in Eq.\ (\ref{equation:w_m}). Here, $M=10^4$ for the KPM and $M=2000$ for the FTM. Note that the $x$ and $y$ axes differ by a factor of $10^6$ in magnitude.}
\label{figure:damping_2}
\end{center}
\end{figure}

In Fig.\ \ref{figure:damping_1} we plot the Jackson and Lorentz damping factors along with the Hann window function, where the expansion order is chosen as $M=10^4$. To demonstrate the performance of the different damping factors and the window function, we use them to approximate the function $\delta(x)$. The results obtained by using the KPM with different damping factors are shown in Fig.\ \ref{figure:damping_2}. Also shown are the results obtained with the Fourier expansion and the CPGF method \cite{ferreira2015prl}. For the same value of $M=10^4$, the Jackson damping gives a narrower shape compared to the Lorentz damping and therefore has finer resolution, while the CPGF method is essentially equivalent to the KPM with the Lorentz damping ($\lambda=4$) when the resolution parameter in the CPGF method is chosen as $4/M$. Although CPGF and KPM with a Lorentz kernel behave similarly near the band center ($\widetilde{E}=0$), it is important to note that CPGF provides a uniform energy resolution, contrary to KPM with a Lorentz kernel, whose resolution changes with energy, leading to overshooting
away from the band centre \cite{Schubert2010,Joao2019}. Furthermore, CPGF allows for approximating the Green's function up to an arbitrary precision, and is thus the ideal choice when the physical origin of the $\delta$-function is a Green's function. We also note that while the Gibbs oscillations can be effectively suppressed using the KPM, they persist in the case of the FTM. Apart from being less effective in suppressing Gibbs oscillations, the FTM has also been shown to be less computationally efficient as compared to the KPM \cite{fan2014cpc}. This comparison and the comparison between the KPM and the LRM \cite{silver1996jcp, weisse2006rmp} indicate that the KPM with the Jackson damping factor is the optimal approach for approximating the quantum projection operator $\delta(E-\hat{H})$.

\begin{algorithm}[H] % must use [H]
\caption{Evaluating the Chebyshev moments $C_m = \langle\phi_\text{L}|T_m(\widetilde{H})|\phi_\text{R}\rangle$}
\label{algorithm:kpm1}
\begin{algorithmic}[1]
\State $|\phi_{0}\rangle \gets |\phi_\text{R}\rangle$
\State $C_{0} \gets \langle\phi_\text{L}|\phi_0\rangle$
\State $|\phi_{1}\rangle \gets \widetilde{H} |\phi_{0}\rangle$
\State $C_{1} \gets \langle\phi_\text{L}|\phi_1\rangle$
\For {$m$ = 2 to $M$}
   \State $|\phi_{2}\rangle \gets 2\widetilde{H} |\phi_{1}\rangle - |\phi_{0}\rangle$
   \State $C_{m} \gets \langle\phi_\text{L}|\phi_2\rangle$
   \State $|\phi_0\rangle \gets |\phi_1\rangle$
   \State $|\phi_1\rangle \gets |\phi_2\rangle$
\EndFor
\end{algorithmic}
\end{algorithm}

We now summarize the procedure of the KPM: (1) construct a set of Chebyshev moments $\{C_m\}$ (see Algorithm \ref{algorithm:kpm1}), and (2) calculate physical properties such as the DOS from the Chebyshev moments through a finite-order Chebyshev polynomial summation as given by Eq.\ (\ref{equation:rho_KPM}). Similar to the case of the LRM and the FTM, the construction of the Chebyshev moments dominates the computation time and the algorithm is parallel in energy. The energy resolution achieved in the KPM is $\delta E \sim \Delta E/M$ \cite{weisse2006rmp}, similar to the case of the FTM.

%% file: secIV_landauer.tex
\section{Landauer-B\"{u}ttiker quantum transport methodology}
\label{section:lb}

Transport properties at the nanoscale in open systems (with electrodes in a device geometry) are conveniently described by the Landauer-B\"{u}ttiker \cite{landauer1957ibm, landauer1970pm, buttiker1985prb} and the nonequilibrium Green's function formalisms \cite{kadanoff1962book, keldysh1965jetp, rammer1986rmp, haug1996book}. The Landauer-B\"{u}ttiker formalism expresses the current response of a multi-port conductor in terms of transmission matrices, and is also derivable straightforwardly from linear response theory \cite{StoneIBM1988, Baranger89}. 

In the Landauer-B\"{u}ttiker (LB) formalism, efficient numerical methods based on recursive Green's functions \cite{ferry1997book} have been developed and are routinely used. As matrix inversion is at the heart of this approach, the computational cost generally scales cubically with respect to the cross-sectional area of the system, making it computationally prohibitive for large and disordered two-dimensional and three-dimensional systems. Despite this limitation, it is still possible to implement density-functional methods in the nonequilibrium transport formalism \cite{Brandbyge2002}, and to investigate low-dimensional nanostructures such as disordered semiconducting nanowires \cite{persson2008nl,markussen2017prb} or chemically functionalized nanotubes with lengths reaching the micrometer scale \cite{Lopez2009}. A recent study of quantum transport in carbon nanotubes confirms for instance the universality of the single-parameter scaling Anderson localization for realistic models of disordered systems \cite{Lopez2019}

In tight-binding calculations of the LB method, the recursive Green's function formalism \cite{lewenkopf2013jce} is usually used. The contacts are modeled as ballistic semi-infinite leads and the conductance $g(E)$ is obtained from the transmission function $\mathcal{T}(E)$, 
\begin{equation}
g(E)=\frac{2e^2}{h} \mathcal{T}(E).
\end{equation}
For a single-mode system, the transmission function equals the probability of a charge carrier to transmit from one contact to another. If there are several transport modes involved, the transmission function equals the sum of the transmission probabilities for the different modes. There are many equivalent forms for the transmission function, and here we adopt the Caroli form \cite{caroli1971jpc},
\begin{equation}
\mathcal{T}(E) = \text{Tr} [G(E) \Gamma_\text{L} G^{\dagger}(E) \Gamma_\text{R}],
\label{equation:transmission}
\end{equation}
where $G(E)$ is the advanced Green's function of the device, $G^{\dagger}(E)$ is the retarded Green's function, and $\Gamma_\text{L/R}$ describe the coupling of the device to the left and right leads. The advanced Green's function for a system attached to two leads is 
\begin{equation}
G(E) = \frac{1}{ E - \hat{H} - \Sigma_\text{L}(E_\text{L}) - \Sigma_\text{R}(E_\text{R}) },
\end{equation}
where $\Sigma_\text{L}(E_\text{L})$ is the self-energy of the left lead at Fermi energy $E_\text{L}$ and $\Sigma_\text{R}(E_\text{R})$ is the self-energy of the right lead at Fermi energy $E_\text{R}$. The Fermi energies $E_\text{L}$ and $E_\text{R}$ of the leads can be set to the same value as in the device, $E$, or to an arbitrary value. In the calculations in the rest of this review, we set $E_\text{L} = E_\text{R} = E$. The self-energy matrices can be obtained through different methods, e.g., using an iterative method \cite{sancho1985jpf}. The coupling matrices $\Gamma_\text{L}$ and $\Gamma_\text{R}$ are the imaginary part of the self energies,
\begin{equation}
\Gamma_\text{L/R}=i\left(\Sigma_\text{L/R}-\Sigma_\text{L/R}^{\dagger}\right)=-2\textmd{Im}\left[\Sigma_\text{L/R}\right].
\end{equation}

Beyond usual decimation techniques, the modified Lanczos algorithm to compute off-diagonal Green's functions for non-symmetric Hamiltonian matrices, described at the end of Sec.\ \ref{sec:lrm}, can be used for the calculation of the Landauer--B\"uttiker conductance \cite{Triozon2005}. In this situation, to cope with device geometries including open boundaries, the Green's function is obtained from an effective Hamiltonian $\hat{H} = \hat{H}_{0} + \Sigma_\text{L}+\Sigma_\text{R}$, where the effect of left and right leads are introduced through the self-energy operators. The technical approach is to rewrite the transmission function of Eq.\ (\ref{equation:transmission}) as
\begin{equation}
\mathcal{T}(E) = \sum_{\alpha,\beta,\alpha',\beta'} \langle \beta | \Gamma_\text{R} | \alpha \rangle \langle \alpha | G^\dagger | \alpha' \rangle \langle \alpha' | \Gamma_\text{L} | \beta' \rangle \langle \beta' | G | \beta \rangle,
\label{equation:transmission2}
\end{equation}
where $\alpha,\beta$ ($\alpha',\beta'$) are the interface states that run over the orbitals coupled to the left (right) electrode. The self-energies of the leads can be calculated efficiently using standard recursion techniques \cite{sancho1985jpf}, while an $\mathcal{O}(N)$ method is needed for evaluating the Green's functions. The off-diagonal elements of the Green's functions can be expressed as a sum of three diagonal elements
\begin{align}
\langle \alpha | G^\dagger | \alpha' \rangle = \frac{1}{2}\left[
(1+i)\langle\psi_+|G^\dagger|\psi_+\rangle \right. \nonumber \\
\left. + (i-1)\langle\psi_-|G^\dagger|\psi_-\rangle
- 2i\langle\psi_i|G^\dagger|\psi_i\rangle \right],
\end{align}
where $|\psi_\pm\rangle = \left( |\alpha\rangle \pm |\alpha'\rangle \right) / \sqrt{2}$ and $|\psi_i\rangle = \left( |\alpha\rangle + i|\alpha'\rangle \right) / \sqrt{2}$. The problem thus reduces to the $\mathcal{O}(N)$ evaluation of $\langle\psi|G^\dagger|\psi\rangle$, described at the end of Sec.\ \ref{sec:lrm}.

This method was tested on carbon nanotube-based heterojunctions \cite{Triozon2005}, with perfect agreement with standard decimation techniques. As the computational cost of a single off-diagonal Green’s function element is $\mathcal{O}(N)$, this approach is particularly well-suited for systems which have large transverse dimensions, since the conventional decimation or transfer matrix techniques require matrix inversion of a layer Hamiltonian. The number of Green’s function elements to calculate for the conductance is $N_\text{R}\times N_\text{L}$, where $N_\text{R}$ ($N_\text{L}$) is the number of orbitals coupling the right (left) lead to the channel. Usually the number of such Green’s function elements is relatively small, and the total numerical cost is essentially proportional to $N$, regardless of the geometry and topological complexity of the system \cite{Triozon2005}.

It is appropriate at this point to mention a computationally efficient wave function formulation of the quantum scattering problem in the Landauer-B\"{u}ttiker formalism -- available in the KWANT code (https://kwant-project.org/) -- that can reduce the computational time compared to the recursive Green's function approach, but at the cost of increased memory footprint \cite{groth2014njp,GAURY20141}. For nanostructures, the computational cost of this wave function matching method has been shown to scale linearly with the number of sites $N_\text{s}$ for large $N_\text{s}$ (or more precisely with $N_\text{s}\times N_\text{c}$, where $N_\text{c}$ is the number of open channels in the leads) \cite{Santos2019}. Finally, Istas, Groth and Waintal have recently proposed an approach to cope with ``mostly translationally invariant systems'' \cite{Istas2018}, i.e., systems with weak disorder. With this method, systems of complex geometries are decomposed into two parts; one fully periodic part that is stitched with another part containing the disorder potential and electrodes. For weak disorder this approach becomes truly $\mathcal{O}(N)$, which makes it promising to study, in particular, quantum transport at the surfaces of large systems such as 3D topological insulators or Weyl semi-metals.

%% file: secV_numerical_examples.tex
\section{Numerical examples}
\label{section:numerical_examples}

In this section, we use some numerical examples to illustrate the formalisms and techniques discussed above. We consider the Anderson model \cite{anderson1958pr}, implemented as a nearest-neighbor tight-binding model defined on a cubic lattice with lattice constant $a$ and dimension $N=N_x\times N_y\times N_z$. The Hamiltonian is written as
\begin{equation}
\hat{H} = \sum_{ij} (-\gamma) c_i^{\dagger} c_j + \sum_i U_i c_i^{\dagger} c_i,
\end{equation}
where $-\gamma$ is the hopping integral between neighboring sites and $U_i$ are the on-site potentials. The on-site potentials are uniformly distributed in the interval $[-W/2,W/2]$, where $W$ is called the Anderson disorder strength. Without loss of generality, we consider transport in the $x$ direction, which has periodic boundary conditions. The boundary conditions in the other directions will be chosen according to the specific application. Note that the two-fold spin degeneracy in this model is not included in the equations but is considered in the results shown in the relevant figures. 

\subsection{Formalisms to be compared}

\begin{table*}[htb]
\caption{Summary of the explicit formulas and computational costs of different LSQT approaches for calculating the dissipative electrical conductivity. $N$ is the Hamiltonian size. For the VAC-KPM and MSD-KPM methods, $M$ is the order of the Chebyshev polynomial expansion of the quantum projection operator, $t_{\rm max}$ is the maximum evolution time with $N_t$ intervals (not necessarily uniform), and $\alpha$ is a numerical factor of the order of 1. In the KG-CGPF method, $N_{e}$ is the number of energy points, $N_{\eta}$ is the number of energy resolution values chosen for each energy, and $\tau_{\phi}$ is the dephasing time.}
\begin{center}
\begin{tabular}{ c c c c }
\hline
\hline
LSQT method  & VAC-KPM & MSD-KPM & KG-CPGF    \\
\hline
Explicit formulas    &
Eqs.\ (\ref{equation:sigma_t_vac_t}), (\ref{equation:VAC_KPM}), (\ref{equation:VAC_moments})   & 
Eqs.\ (\ref{equation:sigma_t_msd_t}), (\ref{equation:MSD_KPM}), (\ref{equation:MSD_moments})   & 
Eqs.\ (\ref{equation:sigma_kg_single_energy})-(\ref{equation:sigma_kg_single_energy_R})  \\
Computational cost  & $\sim$$N(MN_t + 3\alpha\omega_0 t_{\rm max})$    & $\sim$$N(MN_t + 3\alpha\omega_0 t_{\rm max})$   & $\sim$$N(8 N_\text{e} N_{\eta} \omega_0 \tau_{\phi})$  \\ 
\hline
\end{tabular}
\end{center}
\label{table:lsqt}
\end{table*}

We compare three representations of the Kubo conductivity, including the VAC representation of Eq.\ (\ref{equation:sigma_t_vac_t}), the MSD representation of Eq.\ (\ref{equation:sigma_t_msd_t}), and the Kubo-Greenwood representation of Eq.\ (\ref{equation:sigma_KG_T0}). See Table \ref{table:lsqt} for a summary of the explicit formulas and the computational cost for each method. For the VAC and MSD representations, we use the KPM with Jackson damping for the quantum projection operator. The quantity to be calculated in the VAC representation is the product of the DOS and the VAC
\begin{align}
\label{equation:VAC_KPM}
\rho(E)C_\text{vv}(E, t) &= \frac{1}{\pi\Omega\Delta E\sqrt{1-\widetilde{E}^2}} \sum_{m=0}^{M} (2-\delta_{m0})\nonumber \\
&\times  g_m T_m(\widetilde{E})C^{\rm vac}_{m}(t),
\end{align}
where 
\begin{equation}
\label{equation:VAC_moments}
C^{\rm vac}_{m}(t) =\textmd{Re}\left[ \langle\phi_\text{L}^{\rm vac}(t) | T_m(\widetilde{H}) |\phi_\text{R}^{\rm vac}(t)\rangle\right]
\end{equation}
are the Chebyshev moments of $\rho(E)C_\text{vv}(E, t)$. 
The quantity to be calculated in the MSD representation is the product of the DOS and the MSD
\begin{align}
\label{equation:MSD_KPM}
\rho(E)\Delta X^2(E, t) &= \frac{1}{\pi\Omega\Delta E\sqrt{1-\widetilde{E}^2}} \sum_{m=0}^{M} (2-\delta_{m0}) \nonumber \\
&\times  g_m T_m(\widetilde{E})C^{\rm msd}_{m}(t),
\end{align}
where 
\begin{equation}
\label{equation:MSD_moments}
C^{\rm msd}_{m}(t) = \langle\phi_\text{L}^{\rm msd}(t) | T_m(\widetilde{H}) |\phi_\text{R}^{\rm msd}(t)\rangle
\end{equation}
are the Chebyshev moments of $\rho(E)\Delta X^2(E, t)$. The vectors $|\phi_\text{L,R}^{\rm vac}(t)\rangle$ and $|\phi_\text{L,R}^{\rm msd}(t)\rangle$ are defined in Eqs.\ \eqref{equation:phi_vac} and \eqref{equation:phi_msd}, respectively. We call these the VAC-KPM and MSD-KPM methods.

For the Kubo-Greenwood formalism, we consider a numerical implementation based on the Chebyshev polynomial expansion of the Green's function according to Eq.\ (\ref{equation:CPGF-1}), which we call the KG-CPGF method \cite{ferreira2015prl}. Following Ferreira and Mucciolo, we change \emph{both} of the Dirac delta functions in the Kubo-Greenwood formula to the regularized Green's function and rewrite the Kubo-Greenwood conductivity in Eq.\ (\ref{equation:trace_KG}) as
\begin{equation}
\label{equation:KG_Green}
\sigma(E,\eta) = \frac{\hbar e^2}{\pi\Omega} \bra{\phi} V \text{Im} [G^{+}(E)] V \text{Im} [G^{+}(E)] \ket{\phi}.
\end{equation}
Here we have highlighted the $\eta$-dependence ($\eta=\hbar/\tau_{\phi}$) of the conductivity.
Using the Chebyshev expansion of the Green's function in Eq.\ (\ref{equation:CPGF-1}), we then have
\begin{equation}
\label{equation:KG_CPGF}
\sigma(E,\eta) = \frac{\hbar e^2}{\pi\Omega(\Delta E)^2} \sum_{m=0}^{M} \sum_{n=0}^M \text{Im} [\bar{G}_m^{+}(z)] \text{Im} [\bar{G}_n^{+}(z)] C^{\rm kg}_{mn} 
\end{equation}
where $\bar{G}_m^{+}(z)$ is given in Eq.\ (\ref{equation:CPGF}) and
\begin{equation}
\label{equation:KG_moments}
C^{\rm kg}_{mn} = \bra{\phi} VT_m(\widetilde{H})VT_n(\widetilde{H}) \ket{\phi}.
\end{equation}
An efficient algorithm for evaluating this conductivity at a single energy has also been derived \cite{ferreira2015prl}:
\begin{equation}
\label{equation:sigma_kg_single_energy}
\sigma(E,\eta) = \frac{\hbar e^2}{\pi\Omega(\Delta E)^2}
\bra{\phi_\text{L}}\ket{\phi_\text{R}},
\end{equation}
\begin{equation}
\label{equation:sigma_kg_single_energy_L}
\ket{\phi_\text{L}} = \sum_{m=0}^{M}\text{Im} [g_m(z)] T_m(\widetilde{H}) \hat{V}\ket{\phi},
\end{equation}
\begin{equation}
\label{equation:sigma_kg_single_energy_R}
\ket{\phi_\text{R}} = \sum_{n=0}^M  \text{Im} [g_n(z)] \hat{V} T_n(\widetilde{H}) \ket{\phi}.
\end{equation}

In addition to algorithmic improvements, increasing computing power has played an important role in advancing quantum transport simulations. If computational nodes with a large amount of memory ($\geq$ 16 GB RAM) are available, high-resolution spectral calculations of DOS and DC conductivity can be carried out in giant systems with $N$ exceeding $10^9$ \cite{ferreira2015prl}. This RAM-intensive approach inspired the recent open-source KITE initiative ({\it https://quantum-kite.com/}) for accurate real-space calculations of electronic structure and quantum transport with multibillions
of orbitals \cite{Joao2019}. Another implementation combining KPM and this RAM-intensive approach is available in the KWANT simulation package ({\it https://kwant-project.org/}). Finally, all algorithms presented in this review will be distributed open-source within the C++ package TB-TK\\ ({\it http://second-tech.com/wordpress/index.php/tbtk/}).

In massively parallel computing, the use of graphics processing units (GPUs) is playing an increasingly important role in various simulation methods used in computational physics \cite{harju2013lncs}, and the linear scaling techniques presented in this review perform particularly well on the GPU \cite{fan2014cpc,garcia2015prl}. The LSQT calculations shown in this subsection were obtained by using an open source code named GPUQT \cite{fan2018cpc}, which is fully implemented on the GPU. Moreover, a pedagogical Python implementation of the VAC-KPM and MSD-KPM methods using a Jupyter notebook is also available ({\it https://github.com/brucefan1983/LSQT-Jupyter}).

Finally, in this section we compare the above LSQT methods with the LB method \cite{datta1995,ferry1997book}, when appropriate.

\subsection{Ballistic regime}

As discussed in Sec.\ \ref{section:transport_regimes}, the VAC and MSD formalisms capture the essential physics of ballistic transport. To illustrate this, we consider a narrow ribbon with $N_y=2$ and $N_z=1$, and hard-wall boundary conditions in the $y$ and $z$ directions. To achieve high accuracy in the random vector approximation, we set $N_x=5\times10^6$ in the transport direction and average the results over $N_r=10$ random vectors. The total number of tight binding orbitals is thus $N=N_xN_yN_z=10^7$. We use the KPM with the Jackson kernel and $M=3000$.

\begin{figure}[htb]
\begin{center}
\includegraphics[width=\columnwidth]{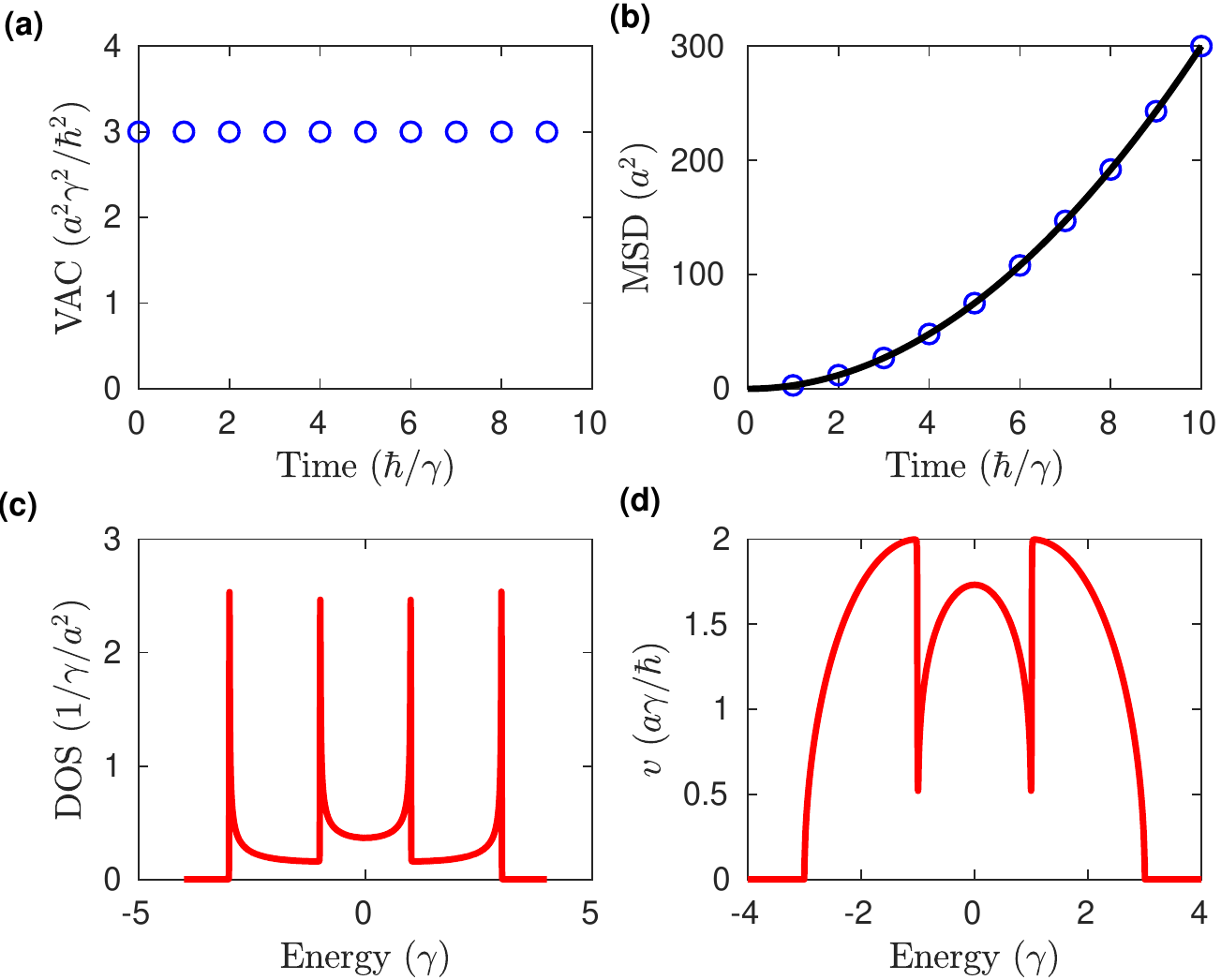}
\caption{Electronic and transport properties of a disorder-free 2D square lattice ($N_z=1$) ribbon with width $N_y=2$ and periodic length $N_x=5\times10^6$ in the transport direction. (a) VAC and (b) MSD at $E=0$ as a function of correlation time. (c) DOS and (d) group velocity as a function of Fermi energy.}
\label{figure:example_ballistic_2a}
\end{center}
\end{figure}

\begin{figure}[htb]
\begin{center}
\includegraphics[width=\columnwidth]{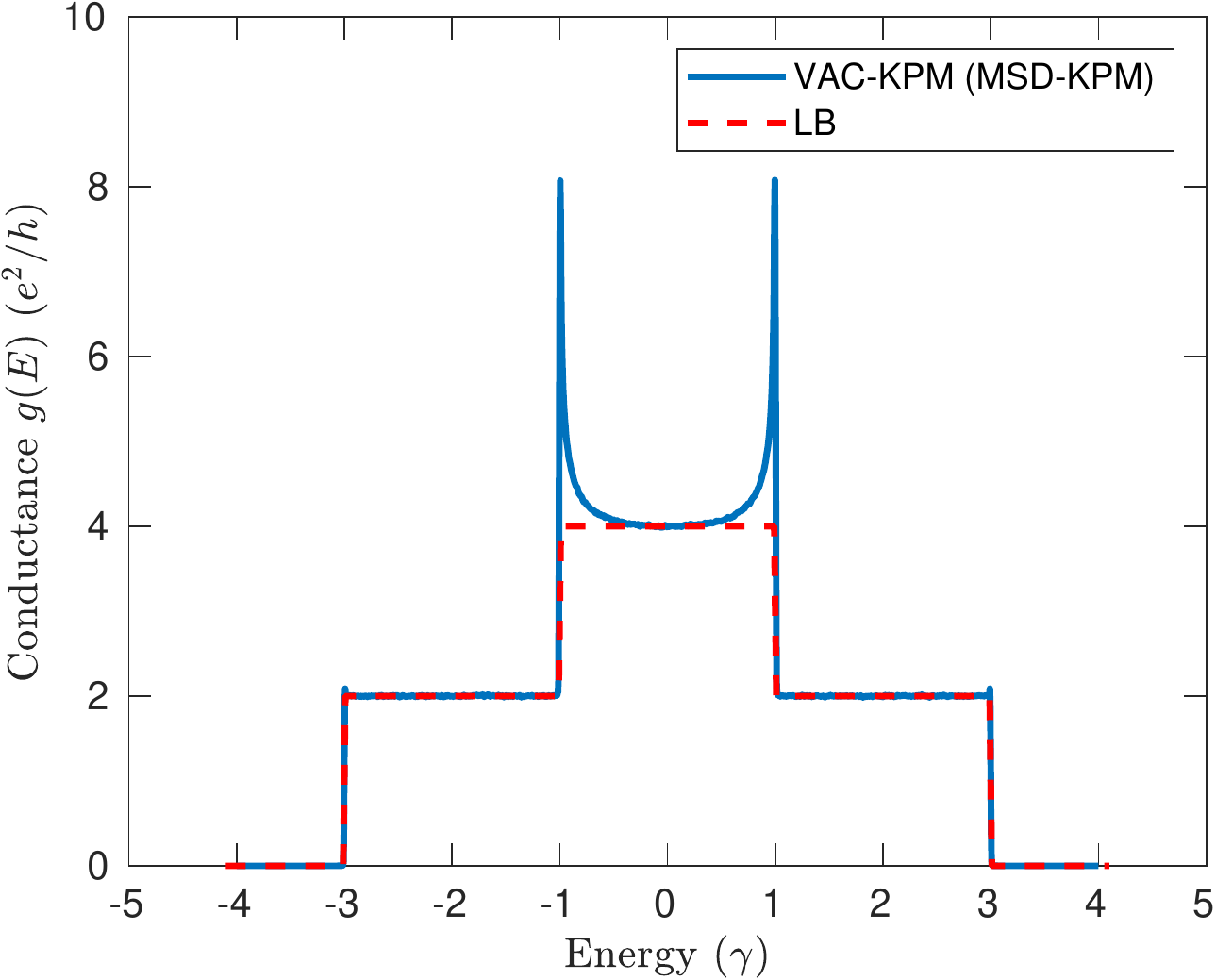}
\caption{Ballistic conductance in a disorder-free 2D square lattice ($N_z=1$) ribbon with width $N_y=2$ and periodic length $N_x = 5 \times 10^6$ in the transport direction, obtained by using the VAC-KPM (or equivalently MSD-KPM) method (solid line) and the LB method (dashed line).}
\label{figure:example_ballistic_2b}
\end{center}
\end{figure}

The VAC at the band center $E=0$, calculated from Eq.\ (\ref{equation:VAC_KPM}) and dividing by the DOS $\rho(E=0)$, is a constant in time with $v_{\rm F}^2=3a^2\gamma^2/\hbar^2$, as shown in Fig.\ \ref{figure:example_ballistic_2a}(a). Accordingly, the MSD in Fig.\ \ref{figure:example_ballistic_2a}(b), calculated from Eq.\ (\ref{equation:MSD_KPM}) and dividing by $\rho(E=0)$, is a quadratic function of the correlation time, $\Delta X^2(E, t)=v_{\rm F}^2(E)t^2$. In other words, the electrons are propagating at a constant velocity without scattering. The DOS $\rho(E)$ is shown Fig.\ \ref{figure:example_ballistic_2a}(c), and the group velocity $v_{\rm F}(E)$, which is the square root of the VAC at $t=0$, is shown in Fig.\ \ref{figure:example_ballistic_2a}(d). From these we can calculate the ballistic conductance according to Eq.\ (\ref{equation:ballistic_conductance}), as given by the solid line in Fig.\ \ref{figure:example_ballistic_2b}. For comparison, we also show the conductance calculated with the LB method, which is represented by the dashed line. The VAC-KPM and MSD-KPM methods clearly produce the correct conductance plateaus. Around the Van Hove singularity points at $E=\pm \gamma$, however, these methods overshoot the conductance plateau, as has been noticed in a variety of studies \cite{charlier2007rmp, markussen2006prb, fan2014cpc}. The overshooting originates from a mixing of the densities of states from different bands around the band edges, which results in an overestimation of the group velocity, as clearly demonstrated by Markussen \textit{et al} \cite{markussen2006prb}. When the system contains some disorder, deviating the conduction regime from purely ballistic motion, the MSD formalism becomes extremely suitable for calculating length-dependent conductance, as largely illustrated in applications to carbon nanotubes \cite{roche2001prl,roche2001prb,charlier2007rmp}. The KG-CPGF method has so far not been used in the ballistic regime.

\subsection{Diffusive regime}
\label{sec:diffusive}

We next consider a disordered system and closely compare the different LSQT methods as well as the LB method in the ballistic-to-diffusive crossover regime. As a generic case, and to make the computation feasible for the LB method, we consider a 2D square lattice ($N_z=1$) with width $N_y=50$ and an Anderson disorder strength $W=\gamma$. In the LSQT calculations, $N_x=2\times10^5$, $N_r=10$, and $M=3000$. In the LB method, we increase the system length from $L=a$ to $100a$ and calculate the conductance $g(E,L)$ iteratively. We average over $100$ disorder realizations in the LB calculations.

\begin{figure}[htb]
\begin{center}
\includegraphics[width=\columnwidth]{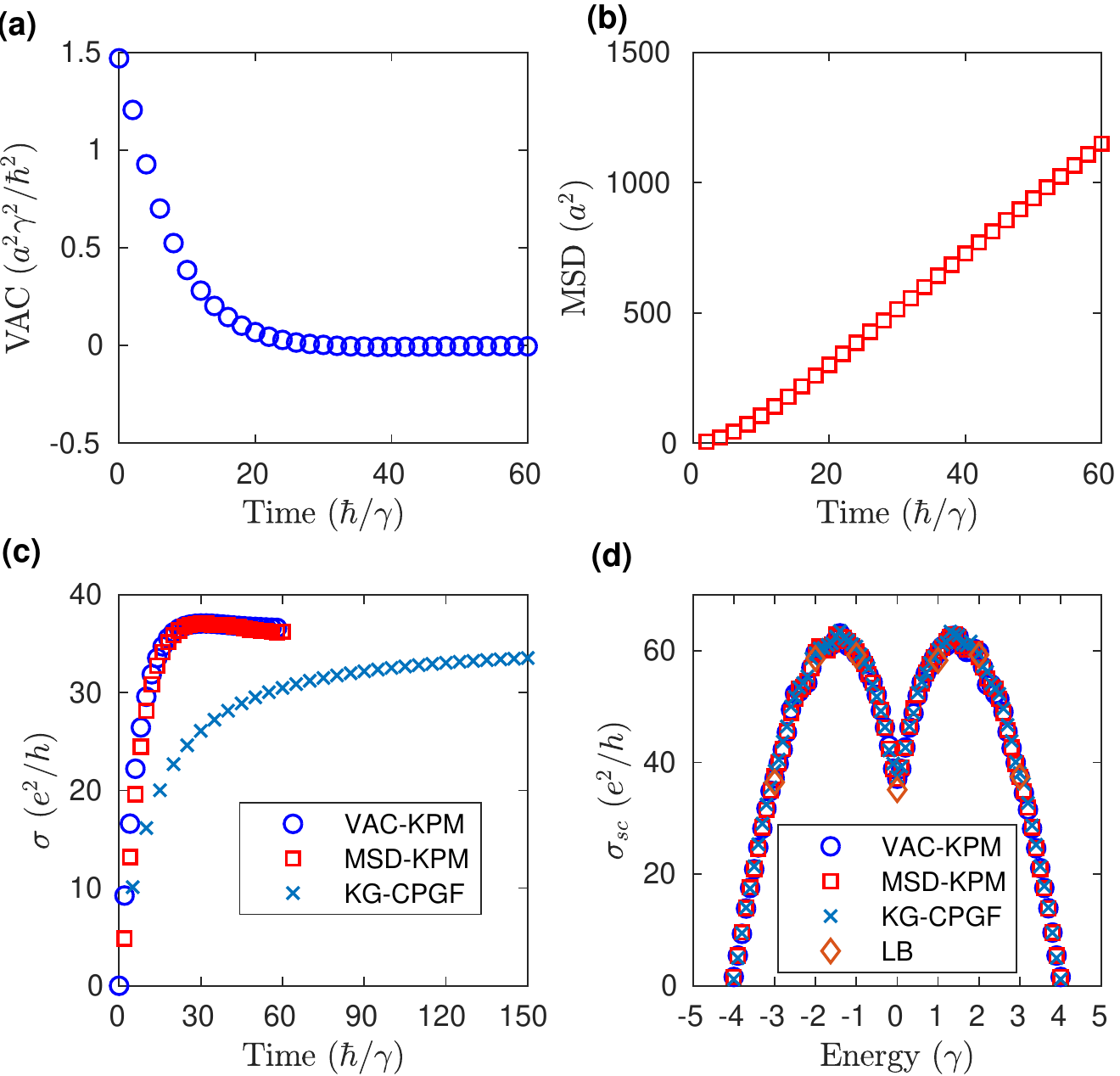}
\caption{Diffusive transport at the $E=0$ energy point in a 2D square lattice ($N_z=1$) with width $N_y=50$, periodic length $N_x=2\times 10^5$ in the transport direction, and disorder strength $W=\gamma$. (a) VAC, (b) MSD, and (c) time-dependent electrical conductivity as a function of the evolution time $t$ for the VAC-KPM and MSD-KPM methods or the dephasing time $\tau_{\phi} = \hbar/\eta$ for the KG-CPGF method. (d) Semiclassical conductivity $\sigma_{\rm sc}$ as a function of energy $E$ from the various methods. Because the LB method is not parallel in energy, only a few energy points were considered.}
\label{figure:example_diffusive_a}
\end{center}
\end{figure}

Results for the $E=0$ energy point are shown in Fig.\ \ref{figure:example_diffusive_a}. As expected, the VAC decays exponentially with increasing evolution time $t$ [Fig.\ \ref{figure:example_diffusive_a}(a)] and the MSD changes from a quadratic to a linear function of $t$ [Fig.\ \ref{figure:example_diffusive_a}(b)]. The time-dependent electrical conductivities calculated from the VAC (\textit{via} a time integration) and the MSD (\textit{via} a time derivative) are equivalent, as can be seen in Fig.\ \ref{figure:example_diffusive_a}(c).

Figure \ref{figure:example_diffusive_a}(c) also shows the evolution of the electrical conductivity calculated using the KG-CPGF method as a function of the dephasing time $\tau_{\phi}=\hbar/\eta$. As we have discussed in the previous section, the dephasing time $\tau_{\phi}$ and the evolution time $t$ are associated with different regularizations of the Green's function, resulting in different time dependence, as shown previously in Fig.\ \ref{figure:transport_regimes}(a) and here in Fig.\ \ref{figure:example_diffusive_a}(c). However, they lead to the same diffusion constant and semiclassical conductivity $\sigma_{\rm sc}(E)$ in the diffusive regime. This is shown in Fig.\ \ref{figure:example_diffusive_a}(d). Here, $\sigma_{\rm sc}(E)$ is taken as the maximum value of the scale-dependent electrical conductivity, which is the conductivity attained in the system before coherent backscattering and quantum interference come into play. In the LB method, we calculate the semiclassical conductivity based on the ballistic-to-diffusive transition formula \cite{datta1995}
\begin{equation}
\frac{1}{g(E,L)} = \frac{1}{\sigma_{\rm sc}(E)} \frac{L}{N_y a} + \frac{1}{g_0(E)},
\end{equation}
where $g_0(E)$ is the ballistic conductance at energy $E$. As shown in Fig.\ \ref{figure:example_diffusive_a}(d), this coincides with each of the other polynomial expansion-based methods.

\subsection{Weak localization regime}
\label{section:wl}

We next analyze the behavior of the conductivity scaling in the 2D square lattice beyond the diffusive regime, when disorder is strong enough to induce weak localization effects. Standard weak localization theory predicts that the conductivity will decay logarithmically with system size \cite{abrahams1979prl},
\begin{equation} \label{eq_wl}
\sigma(L) = \sigma_\text{sc} - \frac{2}{\pi} \frac{e^2}{h} \ln \left( \frac{L}{l_\text{e}} \right),
\end{equation}
where $\sigma_\text{sc}$ is the semiclassical conductivity, $L$ is the system size, and $l_\text{e}$ is the mean free path. In the derivation of Eq.\ (\ref{eq_wl}), $l_\text{e}^{-1}$ is typically chosen as an upper cutoff in the momentum integration, because $l_\text{e}$ is defined as the length when diffusive behavior sets in. However, it is generally known that several mean free paths are needed before a system leaves the quasiballistic regime and enters the fully diffusive regime. Thus, in numerical simulations, $l_\text{e}$ should be practically replaced by a more general ``semiclassical length'' $l_\text{sc}$,
\begin{equation} \label{eq_wl_new}
\sigma(L) = \sigma_\text{sc} - \frac{2}{\pi} \frac{e^2}{h} \ln \left( \frac{L}{l_\text{sc}} \right).
\end{equation}

\begin{figure}[htb]
\begin{center}
\includegraphics[width=\columnwidth]{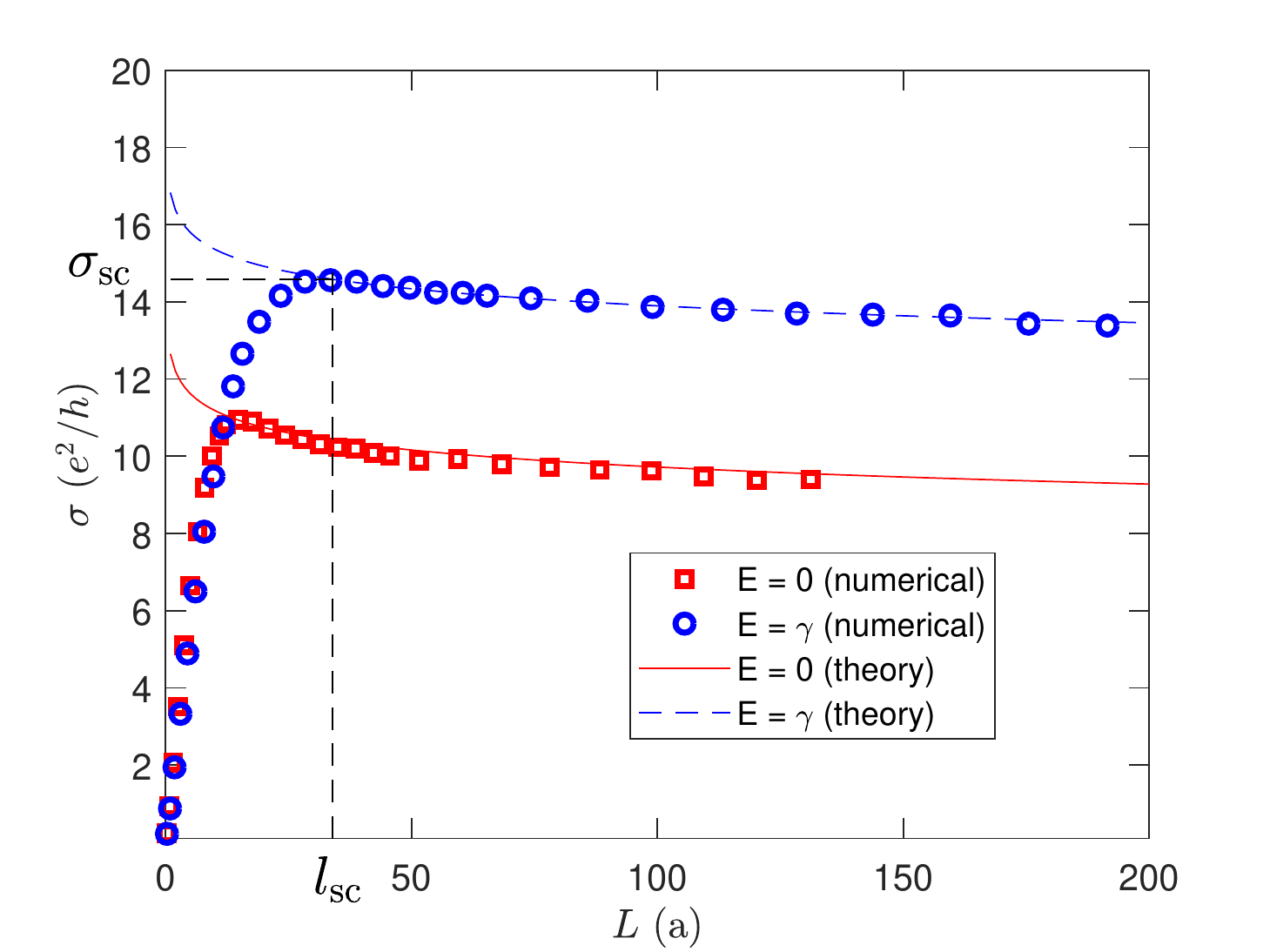}
\caption{Conductivity $\sigma(E,L)$ as a function of length $L$ for $E=0$ and $E=\gamma$ in a 2D square lattice ($N_z=1$) with width $N_y=50$, periodic length $N_x=2\times 10^5$ in the transport direction, and disorder strength $W=\gamma$. The markers are numerical calculations with the MSD-KPM method ($N_r=6$ and $M=3000$) and the solid and dashed lines represent the predictions from weak localization theory, Eq. (\ref{eq_wl_new}), without any fitting parameters. The black dashed lines show the definitions of $\sigma_\text{sc}$ and $l_\text{sc}$ for $E = \gamma$.}
\label{figure:example_wlocalized}
\end{center}
\end{figure}

By definition, $l_\text{sc}$ is the length at which the time-dependent conductivity reaches its maximum value $\sigma_\text{sc}$, which identifies the crossover from the diffusive to the localized regime. As seen in Fig.\ \ref{figure:example_wlocalized}, the decay of the conductivity is perfectly described by Eq.\ (\ref{eq_wl_new}), without using any fitting parameters. As the system length $L$ increases, the quantum correction will eventually grow to the order of the semiclassical conductivity; at this point the electronic system will undergo a transition to the strong (Anderson) localization regime, discussed in the next section.

\subsection{Strong localization regime}

It is well known that any amount of disorder is sufficient to localize electrons in the low-dimensional Anderson model \cite{anderson1958pr}, so that for large enough sample size all electronic states will be exponentially localized. We study the transition to this regime by considering a square lattice with the same sample size as before ($N_x=2 \times 10^5$, $N_y=50$, $N_z=1$), but with a much larger disorder strength, $W=5\gamma$. This permits the simulations to enter the strong localization regime within the chosen system size.

Among the different LSQT methods, only the MSD-KPM method has been compared quantitatively to the LB method in the localized regime. Here we choose $N_r=10$ and $M=3000$ in the MSD-KPM method and convert the computed conductivity to conductance using the standard definition given by Eq.\ (\ref{equation:sigma_to_g}). The length $L$ is calculated using Eq.\ (\ref{equation:L}). In the LB method, we average over $5000$ disorder realizations to obtain the typical conductance \cite{anderson1980}
\begin{equation}
g_{\rm typ}(E,L) = \exp[\langle \ln g(E,L)\rangle].
\end{equation}

\begin{figure}[htb]
\begin{center}
\includegraphics[width=\columnwidth]{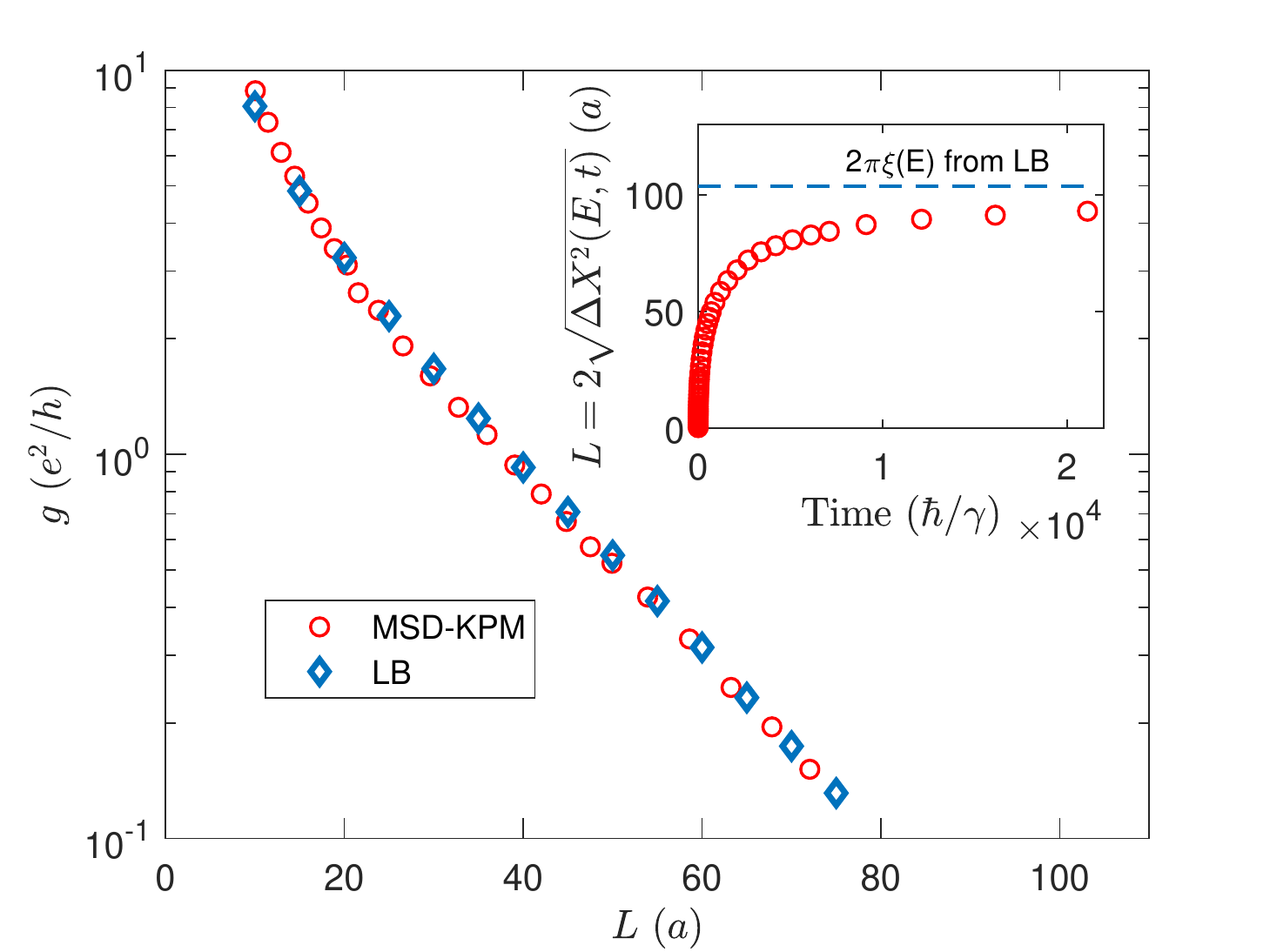}
\caption{(Main panel) Conductance as a function of length for $E=0$ in a 2D square lattice ($N_z=1$) with width $N_y=50$, length $N_x=2 \times 10^5$ in the transport direction, and disorder strength $W=5\gamma$. In the MSD-KPM method, the conductance is determined from the conductivity using Eq.\ (\ref{equation:sigma_to_g}). (Inset) Propagation length $L$ defined in Eq.\ (\ref{equation:L}) as a function of the evolution time in the MSD-KPM method. The horizontal dashed line indicates the value $2\pi\xi(E)$, where $\xi(E)$ is calculated from the LB conductance using Eq.\ (\ref{equation:xi_LB}). The propagation length approaches $2\pi\xi(E)$ in the long-time limit, in accordance with Eq.\ (\ref{equation:xi_from_MSD}).}
\label{figure:example_localized}
\end{center}
\end{figure}

Figure \ref{figure:example_localized} shows the conductance at $E=0$, which decays exponentially in the large length limit for both the MSD-KPM and the LB formalism. This provides a definition of the localization length $\xi(E)$,
\begin{equation}
g_{\rm typ}(E,L) \propto \exp[-L/\xi(E)].
\label{equation:xi_LB}
\end{equation}
The localization length at $E=0$ is fitted to be $\xi \approx 16a$. The conductance calculated from the MSD is equivalent to the LB conductance down to $g \approx 0.1e^2/h$, highlighting this method's ability to capture the strong localization regime.  As shown in the inset of Fig.\ \ref{figure:example_localized}, the propagation length defined in Eq.\ (\ref{equation:L}) approaches $2\pi\xi(E)$ at long times, as expected from Eq.\ (\ref{equation:xi_from_MSD}). This definition has been discussed in Refs.\ \cite{uppstu2014prb,fan2014prb} and shown to be equivalent to that given by Eq.\ (\ref{equation:xi_LB}). Note that such saturation of the propagation length is expected from the absence of diffusion in the Anderson localization regime \cite{anderson1958pr}.

In principle, the VAC-KPM method can also be used in the localized regime. However, it is less practical than the MSD-KPM method because the time integration in the VAC-KPM method requires small time intervals and thus a large number of time steps $N_t$. In contrast, the time derivative in the MSD-KPM method allows the use of large time intervals in the localized regime. The method by Yuan \textit{et al.}\ \cite{yuan2010prb, yuan2010prb_b, zhao2015prb} is based on the VAC formalism and the Fourier transform method for approximating both the quantum resolution operator and the time evolution operator. Therefore, a fixed time step of $\Delta t = \pi/\omega_0$ (determined by the Nyquist sampling theorem) is chosen together with a certain value of $N_t$. However, we note that using a fixed $N_t$ for the whole spectrum might be insufficient for a quantitative study of quantum transport when different energy states exhibit different transport timescales. Finally, we note that a quantitative extraction of the localization length has never been carried out with the KG-CPGF formalism.

\subsection{Convergence, computational cost, and method comparison}

In this section we delve more deeply into some of the practical issues that are faced when using the LSQT methods described above. In particular, we discuss aspects related to the numerical convergence and error associated with these methods. We then address a couple of situations where the time-dependent methods give spurious results in the long-time limit, and give guidelines for avoiding these pitfalls. Finally, we give an analysis of the computational cost of each method, and then we briefly summarize and recap the similarities and differences between the various LSQT methods presented in this review.

\subsubsection{Convergence with the number of moments}

In real-space numerical simulations of bulk materials, periodic boundary conditions are often used to avoid the effect of edges. However, independent of the boundary conditions, all numerical simulations involve systems with a finite number of atoms, meaning the Hamiltonian has a discrete set of eigenvalues at a discrete set of momenta. For example, in a 2D lattice with $\sqrt{N} \times \sqrt{N}$ orbitals, the separation in momentum space of two consecutive states is proportional to $1/\sqrt{N}$. For systems with a linear energy dispersion near the band center, such as graphene, the energy spacing is also proportional to $1/\sqrt{N}$, while in systems with a parabolic dispersion it is proportional to $1/N$. Therefore, when running numerical simulations, one should always be aware of the energy resolution of the numerical method with respect to the energy level spacing of the finite system.

In all the discussed methods based on the KPM, the energy resolution $\delta E$ is inversely proportional to the number of moments $M$, $\delta E = \alpha \Delta E / M$, where $\Delta E$ is the half bandwidth of the energy spectrum and $\alpha \approx 3-5$ is a factor depending on the choice of kernel (see Sec.\ \ref{subsection:kpm}). Meanwhile, in the Lanczos and KG-CPGF methods, the resolution is fixed by a user-defined parameter $\eta$, and a sufficient number of moments is then used to reach this level of broadening. For the KG-CPGF method this is reached for $M \approx 4 \Delta E / \eta$, with $\eta = \hbar / \tau_\phi$.

\begin{figure}[htb]
\begin{center}
\includegraphics[width=\columnwidth]{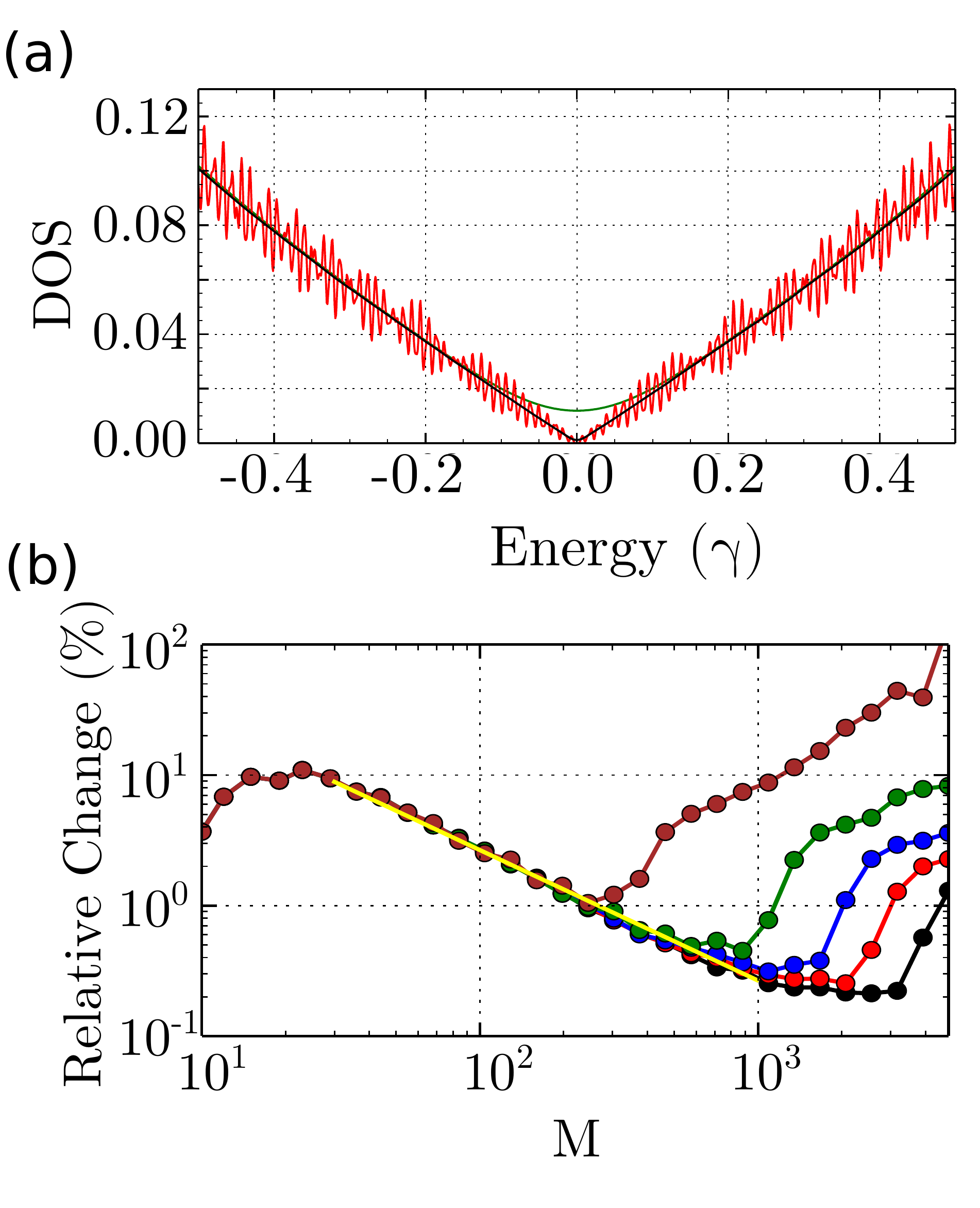}
\caption{(a) Density of states of the tight-binding model of graphene with $2 \times N_1 \times N_2 = 2 \times 10^6$ atoms, approximated by the KPM with the Jackson kernel for different energy resolutions of $\delta E = 0.19\gamma_0$ (green line), $0.019\gamma_0$ (black line), and $0.005\gamma_0$ (red line). (b) Maximum relative difference between two consecutive approximations of the DOS as a function of the number of moments for different system sizes $N_1 = N_2$ = 100 (brown line), 300 (green line), 500 (blue line), 700 (red line), and 1000 (black line).}
\label{figure:convergence_kpm}
\end{center}
\end{figure}

When simulating the properties of bulk materials in a finite system, it is important that the energy resolution of the numerical method be larger than the discrete energy level spacing in order to avoid spurious effects. An example of this is shown in Fig.\ \ref{figure:convergence_kpm}(a), where we calculate the DOS of graphene using the KPM with the Jackson kernel and different numbers of moments. This system has $2 \times 10^6$ atoms with a discrete energy spacing of $0.01\gamma_0$ near $E=0$, where $\gamma_0$ is the hopping energy between nearest-neighbor carbon atoms. The DOS calculated with a relatively large energy resolution, $\delta E = 0.19 \gamma_0$ is linear at large energies but becomes parabolic around $E=0$, which is a consequence of the broadening of states at positive and negative energies. For the finest resolution of $0.005 \gamma_0$, the discrete energy level spacing is resolved, resulting in a strongly oscillatory DOS. Finally, for an intermediate resolution of $0.019\gamma_0$, twice the discrete level spacing, the system shows the expected linear dispersion for a broad range of energies and minimal broadening at $E=0$.

This intermediate value indicates the existence of an optimal resolution. This is better illustrated in Fig.\ \ref{figure:convergence_kpm}(b), where we plot the quantity 
\begin{equation}
\Delta\rho_M = \max\limits_E \left| \frac{ \rho_{M+1}(E) - \rho_{M}(E) }{ \rho_{M+1}(E) } \right|,
\end{equation}
which measures the relative difference between the DOS calculated with two successive numbers of moments $M$. The maximum is taken over the energy range depicted in Fig.\ \ref{figure:convergence_kpm}(a), and we consider this quantity for systems with different numbers of carbon atoms. The relative difference initially decreases with increasing $M$, which is a consequence of the reduced broadening around $E=0$. Eventually, when $M$ is large enough, $\Delta\rho_M$ exhibits a sharp increase owing to the resolution of the discrete energy levels in the system. The optimal value of $M$, corresponding to the minimum value of $\Delta\rho_M$, increases linearly with $\sqrt{N}$ as expected from the discussion above. In general, the optimal value of $M$ depends on the energy spectrum of the system being studied, and the procedure outlined in Fig.\ \ref{figure:convergence_kpm} illustrates how this may be determined.

\subsubsection{Convergence of the stochastic trace approximation}

In Sec.\ \ref{section:trace}, we discussed a powerful tool for evaluating the trace of large matrices. Here we will discuss the error introduced by this approximation and its convergence to zero. We will focus only on the random phase implementation, where the elements of the random vector in Eq.\ \eqref{equation:rp_state} are given by $\xi_{rn} = {\rm e}^{i \phi_{rn}}$, with $\phi_{rn}$ a random number homogeneously distributed between 0 and $2\pi$, but the analysis can be extended to any random vector satisfying the conditions given in section \ref{section:trace}. Let us compute explicitly the right-hand side of Eq.\ (\ref{equation:trace_A}) for this choice of random vectors,
\begin{equation}
\frac{1}{N_{r}}\sum_{r=1}^{N_r}\bra{\phi_r}\hat{A}\ket{\phi_r} = \frac{1}{N}{\rm{Tr}}[ \hat{A}] +\Delta E_r,
\end{equation}
where 
\begin{equation}
\Delta E_r = \frac{1}{N}\sum_{i \neq j=1}^N A_{ij} \,\left(\frac{1}{ N_r}\sum_{r=1}^{N_r} \chi_{rij}\right)
\label{random_phase_errors}
\end{equation}
is the error of the approximation, with $\chi_{rij} \equiv {\rm e}^{i (\phi_{ri}-\phi_{rj})}$.  The first important result is that the error vanishes for diagonal matrices with this choice of random vectors. This result guarantees that one has optimal convergence when choosing the eigenvectors as the basis of the matrix. The quantity inside the parentheses is the average of a random variable $\chi_{rij}$, which has the probability distribution
\begin{equation}
\rho_\chi(x) = \frac{1}{\pi\sqrt{1-x^2}},\quad x\in(-1,1)
\end{equation}
for both its real and imaginary part. This distribution has vanishing mean, which guarantees the convergence of the random phase approximation upon averaging over enough random vectors. For a system with homogeneous off-site terms $A_{ij}= A_{\rm off}$, the error converges to zero as
 \begin{equation}
\Delta E_r = \frac{ A_{\rm off}}{N_r}.
\end{equation}
For a system with an arbitrary number of different off-site terms, the error becomes a sum of random variables with different weights, and converges to a Gaussian distribution via the central limit theorem. In this case, the error converges to zero as 
\begin{equation}
\Delta E_r =\frac{\sigma_{A}}{\sqrt{N N_r}},
\label{eq:conv_rp}
\end{equation}
where $N$ is the total number of orbitals, $N N_r$ is the number of observations of the random variable, and $\sigma_{A}$ is the standard deviation of these observations, which depends on the spectrum of the matrix.

\begin{figure}[htb]
\begin{center}
\includegraphics[width=\columnwidth]{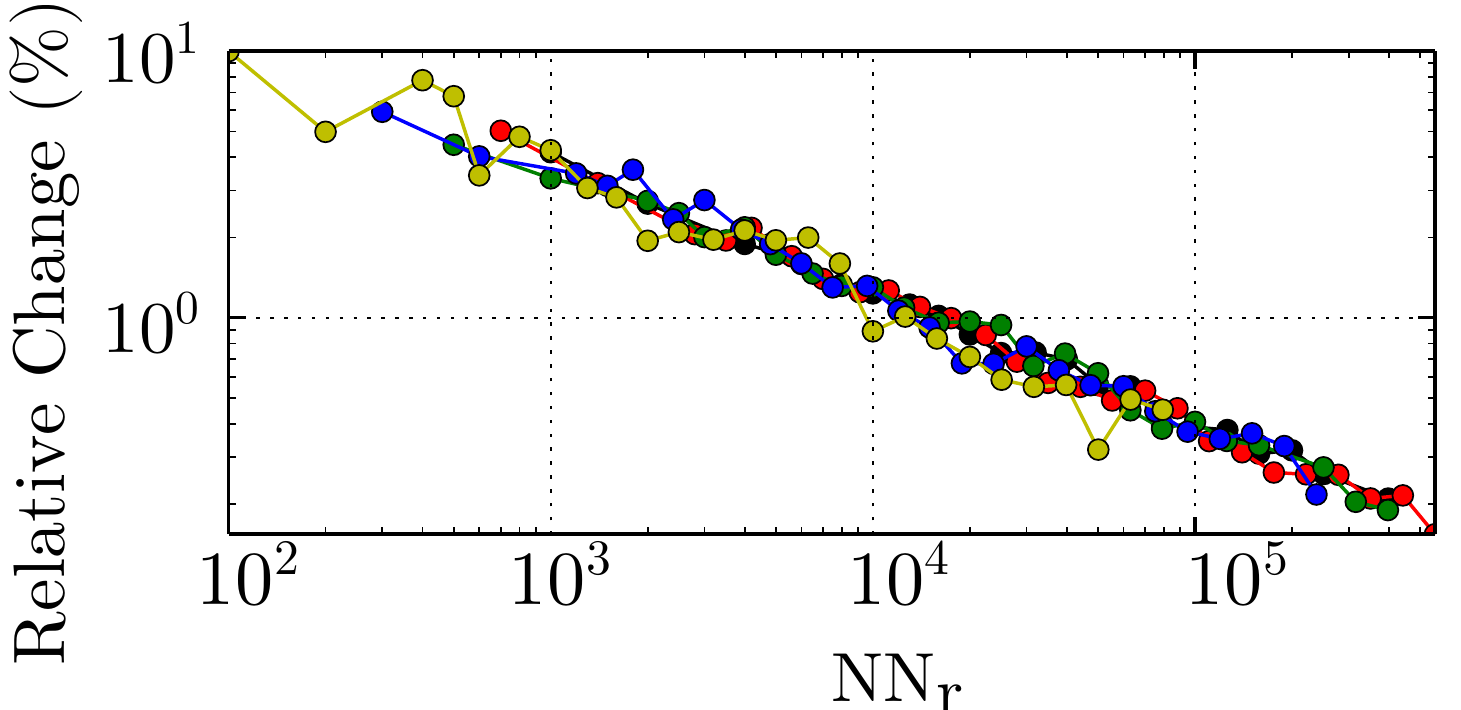}
\caption{Convergence of the graphene density of states with respect to the number of random vectors $N_r$ and the system size $N$ for a honeycomb lattice with different number of orbitals $N = 2 \times N_1 \times N_2$ with $N_1 = N_2 = 1000$ (black), $700$ (red), $500$ (blue), $300$ (green), and $100$ (yellow). All data follow the trend $\Delta \rho_R \propto 1/\sqrt{NN_r}$.}
\label{figure:convergence_randomphase}
\end{center}
\end{figure}

Let us recall that the traces appearing in the Kubo formalism consist of a product of several operators and Green's functions, which means that they are typically neither diagonal nor regular. Therefore, one expects a convergence of the traces following the central limit theorem, as in Eq.\ \eqref{eq:conv_rp}. In Fig.\ \ref{figure:convergence_randomphase} we show this for the graphene DOS, where we calculate the quantity
\begin{equation}
\Delta\rho_{N_r} = \max\limits_E \left| \frac{\rho_{N_r+1} - \rho_{N_r}}{\rho_{N_r+1}} \right|,
\end{equation}
which, similar to $\Delta \rho_M$ above, is the relative difference between the DOS calculated with two successive numbers of random phase states. As before, the maximum is taken over the energy range in Fig.\ \ref{figure:convergence_kpm}(a), and $\Delta \rho_{N_r}$ is calculated for different numbers of orbitals $N$. This quantity measures the convergence of the stochastic trace approximation with respect to the number of random vectors and decays as the inverse of $\sqrt{NN_r}$, as expected from the above discussion.

These results indicate that even in the simple case of the density of states, which consists only of the imaginary part of the Green's function, there are enough off-site elements to apply the central limit theorem. It is important to highlight that except for the inhomogeneity of the off-site elements, the discussion above made no emphasis on any specific feature of the matrix of which we are computing the trace, and the results presented here are applicable to any other transport quantity.

From a practical point of view, in typical simulations we choose the product $NN_r$ such that the error is of the order of $1\%$ of the trace. For the case of the conductivity in a system with $N \sim 10^7$, this convergence is sometimes achieved even with $N_r = 1$. Nevertheless, a convergence analysis should be performed for each system.

\subsubsection{Disorder and convergence}

Until this point, we have discussed the convergence of the DOS of a system in the absence of disorder, with well-defined energy levels. Now we illustrate the convergence of the electrical conductivity in the presence of disorder. When disorder is incorporated, the position of these energy levels is redistributed, with a different distribution for each realization of disorder. This means that averaging over disorder configurations will induce a broadening which is related to the disorder's self-energy, and can mask the discreteness of the spectrum. Under such conditions, the convergence of the conductivity with respect to the number of moments can occur more quickly than in clean systems, and can also exhibit saturation once the numerical broadening becomes smaller than the disorder-induced broadening. To illustrate this effect, we consider a 3D cubic lattice with periodic boundaries in all three directions. We set $N_x=250$, $N_y=N_z=200$ and consider transport in the $x$ direction. The Anderson disorder strength is chosen as $W = 2 \gamma$. As we have demonstrated the equivalence of all the LSQT methods for diffusive transport, we consider only the MSD-KPM with the Jackson kernel in this example.

\begin{figure}[htb]
\begin{center}
\includegraphics[width=\columnwidth]{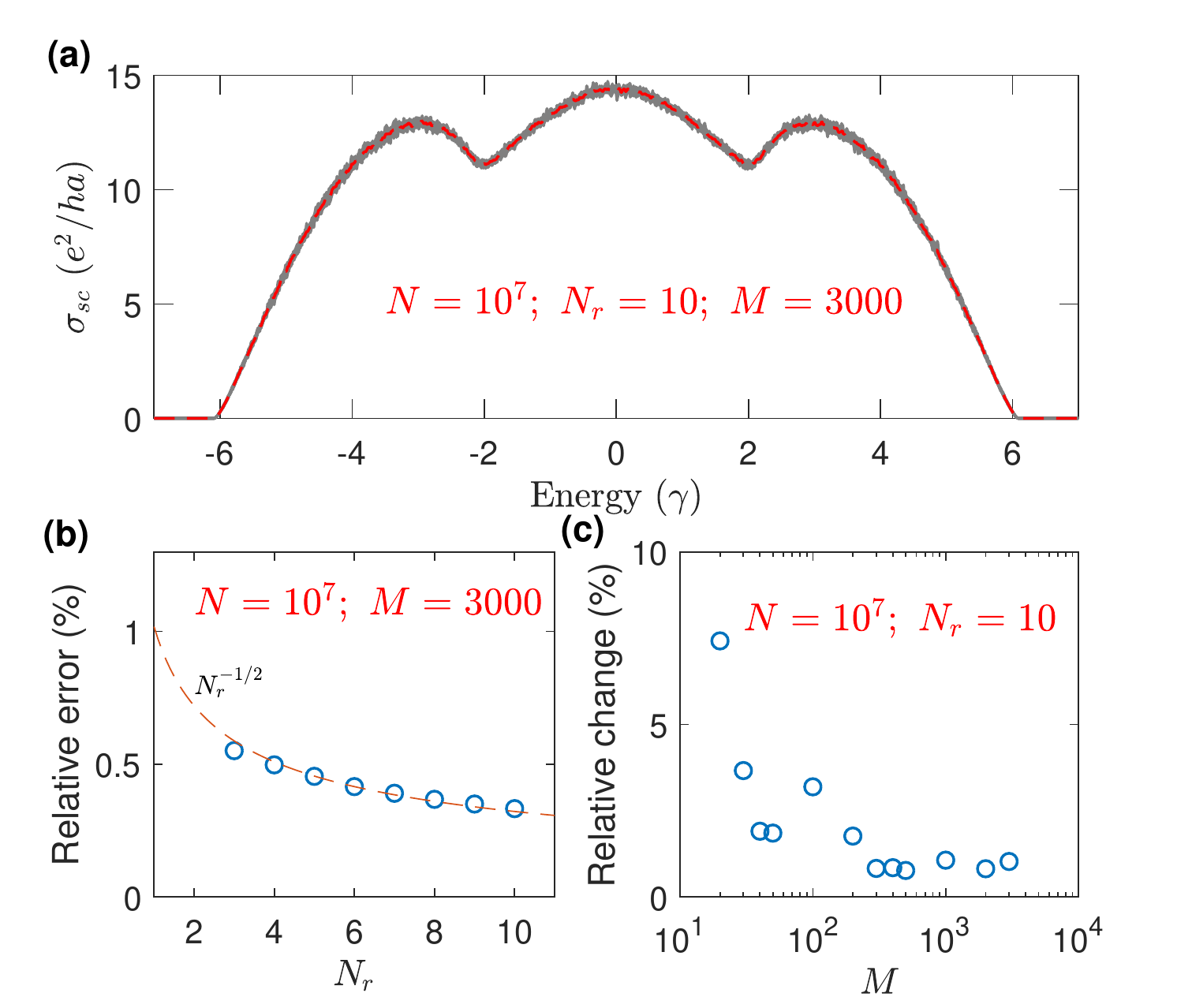}
\caption{(a) Semiclassical conductivity $\sigma_{\rm sc}$ for the whole energy spectrum in a disordered ($W=2\gamma$) 3D cubic lattice. The gray lines represent results from ten independent calculations (each with a different random vector) and the dashed line is their average. Here, the number of Chebyshev moments is $M=3000$. (b) Relative error in $\sigma_{\rm sc}$ (averaged over the energy spectrum) shown in (a) as a function of the number of random vectors $N_r$, which follows the expected scaling $\sim N_r^{-1/2}$. (c) Convergence of the relative change in $\sigma_{\rm sc}$ with increasing $M$. In all the calculations, the number of sites is $N=10^7$.}
\label{figure:example_cubic}
\end{center}
\end{figure}

Figure \ref{figure:example_cubic}(a) shows the semiclassical conductivity over the entire energy spectrum. We have performed $N_r=10$ independent simulations (with $M=3000$ moments), each with a different random vector and disorder configuration. These independent simulations are shown by the gray lines, while the dashed line shows the average. In this figure it is clear that the spread of results is quite small. This is quantified in Fig.\ \ref{figure:example_cubic}(b), where we show the relative error of the conductivity, averaged over the entire energy spectrum, as a function of the number of random vectors $N_r$. As in the previous section this error scales as $1/\sqrt{N_r}$. For this system size, $N=10^7$, even one random vector is sufficient to achieve an error of $\sim$$1\%$.

In Fig.\ \ref{figure:example_cubic}(c) we plot the relative change in the conductivity as a function of the number of moments $M$, calculated in the same way as in Fig.\ \ref{figure:convergence_kpm}(b) above. Here we see that the calculated conductivity converges rapidly with the increase of $M$, saturating when $M > 300$. This quick saturation, despite the large system size, is a consequence of the presence of disorder, as discussed above. In general, the convergence of the conductivity with $M$ depends on the material and the type of disorder, and this should be performed for each system being studied. For relatively ``smooth" systems such as this one, convergence can occur quickly. For systems showing singular behavior, such as for example graphene with vacancies, numerical convergence becomes a much more challenging task, as will discussed further in Sec.\ \ref{section:graphene_point}.

\subsubsection{Spurious effects in the long-time limit}
\label{sec:long_time_limit}
\begin{figure}[htb]
\begin{center}
\includegraphics[width=\columnwidth]{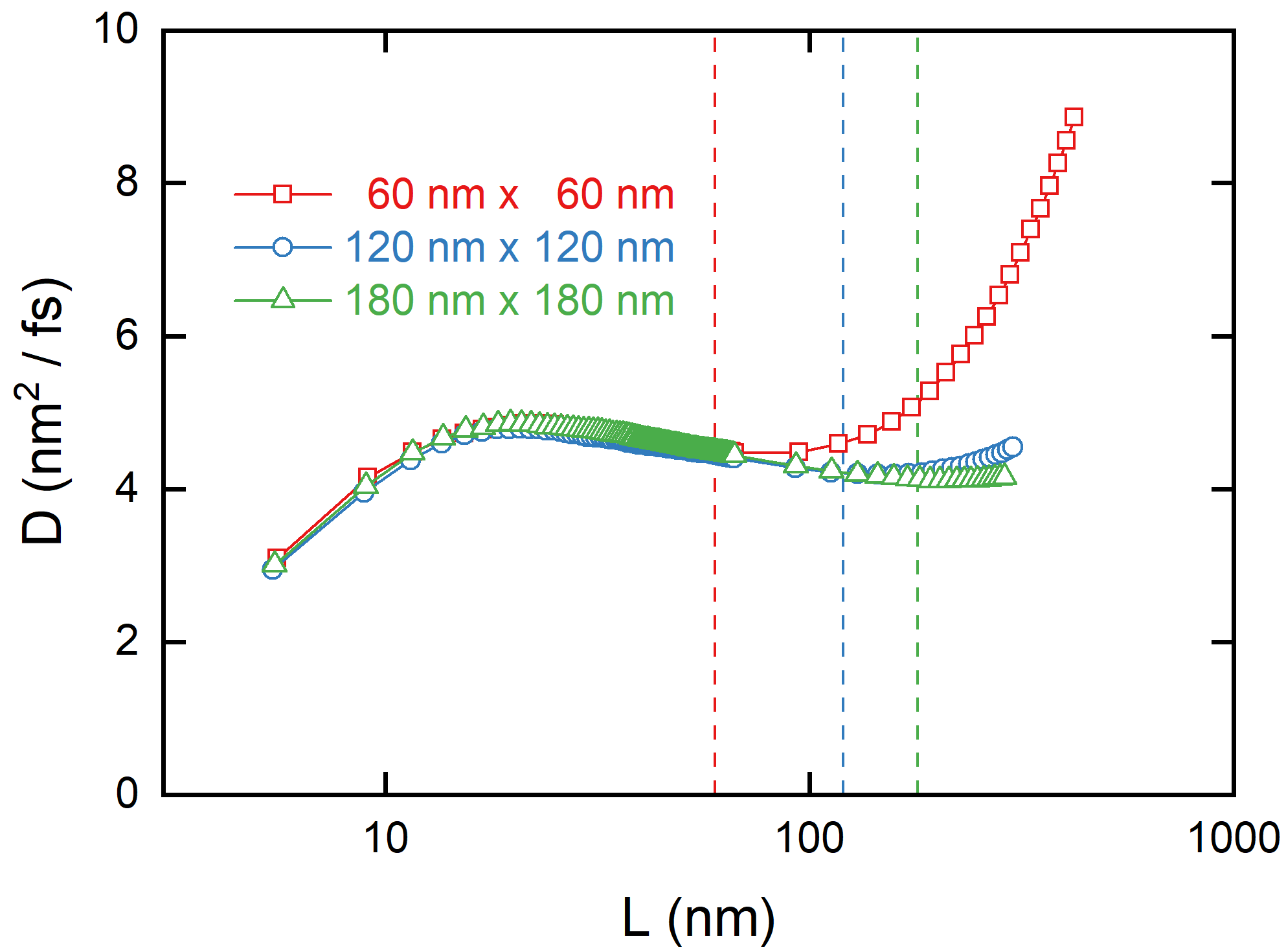}
\caption{Diffusion coefficient $D(E=0,L)$, as a function of transport length $L$, in three samples of polycrystalline graphene with different sizes. The dashed vertical lines indicate the point where $L$ equals the size of each sample, after which $D$ diverges. The samples contain $N = 1.4 \times 10^5$, $5.6 \times 10^5$, and $1.26 \times 10^6$ atoms, and $D$ was calculated using MSD-KPM with the Lorentz kernel and $M = 1000$ moments.}
\label{figure:polyG_DvsL}
\end{center}
\end{figure}

In this section we point out a couple of spurious effects that can arise in the time-dependent LSQT methods at long simulation times, and we discuss how to control such unwanted artifacts. As mentioned above, it is customary in the transport simulations to use periodic boundary conditions to avoid edge effects, hence enabling the calculation of bulk material properties. However, while this eliminates scattering off the edges, it permits the initial electronic state to propagate beyond the size of the simulation cell. When this happens, superperiodicity begins to play a role in transport, leading to a divergence of the conductivity. This phenomenon is illustrated in Fig.\ \ref{figure:polyG_DvsL}, where we plot the diffusivity $D(E,L)$ as a function of the propagation length $L$ for three polycrystalline graphene samples of different size. In each case, $D$ follows the expected transition through the ballistic, diffusive, and weakly-localized regimes. However, when the propagation length exceeds the sample size, the diffusivity diverges rapidly. The exact details of this divergence, such as when it begins and the rate of divergence, depend on the details of the system, but in general simulation results extracted from the time-dependent LSQT methods should only be considered valid when $L$ is less than the diameter of the simulation cell.

\begin{figure}[htb]
\begin{center}
\includegraphics[width=\columnwidth]{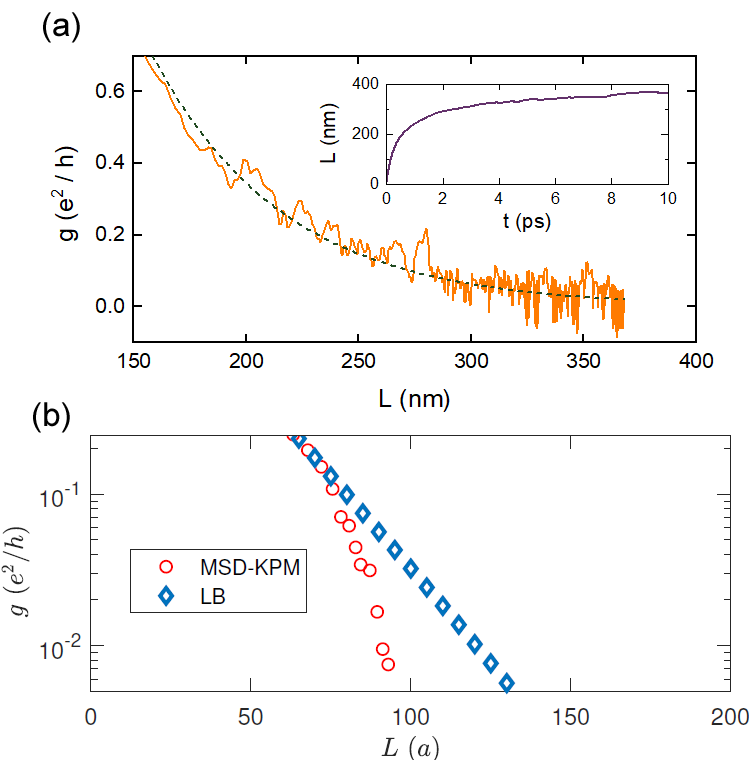}
\caption{(a) Length-dependent conductance at $E=0$ of an armchair graphene nanoribbbon with $N = 9.4 \times 10^6$ atoms, a width of 20.1 nm and a periodic length of 125 $\upmu$m, with Anderson disorder strength $W=4\gamma_0$, calculated with MSD-KPM using the Lorentz kernel and $M=3400$ polynomials. The inset shows the saturating propagation length $L$. A single random-phase state and disorder configuration have been used to highlight the noise present in each individual simulation. The dashed line shows the trend $g \propto \exp(-L/\xi)$, with $\xi$ given by Eq.\ \eqref{equation:xi_from_MSD}. (b) Extension of Fig.\ \ref{figure:example_localized} to smaller conductances, highlighting the transition to superexponential behavior as $L \rightarrow 2\pi\xi$.}
\label{figure:localize_noise}
\end{center}
\end{figure}

Next we discuss the calculation of the conductance in the strong localization regime of transport. As shown in Fig.\ \ref{figure:example_localized}, the conductance $g$ calculated from MSD-KPM exactly matches that from LB simulations, and the localization length $\xi$ extracted from the exponential decay of $g$ also matches the value of $\xi$ derived from the saturated value of the MSD, via Eq.\ \eqref{equation:xi_from_MSD}. However, the calculation of $g$ can become problematic numerically as $L$ approaches $2\pi\xi$. This can be understood from the inset of Fig.\ \ref{figure:example_localized}. As the MSD saturates, even small fluctuations lead to large fluctuations in the conductivity, which is calculated as the derivative of the MSD. The conductivity thus becomes incredibly noisy and can even become negative due to fluctuations in the MSD, as shown in Fig.\ \ref{figure:localize_noise}(a) for a different system.

In this regime, a large number of averages over random phase states and disorder configurations is needed to achieve a converged value of $g$, but even so, spurious behavior can arise. This was seen for example in \citet{uppstu2014prb}, where an anomalous super-exponential decay of the typical conductance $g_\text{typ}$ appeared when $L$ approached $2\pi\xi$. This behavior also appears if we extend the simulations of Fig.\ \ref{figure:example_localized} to lower values of conductance. This is shown in Fig.\ \ref{figure:localize_noise}(b): when $L$ is within $\sim$$20\%$ of its upper bound of $2\pi\xi$, the typical conductance transitions from the expected exponential decay to a super-exponential decay. Because of these issues, extreme care must be taken when calculating conductivity and conductance near the saturated value of the MSD, as their exact values can be overwhelmed by noise.

However, in the Anderson localization regime of transport, what is of more interest than the exact value of the conductance is the localization length $\xi$, which describes the scaling of the conductance to any length scale. This quantity can be calculated via the exponential decay of $g$ prior to the onset of problems with noise. Additionally, if this exponential decay is difficult to obtain numerically, $\xi$ is easily extracted from the MSD-KPM approach via the saturated value of the MSD. This is exemplified by the dashed line in Fig.\ \ref{figure:localize_noise}(a), which shows the trend $g \propto \exp(-L/\xi)$, with $\xi$ given by Eq.\ \eqref{equation:xi_from_MSD}. This approach provides an unambiguous calculation of the localization length, and is much more robust to the noise issues mentioned above.

\subsubsection{Computational cost}

In the MSD-KPM method, the evaluations of the time evolution operator and the quantum projection operator are decoupled. According to Fig.\ \ref{figure:test_evolution}(b), it takes about $\alpha\omega_{0}t_{\rm max}$ iterations to evaluate the time evolution operator $[X, U(t_{\rm max})]$, where $t_{\rm max}$ is the maximum correlation time and $\alpha$ is a numerical factor of the order of 1. According to Algorithm \ref{algorithm:evolvex}, the number of matrix-vector multiplies (MVMs) for each iteration is $3$ and the number of MVMs for the time evolution part is thus $3\alpha\omega_{0}t_{\rm max}$. According to Algorithm \ref{algorithm:kpm1}, the number of MVMs for evaluating the quantum projection operator using the KPM is $M N_t$, where $N_t$ is the number of time intervals. Each MVM costs $\sim$$WN$ multiplication operations, where $W$ is the number of nonzero diagonals of the Hamiltonian. For simplicity, we omit the factor $W$ that is common to all the methods. Then the total computational cost of the MSD-KPM method can be written as $\sim$$N(MN_t + 3\alpha\omega_0 t_{\rm max})$. The computational cost in the VAC-KPM method can be shown to be the same. In the KG-CPGF method, the number of MVMs for a fixed energy $E$ and a fixed energy resolution $\eta$ is $2M$. If one considers $N_\text{e}$ energy points and $N_{\eta}$ energy resolution values for each energy point (needed for checking the $\eta$-dependence of the conductivity), the overall computational cost is $\sim$$N(2 N_\text{e} N_{\eta} M)$. Because the maximum dephasing time $\tau_{\phi}$, or equivalently, the minimum energy resolution $\eta=\hbar/\tau_{\phi}$ achievable with a given $M$ in the KG-CPGF method \cite{ferreira2015prl} is $\eta = 4\Delta E/M$, we can transform $M$ to $4 \omega_0 \tau_{\phi}$, where $\omega_0=\Delta E/\hbar$, and write the computational cost in this method as $\sim$$N(8 N_\text{e} N_{\eta} \omega_0 \tau_{\phi})$. The computational cost of each method is summarized in Table \ref{table:lsqt}.

\subsubsection{Comparison of methods}
\label{sec:method_comparison}
At this point, it is important to point out the fundamental difference between the MSD-KPM (or VAC-KPM) method and the KG-CPGF method. In the MSD-KPM method, the two Dirac delta functions in the Kubo-Greenwood formula are treated differently, owing to their different origins in Eq.\ \eqref{equation:sigma_mu_nu_Greenwood}. One is treated as a quantum projection operator expanded using the KPM, which fixes the energy resolution $\delta E$, and the other is treated as a Green's function in the limit $\tau_\phi \rightarrow \infty$ and regularized by time evolution, which controls the transport regime. Meanwhile, in the KG-CPGF method both Dirac delta functions are treated on an equal footing: as a Green's function in the limit $t \rightarrow \infty$ and regularized by the dephasing time, which thus controls simultaneously the transport regime via $\tau_{\phi}$ and the energy resolution via $\eta = \hbar / \tau_{\phi}$.

As illustrated in Fig.\ \ref{figure:transport_regimes} in Sec.\ \ref{section:transport_regimes}, and Fig.\ \ref{figure:example_diffusive_a} in Sec.\ \ref{sec:diffusive}, this difference appears not to matter in most cases, and when taking the appropriate limits both methods can achieve equivalent results in the ballistic, diffusive, and strong localization regimes. However, as will be discussed further in Sec.\ \ref{section:graphene_point} below, a difference does appear for situations with a diverging DOS. In this case, the different regularization of the delta functions used in each method appear to make a qualitative difference in the numerical results.

Finally, we would like to note that while the focus of this review is on the calculation of zero-frequency (DC) transport, one can also derive an expression for the AC electrical conductivity $\sigma(\omega)$ in the spirit of Sec.\ \ref{section:kubo_formulas}. Several works have used KPM or FTM methods to study optical conductivity numerically \cite{Weisse2004, Yuan2011, Cysne2016}, and in principle it should reduce to the DC conductivity $\sigma_\text{DC}$ in the limit $\omega \rightarrow 0$. To the best of our knowledge, to date there has been no direct comparison of the numerical implementations of DC and AC conductivities using $\mathcal{O}(N)$ techniques, but one group has published two independent papers in which they calculate these quantities for graphene with $0.4\%$ of vacancy defects \cite{ferreira2015prl, Cysne2016}.

A comparison of these papers shows that the DC and AC implementations give identical results, but convergence $\sigma(\omega \rightarrow 0)$ at the band centre (E=0) can require a fine integration mesh. Meanwhile, in Figs.\ \ref{figure:example_ballistic_2b} and \ref{figure:example_localized} we have shown good agreement between the time-dependent methods and the LB method, while early work by Wei\ss{}e \textit{et al.}\ demonstrated good agreement between LB calculations and calculations of $\sigma(\omega \rightarrow 0)$ \cite{Weisse2005}. Altogether, these results indicate general consistency between $\mathcal{O}(N)$ calculations of $\sigma_\text{DC}$, $\mathcal{O}(N)$ calculations of $\sigma(\omega \rightarrow 0)$, and LB calculations.

%% file: secVI_applications.tex
\section{Applications to dissipative transport in disordered materials }
\label{section:applications}
After presenting the LSQT methodologies for dissipative electronic transport, we are now in a position to discuss various applications made during the last two decades. The LSQT method based on the MSD was first developed to study electronic transport in quasicrystals \cite{roche1997prl}, structures with a fivefold symmetry in the absence of translational invariance \cite{shechtman1984prl}. In such aperiodic systems with additional weak disorder, it was demonstrated that the scaling behavior of the quantum conductivity deviates significantly from that predicted using the semiclassical Bloch-Boltzmann approach \cite{roche1997prl}. After this initial work, additional applications were focused on low-dimensional materials, such as silicon nanowires \cite{markussen2006prb, persson2008nl}, carbon nanotubes (CNTs) \cite{roche2001prl, roche2001prb, triozon2004prb, latil2004prl, roche2005prl, avriller2006prb, charlier2007rmp}, and very extensively graphene-based materials \cite{ROZHKOV201177}. We will first illustrate applications of the LSQT methods to graphene-based materials with various types of static disorder, then we will discuss 3D metals and Dirac semimetals, and finally we will cover electronic transport in CNTs and crystalline organic semiconductors considering electron-phonon scattering.

\subsection{Applications to disordered graphene}

Ever since its discovery \cite{novoselov2004science, novoselov2005pnas}, graphene research has included an intense focus on the impact of disorder on its transport properties. Many studies have considered either realistic or simplified theoretical models, and have been inspired by the plethora of observed defects generated during material fabrication and integration into practical devices. Studying quantum transport in graphene and two-dimensional disordered materials is of particular interest given the large variety of physical scattering sources such as long-range charged impurities and screening effects (electron-hole puddle formation), short-range static defects, thermal disorder, as well as many-body effects. The quantum transport theory of massless Dirac fermions in the presence of such disorder is extremely rich in novel phenomena such as Klein tunneling, the minimum conductivity at the Dirac point, weak antilocalization, and the anomalous quantum Hall effect, all of which have been widely studied and presented in excellent reviews \cite{castro2009rmp, peres2010rmp, das2011rmp}. Here we will focus on a few representative cases of defects and their implication for electrical transport. For further background and reviews on the electronic and transport properties of graphene, see \cite{castro2009rmp, peres2009jpcm, peres2010rmp, mucciolo2010jpcm, das2011rmp, torres2014book}.
 
\subsubsection{Anderson disorder}

Anderson disorder \cite{anderson1958pr}, as introduced in the last section, is the canonical disorder model for studying quantum transport in different materials. Although this is not a very realistic disorder model for graphene, it is still of theoretical importance. One advantage of this disorder model is that analytical results \cite{shon1998jpsj, ostrovsky2006prb} can be obtained in the weak-disorder limit based on perturbation theory such as the self-consistent Born approximation (SCBA). 
LSQT calculations of the transport properties of graphene with Anderson disorder were first performed by Lherbier \textit{et al} \cite{lherbier2008prl}. With the presence of Anderson disorder, the electronic DOS at the charge neutrality point is enhanced and the Van Hove singularities are smoothed, which is consistent with the prediction from SCBA \cite{shon1998jpsj}. The semiclassical conductivity from LSQT calculations has a minimum at the charge neutrality point, which approaches the so-called minimum conductivity $4e^2/\pi h$ in the strong-disorder limit but remains generally larger in the weak-disorder limit. In the weak-disorder limit, the LSQT results can be well fitted by the SCBA prediction \cite{roche2012ssc}. Meanwhile, in the strong-disorder limit the SCBA fails to quantitatively describe the conductivity because of the neglect of quantum interference and hence localization effects. In both cases, the semiclassical Boltzmann transport equation approach fails to capture the energy dependence of the conductivity. This comparison highlights the necessity of employing fully quantum mechanical and nonperturbative calculations for a complete description of the transport physics of disordered graphene and related materials.

Beyond the diffusive regime, the conductivity decreases with increasing time or length \cite{lherbier2008prl, fan2014prb}, experiencing weak and strong localization effects consecutively. The weak localization regime is characterized by a logarithmic decay of the conductivity with respect to the length $L(E)$, according to Eq.\ (\ref{eq_wl_new}). This has been confirmed numerically with LSQT calculations \cite{fan2014prb}. In the scaling theory of Anderson localization \cite{abrahams1979prl, lee1985rmp}, $L(E)$ reaches the localization length $\xi(E)$ when the weak localization correction equals the semiclassical conductivity $\sigma_{\rm sc}(E)$. This gives an expression of the two-dimensional localization length
\begin{equation}
\label{equation:xi_from_mfp}
\xi(E)=l_\text{sc}(E) \exp \left[ \frac{\pi h\sigma_{\rm sc}(E)}{2e^2} \right].
\end{equation}
It has been demonstrated \cite{fan2014prb} that the two-dimensional localization length calculated in this way is consistent with that calculated based on the one-parameter scaling of localization in quasi-one-dimensional systems \cite{mackinnon1981prl,kramer1993rpp}. Graphene with Anderson disorder fully follows the one-parameter scaling theory of localization \cite{abrahams1979prl} and there is no extended state in the absence of decoherence. However, as remarked initially, the Anderson disorder model is not a satisfactory description of defects in real materials, and the study of more realistic disorder models is fundamental for any quantitative analysis of experimental measurements.

\subsubsection{Charged impurities}

One realistic disorder model is a long-range electrostatic potential accounting for the effects of charged impurities trapped in the substrate beneath graphene. It has been argued \cite{rycerz2007epl} that the bare Coulomb potential is not suitable for describing the potential induced by charged impurities. A standard model considering screening effects is obtained by replacing the bare Coulomb potential with a smoother Gaussian function, although more complex charged impurity models have been studied using the LSQT approach \cite{radchenko2012prb, radchenko2013prb}. Under this disorder model, the electrostatic potential energy at position $\bm{r}$ is given in real space by 
\begin{equation}
U(\bm{r}) = \sum_{k=1}^{N_{\rm imp}} U_k \exp\left[-\frac{|\bm{r}-\bm{r}_k|^2}{2\xi^2}\right],
\end{equation}
where $N_{\rm imp}$ is the number of screened charge impurities, $U_k$ is the strength of the $k$th impurity located at $\bm{r}_k$, and $\xi$ is the effective range of the potential (here we distinguish this from the localization length discussed in previous sections). The ratio $n_{\rm imp}=N_{\rm imp}/N$, with $N$ being the number of atoms, defines the impurity concentration. The potential heights $U_k$ are assumed to be uniformly distributed in the interval $[-W/2,W/2]$. $W$ is the strength of the potential, which plays a similar role as in the Anderson disorder model. Actually, this charged impurity model reduces to the Anderson disorder model in the limit $\xi \to 0$ and $n_{\rm imp}\to 1$. By tuning the value of $\xi$ across the lattice constant, both short-range and long-range potentials can be realized. A dimensionless quantity which is frequently used to quantify the disorder strength when $n_{\rm imp} \ll 1$ is given by \cite{rycerz2007epl}
\begin{equation}
K_0 \approx 40.5 \times n_{\rm imp} 
\left(\frac{W}{2\gamma_0}\right)^2 
\left( \frac{\xi}{a} \right)^4,
\label{equation:K_0}
\end{equation}
where $a \approx 2.46$ $\text{\AA}$ is the lattice constant of graphene and $\gamma_0 \approx 2.7$ eV is the hopping energy between neighboring carbon atoms.

\begin{figure}[htbp]
\hspace{0cm}\centering
\includegraphics[width=1\linewidth]{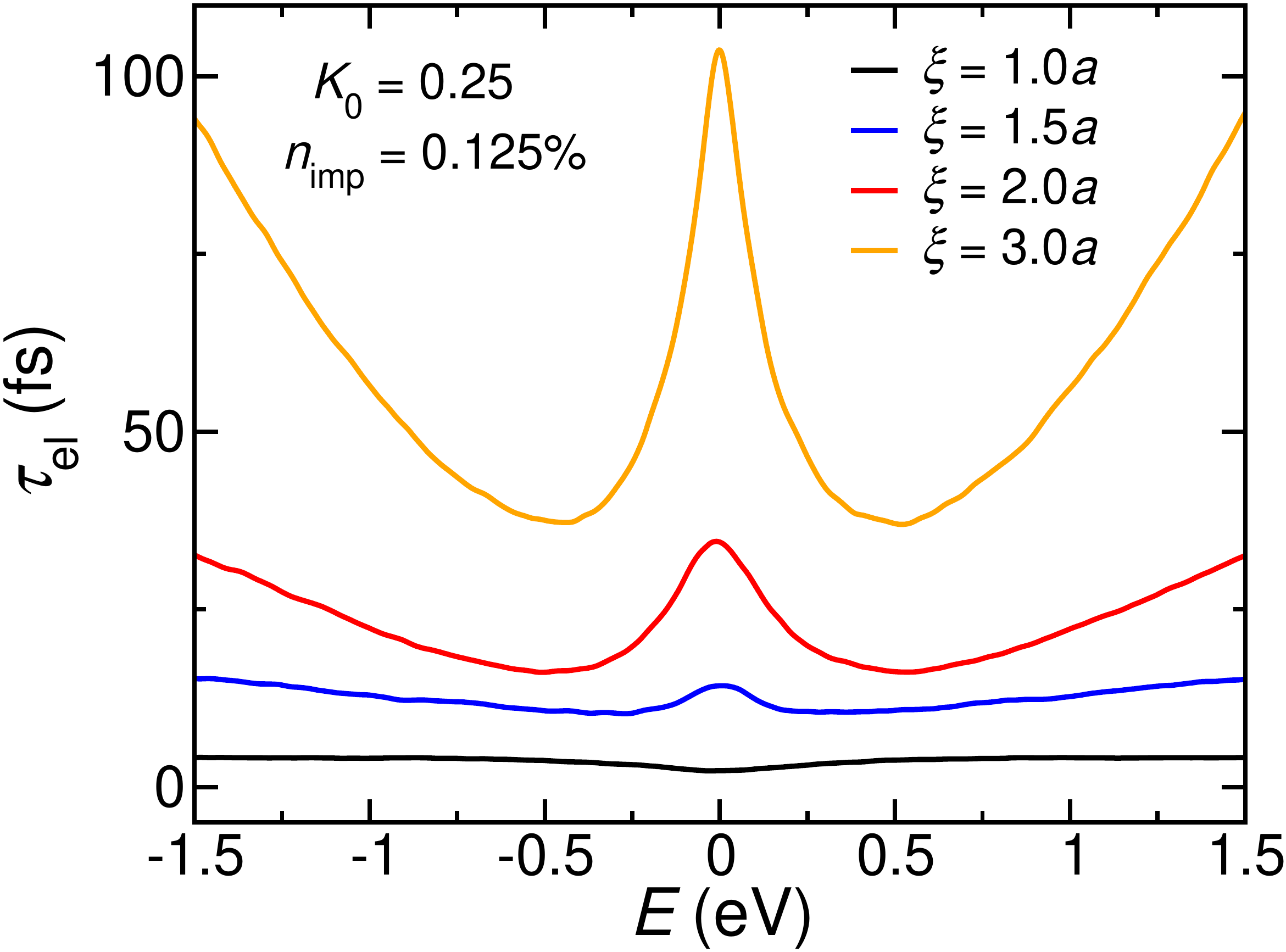}
\caption{Energy-depedent momentum relaxation time in graphene, highlighting a transition between intervalley and intravalley dominated scattering. Results were obtained for electrostatic impurities with a varying range $\xi$ at a concentration $n_{\text{imp}}=0.125\%$ with constant $K_0 = 0.25$. Simulations were carried out using the Lanczos method with 1000 recursion steps and $M=125$ Chebyshev polynomials for the time evolution in a sample of $N=10^7$ carbon atoms.}
\label{fig:Intervalley_intravalley_scattering_time_v4}
\end{figure}

Graphene with this type of disorder shows diverse transport regimes, as the Gaussian-shaped potential can induce two kinds of scattering: intervalley scattering which mixes the states in the two valleys of reciprocal space, and intravalley scattering which does not. The dependence of these two scatterings on the disorder strength has been studied numerically \cite{zhang2009prl}. To explore this transition between intervalley and intravalley dominated transport with the LSQT methodology, one may investigate the energy-resolved momentum relaxation time $\tau_\text{p}(E)$, defined in Eq.\ (\ref{equation:diffusivity}), with $D_\text{sc}(E)$ taken as the maximum value of the time-dependent diffusivity $D(E,t)$. In Fig.\ \ref{fig:Intervalley_intravalley_scattering_time_v4} we plot $\tau_\text{p}(E)$ for varying impurity height and width while fixing $K_0 = 0.25$, with $\xi$ varying from $1a$ to $3a$.

\begin{figure}[htb]
\centering
\includegraphics[width=1.0\linewidth]{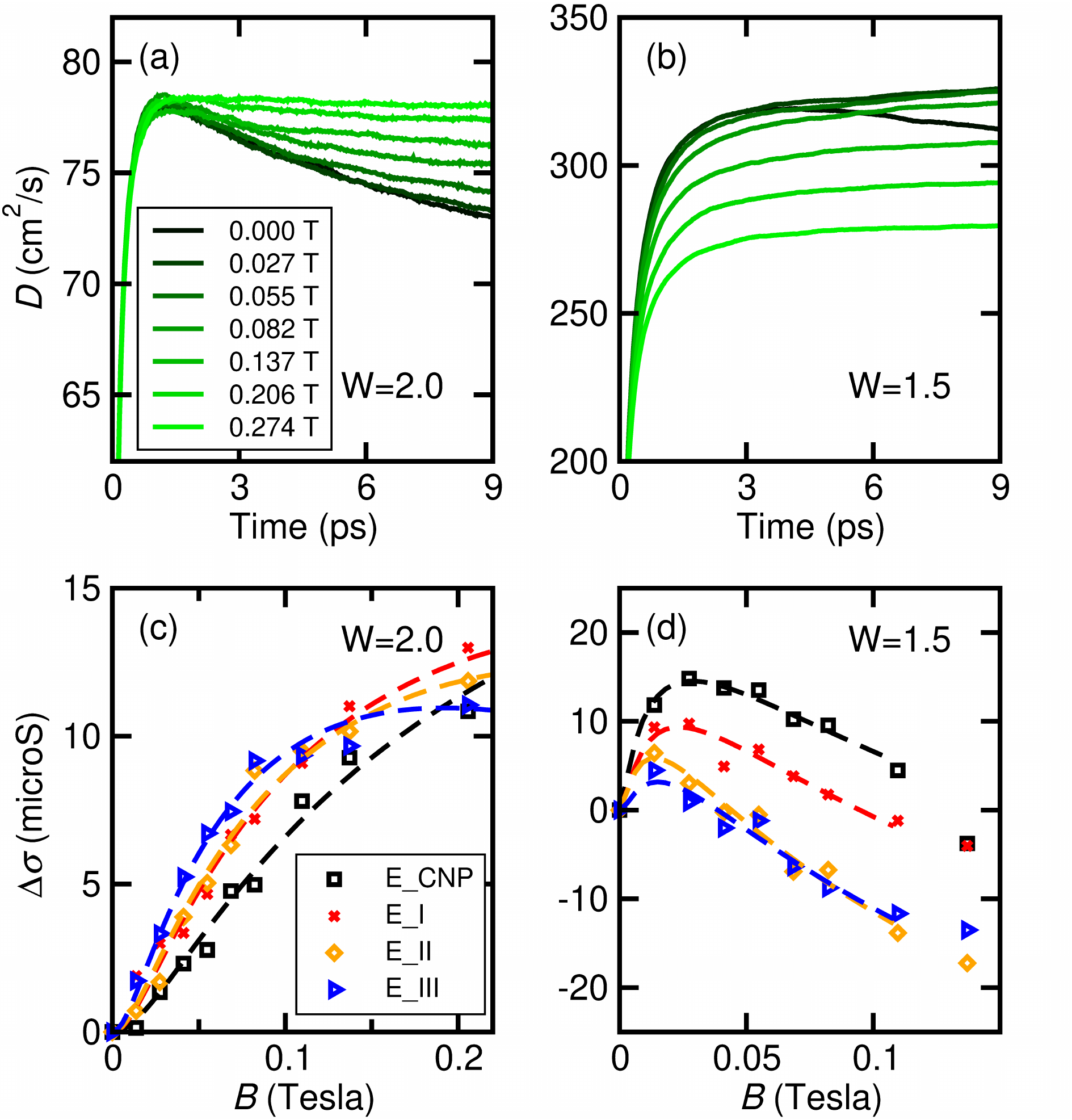}
\caption{Top panels: diffusion coefficient of graphene at the Dirac point for various magnetic fields and $n_\text{imp}=0.125\%$,   $\xi=\sqrt{3}a$, (a) $W=2\gamma_0$, and (b) $W=1.5\gamma_0$. Bottom panels: the corresponding $\Delta\sigma(B)$ for four different Fermi level positions ($E_{\text{CNP}}=0$ eV, $E_{\text{I}}=0.049$ eV, $E_{\text{I}}=0.097$ eV and $E_{\text{III}}=0.146$ eV). Dashed lines are fits as explained in the text. Simulations were run for a system size of $1.15 \times 10^7$ carbon atoms. Adapted from \cite{ortmann2011epl}.}
\label{fig:Charged_Impurity_weak_antilocalization_tuning}
\end{figure}

At $\xi=3a$, $\tau_\text{p}(E)$ strongly increases at low energies; this is a signature of transport limited to a single Dirac cone, where pseudospin conservation prohibits back scattering near the Dirac point, yielding long scattering times. As $\xi$ decreases to $1a$, intervalley scattering begins to play a role, allowing for back scattering processes and a subsequent reduction of $\tau_\text{p}(E)$ around the graphene Dirac point. Here we have tuned the impurity height and width simultaneously at a fixed $K_0$, but we will show in Fig.\ \ref{fig:Charged_Impurity_weak_antilocalization_tuning} below that a similar transition can be induced by tuning only the potential height $W$.

When intervalley scattering is completely excluded by considering a single-valley Dirac Hamiltonian, the conductivity follows a one-parameter scaling, either with \cite{ostrovsky2007prl} or without \cite{bardarson2007prl, nomura2007prl} an unstable fixed point. In this case, the $\beta$ function $\beta(\sigma) = d\ln \sigma/d \ln L$ is positive (metallic), indicating weak antilocalization \cite{CHAKRAVARTY1986193}. On the other hand, when using a disoder model that induces intervalley scattering, numerical simulations reveal that the conductivity follows a one-parameter scaling with a negative $\beta$ function \cite{zhang2009prl}, which is associated with the weak localization regime.

As shown in Sec.\ \ref{section:wl}, the MSD-KPM approach fully captures weak localization in the square lattice with Anderson disorder. Here we show that this approach can capture both weak localization and weak antilocalization in graphene, with the transition driven by the disorder strength. These results are fully in agreement with the theory developed for massless Dirac fermions \cite{mccann2006prl, kechedzhi2007epjst, falko2007ssc}.

\begin{figure}[htb]
\begin{center}
\includegraphics[width=\columnwidth]{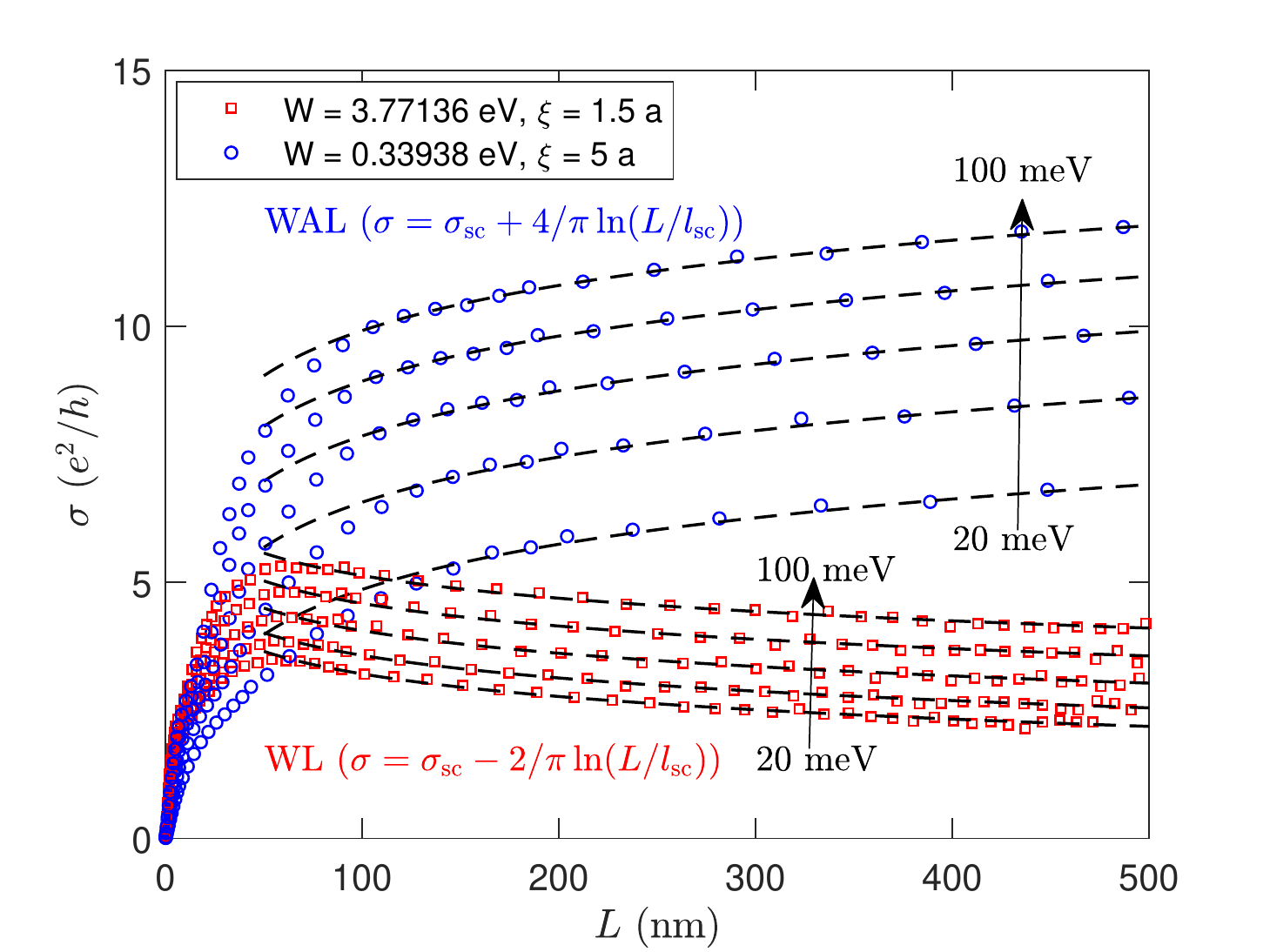}
\caption{Length-dependent conductivity $\sigma(E,L)$ in graphene with Gaussian-shaped disorder, at Fermi energies from $E=20$ to $100$ meV. The impurity concentration is fixed to $n_{\rm imp}=1\%$. Two disorder profiles are considered, one leading to strong valley mixing and weak localization (WL), the other leading to weak valley mixing and weak antilocalization (WAL). Symbols are numerical simulations and dashed lines correspond to Eq.\ (\ref{eq_wl_new}) or (\ref{equation:wal}). Simulations were performed using the MSD-KPM method with $M=3000$ moments, $N_r=10$ random vectors, and  $N=4\times 10^7$ orbitals}.
\label{figure:wal}
\end{center}
\end{figure}

Figure \ref{figure:wal} shows the length-dependent quantum conductivity $\sigma(E,L)$ of graphene with Gaussian-shaped disorder. Two sets of disorder parameters are considered, describing either a weak and relatively long-ranged disorder ($\xi=5a$ and $W=0.33938$ eV) that mainly induces intravalley scattering, or a strong and relatively short-ranged disorder ($\xi=1.5a$ and $W=3.77136$ eV) that induces significant intervalley scattering. In both cases the impurity concentration is fixed to $n_{\rm imp}=1\%$, such that $K_0=1$, and we consider Fermi energies ranging from $E=20$ to $100$ meV. For disorder large enough to induce intervalley scattering, the conductivity first increases from zero to a plateau, corresponding to the ballistic-to-diffusive transition, followed by a logarithmic decay given by Eq.\ (\ref{eq_wl_new}), as already discussed for the disordered square lattice. Similar results have been obtained for graphene with structural defects \cite{lherbier2012prb}. In contrast, when the disorder does not mix the valleys, weak antilocalization appears with a change of sign of the quantum correction and thus an increase of the conductivity with increasing propagation length, in full agreement with the theory developed for a single-valley Dirac Hamiltonian \cite{bardarson2007prl, nomura2007prl, ostrovsky2007prl},
\begin{equation}
\label{equation:wal}
\sigma(L) = \sigma_\text{sc} + \frac{4}{\pi} \frac{e^2}{h} \ln \left( \frac{L}{l_\text{sc}} \right).
\end{equation}
Similar conductivity scaling in the presence of Gaussian-shaped disorder has also been obtained using the Landauer-B\"{u}ttiker approach \cite{lewenkopf2008prb}.

Experimentally, quantum corrections to the conductivity can be explored by measuring the low-temperature magnetoresistance, or equivalently $\Delta\sigma(B) = \sigma(B) - \sigma(B=0)$, where $\sigma(B)$ is the magnetoconductivity ($B$ the external magnetic field). A diagrammatic theory of quantum interference in disordered graphene \cite{mccann2006prl, kechedzhi2007epjst, falko2007ssc} has been developed and provides a possible quantitative analysis of magnetoconductivity data. Both positive (weak localization) and negative (weak antilocalization) magnetoconductivity can be obtained, depending on the relative strength between the intravalley scattering time, the intervalley scattering time, and the coherence time. A transition from localization to antilocalization has been demonstrated experimentally \cite{tikhonenko2009prl}.

Here we show that similar results can be obtained from numerical calculations based on the LSQT method in the MSD formalism \cite{ortmann2011epl}. As above, we use the Gaussian charged impurity model to tune the intervalley scattering strength. In this case we fix the potential range at $\xi = \sqrt{3}a$ and the impurity concentration at $n_\text{imp} = 0.125\%$, and we choose impurity strengths of $W = 1.5\gamma_0$ and $2\gamma_0$ (corresponding to $K_0 = 0.26$ and $0.46$). A summary of the results is shown in Fig.\ \ref{fig:Charged_Impurity_weak_antilocalization_tuning}.

Figure \ref{fig:Charged_Impurity_weak_antilocalization_tuning}(a) shows the time-dependent diffusion coefficient $D(E,t)$ at the graphene Dirac point ($E=0$) for $W = 2\gamma_0$, with the decay at long times indicating the presence of weak localization induced by the strong disorder. Turning on the magnetic field in the millitesla range suppresses the localization, as indicated by the progressively slower decay of $D(E,t)$ with increasing $B$. Figure \ref{fig:Charged_Impurity_weak_antilocalization_tuning}(b) shows the case for weaker disorder, with $W = 1.5\gamma_0$. This disorder is still strong enough to induce weak localization at longer times, as seen for the black curve with $B=0$. However, a small magnetic field ($B = 27$ mT) is sufficient to completely suppress this localization in the simulated time scale, leaving behind pure antilocalization behavior that is then suppressed by progressively higher magnetic fields. Figures \ref{fig:Charged_Impurity_weak_antilocalization_tuning}(c) and (d) show the corresponding magnetoconductivities at different Fermi energies, where in all cases the value of $\sigma(B)$ was taken at the maximum simulation time ($t = 9$ ps). These figures highlight the pure WL behavior induced by the stronger disorder, and a WL-WAL transition for the case of weaker disorder.

To interpret these results in terms of the Hikami-Larkin-Nagaoka model \cite{mccann2006prl,HikamiLarkinNagaoka} the magnetoconductivity can be expressed in a simplified form,
\begin{align}
\Delta\sigma(B)=\frac{e^2}{\pi h}\left[\mathcal{F}\left(\frac{\tau_B^{-1}}{\tau_{\varphi}^{-1}}\right)-3\mathcal{F}\left(\frac{\tau_B^{-1}}{\tau_{\varphi}^{-1}+2\tau_*^{-1}}\right)\right], 
\label{Magnetoresisntance}
\end{align}
where $\mathcal{F}(z)=\ln z +\psi\left(\frac{1}{2}+z^{-1}\right)$, $\psi(x)$ is the digamma function, and the magnetic time scale $\tau_B^{-1} = 4D/l_B^2$ is defined \textit{via} the diffusion coefficient $D$ and the magnetic length $l_{B}=\sqrt{\hbar/(eB)}$. In addition to the magnetic time scale $\tau_B$ and the coherence/dephasing time $\tau_{\varphi}$, this model assumes a phenomenological elastic scattering time $\tau_*$ that contains both intra- and intervalley scattering \cite{mccann2006prl}. The numerical data in Figs.\ \ref{fig:Charged_Impurity_weak_antilocalization_tuning}(c) and (d) are fitted with Eq.\ (\ref{Magnetoresisntance}) and clearly follow the dependence on the magnetic field $B$ for both the WL and the WAL regimes. For short intervalley scattering times, $\tau_\text{iv}<\tau_{\varphi}$, for the lowest magnetic fields WL is described in full consistency with \cite{mccann2006prl} and $\tau_*$ ranging from $1.1$ to $2.3$ ps. In case of WAL the effective scattering time $\tau_*$ is increased to $1.5$ to $6.3$ ps.

Despite the observation of WAL for $W=1.5\gamma_0$, which is more pronounced at energies away from the Dirac point (see Fig.\ \ref{fig:Charged_Impurity_weak_antilocalization_tuning}(d)), experiments suggest an opposite behavior, namely that WAL is stronger close to the Dirac point \cite{tikhonenko2009prl}. This discrepancy in the interpretation of WAL may be caused by as-yet-unknown differences in the disorder potential for the charged impurities between the model and the graphene in the experiments. A further step would be a better understanding of the time scales for intra- and intervalley scattering as well as dephasing in the experiments, and further numerical tests with different disorder models.

\subsubsection{Point-like defects\label{section:graphene_point}}

Point-like structural defects have also been observed in graphene and have been shown to greatly affect transport properties \cite{cresti2008nr, roche2012ssc}. Beyond those induced by material and device fabrication, point-like defects can also be deliberately created using ion irradiation or chemical treatments for tailoring the conduction regime \cite{nakaharai2013acsnano}. Chemical substitution of carbon with nitrogen or boron atoms has been experimentally observed \cite{zhao2011science}, and numerical studies using the LSQT approach have discovered the emergence of mobility gaps \cite{lherbier2008prl_chemical, biel2009nl, lherbier2013nl}, which can help in fabricating p-type or n-type graphene-based transistors \cite{marconcini2012acsnano}.

A generic and common defect that is found in any material is the missing lattice atom. Single vacancies in graphene have been produced and characterized by transmission electron microscopy \cite{meyer2008nl} and scanning tunneling microscopy \cite{ugeda2010prl}. This type of disorder has a dramatic impact on the electronic structure of graphene, with the formation of low-energy impurity resonances (also called zero-energy modes) which are localized at the Dirac point and which display a wavefunction decay following a power law \cite{pereira2008prb}. The impact of such anomalous localization behavior on quantum transport at the graphene Dirac point has been the subject of an intense debate in the literature. Meanwhile, away from the graphene Dirac point, all studies have found that electronic states are localized \cite{TramblyPRL2006,TramblyMPL2011,cresti2013prl,laissardiere2013prl,fan2014prb,ferreira2015prl}.

\begin{figure}[htb]
\begin{center}
\includegraphics[width=\columnwidth]{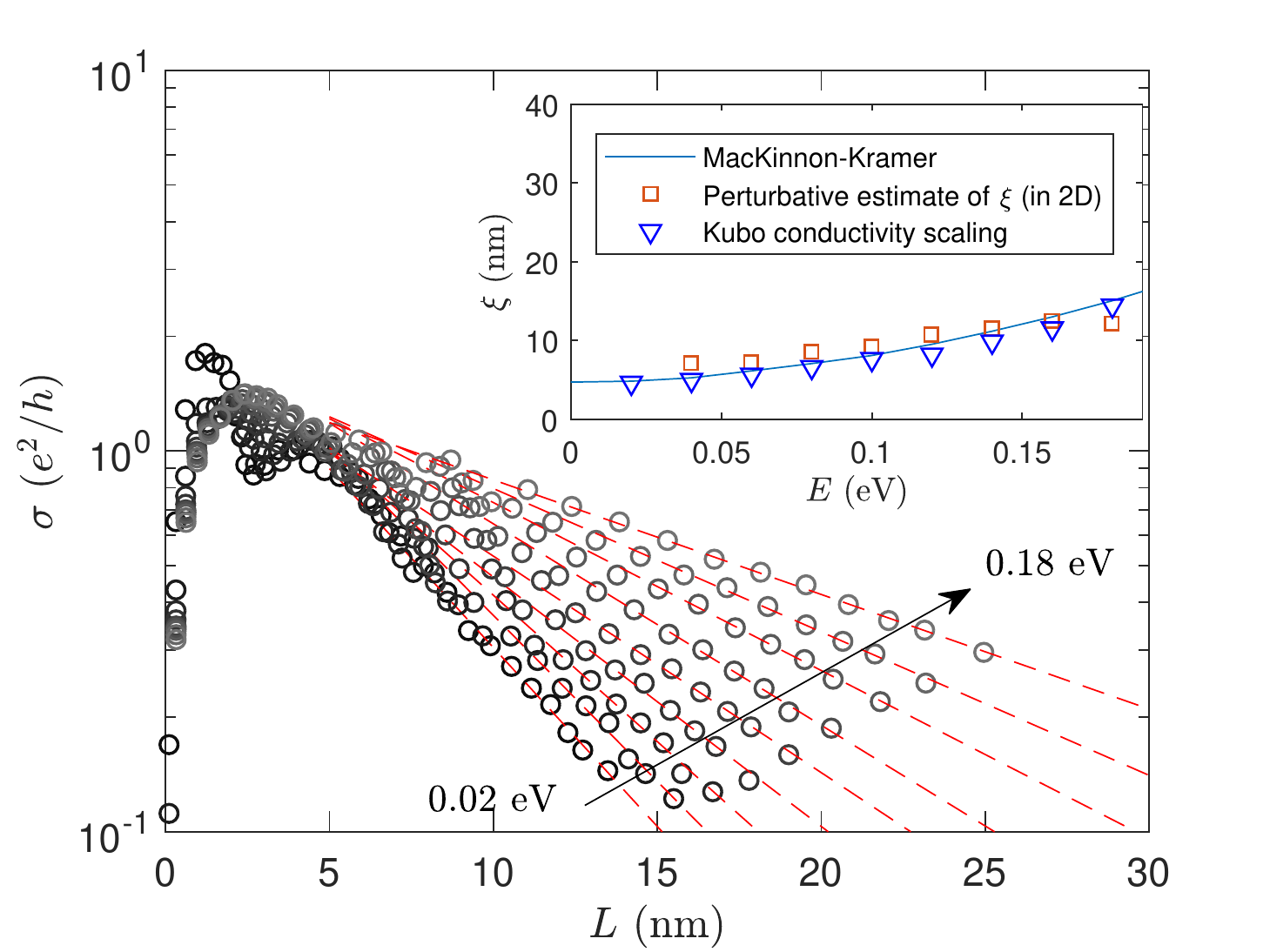}
\caption{Length-dependent conductivity $\sigma(E,L)$ in graphene with $1\%$ of vacancy defects at Fermi energies from $E = 0.02$ eV to $0.18$ eV. The inset shows the localization lengths, as a function of Fermi energy, extracted from the exponential decay of conductivity data (triangles), from the perturbative estimate using the mean free path and semiclassical conductivity (squares) \cite{lee1985rmp}, and from the one-parameter scaling method by MacKinnon and Kramer (solid line) \cite{kramer1993rpp,mackinnon1981prl}. The results were obtained from the MSD-KPM method with $M=3000$ moments, $N_r=10$ random vectors, and $N=2\times 10^7$ orbitals. }
\label{figure:vacancy}
\end{center}
\end{figure}

For a low density of vacancies distributed roughly equally on both sublattices in a random fashion, one expects short-range scattering and localization effects to emerge. This was numerically studied using linear scaling methods by Cresti \textit{et al.}\ \cite{cresti2013prl}, Fan \textit{et al.}\ \cite{fan2014prb} and Ferreira \textit{et al.}\ \cite{ferreira2015prl}, and all simulations obtain localization for energies away from the Dirac point. Figure \ref{figure:vacancy} shows the results of MSD-KPM simulations of graphene with $1\%$ of randomly-distributed vacancies. Here one can see an unambiguous exponential decay of the conductivity for a wide energy range away from the Dirac point, with extracted localization lengths in good agreement with Eq.\ \eqref{equation:xi_from_mfp} as well as the one-parameter scaling theory of localization \cite{kramer1993rpp, mackinnon1981prl}.

At the Dirac point, simulations from \cite{fan2014prb} and \cite{laissardiere2013prl} also indicated the presence of Anderson localization, while the authors of \cite{cresti2013prl} proposed an alternative conductivity scaling following $\sigma \propto 1/L^{\beta}$ with $\beta \approx 2$. However, the presence of zero-energy modes means that at the Dirac point numerical results are extremely sensitive to the chosen parameters, and difficult to converge numerically. In contrast to these results, field-theoretical calculations predict the absence of localization corrections at the band center of 2D disordered systems with chiral symmetry \cite{gade1991npb, gade1993npb}, suggesting that the localization length diverges at $E = 0$ in chiral-symmetric disordered graphene (class BDI, which is the case of graphene with vacancies) \cite{ostrovsky2006prb, ostrovsky2010prl, Ostrovsky2014}. Such a critical point is predicted to yield a finite value of the Dirac point quantum conductivity on the order of $e^2/h$ \cite{ostrovsky2006prb}. This behavior was reproduced by Ferreira and Mucciolo, who used the KG-CPGF method to obtain delocalized transport with a conductivity of $4e^2/\pi h$ at the Dirac point over a wide range of vacancy concentrations with energy resolutions down to 1 meV \cite{ferreira2015prl}.

As discussed in previous sections, the difference between the KG-CPGF and KPM-MSD methods is their treatment of the two Dirac delta functions. In KG-CPGF both are treated as Green’s functions regularized by the dephasing time $\tau_\phi$. When approaching the adiabatic limit this simultaneously modulates the transport regime ($\tau_\phi \rightarrow \infty$) and the energy resolution ($\eta = \hbar/\tau_\phi \rightarrow 0$). With this method the conductivity at the Dirac point is delocalized and equal to $4e^2/\pi h$ down to $\eta = 1$ meV. Meanwhile, in the KPM-MSD method the energy resolution is fixed by the KPM expansion of one delta function, and the transport regime is then modulated by the time evolution expansion of the other delta function. The best calculations with this methodology \cite{fan2014prb} showed identical Anderson localization for all energy resolutions down to ${\sim}1$ meV in the KPM expansion for simulation times $t > 1$ ps ($\hbar / t < 0.7$ meV). Achieving better energy resolution with either method may yet reveal different behavior, but this is a highly demanding computational task, and at this point the difference appears to depend on the numerical approximations of delta functions in the adiabatic limit. Importantly this only becomes problematic in the presence of a highly diverging DOS, whereas in all other situations with smooth spectra, all linear scaling methods lead to equivalent results and are able to accurately capture all regimes of quantum transport.

\subsubsection{Large-scale structural defects}

Beyond point-like or electrostatic disorder, the methods presented in this review are also readily applicable to large-scale lattice defects such as grain boundaries (GBs) or graphene antidots \cite{LherbierPRL2011,pedersen2008prl, bai2010nnt}. Grain boundaries are a natural result of chemical vapor deposition (CVD), which is the best approach for the large-scale production of graphene \cite{2013zhang_cvdreview}. During the CVD growth process, graphene grains nucleate and grow at random positions and orientations, resulting in a polycrystalline structure when growth is complete \cite{Arjmandi_Tash_2018}. The grain boundaries that form at the interface of the graphene grains typically consist of disordered arrays of carbon pentagons, heptagons, and octagons. Experiments based on scanning tunneling microscopy or quantum transport have shown that GBs are strong charge scatterers, and can thus limit the transport properties of large-area CVD-grown graphene \cite{Isacsson2017_2dmater}.

\begin{figure*}[htbp]
\begin{center}
\includegraphics[width=\textwidth]{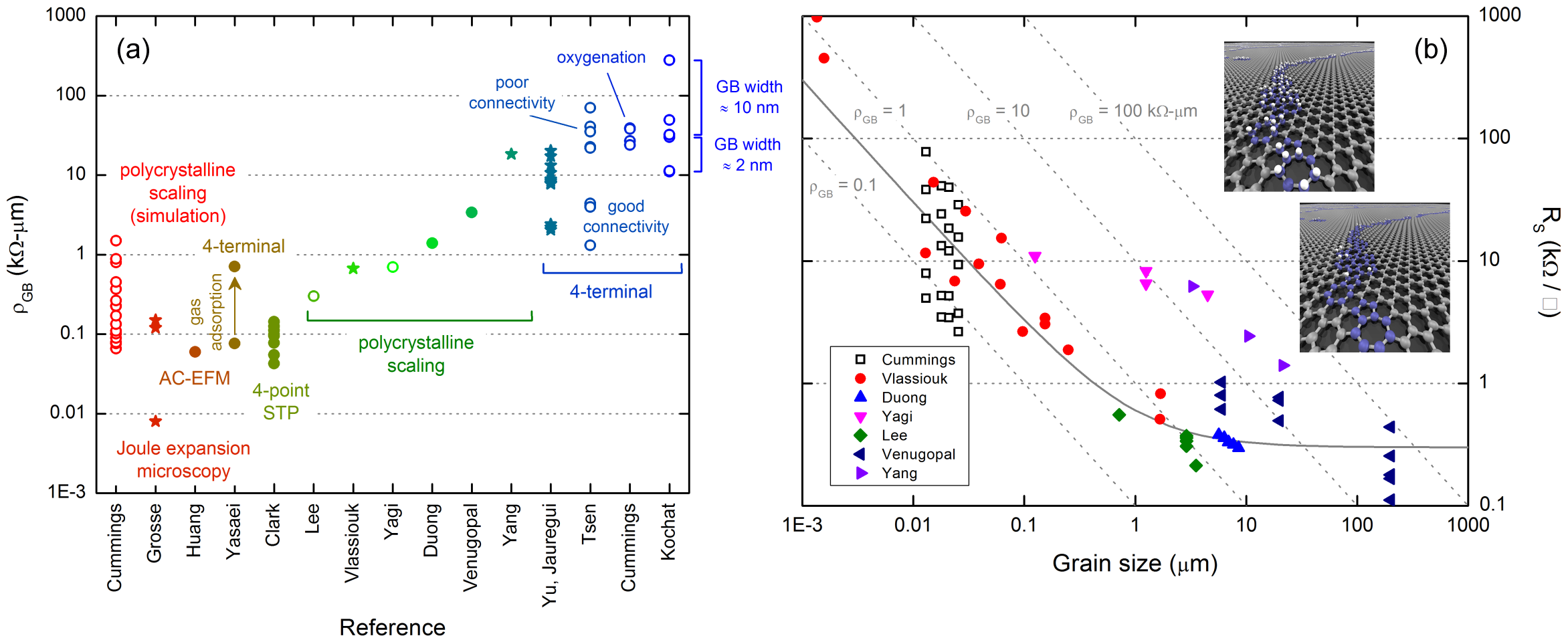}
\caption{(a) Summary of the values of grain boundary resistivity ($\rho_{\rm GB}$) extracted from the literature. Open circles are measurements at the charge neutrality point, closed circles are measurements far from the charge neutrality point, and stars are for measurements where the position of the Fermi level is unknown. (b) Summary of the values of graphene sheet resistance, as a function of grain size, extracted from the literature. The solid gray line illustrates the scaling law of Eq.\ (\ref{equation:polyG_scaling}), assuming $R_{\rm s}^{\rm G} = 300$ $\Omega/\Box$ and $\rho_{\rm GB} = 0.3$ k$\Omega$ $\upmu$m. In both panels, the spread of simulation results is due to the impact of chemical functionalization of the grain boundaries, as depicted in the insets. In the simulations, system sizes ranged from $\sim$140,000 to $\sim$280,000 atoms, the semiclassical conductivity was calculated using the Lanczos expansion of the MSD with 1000 moments and an energy broadening of $\sim$100 meV, and the diffusive regime was reached after a simulation time of 0.1-0.3 ps. Adapted from \cite{Isacsson2017_2dmater}.}
\label{figure:pGB}
\end{center}
\end{figure*}

A variety of numerical simulations, based on the methods presented in this review, have been carried out to quantify the impact that GBs have on charge transport in CVD-grown graphene \cite{vantuan2013nl, cummings2014am, cummings2014prb, seifert2015twodm, barrios-vargas2017nl}. By applying Eq.\ (\ref{equation:sigma_t_msd_t}) to realistic models of polycrystalline graphene generated by molecular dynamics simulations, Van Tuan \textit{et al.}\ showed that the semiclassical conductivity $\sigma_{\rm sc}$ of these materials scales linearly with the average grain size \cite{vantuan2013nl}. Subsequent work quantified the impact of GBs through the scaling relation \cite{cummings2014am}
\begin{equation}
R_{\rm s} = R_{\rm s}^{\rm G} + \rho_{\rm GB} / l_{\rm G},
\label{equation:polyG_scaling}
\end{equation}
where $ R_{\rm s} \equiv 1/\sigma_{\rm sc}$ is the sheet resistance of the polycrystalline graphene, $R_{\rm s}^{\rm G}$ is the sheet resistance within the graphene grains, $l_{\rm G}$ is the average graphene grain size, and $\rho_{\rm GB}$ is the GB resistivity. By calculating $R_{\rm s}$ for polycrystalline samples with a variety of grain sizes and fitting to Eq.\ (\ref{equation:polyG_scaling}), Cummings \textit{et al.}\ extracted an intrinsic GB resistivity of $\rho_{\rm GB} = 0.07$ k$\Omega$ $\upmu$m \cite{cummings2014am}. This value is on the low end of those obtained experimentally. However, as shown in Fig.\ \ref{figure:pGB}(a), the value of $\rho_{\rm GB}$ depends significantly on the measurement technique, doping level, material quality, and degree of chemical functionalization \cite{Isacsson2017_2dmater}. Indeed, the spread of simulation results indicates that $\rho_{\rm GB}$ can be tuned by more than one order of magnitude by varying the concentration of chemical adsorbates on the GBs.

The impact of GBs on the electrical properties of CVD graphene can be seen in Fig.\ \ref{figure:pGB}(b), where we show a summary of the values of graphene sheet resistance, as a function of grain size, extracted from the experimental literature. Simulation results are shown as open squares, with the spread of values resulting from different degrees of chemical functionalization of the GBs. Overall, the measurements follow the scaling trend described by Eq.\ (\ref{equation:polyG_scaling}), and the crossover between GB-dominated and grain-dominated transport occurs for grain sizes in the range of 1-10 $\upmu$m.

Apart from GBs, large-scale lattice defects can also be intentionally engineered. A graphene antidot lattice \cite{pedersen2008prl}, also called a graphene nanomesh \cite{bai2010nnt}, is a graphene sheet containing a pattern of nanometer-sized holes. These structures have been proposed to create a band gap in otherwise gapless graphene. 
However, deviations from a perfect superlattice structure are usually present in real experimental situations. The effects of geometrical disorder, modeled as fluctuations in the antidot radius and location \cite{power2014prb}, have been studied using the LSQT method. It was shown that the band gap in a perfect antidot lattice vanishes with the introduction of sufficiently strong geometrical disorder, and a transport gap can be induced via Anderson localization \cite{fan2015prb}, in accordance with experimental results \cite{eroms2009njp, giesbers2012prb, zhang2013prl}. The charge carrier mobilities are found to be very small
compared to values found in graphene without antidots, and quantitative agreement with experiments has been obtained \cite{zhang2013prl}. In a model of anisotropic geometrical disorder, a coexistence of ballistic conduction and Anderson localization in different directions have also been predicted using the LSQT method \cite{pedersen2014prb}.

Finally, interestingly the synthesis of wafer-scale two-dimensional amorphous carbon monolayers has
been recently demonstrated experimentally \cite{Jooe1601821, Toh2020}. Such amorphous material presents unrivalled properties as coating of metals, semiconductors or magnetic materials, and is expected to strongly improve
atomic layer deposition of dielectrics, hence fostering the development of ultracompact
technologies. Theoretical studies using LSQT methods have shown that localization lengths in such materials are no longer than a few nanometres, making them prototypes of two-dimensional Anderson insulators \cite{VanTuanPRB2012, LherbierNR2013}, as confirmed experimentally \cite{Toh2020}.

\subsection{3D metals and semimetals}	
\subsubsection{Electrical conductivity in liquid transition metals}

Early studies of electron transport in disordered metallic systems have been conducted for liquid phase $3d$ transition metals such as Cr, Mn, Fe, Co and Ni with the so-called tight-binding linear muffin-tin orbital (TB-LMTO) recursion method \cite{Bose1993}. This method is an illustrative example of the connection between {\it ab initio} electronic structure approaches and linear-scaling quantum transport methods.

The TB-LMTO method divides the simulation space into atomic and interstitial regions with muffin-tin spheres that are centered at the atomic sites $\mathbf{R}$. Within these spheres, orbitals are defined with a collective angular momentum index $L=(l,m)$ through which $s$, $p$ and $d$ orbitals are included. The TB parameters for these orbitals are obtained from DFT calculations, thus providing a self-consistent description of the electronic properties at the Kohn-Sham level, which describes the general wave functions by orbitals
\begin{align}
\chi^{\alpha}_{RL}(\mathbf{r}_{R})=\phi_{RL}(\mathbf{r}_R)+\sum\limits_{R'L'}\dot{\phi}^{\alpha}_{R'L'}(\mathbf{r}_R)h^{\alpha}_{R'L',RL}
\end{align}
in the TB representation $\alpha$. Here $\phi_{RL}(\mathbf{r}_R)$ are reference wave functions inside a sphere of radius $s_R$ centered at $\mathbf{R}$ for a particular reference energy $E_{\nu Rl}$. Inside the sphere the potential is calculated with DFT and additional functions $\dot{\phi}^{\alpha}_{RL}(\mathbf{r}_R)$ related to the energy derivative of $\phi_{RL}(\mathbf{r}_R)$ at the reference energy enter with expansion coefficients $h^{\alpha}_{R'L',RL}$. These expansion coefficients typically vanish for the second neighbor shell in close-packed structures. This short-range nature makes the application of the Lanczos approach particularly efficient. 

The geometry of the liquid metals was modeled by clusters with a size of 600 particles, which were generated with a Monte Carlo method. This leads to a disordered configuration of the metal atoms deviating from their fcc or bcc crystal structure. The strength and type of disorder are reflected in the matrix elements $h^{\alpha}_{R'L',RL}$ that can intermix different angular momentum components of the muffin-tin orbitals.

\begin{figure}[htb]
\centering
\includegraphics[width=1\linewidth]{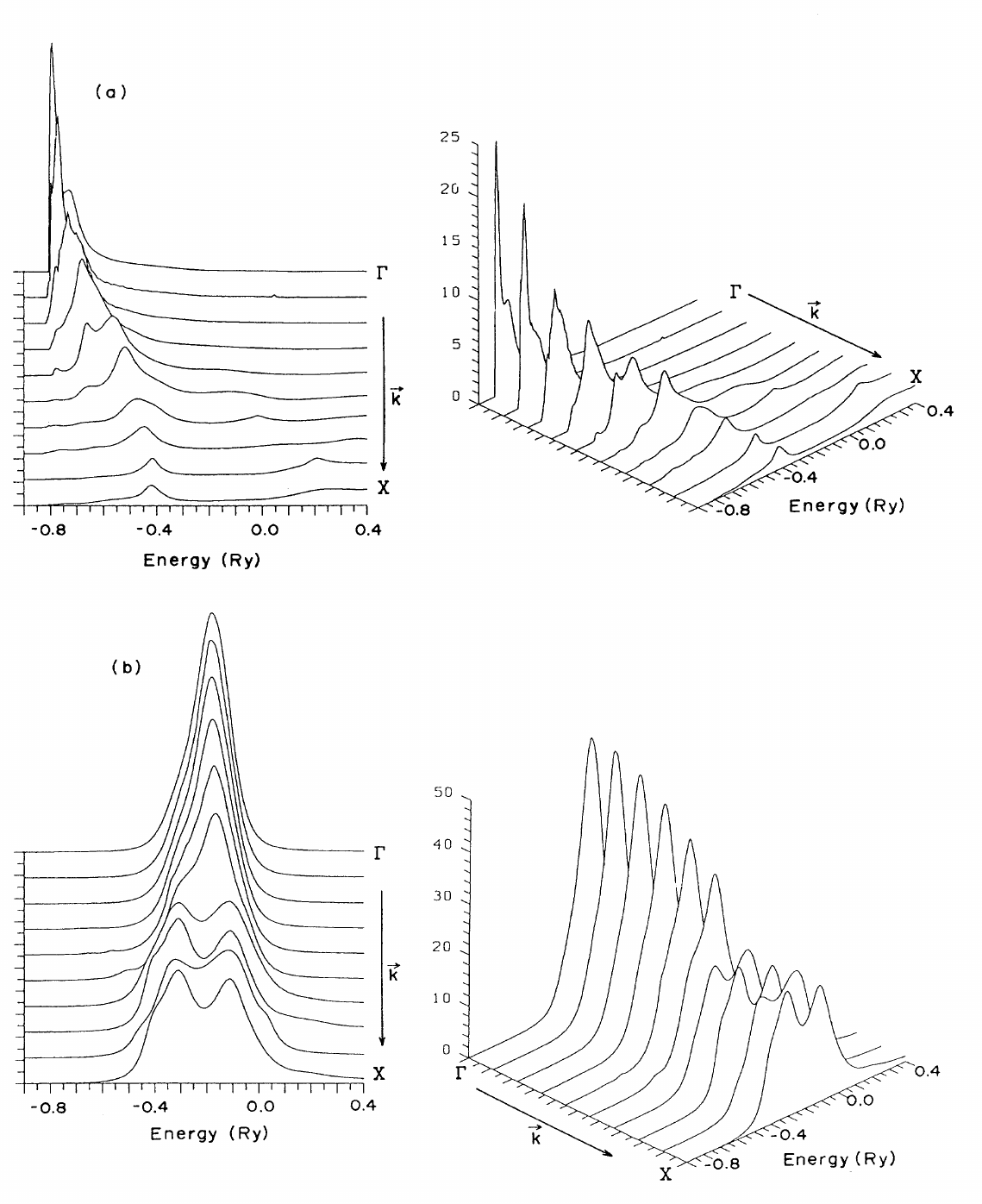}
\caption{Momentum-resolved spectral functions $n_{\mathbf{k}}^l(E)$ for the (a) $s$-orbitals and (b) $d$-orbitals in liquid Fe obtained with the TB-LMTO recursion method. The $k$ values range from zero to $2\pi/a$, with $a$ the lattice parameter of fcc Fe at the density of liquid Fe ($0.0756/\text{\AA}^3$). The results were obtained for a cubic cluster with 600 particles. From \cite{Bose1993}.}
\label{fig:*Bose1993_page7_crop}
\end{figure}

In a preliminary step towards calculating electronic transport, Bose {\it et al.}\ \cite{Bose1993} analyzed the momentum and angular momentum-resolved spectral functions as
\begin{align}
n_{\mathbf{k}}^l(E)=-\frac{1}{\pi}\lim\limits_{\epsilon\rightarrow 0^+}\Im\left\{\sum\limits_m\bra{u_{\mathbf{k}}^{l,m}}G(E+i\epsilon)\ket{u_{\mathbf{k}}^{l,m}}\right\},
\end{align}
with $|u_{\mathbf{k}}^{l,m}\rangle=\sum\limits_je^{i\mathbf{k}\cdot\mathbf{R}_j}|\chi^{\alpha}_{R_j,l,m}\rangle$ being the Fourier transform of the muffin-tin orbitals. In Fig.\ \ref{fig:*Bose1993_page7_crop}, the spectral function $n_{\mathbf{k}}^l(E)$ for liquid Fe obtained with the TB-LMTO recursion method is plotted. The spectral function is used to study residual dispersion and the effect of disorder on the $s$, $p$, and $d$ orbitals in the liquid phase.

Subsequently real-space calculations of the electrical conductivity were performed with the Kubo-Greenwood formula in the form
\begin{align}
\sigma_{jj}=\frac{e^2}{\Omega_a}n(E_\text{F})D(E_\text{F})
\end{align}
with $\Omega_a$ the atomic volume, $n(E_\text{F})$ the density of states at the Fermi -level and $D(E_\text{F})$ the diffusivity at $E_\text{F}$. The latter is given by
\begin{align}
D(E_\text{F})=-\hbar\lim\limits_{\epsilon\rightarrow 0}\Im\left\{\bra{E_m}v_jG(E_\text{F}+i\epsilon)v_j\ket{E_m}\right\}_{E_m=E_\text{F}},
\end{align}
which is equivalent to the zero temperature result of Eq.\ \eqref{equation:sigma_KG_T0} at the Fermi energy.

The transport approach was applied to the $3d$ transition metals Cr, Mn, Fe, Co and Ni. The calculated diffusivity has contributions from the purely diagonal $s$, $p$ and $d$ channel as well as contributions due to the hybridization of the orbitals induced by the off-diagonal matrix elements $h^{\alpha}_{R'L',RL}$. The analysis of the conductivity predicts that the conductivity due to the $d$-channel $\sigma_{dd}$ dominates the contributions $\sigma_{ss}$ and $\sigma_{pp}$ by a factor of five to six. Because the diffusivity $D_d$ is much smaller than $D_s$ and $D_p$ (7-12 times), this dominance is due to the higher weight in the DOS of the $d$-states. Additional mixed-channel contributions to the conductivity occur due to the hybridization of the $s$, $p$, and $d$ orbitals. They turned out to be substantial for these liquid metal systems, reaching the conductivities of the conventional channels but with opposite sign. Hence, the orbital mixing strongly impacts the conductivity and impedes the conduction process.

Finally, electrical resistivities were compared to experimetal results with good quantitative agreement (except for liquid Ni), which validates the TB-LMTO recursion method. Further studies of the electrical conductivity with the Kubo-Greenwood formula extended the application of the TB-LMTO recursion approach to a larger number of systems including liquid La, Hg, and metallic glasses \cite{Bose_1994, Bose1998, Bose_1999}. 

\subsubsection{Localization transitions in disordered Dirac semimetals}
   
In recent years $\mathcal{O}(N)$ approaches have also been used to study higher-dimensional materials such as three-dimensional Dirac semimetals \cite{PixleyPRL2015, DasSarma20162} or disordered Weyl fluids \cite{DasSarma2017, DasSarma20172, DasSarma2018}. In general, a Dirac semimetal is a condensed matter system where twofold Kramers-degenerate conduction and valence bands touch each other. These materials can be described with a massless Dirac equation in the infrared limit. In the undoped case the Fermi level lies exactly at the Dirac point where the bands touch each other.
Examples of Dirac semimetals are $\text{Cd}_2\text{As}_2$,  $\text{Na}_3\text{Bi}$,  $\text{Bi}_{1-x}\text{Sb}_x$,  $\text{Bi}\text{Tl}(\text{S}_{1-x}\text{Se}_{x})_2$,  $(\text{Bi}_{1-x}\text{In}_x)_2\text{Se}_3$, or  $\text{Pb}_{1-x}\text{Sn}_x\text{Te}$. 

Dirac semimetals can be modeled by a massless Dirac Hamiltonian in its non-covariant form that additionally involves a disorder potential. The underlying Dirac Hamiltonian is defined by
\begin{align}
H=\frac{1}{2}\sum\limits_{\mathbf{r},\hat{\mu}}\left(it\psi_{\mathbf{r}}^{\dagger}\alpha_{\mu}\psi_{\mathbf{r}+\mathbf{e}_{\hat{\mu}}}+H.c.\right)+\sum\limits_{\mathbf{r}}V(\mathbf{r})\psi^{\dagger}_{\mathbf{r}}A_W\psi_{\mathbf{r}},
\end{align} 
where $\psi_{\mathbf{r}}=(c_{\mathbf{r},+,\uparrow},c_{\mathbf{r},-,\uparrow},c_{\mathbf{r},+,\downarrow},c_{\mathbf{r},-,\downarrow})^T$ denotes the four components of a Dirac spinor referring to an electron at site ${\mathbf{r}}$ with parity ($\pm$) and spin ($\uparrow$$/$$\downarrow$); $\mathbf{e}_{\hat{\mu}}$ (with $\hat{\mu}=\hat{x},\hat{y},\hat{z}$) refers to a unit vector pointing to the nearest neighbor; and $\alpha_{\mu}=\sigma_{\mu}\otimes\mathds{1}_2$ are the $4\times4$ Dirac matrices that obey the anti-commutation relation $\left\{\alpha_{\mu},\alpha_{\nu}\right\}=2\delta_{\mu\nu}\mathds{1}_4$ according to the anti-commutation relations of the $2\times2$ Pauli spin matrices $\sigma_{\mu}$. 
The type of the disorder (symmetry of the disorder) is given by $A_W$. When it is diagonal in the spinor components ($A_W=\mathds{1}_4$), the disorder potential is just a scalar potential and $V(\mathbf{r})$ describes a random scalar potential at site $\mathbf{r}$ with strength $V(\mathbf{r})\in \left[-W/2,W/2\right]$. Off-diagonal terms are also studied, e.g.\ with an axial chemical potential $A_W=\gamma_5=i\alpha_1\alpha_2\alpha_3$.
 
The model study in \cite{PixleyPRL2015} determined the quantum phase transition of a Dirac semimetal into a conventional diffusive metal, and at larger disorder into an Anderson insulator. These quantum phase transitions can be determined by comparing the average and typical DOS, both of which can be calculated using the KPM. This was shown, e.g., by Wei\ss{}e \textit{et al.}, who used the KPM to demonstrate a metallic to insulating transition in the 3D Anderson model \cite{weisse2006rmp}. The average DOS is given by
\begin{align}
\rho_a(E)=\left<\frac{1}{4N_s}\sum\limits_{i=1}^{N_s}\sum\limits_{\alpha=1}^4\delta(E-E_{i\alpha})\right>,
\end{align}
with site index $i$ and orbital index $\alpha$ at eigenenergy $E_{i\alpha}$. $N_s$ is the size of the system and $\langle\dots\rangle$ represents the average over several realizations of disorder. The typical DOS is calculated as
\begin{align}
\rho_t(E)=\exp\left(\frac{1}{4N_s}\sum\limits_{i=1}^{N_s}\sum\limits_{\alpha=1}^4\left<\log\rho_{i\alpha}(E)\right>\right)
\end{align}
with 
\begin{align}
\rho_{i\alpha}(E)=\sum_{k,\beta}\left|\bra{k,\beta}\ket{i,\alpha}\right|^2\delta(E-E_{k\beta})
\end{align}
the local DOS. The above quantities for the average DOS and the typical DOS were calculated with the KPM. For a system size of $N_s=60^3$, model simulations of the average DOS and the typical DOS were performed with $N_\text{p}=1028$ Chebyshev polynomials for the average DOS and $N_\text{p}=8192$  Chebyshev polynomials for the typical DOS.
\begin{figure}[htb]
\centering
\includegraphics[width=\linewidth]{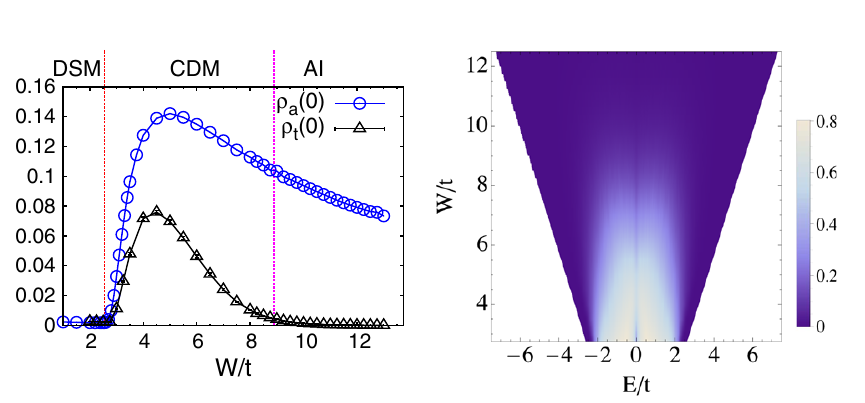}
\caption{Quantum phase transitions of a three-dimensional Dirac semimetal (DSM) to a conventional diffusive metal (CDM) and to an Anderson insulator (AI). Left: average DOS $\rho_a(0)$ and typical DOS $\rho_t(0)$ for a cubic system of size $N_s=60^3$ at the Dirac point $E=0$ as a function of the disorder strength $W$. For the average DOS the KPM was used with 1028 moments whereas the typical DOS required 8192 moments. Right: mobility edge as a function of Fermi energy and disorder strength. The white region represents metallic behavior, the blue region localized behavior, and the blank region is gapped. From \cite{PixleyPRL2015}.}
\label{fig:*DasSarma_2015_page2_crop_no_caption}
\end{figure}

Further analysis involved the evaluation of the inverse participation ratio of the wave function as an indicator of the localization transition. The average participation ratio is defined as
\begin{align}
P_{\text{avg}}=\left<\frac{\left[\sum_{i,\alpha}\left|\psi_{\alpha}(\mathbf{r}_i)\right|^2\right]^2}{\sum_{i,\alpha}\left|\psi_{\alpha}(\mathbf{r}_i)\right|^4}\right>
\end{align}
and was calculated for much smaller system sizes than used for the DOS calculation. From the inverse participation ratio $P_{\text{avg}}^{-1}$ the localization transition was obtained in full accordance with KPM-based results, which corroborates the approach. In essence, the authors in \cite{PixleyPRL2015} studied the transport properties of disordered Dirac semimetals via the density of states and compared the results to the inverse participation ratio, leading to the same findings for the localization transitions.

Subsequent publications were dedicated to further analyze localization transitions in Dirac semimetals \cite{DasSarma20162} and Weyl semimetals \cite{DasSarma20172, DasSarma2018} as well as the investigation of spectral properties of disordered Weyl fluids \cite{DasSarma2017} that could be tested by appropriate ARPES or STM measurements in undoped compounds dominated by neutral defects. We note that these studies do not explicitly calculate transport quantities, but highlight the power of KPM techniques for efficiently examining the behavior of exotic topological materials. They also represent an opportunity for further study of transport in these materials using the methods described in this review.

\subsection{Quantum transport in nanotubes and crystalline organic semiconductors with electron-phonon coupling}

Electron-phonon coupling (EPC) \cite{giustino2017rmp} plays a crucial role in many transport properties, notably in conventional superconductivity \cite{bardeen1958pr} and temperature-dependent electrical resistivity. Although EPC and electrical conductivity can be studied using first principles calculations combined with the Boltzmann transport equation, this method is computationally formidable for complex systems. EPC can also be rigorously taken into account in quantum transport calculations based on the LB method, but various approximations \cite{frederiksen2007prb, luisier2009prb, rhyner2014prb} have to be used in practical calculations and the computation is generally very expensive.

Phonons are lattice vibrations which are associated with deviations of the atom from their equilibrium positions $\bm{R}_i^0$. In the TB formalism, the hopping integral $\gamma_{ij}$ between atoms $i$ and $j$ is affected by the variation of the bond length between two atoms $R_{ij}(t) = |\bm{R}_j(t)-\bm{R}_i(t)|$. A simple relation between $\gamma(R_{ij})$ and $R_{ij}$ is $\gamma_{ij} \propto 1/R_{ij}^2$ \cite{harrison1989book}, although more sophisticated models \cite{porezag1995prb} can be constructed in specific materials. Based on the idea of distance-dependent hopping integrals, Roche \textit{et al.}\ proposed a method to take the EPC into account in the MSD formalism \cite{roche2005prl,roche2005prb}. In this approach, the EPC is encoded in a time-dependent TB Hamiltonian $H(\{\bm{R}_i(t)\})$, where the time dependence of the atom positions $\bm{R}_i(t)$ is induced by phonon modes (labeled by the phonon branch $\nu$ and wave vector $\bm{q}$) with amplitude $A_{\nu}(\bm{q})$, frequency $\omega_{\nu}(\bm{q})$, and polarization $\bm{e}_{\nu}(\bm{q})$,
\begin{equation}
\bm{R}_i(t) = \bm{R}_i^0 + A_{\nu}(\bm{q}) \bm{e}_{\nu}(\bm{q}) \cos\left(\bm{q} \cdot \bm{R}_i^0 + \omega_{\nu}(\bm{q})t \right).
\end{equation}
The total simulation time in the MSD formalism is divided into a number of time intervals which are about one-tenth of the oscillation period of the considered phonon mode. The electron Hamiltonian is kept constant during each time interval and is updated after each time interval according to the updated atomic positions. In this way, the electron wave propagation is coupled to the phonons.

This dynamical off-diagonal disorder for electrons can also be modeled by combining the quantum evolution of the electronic wave function and the classical evolution of the lattice sites \cite{troisi2006prl}. An approach combining the MSD approach and molecular dynamics (MD) simulations has also been developed \cite{ishii2009crp, ishii2010prb}, where the atomic positions are updated according to interatomic forces from an empirical potential. Using these methods, the impact of EPC on quantum decoherence in carbon nanotubes has been quantified \cite{roche2005prl, roche2005prb, roche2007jpcm, ishii2010prl}.

While dynamical disorder from EPC is responsible for decoherence, a static disorder approximation can be used when the purpose is to compute the phonon-limited electron mobility. Based on the Born-Oppenheimer approximation, the electrons essentially experience a static potential profile associated with an instantaneous atomic configuration. In this approximation, one only needs to use MD simulations to generate a few equilibrated configurations and calculate the transport properties for each one separately. When the simulation system is large, the results from different configurations should not differ significantly. This approach has been used to study the phonon-limited electrical conductivity and mobility in suspended single-layer graphene with large-scale ripples spontaneously formed at room temperature \cite{fan2017tdm}. It was found that the semiclassical conductivity is a constant and the mobility is inversely proportional to the carrier density, in good agreement with results obtained by using the many-body perturbative GW approximation \cite{li2013prb}. A similar static disorder approximation has been used in the LB approach, where harmonic lattice dynamics \cite{liu2015prb} or classical MD simulations \cite{markussen2017prb} are used to generate equilibrated configurations at a given temperature, and the electron transmissions in these systems are then calculated by combining DFT and nonequilibrium Green's function calculations. Instead of using MD simulations or stochastic sampling, Gunst \textit{et al.} showed that a single ``special thermal displacement'' (STD) of the atoms in a large supercell can give the correct thermal average of the LB conductance and phonon-assisted current \cite{gunst2017prb}. This STD method would be an optimal way to include EPC in linear scaling quantum transport methods for large systems.

\begin{figure}[htb]
\begin{center}
\includegraphics[width=\columnwidth]{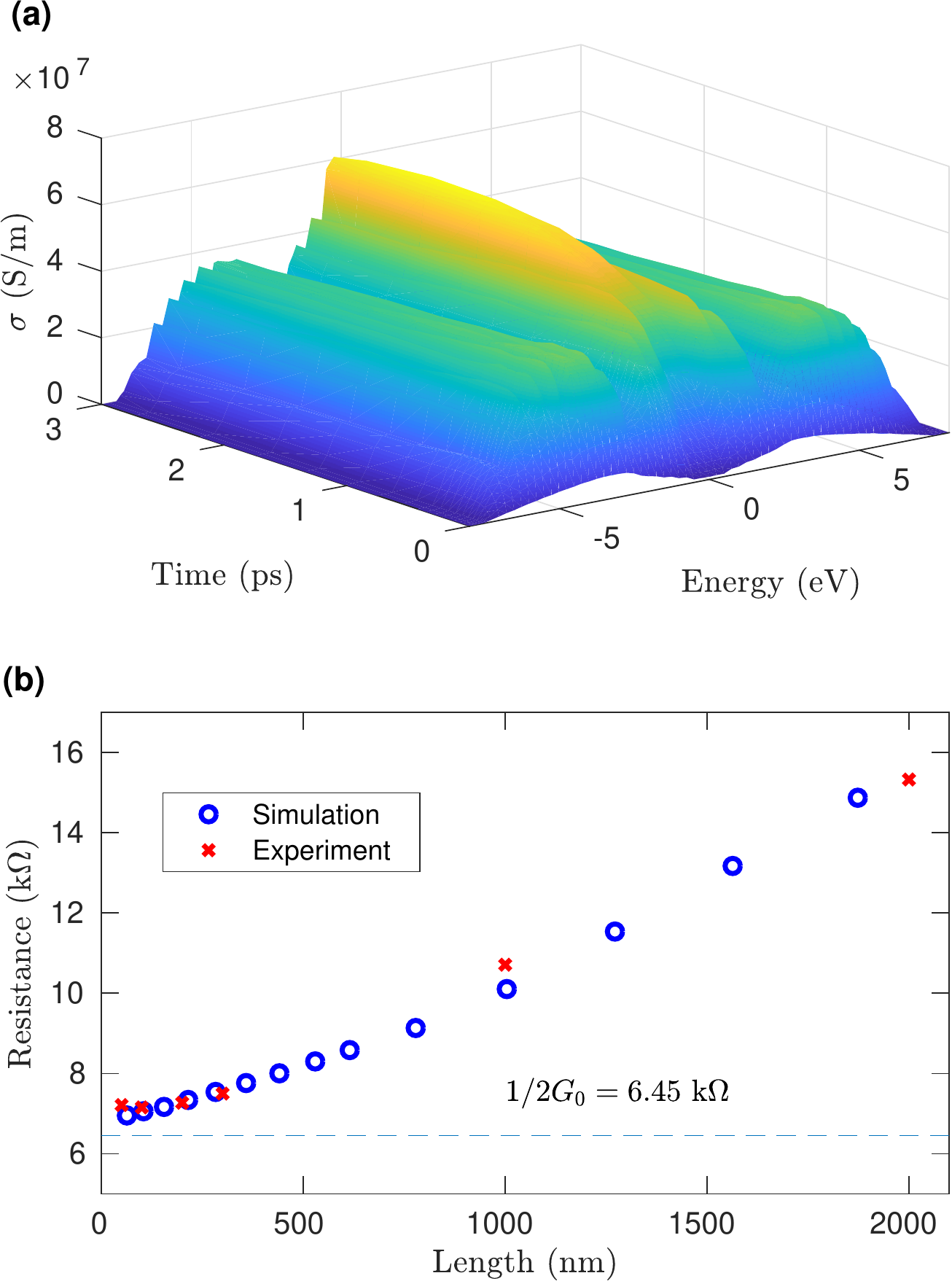}
\caption{(a) Surface plot of the conductivity $\sigma(E,t)$ of a $(14,14)$-CNT as a function of the Fermi energy $E$ and simulation time $t$ calculated using the MSD-KPM method. A thickness of $0.335$ nm for the carbon wall was assumed to calculate the volume of the CNT. (b) Simulated resistance of the CNT as a function of length (circles) compared with the experimental data (crosses) \cite{park2004nl}. A simulation cell $5$ $\upmu$m long (with $1.12 \times 10^6$ carbon atoms) was chosen. First, a classical MD simulation was used to generate a relaxed configuration at 300 K. Then, the electronic Hamiltonian was constructed based on the relaxed configuration and the length-dependent hopping parameter, $\gamma_{ij}=\gamma_0(R_0/R_{ij})^2$, where $\gamma_0=-2.5$ eV. The results were obtained by using the MSD-KPM method with $M=2000$ moments and $N_r=10$ random vectors. Adapted from \cite{fan2018cpc}.}
\label{figure:cnt}
\end{center}
\end{figure}

As an example of the application of the static disorder approximation for EPC, we show results \cite{fan2018cpc} for electron transport in a single-walled metallic $(14,14)$-CNT with a diameter of about $1.8$ nm, which is comparable to that reported in prior experiments \cite{park2004nl}. Figure \ref{figure:cnt}(a) shows $\sigma(E,t)$ calculated using the MSD-KPM method. We see that for the entire energy spectrum, the conductivity converges well up to a simulation time of $3$ ps. The ballistic-to-diffusive transition is clearly seen in Fig.\ \ref{figure:cnt}(b), where the resistance (the inverse of the conductance defined in Eq.\ (\ref{equation:ballistic_conductance})) at $E=0$ (corresponding to the low-bias situation in the experiments) as a function of the channel length is shown. In the short-length limit, the resistance approaches the ballistic value of $1/(2G_0)=h/4e^2=6.45$ k$\Omega$ (there are two conducting channels at the charge neutrality point). In the long-length limit, the resistance scales linearly with the channel length, which is the expected diffusive behavior. The good agreement with experiments demonstrates the applicability of the static disorder approximation in this case and the predictive power of the MSD-based LSQT method.

More complex systems where LSQT approaches have been applied are organic semiconductor crystals, which are used for instance in organic transistors \cite{PodzorovRMP2006}. While charge transport in organic crystals has been studied extensively over the last few decades, the microscopic picture of transport and the crossover between different mechanisms is still not fully clear. The higher complexity in such systems compared to CNTs stems from the large number of molecular vibrations and from the electronic anisotropy. The former influence the electronic properties in different ways depending on temperature, while the latter induces an anisotropic transport behavior. To understand how EPC affects charge transport beyond simple models is a key requirement for predicting the electrical conductivity of crystalline organic semiconductors, and the efficiency of LSQT approaches makes them a useful tool for tackling such complexity in the limit of coherent electronic transport.

Different theoretical approaches exist to include the EPC for intra- and intermolecular vibrational modes in the Kubo transport framework based on the LSQT methods reviewed in this article. The EPC of high-frequency modes can be treated within polaron theories \cite{ortmann2009prb,ortmann2011prb} that take into account their full quantum mechanical nature with a non-adiabatic approach, while other implementations of the EPC use a mixed classical-quantum mechanical description \cite{troisi2006prl, ciuchi2011prb, ishii2012prb}. This mixed description is often referred to as the adiabatic limit as the nuclei are treated semiclassically.

\begin{figure}[htb]
\centering
\includegraphics[width=1.05\linewidth]{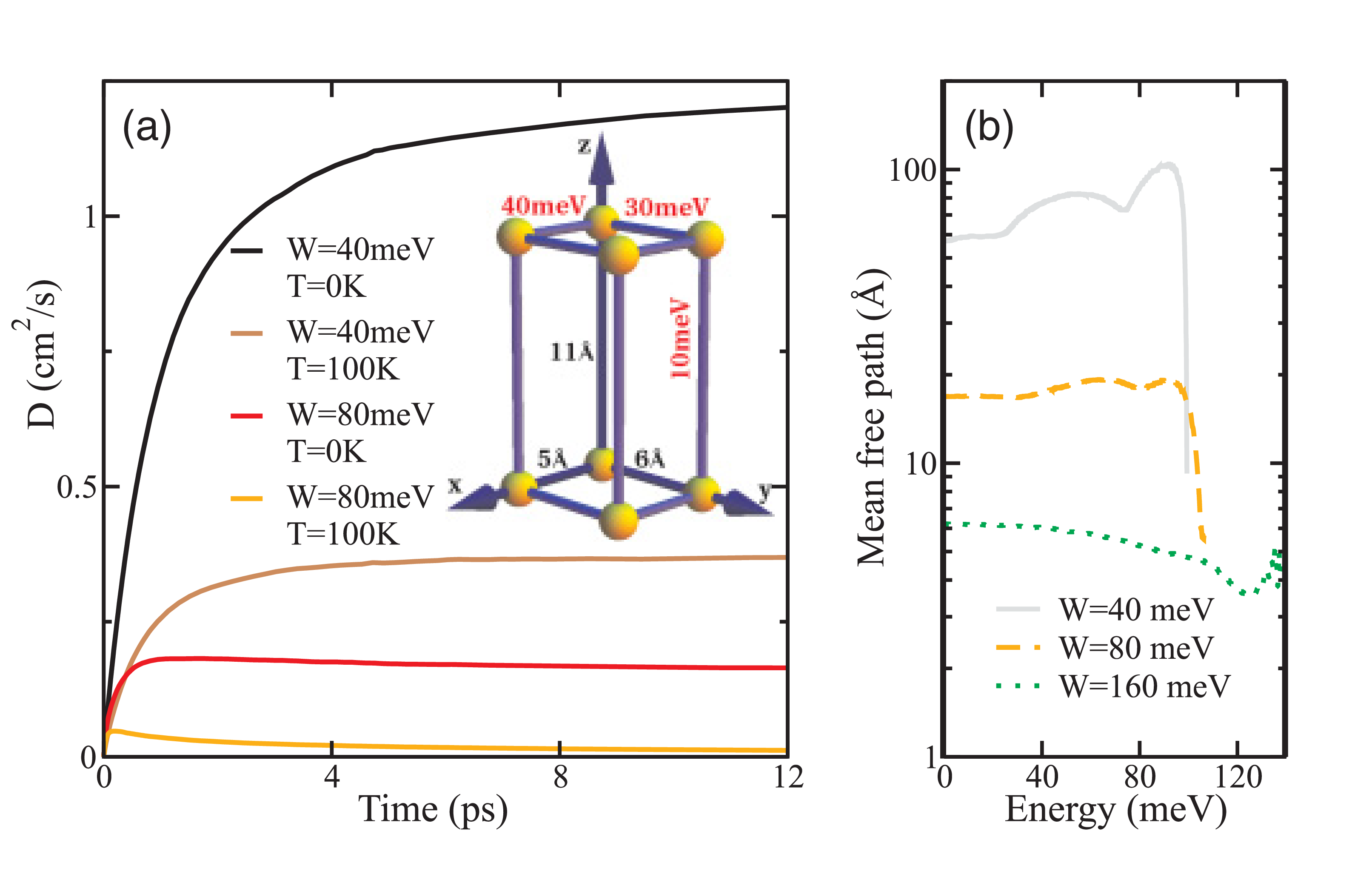}
\caption{Charge transport in a three-dimensional model of an organic crystal, calculated with the MSD-KPM method. (a) Typical time-dependent diffusion coefficient for various strengths of the disorder $W$ and temperatures $T$. Inset: three-dimensional crystal structure where system parameters $\varepsilon_{MN}$ and $R_{MN}$ are transfer integrals and distances to the nearest neighbors. (b) Polaron mean free path vs. energy for various disorder strengths. Only positive energies are shown because of symmetry. The simulations were performed on a system of $(0.24\times 0.22 \times 0.14)$ $\upmu$m$^3$, corresponding to $N_s=4\times10^7$, with a simulation time of 61 ps, a time step of 5$\hbar/(40$ meV) $\approx 82$ fs and 88 Chebyshev polynomials for the time evolution. Adapted from \cite{ortmann2011prb}.}
\label{fig:LSQT_in_Organics}
\end{figure}

The work of Ortmann and Roche \cite{ortmann2011prb} used a non-perturbative description of the EPC in organic crystals via a polaron transformation that takes into account the transfer integrals in all directions. Similar to the original work of Holstein \cite{Holstein1959}, a coherent phonon dressing results in a renormalization of the electronic bandwidth that depends on temperature. By considering the finite electronic bandwidth of the charge carriers, this theory overcomes the limitations of narrow-band transport theories (see e.g.\ \cite{HannewaldBobbertPRB2004, Silbey2008}). For instance, the MSD-KPM method allows studies of localization due to an interplay of electronic transfer integrals and disorder \cite{ortmann2011prb}. By combining polaron dressing and disorder effects, this approach enables access to transport parameters such as mean free paths or diffusion constants during the coherent propagation of polaronic wave packets. 	
	
In Fig.\ \ref{fig:LSQT_in_Organics} we show the time-dependent diffusion constant and the energy-resolved mean free path of a three-dimensional cubic model of an organic crystal, parametrized as shown in the inset. At high enough temperature and disorder strength, the transport regime changes from diffusive to localized, as seen in panel (a). The diffusion constant depends on the polaronic bandwidth, and thus the carrier mobility varies with temperature through the temperature dependence of the bandwidth. In addition, disorder-induced localization is apparent at low temperatures and is reduced with increasing temperature. The combination of both of these effects may induce a transition of the transport regime from band-like to hopping transport with increasing temperature.

Troisi \textit{et al.}\ \cite{troisi2006prl} have presented a transport approach based on a microscopic description of dynamical lattice disorder within the adiabatic regime. In the adiabatic approximation, the electronic transfer integral is assumed to exceed typical vibrational frequencies by one order of magnitude (or at least a large factor) and thus leads to a semiclassical treatment of the vibrational modes and the EPC. The carrier's MSD and the diffusion constant for a coherently-propagated electronic wave packet can then be calculated based on a mixed quantum-classical description employing Ehrenfest equations.

A related numerical approach utilizing LSQT methods was applied to organic semiconductors by other authors \cite{ishii2012prb}, where several transport scenarios were investigated in a pentacene model system including intra- and intermolecular EPC as well as static disorder. Their numerical approach (referred to as time-dependent wave packet diffusion, TD-WPD) is an extension of an approach that was successfully applied to CNTs and graphene nanoribbons -- during the propagation of the electronic wave packet the system Hamiltonian is updated at each time step according to the molecular dynamics of the vibrational modes. As described above the approach considers the modes dynamically, leading to polaronic effects on the initial electronic wave packet. Additional static disorder effects and the interplay with the EPC were also studied in this work.
	
In the same spirit, Ciuchi \textit{et al.}\ \cite{ciuchi2011prb} demonstrated that the lattice dynamics of low-frequency intermolecular modes lead to a localization of the charge carriers on the time scale below a vibration period, at which the lattice is assumed to be frozen. Here rubrene is taken as the reference system, which has been intensively studied as a prototype compound in recent years \cite{SundarScience2004, PodzorovPRL2004, TroisiAdvMat2007, KeraPRL2010, MasinoPRB2010, OrdejonPRB2017}. The results from their LSQT-based method suggest that charge carriers (after an initial localization) remain in a diffusive transport regime with diffusion constant $D$ leading to finite carrier mobilities according to $\mu(T)=eD/k_{\text{B}}T$. Implementing this idea, they proposed an exponential decay of the velocity correlation function over time with an inelastic scattering time $\tau_{\text{in}}$ that is on the order of the vibrational period of a typical inter-molecular mode, i.e., $\tau_{\text{in}}\propto\omega_{\text{inter}}^{-1}$.

This relaxation time approach is designed to counteract the localization phenomenon, which eventually results in finite mobilities, in contrast to the semiclassical Ehrenfest method proposed earlier. Indeed, it has been shown that the latter suffers from an increase of the velocity correlation function, leading to an increase of the time-dependent diffusion constant at time scales above the period of the inter-molecular vibrations, which results in diverging carrier mobilities. Employing the relaxation time approach, the diffusion constant is obtained from the so called transient localization length and the inelastic scattering time via $D=L^2(\tau_{\text{in}}) / 2\tau_{\text{in}}$. Recently this approach has been applied to charge transport properties in two-dimensional herringbone structures \cite{FratiniNatMat2017}. The anisotropy of the electronic coupling (distribution of transfer integrals) was studied and connected to the localization behavior and the carrier mobility.
	
The emerging picture from these various studies using LSQT approaches is that there is a partial localization of charge carriers induced by disorder that can have vibrational or static origin. The spatial extent, or localization length, is still difficult to predict, since it is influenced both by high-frequency molecular vibrations leading to polaronic effects and by semiclassical dynamical disorder leading to localization. Since each can enhance the other, a combination of different approaches including those for high frequencies and low frequencies is desirable. The present success of the efficient numerical approaches in these studies suggests that future developments might emerge based on similar methods \cite{Panhans2020NatCommun} and that the material classes to which they are applied will be extended \cite{Mayou2020PRL}.

%% file: secVII_hall_spins.tex
\section{Hall and spin transport }
\label{section:hall_spin}

\subsection{Topological and Fermi surface contributions}

In the previous sections, we explained how to combine different numerical techniques to compute the diagonal conductivity from different representations of the Kubo-Greenwood formula in a linear scaling way. However, it is nontrivial to extend this approach to study other transport properties such as the Hall conductivity. The reason is that the Kubo-Greenwood formula only captures the Fermi level properties of the system, but as shown by Thouless \textit{et al.}\ in their seminal work \cite{ThoulessPRL1982}, some quantities are defined in terms of the topology of the electronic structure and therefore depend on the whole energy spectrum. This means that in order to compute a general observable, one should first determine whether the topological contributions are negligible and choose the appropriate methodology accordingly.

\begin{figure}[htb]
\centering	\includegraphics[width=0.9\linewidth]{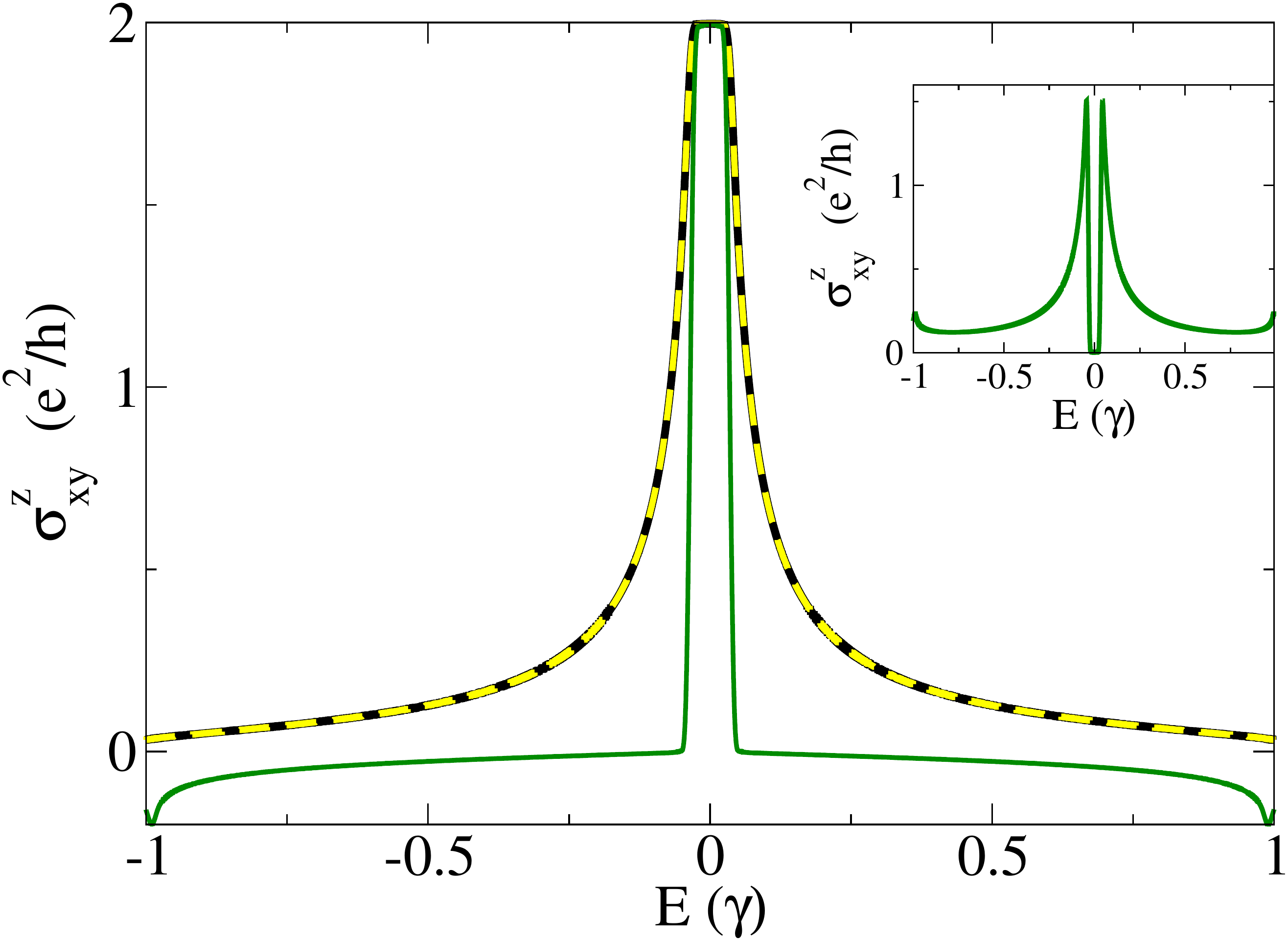}
	\caption{ Comparison between the different decompositions of the spin Hall conductivity applied to the Kane-Mele model considering a SOC strength $\lambda_\text{I} = 0.178\gamma$, where $\gamma$ is the nearest-neighbor transfer integral. The full spin Hall conductivity obtained from the Kubo-Bastin formula (black solid line) matches inside the gap with both Streda's second component $\sigma_{xy}^{z,\rm II}$ (green solid line) and the topological contribution $\sigma_{xy}^{z,\rm T}$ (yellow dashed line) yielding to the same quantized value. 
	However, outside the gap, $\sigma_{xy}^{z,\rm II}$ differs from the Kubo-Bastin formula due to a finite contribution of $\sigma_{xy}^{z,\rm I}$ which is shown as in inset. We verified that summing both contributions in each decomposition lead to the full Kubo-Bastin result. The simulation was performed on a system containing $4 \times 10^6$ atoms, and the Green's functions were approximated using the KPM with 2000 moments.}
\label{fig:sea-vs-level}
\end{figure}

The Hall conductivity is a quantity for which topological effects are prominent. In Sec.\ \ref{section:kubo_formulas} we demonstrated that the Kubo-Bastin formula, Eq.\ (\ref{equation:sigma_mu_nu_Bastin}), is the single-particle approximation of the general Kubo formula, and as such, should contain both the topological and the Fermi level contributions. Therefore, it is reasonable to assume that the electrical response $A\equiv\langle \hat{A}\rangle$ of an arbitrary operator $\hat{A}$ can be separated into two different contributions 
\begin{equation}
A = A^{\rm FS} + A^{\rm T}.\label{streda_separation}
\end{equation}
where $A^{\rm FS}$ and $A^{\rm T}$ representing the Fermi level and topological contribution respectively. 

The first attempt of such decomposition was presented in the seminal work of Streda \cite{Streda1982PRC,crepieux2001prb}, which he  developed for the Hall conductivity and was later extended to an arbitrary operator \cite{cresti2016rnc}. He showed that by using algebraic manipulation and Green's function properties the Kubo-Bastin formula can be broken into two contributions 
\begin{align}
A^{\rm I} = \, &\hbar \Omega \int_{-\infty}^\infty dE' \dv{f(E'-\mu )}{E'}\times \nonumber\\
 &\text{Im}\left(\text{Tr}\left[ \delta(\hat{H}-E')\hat{A}\,G^+(E')\,(\hat{\bm{J}}\cdot \bm{E}_0)\right] \right)
 \label{KuboStredaFermiLevel}
\end{align}
and
\begin{equation}
A^{\rm II} =   \frac{\hbar \Omega}{2\pi}\text{Re}\left(\int_{-\infty}^\infty dE' f(E'-\mu)\text{Tr}[\hat{B}]\right) \label{KuboStredaFermiSea}
\end{equation}
where
\begin{equation}
\hat{B}= \left\{G^+(E')\hat{A}\,\dv{G^+(E')}{E'} -\dv{G^+(E')}{E'} \hat{A}\,G^+(E')  \right\}\hat{\bm{J}} \cdot \bm{E}_0,
\end{equation}
and later demonstrate that {\it{in the band-gap }} they correspond to the Fermi surface term $A^{\rm FS}=A^{\rm I}$ and the topological term $A^{\rm T}=A^{\rm II}$ respectively.  However, it was recently demonstrated by Bonbien \& Manchon \cite{Bonbien202Arxvi} that such separation fails outside the band-gap due to an overlap between $A^{\rm I}$ and $A^{\rm II}$ . The same authors presented a new decomposition of the Kubo-Bastin formula that fixes the overlap problem and also consist in two contributions, the first one 
\begin{align}
A^{\rm FS} = \, & -\pi \hbar \Omega \int_{-\infty}^\infty dE' \dv{f(E'-\mu )}{E'}\times \nonumber\\
 &\text{Re}\left(\text{Tr}\left[ \delta(\hat{H}-E')\hat{A}\delta(\hat{H}-E')\,(\hat{\bm{J}}\cdot \bm{E}_0)\right] \right)
 \label{KuboFermiLevel}
\end{align}
has exactly the same shape as the Kubo-Greenwood formula used for quantum transport calculations, allowing then for exploiting all numerical artillery developed through this review. The second contribution
\begin{align}
A^{\rm T} = \, & - \hbar \Omega \int_{-\infty}^\infty dE' f(E'-\mu )\times \nonumber\\
 &\text{Im}\left(\text{Tr}\left[ \delta(\hat{H}-E')\hat{A}\dv{(G^+(E')+G^-(E'))}{E'}\,(\hat{\bm{J}}\cdot \bm{E}_0)\right] \right)
 \label{KuboFermiSea}
\end{align}
is similar to $A^{\rm II}$ in the sense that depends solely on the Fermi-Dirac distribution and contain derivatives of the Green's functions.

To illustrate how such procedures allows to determine the quantized conductivity arising from the quantum Hall and quantum spin Hall effects, in Fig \ref{fig:sea-vs-level} we show the intrinsic spin Hall conductivity of the Kane-Mele model \cite{KaneMele2005}, this quantity is expected to originate solely from the Berry curvature and should display negligible Fermi level contributions. This model describes the electronic behavior of a system composed of a honeycomb lattice with nearest-neighbor hoppings and strong intrinsic spin-orbit coupling (SOC) characterized by a strength $\lambda_\text{I}$. This system is a topological insulator, with a bulk gap and topological edge states for $|E|<\lambda_\text{I}$, leading to a quantized spin Hall conductivity for the same range of Fermi energies. This calculation was performed using a system of $4 \times 10^6$ orbitals and the KPM, following the methodology developed by Garcia \textit{et al.}\ \cite{garcia2015prl}, which will be discussed in detail in the next subsection. As one can see, both decomposition capture the topological contributions leading to a quantized conductivity in the band-gap. However, deep in the conduction or valence bands, both components of the Streda decomposition yield to sizeable contributions, which is not expected for this system, on the other hand, the decomposition by Bonbien \& Manchon do yield to the expected result, i.e, a vanishing Fermi-level contribution. It should be noted that the both decompositions reduces to Aoki's formula when used for the Hall conductivity \cite{aokisscomm1981, aoki1985prl},
\begin{equation}
\sigma_{\mu\nu} =-\lim_{\eta\rightarrow 0}\frac{i\hbar}{\Omega}\sum_{m,n}f(E_m)\frac{\bra{E_m}\hat{J}_\mu \ket{E_n}\bra{E_n}\hat{J}_\nu\ket{E_m}}{E_m-E_n+i\eta}+\text{h.c.}, \label{aokiEq}
\end{equation}
which is commonly used to compute the topological conductivity through exact diagonalization.

\subsection{Numerical implementations of the Kubo-Bastin formula}
\label{KuboHallNumm}

Previously, we showed how to use the KPM and the time evolution approaches for approximating the Dirac delta function and the Green's function. These approximations can also be applied to the Kubo-Bastin formula. The simplest approach is to expand the Green's function in terms of a polynomial series and regularize it by either using the KPM or by including a finite but small broadening $\eta$. The advantage of this approach is that the energy derivative, present in the Kubo-Bastin formula, only affects the Chebyshev coefficients and therefore can be done analytically. After insertion of the Chebyshev series into Eq.\ (\ref{equation:sigma_mu_nu_Bastin}) and the application of the corresponding derivative to the Chebyshev coefficients in Eq.\ (\ref{KPMGreenFuncEta0}), one obtains the following expression for Kubo-Bastin formula \cite{garcia2015prl},
\begin{equation}
\langle \hat{A} \rangle=\hbar \Omega E_\alpha  \int dE f(E'\mu) \sum_{m,n} \Gamma_{m,n}(\widetilde{E}) \mu_{mn}^{\alpha}, 
\label{HallJose}
\end{equation}
where $\widetilde{H}$ is defined in Eq.\ (\ref{equation:scale}) and $\widetilde{E}$ is defined similarly, and where
\begin{equation}
\mu_{mn}^{\alpha}\equiv g_m g_n  \text{Tr}\left[ \hat{J}_\alpha T_m(\widetilde{H})\,\hat{A}\, T_n(\widetilde{H}) \right]
\end{equation}
are the multi-dimensional version of the Chebyshev moments \cite{weisse2006rmp} and 
\begin{align}
\Gamma_{m,n}(\widetilde{E}) &\equiv\frac{4}{\Delta E^2(1-\widetilde{E}^2)^2(1+\delta_{m0})(1+\delta_{n0})}\nonumber\\  
& \times \Big((\widetilde{E} - i n \sqrt{1- \widetilde{E}})\text{e}^{in \text{arccos}(\widetilde{E})}T_m(\widetilde{E}) \nonumber\\
& + (\widetilde{E} + i m \sqrt{1- \widetilde{E}})\text{e}^{-i m \text{arccos}(\widetilde{E})}T_n(\widetilde{E}) \Big)
\end{align}
are energy-dependent Chebyshev coefficients. 

The second approach is based on Lanczos recursion and the time-dependent Kubo-Bastin formula presented in Eq.\ (\ref{equation:sigma_mu_nu_Bastin}), and has been called the time-evolution Kubo (TEK) approach. Although the simulation time for the transversal components of the DC conductivity is increased compared to the simulation time for the longitudinal components (by a factor of about 500-5000 depending on the number of Lanczos vectors), the time evolution of the studied quantity usually provides more physical insight into the mechanism leading to the stationary state, as already discussed in Sec.\ \ref{section:kubo_formulas}. The core of this method lies in the approximation of the completeness relation by random-phase vectors \cite{ortmann2013prl,ortmann2015prb},
\begin{equation}
\mathds{1}\approxeq\sum\limits_{j=1}^{N_\text{R}}\ket{\phi_j}\bra{\phi_j},
\end{equation}
where $N_\text{R}$ is the number of Lanczos recursion steps and the set $\{\ket{\phi_j}\}$ are random phase vectors as defined in Sec.\ \ref{section:trace}. This identity can then be inserted into Eq.\ (\ref{equation:sigma_mu_nu_Bastin}) in order to obtain an alternative representation of Eq.\ (\ref{HallJose}),
\begin{align}
\begin{aligned}
& \langle \hat{A(\mu)}  \rangle = 4E_0\Omega\lim_{\eta\rightarrow 0^+}  \int\limits_0^{t_\text{c}}\frac{dt}{2\pi}\,{\rm e}^{-\eta t/\hbar}\int\limits_{-\infty}^{\infty}dEf(E,\mu,T) \\
& \times  \sum\limits_{j=0}^{N_\text{R}}\Im\left[\kappa_j(E)\right] \Re \left[\bra{\phi_j} \bm{E}_0 \cdot \hat{\bm{J}}  G^+(E)\hat{A}(t)\ket{\phi_1} \right],
\label{Hall conductivity}
\end{aligned}
\end{align}
where
\begin{equation}
\kappa_j(E) = \bra{\phi_j} G^+(E) \ket{\phi_1}
\end{equation}
are the elements of the first column of the matrix-valued Green's function. This numerical implementation avoids the tedious computation of the eigensystem by using a combination of $\order{N}$ techniques. The conductivity can then be obtained from numerical simulations using the formerly introduced Lanczos algorithm and continued fraction expansion for the calculation of the $\kappa_j(E)$. The $\kappa_j(E)$ are defined recursively with the initial element $\kappa_1(E)=\bra{\phi_1}G^+(E)\ket{\phi_1}$ being related to the DOS $\rho(E)$ of the system via
\begin{equation}
-\frac{1}{\pi} \text{Im} \left[ \kappa_1(E) \right] = \rho(E).
\end{equation}
 In the second step, $\kappa_2(E)$ is
 \begin{align}
  \kappa_2(E)=\frac{1}{b_1}\left(-1+(E-a_1+i\eta)\kappa_1(E)\right).
  \label{equation:kappa_2}
 \end{align} 
For $n>2$, we find the recursion relation
\begin{align}
\kappa_{n+1}(E)=\frac{1}{b_n}\left(-b_{n+1}\kappa_{n-1}(E)+(E-a_n+i\eta)\kappa_n(E)\right).
  \label{equation:kappa_n}
\end{align}
In Eqs.\ (\ref{equation:kappa_2}) and (\ref{equation:kappa_n}), the coefficients $a_n$ and $b_n$ are the matrix elements of the tridiagonal matrix obtained from the Lanczos algorithm for the initial random phase vector $\ket{\phi_1}$. In addition, the Chebyshev polynomial expansion method is used for the time evolution operator $\hat{U}(t)$ as explained in Sec.\ \ref{section:numerical_techniques}. This approach will be illustrated in the next subsection for calculations of the quantum Hall effect in graphene.
 
\subsection{Quantum Hall effect}

One canonical example where the topological contribution plays a dominant role is the quantum Hall effect. When a two-dimensional system is subjected to a perpendicular homogeneous magnetic field, under appropriate conditions the electrons will move in degenerate orbitals which for certain Fermi energies will produce bulk insulating behavior and quantized edge currents, both originating from the topology of the band structure. This effect, and the interaction of the topological states with disorder, has been studied numerically using the two implementations of the Kubo-Bastin formula presented in the above subsection \cite{garcia2015prl,ortmann2015prb}.

As an example, we discuss the quantum Hall effect in disordered graphene. The Hamiltonian for this system is
\begin{align}
\hat{H}=\sum\limits_iV_i\hat{c}_i^{\dagger}\hat{c}_i-\sum\limits_{ij}\gamma \text{e}^{-i\phi_{ij}}\hat{c}_i^{\dagger}\hat{c}_j \label{Hamiltonian Graphene oniste potential}
\end{align}
with the nearest neighbor transfer integral $\gamma=\gamma_0=2.7$ eV. To include disorder, we use an uncorrelated Anderson model with matrix elements $V_i$ taken at random from the interval $\left[-W\gamma_0/2,W\gamma_0/2\right]$. The strength of the disorder in units of the nearest neighbor transfer integral is given by $W$. 

The constant magnetic field $\bm{B}=\nabla\times\bm{A}$ is implemented via a Peierls phase \cite{Luttinger1951PRL}, leading to an additional phase evolution $\phi_{ij}$ that modifies the transfer integral between the sites $i$ and $j$ as
\begin{align}
\phi_{ij}=\frac{e}{\hbar}\int\limits_{\bm{r}_i}^{\bm{r}_j}d\bm{r}\cdot\bm{A}.
\end{align}
\begin{figure}[htb]
\centering
\includegraphics[width=1\linewidth]{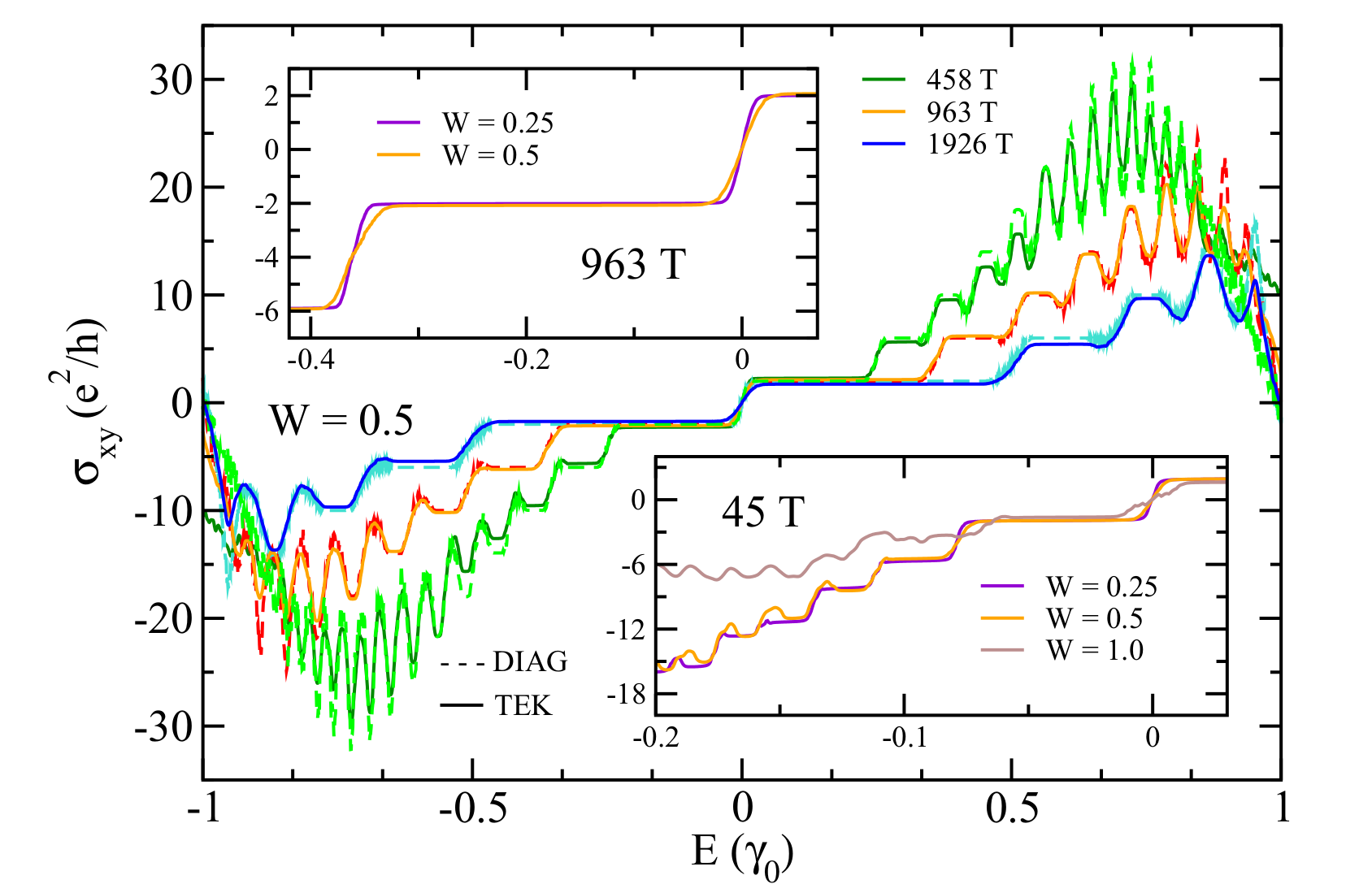}
\caption{Hall conductivity of graphene with Anderson disorder (dashed curves, exact diagonalization; solid curves, TEK) at different magnetic fields (main frame). Effect of increasing disorder for high field 964 T (upper inset) and intermediate field 45 T (lower inset). The simulation was performed in a system of 10 million atoms, using one random vector and the Lanczos algorithm with a broadening of 0.002$\gamma_0$ and at least 1000 Lanczos recursion steps. Adapted from \cite{ortmann2015prb}.}
\label{fig:FO_NBL_SR_PRB_2015_fig3}
\end{figure}

The numerical results for the Hall conductivity $\sigma_{xy}$, which is obtained by replacing $\hat{A}=\hat{{J}}_y$ in the Kubo-Bastin formula, are shown in Fig.\ \ref{fig:FO_NBL_SR_PRB_2015_fig3}. The quantization of the Hall conductivity, following the sequence of steps according to $\sigma_{xy}=\pm4\left(\frac{1}{2}+n\right)\frac{e^2}{h}$, reproduces experimental measurements \cite{Novoselov2005Nature,Zhang2005}. The results are plotted for large and intermediate magnetic field strengths. A comparison of our method with the results from exact diagonalization yields high quantitative agreement.

\begin{figure}[htb]
\centering
\includegraphics[width=\columnwidth]{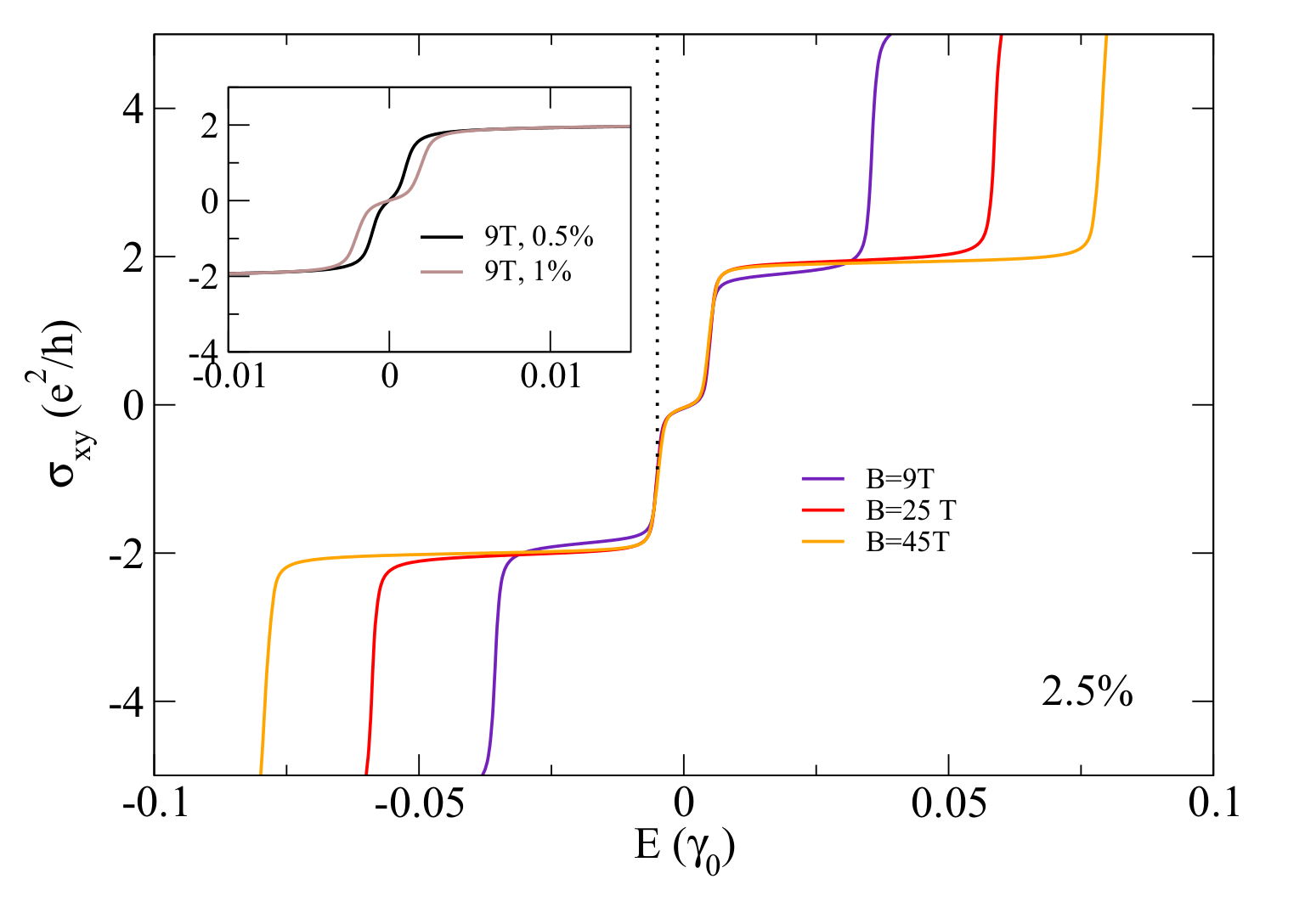}
\caption{Hall conductivity of graphene for $p=2.5\%$ of AB sublattice-breaking defects (strength $V_s=0.2\gamma_0$) distributed randomly in space. The vertical dotted line indicates the nominal gap of a correspondingly homogeneous AB potential of strength $pV_{AB}=0.005\gamma_0$. The simulation was performed in a system of about $10^6$ atoms, with one random vector and the Lanczos algorithm with a broadening of $0.0004\gamma_0$ and 3600 Lanczos recursion steps. Adapted from \cite{ortmann2015prb}.}
\label{fig:FO_NBL_SR_PRB_2015_fig4}
\end{figure}

Using LSQT methods, Ortmann \textit{et al.}\ also examined the impact that sublattice-dependent disorder can have on the quantum Hall effect in graphene \cite{ortmann2015prb}. This was done by including an additional sublattice-symmetry breaking potential according to $V_{i}\rightarrow \left(V_{i}+V_{AB}\left(\delta_{iA}-\delta_{iB}\right)\right)$ with $V_{AB}=0.2\gamma_0$ and this modification applied randomly to $p=2.5\%$ of the sites in the sample. As shown in Fig.\ \ref{fig:FO_NBL_SR_PRB_2015_fig4}, a zero-energy Landau level splitting is clearly visible and corresponds to a plateau onset energy of $pV_{AB}=0.005\gamma_0$ (indicated by the dotted vertical line). Finally the emergence of quantum Hall effect in polycrystalline graphene was shown to be strongly dependent on the ratio between the average grain size and the magnetic length defined by the magnetic field \cite{cummings2014prb}.

\subsection{Quantum valley Hall effect}

Another phenomenon where topology plays an important role is the quantum valley Hall effect. Honeycomb structures are characterized by a linear energy dispersion centered at two inequivalent Brillouin zone points, usually denoted as $\bm{K}^+$ and $\bm{K}^-$, or equivalently, $K$ and $K'$. However, when these systems become massive an anomalous Hall effect is predicted to occur, but with opposite sign in each valley as imposed by the system's inversion symmetry \cite{Sinitsyn2006}. Moreover, similarly to the previously discussed quantum Hall effect, a periodically strained system will behave as if it were subjected to a pseudomagnetic field, producing Landau levels \cite{Levy2010} and a valley quantum Hall effect \cite{settnes2017tdm}. Here we show that by using the Kubo-Bastin formula and an adequate definition of the valley current operator, one can obtain appropriate transport coefficients for graphene under uniform strain and propose an experimental way to detect valley polarized currents. 

For modeling graphene under uniform strain we use a first-nearest-neighbor TB model where strain is included through a modification of the hopping parameters, while the external magnetic field is added using the standard Peierls substitution described in the previous subsection. In the Dirac approximation the strain is described by a gauge field $\pm \bm{A}_{\rm S}$, where $\pm$ denote the two valleys. This gauge field is related to the strain tensor $\epsilon_{ij}$ through $\bm{A}_{\rm S} \propto \big(\epsilon_{xx} -\epsilon_{yy},-2\epsilon_{xy}\big)$ \cite{Guinea2010, Vozmediano2010, Fujita2011}, and the pseudomagnetic field becomes $\bm{B}_{\rm S} = \nabla\times \bm{A}_{\rm S}$. From this, it is straightforward to show that a triaxial deformation $\bm{u}(x,y)=u_0 \big(2xy,x^2-y^2\big)$ induces a constant pseudomagnetic field. Uniaxial tensile strain has also been shown to generate a constant pseudomagnetic field \cite{Zhu2015}.

In order to resolve each valley, one needs to remember that in linear response theory one is computing the average of a microscopic operator, which for charge transport is the current operator. Therefore, we need to find an appropriate microscopic valley current operator. This can be achieved by taking inspiration from the spin current operator, which is in general defined as
\begin{equation}
\hat{J}_\alpha^z \equiv \frac{1}{2}\{\hat{J}_\alpha,s_z\},
\end{equation}
where $\hat{J}_\alpha$ is the single-particle current operator in the $\alpha$ direction as defined in Sec.\ \ref{section:kubo_formulas}, and $s_z$ is the spin operator in the $z$ direction. Then, by expressing $s_z$ in terms of its eigenvector projectors $P_s^\pm=\ket{\pm}\bra{\pm}$, we have
\begin{equation}
\hat{J}_\alpha^z = \frac{1}{2}(\,P_s^+\hat{J}_\alpha P_s^+-P_s^-\hat{J}_\alpha P_s^-),
\end{equation}
and from this expression we conclude that the spin current operator is nothing but the difference between the projections of the current operator in each spin subspace. From here the extension is obvious; we define the valley projector operators as $P^\pm_v=\ket{\bm{K}^\pm}\bra{\bm{K}^\pm}$, and define the valley current operator as
\begin{equation}
\hat{J}_\alpha^v = \frac{1}{2}(\,P^+_v \hat{J}_\alpha P^+_v - P^-_v\hat{J}_\alpha P^-_v\,).
\end{equation}
Different from the case of spin, there is no valley operator in the full tight-binding Hamiltonian, and therefore it is in general impossible to find $P^\pm$ using the same approach. However, from a numerical perspective one can consider the projector as a filter of electrons with momentum not belonging to the $\bm{K}^\pm$ valley, or in explicit terms
\begin{equation}
P^\pm_v = \sum_{\bm{k}} \theta(|\bm{k}-\bm{K}^\pm|-R)\ket{\bm{k}}\bra{\bm{k}} 
\end{equation}
where $\theta(x)$ the Heaviside function and $R$ is a valley cutoff which in general is defined by the disorder energy scale and can be chosen to be for example $R=|\bm{M}-\bm{K}^\pm|$.

\begin{figure}[htb]
\centering
\includegraphics[width=\columnwidth]{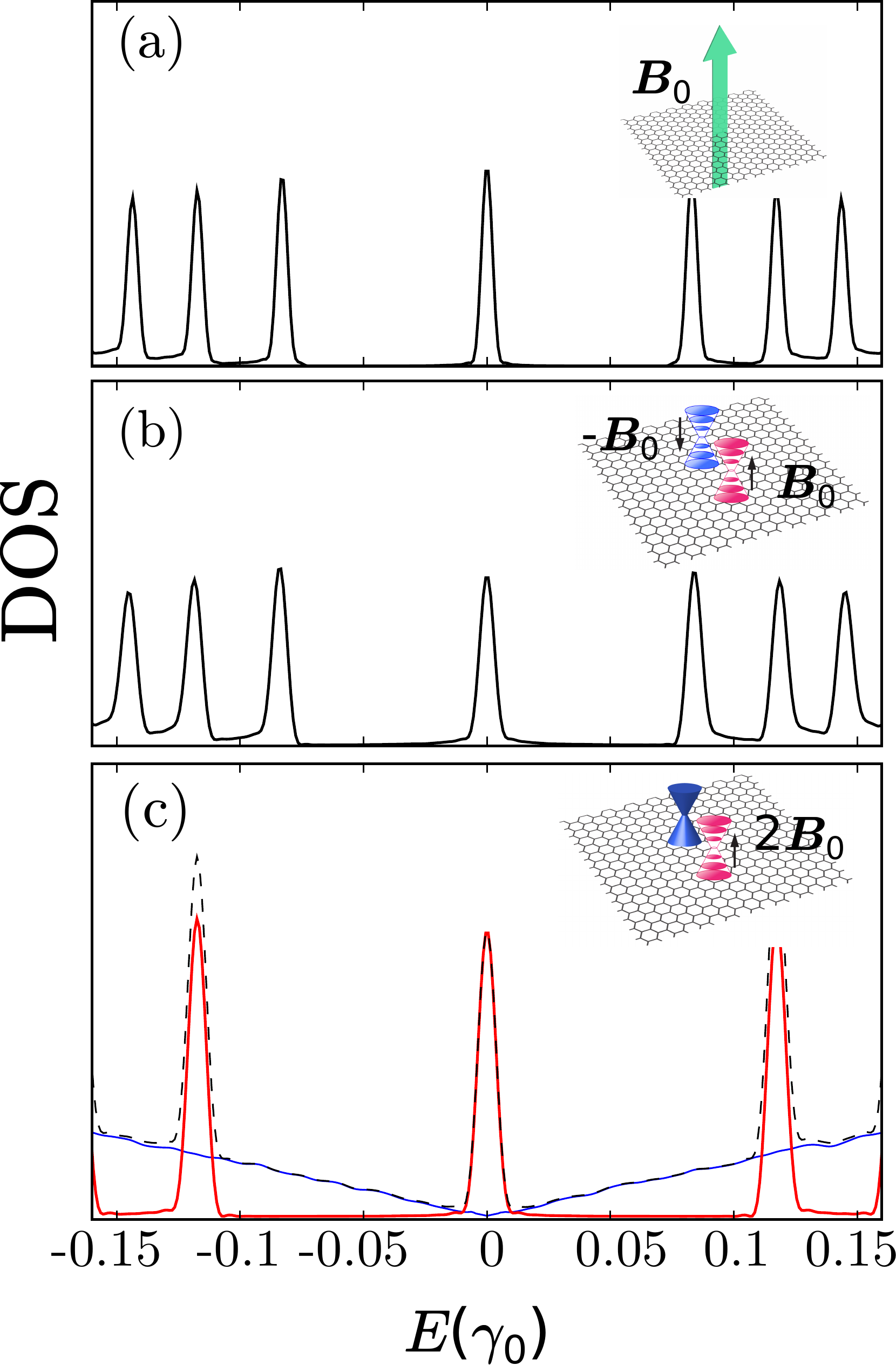}
\caption{DOS of graphene in presence of (a) an external magnetic field $B_\text{M} = B_0$, (b) a strain-induced pseudomagnetic field $B_\text{S} = B_0$, positive and negative for the $K$ and $K'$ valleys respectively. (c) Valley polarized DOS for graphene with both a strain-induced pseudomagnetic field and external magnetic field with $B_\text{S} = B_\text{M} = B_0$, thus cancelling and doubling the field in the $K'$ (Blue) and $K$ (Red) points respectively. We take $B_0 = 50$ T. The simulation was performed in a system of $5 \times 10^5$ atoms, using 200 random vectors and the KPM for expanding the Green's functions with 4000 moments. Adapted from \cite{settnes2017tdm}.}
\label{fig:vhe-fig1}
\end{figure}

In the following, we present calculations using a 100 nm $\times$ 100 nm graphene sample ($\sim 4\times10^5$ atoms) with a maximum strain of $\Delta_m \approx 8 \%$ corresponding to a pseudomagnetic field of 50 T. The maximum strain is obtained along the edge of the sample, so all results can be rescaled such that by keeping $\Delta_m =8$ \% we find a pseudomagnetic field of 5 T for a 1 $\upmu$m $\times$ 1 $\upmu$m sample. The sample choice also implies that not all parts of the sample experience a uniform pseudomagnetic field. This happens along the edge of the samples where nonuniformity of the pseudomagnetic field will act as a scatterer that can mix valleys. The presented results are robust against this type of valley mixing as we only consider a bulk effect in the part of the sample with a constant pseudomagnetic field. The results remain qualitatively unchanged as long as a sufficiently large part of the sample experiences a uniform field.

\begin{figure}[htb]
\centering
\includegraphics[width=1.0\columnwidth]{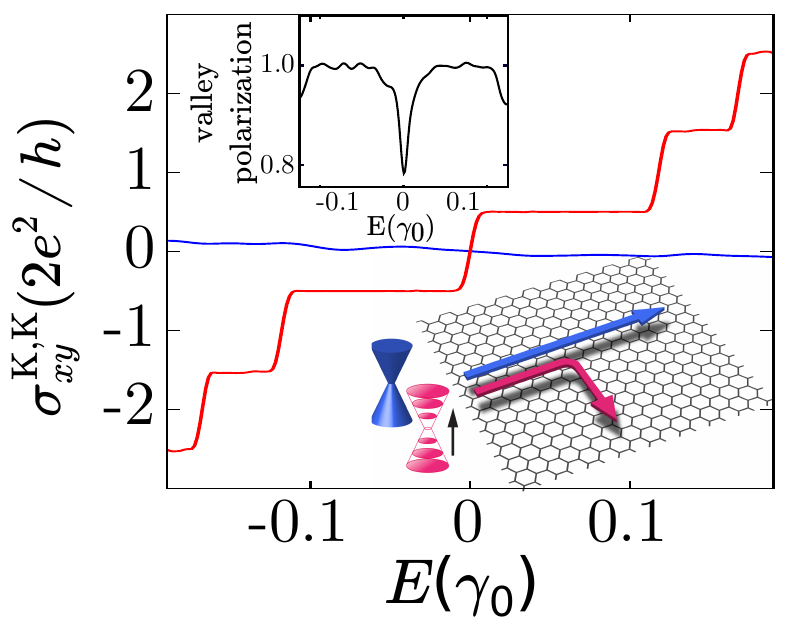}
\caption{Transport in graphene with compensating pseudo- and real magnetic fields. When the strain pseudomagnetic field compensates with a external magnetic field, the Hall conductivity shows the quantum Hall effect in one valley only, while it is suppressed in the other. Another manifestation of such a cancellation is the valley polarized current density (computed as  ($\sigma^K_{xx}-\sigma^{K'}_{xx} )/\sigma_{xx}$), which appears as a consequence of the different densities of states at each valley and is shown in the inset. The simulation was performed in a system of $5 \times 10^5$ atoms, with 200 random vectors and the KPM for expanding the Green's functions with 4000 moments. Adapted from \cite{settnes2017tdm}.}
\label{fig:vhe-fig2}
\end{figure}

In Fig.\ \ref{fig:vhe-fig1} we show the density of states of graphene for the cases with magnetic field, pseudomagnetic field, and a combination of both with the same strength. We see the formation of Landau levels in the first two cases, and for the third one we see a perfect cancellation of fields for one of the valleys, leading to the typical metallic state. This is because due to inversion symmetry, the pseudomagnetic field has opposite sign in each valley and will add to or subtract from the real magnetic field.

Next we compute the Hall conductivity in the situation where the pseudomagnetic field compensates the real magnetic field. This is shown in Fig.\ \ref{fig:vhe-fig2}. In this scenario, we can see that the system behaves exactly as expected for the quantum Hall effect discussed previously, but with a Hall conductivity reduced by half because only one of the valleys is carrying the current. Moreover, because the system is metallic the longitudinal current is fully valley polarized, which is key for valleytronic applications.

\subsection{Spin transport physics}

Spintronics, or spin electronics, involves the study of spin information transfer as well as the manipulation of spin degrees of freedom in solid-state systems \cite{ZuticRMP2004}. Spin transport differs from charge transport in that spin is generally a nonconserved quantity in solids due to spin-orbit and hyperfine coupling. An essential metric to characterize spin transport is given by the upper limits of time or distance over which spin signals can be measured or manipulated. To evaluate the corresponding spin lifetime (or relaxation time) and spin diffusion length in disordered materials, one can either use a semiclassical spin Bloch transport equation \cite{Fabian2007}, or compute numerically the time evolution of the spin polarization of propagating wavepackets. As shown below, real-space $\order{N}$ methods provide a new tool for exploring spin dynamics, spin relaxation and spin transport phenomena (such as the spin Hall effect) in complex materials.

\subsubsection{Spin relaxation time}

To study spin dynamics and spin relaxation using the numerical methods presented above, it suffices to calculate the energy- and time-dependent spin polarization
\begin{equation}
\bm{S}(E,t) = \frac{1}{2} \frac{\left\langle \phi(t) \right|  \hat{\bm{s}} \delta(E-\hat{H})  \left| \phi(t) \right\rangle + {\rm h.c.}}{\left\langle \phi(t) \right|  \delta(E-\hat{H})  \left| \phi(t) \right\rangle},
\label{equation:spin_polarization}
\end{equation}
where $\hat{\bm{s}}$ are the spin Pauli matrices, ``h.c.'' is the Hermitian conjugate, and $\left| \phi(t) \right\rangle = \hat{U}(t) \left| \phi(0) \right\rangle$ is the time-evolved initial state of the system. This initial state is spin polarized along axis $\bm{j}$ according to
\begin{equation}
\left| \phi(0) \right\rangle = \frac{1}{2} (\mathds{1}_{2N} + {\bm j}\cdot \hat{\bm{s}}) \left| \phi_{\rm r} \right\rangle,
\label{equation:spin_initial}
\end{equation}
where $\mathds{1}_{2N}$ is the $2N \times 2N$ identity matrix and $\left| \phi_{\rm r} \right\rangle$ is the random-phase state defined in Eq.\ (\ref{equation:rp_state}) with the replacement $N \rightarrow 2N$ to account for spin.

With a bit of knowledge about spin relaxation mechanisms and the nature of the system under investigation, the spin relaxation time can be extracted from the time-dependent spin polarization. For example, the typical Elliott-Yafet (EY) and D'yakonov-Perel' (DP) spin relaxation mechanisms give $\bm{S}(t) = \bm{S}(0) \exp(-t / \tau_{\rm s})$, where $\tau_{\rm s}$ is the spin relaxation time \cite{Elliott1954, Yafet1963, dyakonov1971spin}. When outside the motional narrowing regime or in the presence of a uniform magnetic field, the DP mechanism changes to $\bm{S}(t) = \bm{S}(0) \exp(-t / \tau_{\rm s}) \cos(\omega_{\rm s}t)$, where $\omega_{\rm s}$ is the spin precession frequency \cite{Gridnev2001}. Meanwhile, more complicated dephasing mechanisms can lead to different behaviors \cite{vantuan2014np,cummings2016prl}.

\begin{figure}[htb]
\begin{center}
\includegraphics[width=\columnwidth]{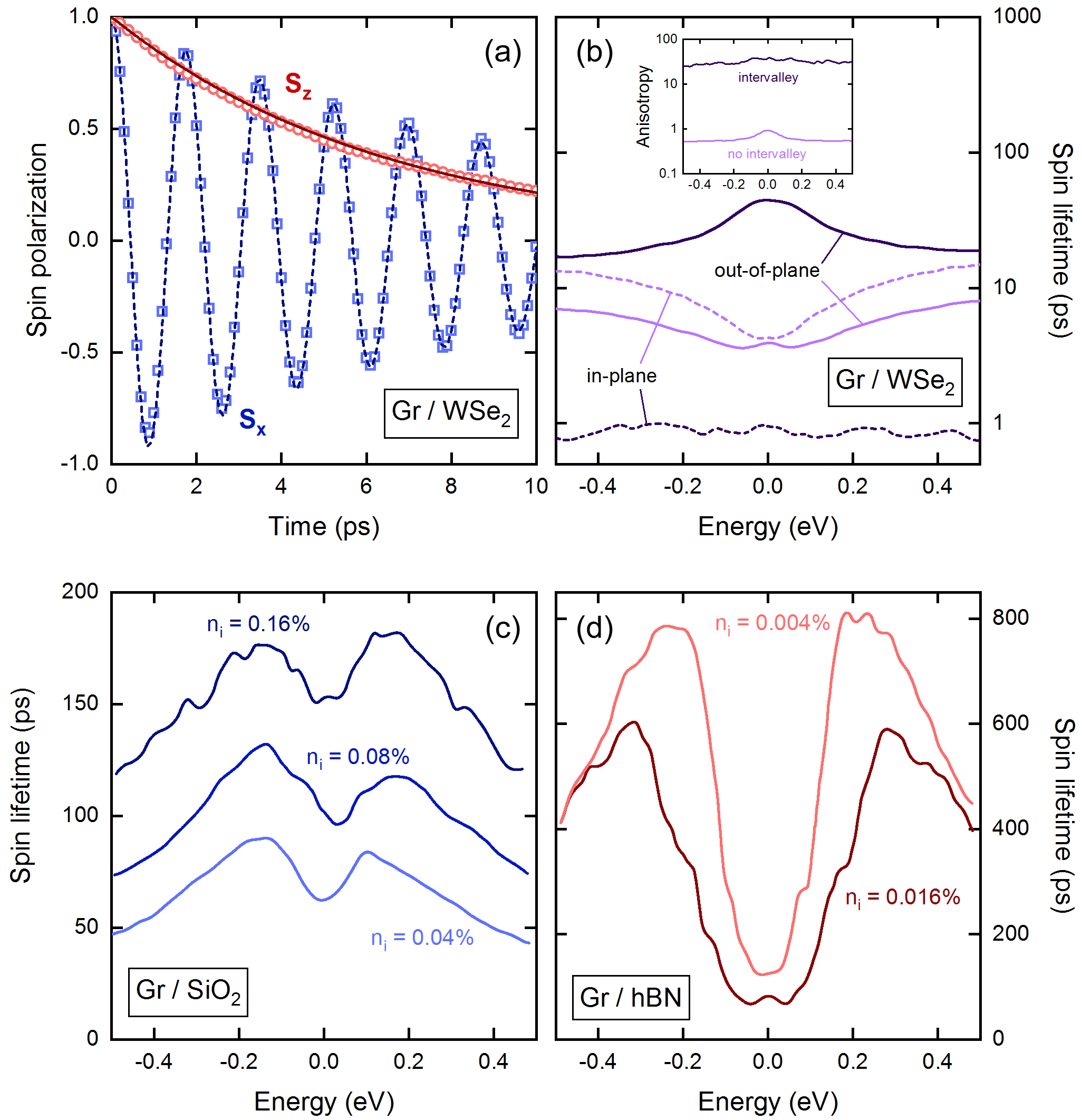}
\caption{(a) Example of spin dynamics and relaxation in graphene on a WSe$_2$ substrate with weak electron-hole puddle disorder. The red (blue) symbols are numerical simulations of the spin polarization along the $z$ ($x$) axis, and the lines are fits. (b) Simulation of anisotropic spin relaxation in graphene on a WSe$_2$ substrate. Light (dark) purple curves show the spin lifetimes in the absence (presence) of intervalley scattering. The inset shows the ratio of the out-of-plane and in-plane spin lifetimes. (c) Energy-dependent spin lifetime in graphene on a SiO$_2$ substrate, which increases with increasing defect density. Opposite scaling can be seen in panel (d), for graphene on an hBN substrate. Simulations of graphene on WSe$_2$ were done for a system size of $9.2\times10^6$ carbon atoms, and the spin polarization was calculated using KPM and the Lorentz kernel, with $M=3400$ moments corresponding to an energy broadening of $\sim$20 meV. Simulations of graphene on SiO$_2$ and hBN were done for a system size of $2\times10^6$ atoms, and the spin polarization was calculated using a Lanczos recursion with an energy broadening of 13.5 meV. Panels (a,b) are adapted from \cite{Cummings2017prl}, while (c,d) are adapted from \cite{vantuan2016scirep}.}
\label{figure:svst}
\end{center}
\end{figure}

An example of spin dynamics and relaxation is shown in Fig.\ \ref{figure:svst}(a). Here we plot the time dependence of spins oriented in (blue symbols) or out of (red symbols) the graphene plane, for graphene on a WSe$_2$ substrate in the presence of weak electron-hole puddles \cite{Cummings2017prl}. Here we see that the in-plane spins undergo precession plus relaxation, while the out-of-plane spins undergo simple exponential decay. Lines show the fits to these numerical results.

The methodology presented in Eqs.\ (\ref{equation:spin_polarization}) and (\ref{equation:spin_initial}) has been applied to the study of spin dynamics and relaxation in a wide variety of graphene-based systems. The first studies using this methodology revealed the role that spin-pseudospin entanglement has on spin relaxation in graphene with gold impurities \cite{vantuan2014np}, graphene on typical SiO$_2$ or hBN substrates \cite{vantuan2016scirep, cummings2016prl}, or graphene functionalized with fluorine adatoms \cite{vantuan2016prl}. An example of this can be seen in Figs.\ \ref{figure:svst}(c) and (d), which show calculations of the spin lifetime in graphene on a SiO$_2$ or hBN substrate for different defect densities. For graphene on SiO$_2$, the spin lifetime increases with increasing defect density, indicating the presence of DP spin relaxation. Meanwhile, graphene on hBN shows the opposite scaling behavior, indicating a transition out of the motional narrowing regime of spin dynamics due to the much weaker scattering induced by the hBN substrate. In all cases, a minimum in the spin lifetime at the charge neutrality point is a signature of spin-pseudospin entanglement in graphene systems dominated by Rasha spin-orbit coupling \cite{vantuan2014np, vantuan2016scirep, cummings2016prl}. We also mention that one can access the spin lifetime by studying the scaling behavior of nonlocal resistance using methods presented in \ref{section:lb}. The advantage is a possible direct comparison with experiments, and the exploration of spin transport in the crossover from diffusive to ballistic motion \cite{Vila2020}.

Recent work investigated spin relaxation in graphene on transition metal dichalcogenide (TMDC) substrates, and predicted the presence of giant spin lifetime anisotropy, with in-plane spins relaxing much faster than out-of-plane spins \cite{Cummings2017prl}. This is depicted in Fig.\ \ref{figure:svst}(b), which shows the simulated spin lifetime in a graphene/WSe$_2$ system. In the presence of intervalley scattering the spin lifetime anisotropy can reach values of several tens, while for graphene on typical SiO$_2$ substrates this value is on the order of one \cite{Raes2016ncomm}. These results have been supported and generalized using a time-dependent perturbative treatment to derive the spin Bloch equations governing the spin dynamics at high electronic density \cite{Offidanispin2018}. The predicted giant spin lifetime anisotropy has also been verified experimentally \cite{Benitez2018np, ghiasi2017nl}, confirming the strong impact that TMDC substrates can have on spin transport in graphene.

Beyond the aforementioned examples, spin relaxation in graphene functionalized with thallium or hydrogen atoms has also been studied with these methods \cite{cresti2014prl,soriano2015twodm}, as has the impact of local magnetism coupled with electron-hole puddles \cite{vierimaa2017prb}.

\subsubsection{Spin Hall effect}

The spin Hall effect (SHE) is another phenomenon where the Fermi sea contribution is highly relevant. It consists of the generation of a spin current that is transverse to an applied electric field due to the presence of spin-orbit coupling \cite{dyakonov1971spin, Hirsch1999,PRLMilletari2017}. There are two mechanisms behind the emergence of SHE. The first is named the intrinsic SHE since it occurs solely due to the spin-orbit coupling encoded in the band structure of the materials, whereas the extrinsic SHE stems from an interplay between disorder and the states at the Fermi level \cite{Sinova2015}. In general, the spin Hall effect measured experimentally is usually a combination of both, and there are even situations where these two effects exactly cancel \cite{PRLMishchenko2004, PRBInoue2004, PRLMilletari2017}. The intrinsic SHE can be considered as the time-reversal generalization of the quantum Hall effect, in the sense that it is the sum of the Berry curvature of each band that determines the behavior of the system \cite{Sinova2015}.

\begin{figure}[htb]
\centering
\includegraphics[width=\columnwidth]{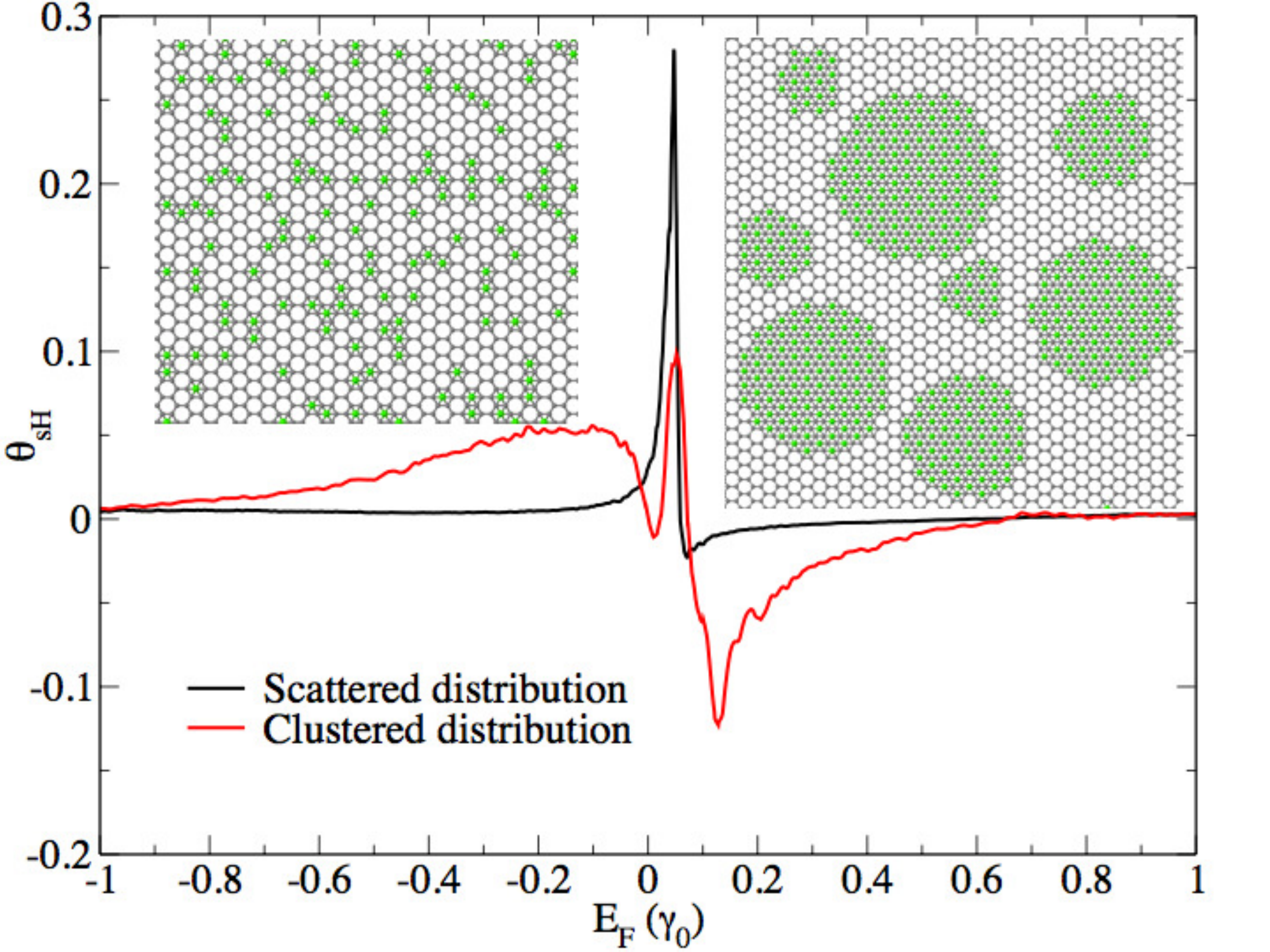}
\caption{Spin Hall angle $\theta_\mathrm{sH}$ for two cases of 15\% gold adatoms distributed onto graphene: scattered (black) and clustered distributions (red), as illustrated in the insets. The simulation was done in a system consisting of 4 million atoms, with one random vector and the KPM for expanding the Green's functions using 1500 moments and the Jackson kernel. Adapted from \cite{DinhPRL2016c}.}
\label{FigSHE}
\end{figure}

The Kubo formula for bulk conductivity allows one to define the main figure of merit of the SHE, namely the spin Hall angle (SHA), which measures how much pure spin current is produced by a charge current, and is connected to transport coeffficients through \cite{cresti2016rnc}
\begin{equation} \label{eq:thsH2}
\theta_{\rm sH}=\frac{\sigma_{xy}^{z}}{\sigma_{xx}},
\end{equation}
where $\sigma_{xy}^{z}$ is the SH conductivity and $\sigma_{xx}$ is the longitudinal charge conductivity. The formal expression of the spin Hall conductivity $\sigma_\mathrm{sH}$ used in numerical simulations is \cite{Sinova2015}
 \begin{equation}
 \sigma_\mathrm{sH}=\frac{e\hbar}{\Omega}\sum_{m,n}\frac{f(E_m)-f(E_n)}{E_m-E_n}\frac{\text{Im}[\left\langle m\left|J_x^z\right|n \right\rangle \left\langle n\left|   v_y\right|m \right\rangle]}{E_m-E_n+i\eta},
\label{EQSC}
 \end{equation}
where $J_x^z=\frac{\hbar}{4}\{s_z,v_x\}$ is the spin current operator and $s_z$ is the $z$-component of the spin Pauli matrices. This formula can be understood as a generalization of Aoki's formula \cite{aokisscomm1981, aoki1985prl}, presented in Eq.\ (\ref{aokiEq}). This formula becomes computationally prohibitive for large systems given that it requires the full spectrum of eigenvalues and eigenvectors of the Hamiltonian. However, the Kubo-Bastin formula and its variants remain valid given that it is derived for an arbitrary Hermitian operator, a condition that the spin current operator satisfies. This approach has been used used to determine the SHA of spin-orbit-enhanced graphene in recent years  \cite{Berg2011, garcia2015prl, DinhPRL2016c, garcia2016tdmat,savero2020}. One illustrative example of the SHA computed for graphene with random adsorbed gold adatoms is shown in Fig.\ \ref{FigSHE}, where a large SHA is observed when gold adatoms are deposited randomly on the graphene surface, while atomic segregation into clusters affects its energy dependence substantially \cite{DinhPRL2016c}.

\begin{figure}[htb]
\centering
\includegraphics[width=\columnwidth]{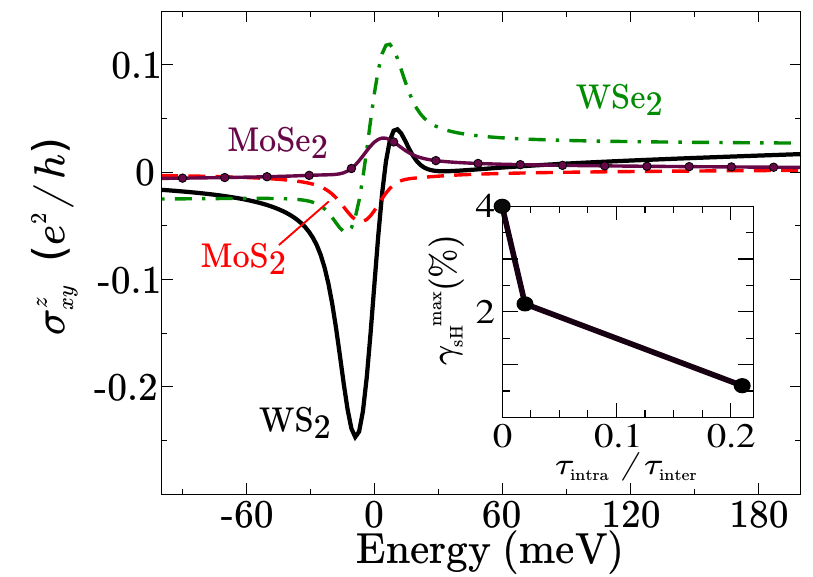}
\caption{Spin Hall conductivity for pristine graphene on different TMDCs. The inset shows the scaling of the spin Hall angle with intervalley scattering strength. The simulation was performed on a system containing $20 \times 10^6$ atoms, using at least 10 random vectors and the KPM for expanding the Green's functions with 3000 moments and the Jackson kernel. Adapted from \cite{garcia2018csr}.}
\label{FigSHEGrTMD}
\end{figure}

An additional example is shown in Fig.\ \ref{FigSHEGrTMD}, where the intrinsic spin Hall conductivity is computed for graphene on different graphene/TMDC substrates. In this particular work it was shown that this methodology can capture both the intrinsic and extrinsic contributions, because the intrinsic SHE is effectively canceled by an opposite extrinsic SHE originating from disorder-induced intervalley scattering. This suppression was studied as a function of the intervalley scattering rate in \cite{garcia2017nanolet,garcia2018csr}. The recent experimental confirmation of the SHE induced by proximity effects in graphene/TMDC heterostructures \cite{Safeer2019,Ghiasi2019,Benitez2020} or bilayer graphene/insulator \cite{Safeer2020} open a new playground to search for the upper limit of SHE efficiency, a task which can be supported by the simulation methods presented here.

Others studies combining $\order{N}$ bulk Kubo approaches with multiterminal Landauer-B\"{u}ttiker quantum transport methods have revealed more complexity in understanding the physics of the SHE than can be obtained from a simple theoretical interpretation of experimental data. For instance, Gregersen and coworkers have demonstrated how geometrical effects allow finite samples to display transverse resistances that are reminiscent of the SHE, but which disappear in the bulk limit \cite{Gregersen2018}. Another important finding concerns the parasitic background contributions that appear when calculating the nonlocal resistance of chemically functionalized graphene systems, which can mask spin effects or mislead the interpretation of experiments \cite{DinhPRL2016c}. Importantly, this type of theoretical analysis has  recently refuted the claim that topological valley Hall currents \cite{Song10879, Beconcini2016} carried by the Fermi sea can explain large nonlocal resistance measured at the Dirac point for certain graphene/hBN interfaces \cite{Gorbachev448}. A complete analysis of bulk and multiterminal quantum transport reveals a limit of the direct connection between the valley Hall conductivity and nonlocal resistance, and shows that non-topological dispersive edge states, resilient to (weak) disorder, give a more solid explanation for the large nonlocal resistance \cite{cresti2016rnc, Marmolejo2018}. 

\subsubsection{Quantum spin Hall effect}

In 1988, Haldane \cite{Haldane1988} proposed a curious two-dimensional tight-binding lattice model exhibiting a nonzero quantization of the Hall conductance $\sigma_{xy}$ in the absence of an external magnetic field. Many years after, Kane and Mele extended the concept to the quantum spin Hall effect (QSHE) studying models of graphene nanoribbons with intrinsic spin-orbit coupling \cite{KaneMele2005}. In such model $S_{z}$-spin component is a conserved quantity and the simultaneous presence of SOC-induced gaps and well defined sample boundaries generate topologically robust spin-polarized edge states. As a result, a new state of QSHE manifests as a quantized spin-Hall conductance and a vanishing charge-Hall conductance. 

 The very low micron eV bandgap of the original Kane-Mele model has impeded the observation of QSHE in clean graphene, but the fundamental underlying mechanism has been further shown to be ubiquitous in condensed matter \cite{Hasan2010RMP,ORTMANNti} and the family of two-dimensional topological insulators is being enriched regularly \cite{Ren_2016, Olsen2019}.

There have been several computational studies, using LSQT methods, exploring how QSHE could emerge by reinforcing the spin-orbit coupling in monolayer graphene systems.  The engineering of a robust QSHE in graphene via uniform heavy ad-atoms deposition was proposed theoretically \cite{Weeks2011, JiangPRL2012}. However, the random distribution of ad-atoms and aggregation effects could be only investigated thoroughly using LSQT \cite{cresti2014prl, Liu2015QSHE,Santos2018}. In Ref.\ \cite{cresti2014prl} the change from homogeneous to inhomogeneous distribution of Thallium ad-atoms on graphene surface was found to trigger multiple quantum Phases from QSHE to SHE or even some anomalous metallic states driven by cluster-to-cluster percolation transport mechanism, robust to localization effects, and identified by a scale-independent quantum conductivity $\sigma_{xx}\simeq 4e^{2}/h$. There is certainly plenty of rooms to further analyse how complex and fluctuating spin-orbit coupling fields induce exotic quantum transport regimes in the wide variety of topological matter, and clearly LSQT stands as an enabling tool to unravel such fascinating physical phenomena.

%% file: LSQT_main.bbl
\begin{thebibliography}{349}
\expandafter\ifx\csname natexlab\endcsname\relax\def\natexlab#1{#1}\fi
\providecommand{\url}[1]{\texttt{#1}}
\providecommand{\href}[2]{#2}
\providecommand{\path}[1]{#1}
\providecommand{\DOIprefix}{doi:}
\providecommand{\ArXivprefix}{arXiv:}
\providecommand{\URLprefix}{URL: }
\providecommand{\Pubmedprefix}{pmid:}
\providecommand{\doi}[1]{\href{http://dx.doi.org/#1}{\path{#1}}}
\providecommand{\Pubmed}[1]{\href{pmid:#1}{\path{#1}}}
\providecommand{\bibinfo}[2]{#2}
\ifx\xfnm\relax \def\xfnm[#1]{\unskip,\space#1}\fi
%Type = Article
\bibitem[{Hohenberg and Kohn(1964)}]{hohenberg1964pr}
\bibinfo{author}{P.~Hohenberg}, \bibinfo{author}{W.~Kohn},
\newblock \bibinfo{title}{{Inhomogeneous Electron Gas}},
\newblock \bibinfo{journal}{Phys. Rev.} \bibinfo{volume}{136}
  (\bibinfo{year}{1964}) \bibinfo{pages}{B864--B871}. \URLprefix
  \url{https://link.aps.org/doi/10.1103/PhysRev.136.B864}.
  \DOIprefix\doi{10.1103/PhysRev.136.B864}.
%Type = Article
\bibitem[{Kohn and Sham(1965)}]{kohn1965pr}
\bibinfo{author}{W.~Kohn}, \bibinfo{author}{L.~J. Sham},
\newblock \bibinfo{title}{{Self-Consistent Equations Including Exchange and
  Correlation Effects}},
\newblock \bibinfo{journal}{Phys. Rev.} \bibinfo{volume}{140}
  (\bibinfo{year}{1965}) \bibinfo{pages}{A1133--A1138}. \URLprefix
  \url{https://link.aps.org/doi/10.1103/PhysRev.140.A1133}.
  \DOIprefix\doi{10.1103/PhysRev.140.A1133}.
%Type = Article
\bibitem[{Kohn(1999)}]{kohn1999rmp}
\bibinfo{author}{W.~Kohn},
\newblock \bibinfo{title}{{Nobel Lecture: Electronic structure of matter---wave
  functions and density functionals}},
\newblock \bibinfo{journal}{Rev. Mod. Phys.} \bibinfo{volume}{71}
  (\bibinfo{year}{1999}) \bibinfo{pages}{1253--1266}. \URLprefix
  \url{https://link.aps.org/doi/10.1103/RevModPhys.71.1253}.
  \DOIprefix\doi{10.1103/RevModPhys.71.1253}.
%Type = Article
\bibitem[{Jones(2015)}]{jones2015rmp}
\bibinfo{author}{R.~O. Jones},
\newblock \bibinfo{title}{{Density functional theory: Its origins, rise to
  prominence, and future}},
\newblock \bibinfo{journal}{Rev. Mod. Phys.} \bibinfo{volume}{87}
  (\bibinfo{year}{2015}) \bibinfo{pages}{897--923}.
  \DOIprefix\doi{10.1103/RevModPhys.87.897}.
%Type = Incollection
\bibitem[{Haydock(1980)}]{haydock1980ssp}
\bibinfo{author}{R.~Haydock},
\newblock \bibinfo{title}{The recursive solution of the schrodinger equation},
\newblock volume~\bibinfo{volume}{35} of \textit{\bibinfo{series}{Solid State
  Physics}}, \bibinfo{publisher}{Academic Press}, \bibinfo{year}{1980}, pp.
  \bibinfo{pages}{215 -- 294}. \URLprefix
  \url{http://www.sciencedirect.com/science/article/pii/S0081194708605056}.
  \DOIprefix\doi{10.1016/S0081-1947(08)60505-6}.
%Type = Article
\bibitem[{Haydock et~al.(1975)Haydock, Heine, and Kelly}]{haydock1975jpc}
\bibinfo{author}{R.~Haydock}, \bibinfo{author}{V.~Heine},
  \bibinfo{author}{M.~J. Kelly},
\newblock \bibinfo{title}{{Electronic structure based on the local atomic
  environment for tight-binding bands. II}},
\newblock \bibinfo{journal}{J. Phys. C: Solid State Phys.} \bibinfo{volume}{8}
  (\bibinfo{year}{1975}) \bibinfo{pages}{2591}. \URLprefix
  \url{http://stacks.iop.org/0022-3719/8/i=16/a=011}.
%Type = Article
\bibitem[{Haydock et~al.(1972)Haydock, Heine, and Kelly}]{haydock1972jpc}
\bibinfo{author}{R.~Haydock}, \bibinfo{author}{V.~Heine},
  \bibinfo{author}{M.~J. Kelly},
\newblock \bibinfo{title}{{Electronic structure based on the local atomic
  environment for tight-binding bands}},
\newblock \bibinfo{journal}{J. Phys. C: Solid State Phys.} \bibinfo{volume}{5}
  (\bibinfo{year}{1972}) \bibinfo{pages}{2845}. \URLprefix
  \url{http://stacks.iop.org/0022-3719/5/i=20/a=004}.
%Type = Article
\bibitem[{Tal-Ezer and Kosloff(1984)}]{ezer1984jcp}
\bibinfo{author}{H.~Tal-Ezer}, \bibinfo{author}{R.~Kosloff},
\newblock \bibinfo{title}{{An accurate and efficient scheme for propagating the
  time dependent Schr\"{o}dinger equation}},
\newblock \bibinfo{journal}{J. Chem. Phys.} \bibinfo{volume}{81}
  (\bibinfo{year}{1984}) \bibinfo{pages}{3967--3971}. \URLprefix
  \url{https://doi.org/10.1063/1.448136}. \DOIprefix\doi{10.1063/1.448136}.
%Type = Article
\bibitem[{Leforestier et~al.(1991)Leforestier, Bisseling, Cerjan, Feit,
  Friesner, Guldberg, Hammerich, Jolicard, Karrlein, Meyer, Lipkin, Roncero,
  and Kosloff}]{leforestier1991jcp}
\bibinfo{author}{C.~Leforestier}, \bibinfo{author}{R.~Bisseling},
  \bibinfo{author}{C.~Cerjan}, \bibinfo{author}{M.~Feit},
  \bibinfo{author}{R.~Friesner}, \bibinfo{author}{A.~Guldberg},
  \bibinfo{author}{A.~Hammerich}, \bibinfo{author}{G.~Jolicard},
  \bibinfo{author}{W.~Karrlein}, \bibinfo{author}{H.-D. Meyer},
  \bibinfo{author}{N.~Lipkin}, \bibinfo{author}{O.~Roncero},
  \bibinfo{author}{R.~Kosloff},
\newblock \bibinfo{title}{{A comparison of different propagation schemes for
  the time dependent Schr\''odinger equation}},
\newblock \bibinfo{journal}{J. Comp. Phys.} \bibinfo{volume}{94}
  (\bibinfo{year}{1991}) \bibinfo{pages}{59 -- 80}. \URLprefix
  \url{http://www.sciencedirect.com/science/article/pii/002199919190137A}.
  \DOIprefix\doi{10.1016/0021-9991(91)90137-A}.
%Type = Book
\bibitem[{Petitfor and Weaire(1985)}]{petitfor1985book}
\bibinfo{author}{D.~Petitfor}, \bibinfo{author}{D.~Weaire},
  \bibinfo{title}{Recursion Method and its Applications (Springer Series in
  Solid States Sciences, Vol. {\bf 58})}, \bibinfo{publisher}{Springer Verlag,
  Berlin}, \bibinfo{year}{1985}.
%Type = Article
\bibitem[{Wei\ss{}e et~al.(2006)Wei\ss{}e, Wellein, Alvermann, and
  Fehske}]{weisse2006rmp}
\bibinfo{author}{A.~Wei\ss{}e}, \bibinfo{author}{G.~Wellein},
  \bibinfo{author}{A.~Alvermann}, \bibinfo{author}{H.~Fehske},
\newblock \bibinfo{title}{The kernel polynomial method},
\newblock \bibinfo{journal}{Rev. Mod. Phys.} \bibinfo{volume}{78}
  (\bibinfo{year}{2006}) \bibinfo{pages}{275--306}. \URLprefix
  \url{https://link.aps.org/doi/10.1103/RevModPhys.78.275}.
  \DOIprefix\doi{10.1103/RevModPhys.78.275}.
%Type = Article
\bibitem[{Fehske et~al.(2009)Fehske, Schleede, Schubert, Wellein, Filinov, and
  Bishop}]{fehske2009pla}
\bibinfo{author}{H.~Fehske}, \bibinfo{author}{J.~Schleede},
  \bibinfo{author}{G.~Schubert}, \bibinfo{author}{G.~Wellein},
  \bibinfo{author}{V.~S. Filinov}, \bibinfo{author}{A.~R. Bishop},
\newblock \bibinfo{title}{{Numerical approaches to time evolution of complex
  quantum systems}},
\newblock \bibinfo{journal}{Phys. Lett. A} \bibinfo{volume}{373}
  (\bibinfo{year}{2009}) \bibinfo{pages}{2182 -- 2188}. \URLprefix
  \url{http://www.sciencedirect.com/science/article/pii/S0375960109004927}.
  \DOIprefix\doi{10.1016/j.physleta.2009.04.022}.
%Type = Article
\bibitem[{Jing and Ma(2007)}]{JunPRB2007}
\bibinfo{author}{J.~Jing}, \bibinfo{author}{H.~R. Ma},
\newblock \bibinfo{title}{Polynomial scheme for time evolution of open and
  closed quantum systems},
\newblock \bibinfo{journal}{Phys. Rev. E} \bibinfo{volume}{75}
  (\bibinfo{year}{2007}) \bibinfo{pages}{016701}.
  \DOIprefix\doi{10.1103/PhysRevE.75.016701}.
%Type = Book
\bibitem[{Viswanath and M\"uller(1994)}]{viswanath1994book}
\bibinfo{author}{V.~Viswanath}, \bibinfo{author}{G.~M\"uller},
  \bibinfo{title}{The Recursion Method: Application to Many-Body Dynamics
  (Lectures Notes in Physics, Vol. {\bf 23})}, \bibinfo{publisher}{Springer
  Verlag, Berlin}, \bibinfo{year}{1994}.
%Type = Article
\bibitem[{Boehnke et~al.(2011)Boehnke, Hafermann, Ferrero, Lechermann, and
  Parcollet}]{BoehnkePRB2011}
\bibinfo{author}{L.~Boehnke}, \bibinfo{author}{H.~Hafermann},
  \bibinfo{author}{M.~Ferrero}, \bibinfo{author}{F.~Lechermann},
  \bibinfo{author}{O.~Parcollet},
\newblock \bibinfo{title}{Orthogonal polynomial representation of
  imaginary-time green's functions},
\newblock \bibinfo{journal}{Phys. Rev. B} \bibinfo{volume}{84}
  (\bibinfo{year}{2011}) \bibinfo{pages}{075145}.
  \DOIprefix\doi{10.1103/PhysRevB.84.075145}.
%Type = Article
\bibitem[{Ganahl et~al.(2014)Ganahl, Thunstr\"om, Verstraete, Held, and
  Evertz}]{Ganahl2014}
\bibinfo{author}{M.~Ganahl}, \bibinfo{author}{P.~Thunstr\"om},
  \bibinfo{author}{F.~Verstraete}, \bibinfo{author}{K.~Held},
  \bibinfo{author}{H.~G. Evertz},
\newblock \bibinfo{title}{Chebyshev expansion for impurity models using matrix
  product states},
\newblock \bibinfo{journal}{Phys. Rev. B} \bibinfo{volume}{90}
  (\bibinfo{year}{2014}) \bibinfo{pages}{045144}.
  \DOIprefix\doi{10.1103/PhysRevB.90.045144}.
%Type = Article
\bibitem[{Marinov et~al.(2020)Marinov, Deen, Jiménez-Tejada, and
  Chen}]{2020PhysRep}
\bibinfo{author}{O.~Marinov}, \bibinfo{author}{M.~Deen},
  \bibinfo{author}{J.~Jiménez-Tejada}, \bibinfo{author}{C.~Chen},
\newblock \bibinfo{title}{Variable-range hopping charge transport in organic
  thin-film transistors},
\newblock \bibinfo{journal}{Physics Reports} \bibinfo{volume}{844}
  (\bibinfo{year}{2020}) \bibinfo{pages}{1 -- 105}. \URLprefix
  \url{http://www.sciencedirect.com/science/article/pii/S0370157319304028}.
  \DOIprefix\doi{https://doi.org/10.1016/j.physrep.2019.12.002},
  \bibinfo{note}{variable-range hopping charge transport in organic thin-film
  transistors}.
%Type = Article
\bibitem[{Abrahams et~al.(1979)Abrahams, Anderson, Licciardello, and
  Ramakrishnan}]{abrahams1979prl}
\bibinfo{author}{E.~Abrahams}, \bibinfo{author}{P.~W. Anderson},
  \bibinfo{author}{D.~C. Licciardello}, \bibinfo{author}{T.~V. Ramakrishnan},
\newblock \bibinfo{title}{{Scaling Theory of Localization: Absence of Quantum
  Diffusion in Two Dimensions}},
\newblock \bibinfo{journal}{Phys. Rev. Lett.} \bibinfo{volume}{42}
  (\bibinfo{year}{1979}) \bibinfo{pages}{673--676}. \URLprefix
  \url{https://link.aps.org/doi/10.1103/PhysRevLett.42.673}.
  \DOIprefix\doi{10.1103/PhysRevLett.42.673}.
%Type = Article
\bibitem[{Gor'kov et~al.(1979)Gor'kov, Larkin, and Khmel'nitskii}]{Gorkov1979}
\bibinfo{author}{L.~Gor'kov}, \bibinfo{author}{A.~Larkin},
  \bibinfo{author}{D.~Khmel'nitskii},
\newblock \bibinfo{title}{Particle conductivity in a two-dimensional random
  potential},
\newblock \bibinfo{journal}{JETP Lett.} \bibinfo{volume}{30}
  (\bibinfo{year}{1979}) \bibinfo{pages}{228}. \URLprefix
  \url{http://www.jetpletters.ac.ru/ps/1364/article\_20629.shtml}.
%Type = Article
\bibitem[{Altshuler et~al.(1980)Altshuler, Khmel'nitzkii, Larkin, and
  Lee}]{Altshuler1980}
\bibinfo{author}{B.~Altshuler}, \bibinfo{author}{D.~Khmel'nitzkii},
  \bibinfo{author}{A.~Larkin}, \bibinfo{author}{P.~Lee},
\newblock \bibinfo{title}{Magnetoresistance and hall effect in a disordered
  two-dimensional electron gas},
\newblock \bibinfo{journal}{Phys. Rev. B} \bibinfo{volume}{22}
  (\bibinfo{year}{1980}) \bibinfo{pages}{5142--5153}.
  \DOIprefix\doi{10.1103/PhysRevB.22.5142}.
%Type = Article
\bibitem[{Lee and Ramakrishnan(1985)}]{lee1985rmp}
\bibinfo{author}{P.~A. Lee}, \bibinfo{author}{T.~V. Ramakrishnan},
\newblock \bibinfo{title}{{Disordered electronic systems}},
\newblock \bibinfo{journal}{Rev. Mod. Phys.} \bibinfo{volume}{57}
  (\bibinfo{year}{1985}) \bibinfo{pages}{287--337}. \URLprefix
  \url{https://link.aps.org/doi/10.1103/RevModPhys.57.287}.
  \DOIprefix\doi{10.1103/RevModPhys.57.287}.
%Type = Article
\bibitem[{Rammer and Smith(1986)}]{rammer1986rmp}
\bibinfo{author}{J.~Rammer}, \bibinfo{author}{H.~Smith},
\newblock \bibinfo{title}{Quantum field-theoretical methods in transport theory
  of metals},
\newblock \bibinfo{journal}{Rev. Mod. Phys.} \bibinfo{volume}{58}
  (\bibinfo{year}{1986}) \bibinfo{pages}{323--359}. \URLprefix
  \url{https://link.aps.org/doi/10.1103/RevModPhys.58.323}.
  \DOIprefix\doi{10.1103/RevModPhys.58.323}.
%Type = Article
\bibitem[{Belitz and Kirkpatrick(1994)}]{Belitz94}
\bibinfo{author}{D.~Belitz}, \bibinfo{author}{T.~R. Kirkpatrick},
\newblock \bibinfo{title}{{The Anderson-Mott transition}},
\newblock \bibinfo{journal}{Rev. Mod. Phys.} \bibinfo{volume}{66}
  (\bibinfo{year}{1994}) \bibinfo{pages}{261--380}. \URLprefix
  \url{https://link.aps.org/doi/10.1103/RevModPhys.66.261}.
  \DOIprefix\doi{10.1103/RevModPhys.66.261}.
%Type = Book
\bibitem[{Akkermans and Montambaux(2010)}]{Montambaux}
\bibinfo{author}{E.~Akkermans}, \bibinfo{author}{G.~Montambaux},
  \bibinfo{title}{Mesoscopic Physics of Electrons and Photons},
  \bibinfo{publisher}{Cambridge University Press}, \bibinfo{year}{2010}.
  \DOIprefix\doi{10.1017/CBO9780511618833}.
%Type = Article
\bibitem[{Rurali(2010)}]{rurali2010rmp}
\bibinfo{author}{R.~Rurali},
\newblock \bibinfo{title}{{Colloquium: Structural, electronic, and transport
  properties of silicon nanowires}},
\newblock \bibinfo{journal}{Rev. Mod. Phys.} \bibinfo{volume}{82}
  (\bibinfo{year}{2010}) \bibinfo{pages}{427--449}. \URLprefix
  \url{https://link.aps.org/doi/10.1103/RevModPhys.82.427}.
  \DOIprefix\doi{10.1103/RevModPhys.82.427}.
%Type = Article
\bibitem[{Dasgupta et~al.(2014)Dasgupta, Sun, Liu, Brittman, Andrews, Lim, Gao,
  Yan, and Yang}]{dasgupta2014am}
\bibinfo{author}{N.~P. Dasgupta}, \bibinfo{author}{J.~Sun},
  \bibinfo{author}{C.~Liu}, \bibinfo{author}{S.~Brittman},
  \bibinfo{author}{S.~C. Andrews}, \bibinfo{author}{J.~Lim},
  \bibinfo{author}{H.~Gao}, \bibinfo{author}{R.~Yan},
  \bibinfo{author}{P.~Yang},
\newblock \bibinfo{title}{{25th Anniversary Article: Semiconductor Nanowires -
  Synthesis, Characterization, and Applications}},
\newblock \bibinfo{journal}{Adv. Mater.} \bibinfo{volume}{26}
  (\bibinfo{year}{2014}) \bibinfo{pages}{2137--2184}. \URLprefix
  \url{https://onlinelibrary.wiley.com/doi/abs/10.1002/adma.201305929}.
  \DOIprefix\doi{10.1002/adma.201305929}.
%Type = Article
\bibitem[{Charlier et~al.(2007)Charlier, Blase, and Roche}]{charlier2007rmp}
\bibinfo{author}{J.-C. Charlier}, \bibinfo{author}{X.~Blase},
  \bibinfo{author}{S.~Roche},
\newblock \bibinfo{title}{Electronic and transport properties of nanotubes},
\newblock \bibinfo{journal}{Rev. Mod. Phys.} \bibinfo{volume}{79}
  (\bibinfo{year}{2007}) \bibinfo{pages}{677--732}. \URLprefix
  \url{https://link.aps.org/doi/10.1103/RevModPhys.79.677}.
  \DOIprefix\doi{10.1103/RevModPhys.79.677}.
%Type = Article
\bibitem[{Laird et~al.(2015)Laird, Kuemmeth, Steele, Grove-Rasmussen,
  Nyg\aa{}rd, Flensberg, and Kouwenhoven}]{laird2015rmp}
\bibinfo{author}{E.~A. Laird}, \bibinfo{author}{F.~Kuemmeth},
  \bibinfo{author}{G.~A. Steele}, \bibinfo{author}{K.~Grove-Rasmussen},
  \bibinfo{author}{J.~Nyg\aa{}rd}, \bibinfo{author}{K.~Flensberg},
  \bibinfo{author}{L.~P. Kouwenhoven},
\newblock \bibinfo{title}{{Quantum transport in carbon nanotubes}},
\newblock \bibinfo{journal}{Rev. Mod. Phys.} \bibinfo{volume}{87}
  (\bibinfo{year}{2015}) \bibinfo{pages}{703--764}. \URLprefix
  \url{https://link.aps.org/doi/10.1103/RevModPhys.87.703}.
  \DOIprefix\doi{10.1103/RevModPhys.87.703}.
%Type = Article
\bibitem[{Castro~Neto et~al.(2009)Castro~Neto, Guinea, Peres, Novoselov, and
  Geim}]{castro2009rmp}
\bibinfo{author}{A.~H. Castro~Neto}, \bibinfo{author}{F.~Guinea},
  \bibinfo{author}{N.~M.~R. Peres}, \bibinfo{author}{K.~S. Novoselov},
  \bibinfo{author}{A.~K. Geim},
\newblock \bibinfo{title}{{The electronic properties of graphene}},
\newblock \bibinfo{journal}{Rev. Mod. Phys.} \bibinfo{volume}{81}
  (\bibinfo{year}{2009}) \bibinfo{pages}{109--162}. \URLprefix
  \url{https://link.aps.org/doi/10.1103/RevModPhys.81.109}.
  \DOIprefix\doi{10.1103/RevModPhys.81.109}.
%Type = Article
\bibitem[{Das~Sarma et~al.(2011)Das~Sarma, Adam, Hwang, and Rossi}]{das2011rmp}
\bibinfo{author}{S.~Das~Sarma}, \bibinfo{author}{S.~Adam},
  \bibinfo{author}{E.~H. Hwang}, \bibinfo{author}{E.~Rossi},
\newblock \bibinfo{title}{{Electronic transport in two-dimensional graphene}},
\newblock \bibinfo{journal}{Rev. Mod. Phys.} \bibinfo{volume}{83}
  (\bibinfo{year}{2011}) \bibinfo{pages}{407--470}. \URLprefix
  \url{https://link.aps.org/doi/10.1103/RevModPhys.83.407}.
  \DOIprefix\doi{10.1103/RevModPhys.83.407}.
%Type = Article
\bibitem[{Ferrari et~al.(2015)Ferrari, Bonaccorso, Fal{'}ko, Novoselov, Roche,
  B{\o}ggild, Borini, Koppens, Palermo, Pugno, Garrido, Sordan, Bianco,
  Ballerini, Prato, Lidorikis, Kivioja, Marinelli, Ryh{\"a}nen, Morpurgo,
  Coleman, Nicolosi, Colombo, Fert, Garcia-Hernandez, Bachtold, Schneider,
  Guinea, Dekker, Barbone, Sun, Galiotis, Grigorenko, Konstantatos, Kis,
  Katsnelson, Vandersypen, Loiseau, Morandi, Neumaier, Treossi, Pellegrini,
  Polini, Tredicucci, Williams, Hee~Hong, Ahn, Min~Kim, Zirath, van Wees,
  van~der Zant, Occhipinti, Di~Matteo, Kinloch, Seyller, Quesnel, Feng, Teo,
  Rupesinghe, Hakonen, Neil, Tannock, L{\"o}fwander, and
  Kinaret}]{ferrari2015nanoscale}
\bibinfo{author}{A.~C. Ferrari}, \bibinfo{author}{F.~Bonaccorso},
  \bibinfo{author}{V.~Fal{'}ko}, \bibinfo{author}{K.~S. Novoselov},
  \bibinfo{author}{S.~Roche}, \bibinfo{author}{P.~B{\o}ggild},
  \bibinfo{author}{S.~Borini}, \bibinfo{author}{F.~H.~L. Koppens},
  \bibinfo{author}{V.~Palermo}, \bibinfo{author}{N.~Pugno},
  \bibinfo{author}{J.~A. Garrido}, \bibinfo{author}{R.~Sordan},
  \bibinfo{author}{A.~Bianco}, \bibinfo{author}{L.~Ballerini},
  \bibinfo{author}{M.~Prato}, \bibinfo{author}{E.~Lidorikis},
  \bibinfo{author}{J.~Kivioja}, \bibinfo{author}{C.~Marinelli},
  \bibinfo{author}{T.~Ryh{\"a}nen}, \bibinfo{author}{A.~Morpurgo},
  \bibinfo{author}{J.~N. Coleman}, \bibinfo{author}{V.~Nicolosi},
  \bibinfo{author}{L.~Colombo}, \bibinfo{author}{A.~Fert},
  \bibinfo{author}{M.~Garcia-Hernandez}, \bibinfo{author}{A.~Bachtold},
  \bibinfo{author}{G.~F. Schneider}, \bibinfo{author}{F.~Guinea},
  \bibinfo{author}{C.~Dekker}, \bibinfo{author}{M.~Barbone},
  \bibinfo{author}{Z.~Sun}, \bibinfo{author}{C.~Galiotis},
  \bibinfo{author}{A.~N. Grigorenko}, \bibinfo{author}{G.~Konstantatos},
  \bibinfo{author}{A.~Kis}, \bibinfo{author}{M.~Katsnelson},
  \bibinfo{author}{L.~Vandersypen}, \bibinfo{author}{A.~Loiseau},
  \bibinfo{author}{V.~Morandi}, \bibinfo{author}{D.~Neumaier},
  \bibinfo{author}{E.~Treossi}, \bibinfo{author}{V.~Pellegrini},
  \bibinfo{author}{M.~Polini}, \bibinfo{author}{A.~Tredicucci},
  \bibinfo{author}{G.~M. Williams}, \bibinfo{author}{B.~Hee~Hong},
  \bibinfo{author}{J.-H. Ahn}, \bibinfo{author}{J.~Min~Kim},
  \bibinfo{author}{H.~Zirath}, \bibinfo{author}{B.~J. van Wees},
  \bibinfo{author}{H.~van~der Zant}, \bibinfo{author}{L.~Occhipinti},
  \bibinfo{author}{A.~Di~Matteo}, \bibinfo{author}{I.~A. Kinloch},
  \bibinfo{author}{T.~Seyller}, \bibinfo{author}{E.~Quesnel},
  \bibinfo{author}{X.~Feng}, \bibinfo{author}{K.~Teo},
  \bibinfo{author}{N.~Rupesinghe}, \bibinfo{author}{P.~Hakonen},
  \bibinfo{author}{S.~R.~T. Neil}, \bibinfo{author}{Q.~Tannock},
  \bibinfo{author}{T.~L{\"o}fwander}, \bibinfo{author}{J.~Kinaret},
\newblock \bibinfo{title}{{Science and technology roadmap for graphene{,}
  related two-dimensional crystals{,} and hybrid systems}},
\newblock \bibinfo{journal}{Nanoscale} \bibinfo{volume}{7}
  (\bibinfo{year}{2015}) \bibinfo{pages}{4598--4810}. \URLprefix
  \url{http://dx.doi.org/10.1039/C4NR01600A}.
  \DOIprefix\doi{10.1039/C4NR01600A}.
%Type = Article
\bibitem[{Mucciolo and Lewenkopf(2010)}]{mucciolo2010jpcm}
\bibinfo{author}{E.~R. Mucciolo}, \bibinfo{author}{C.~H. Lewenkopf},
\newblock \bibinfo{title}{Disorder and electronic transport in graphene},
\newblock \bibinfo{journal}{J. Phys. Condens. Matter} \bibinfo{volume}{22}
  (\bibinfo{year}{2010}) \bibinfo{pages}{273201}. \URLprefix
  \url{http://stacks.iop.org/0953-8984/22/i=27/a=273201}.
%Type = Article
\bibitem[{Geim and Grigorieva(2013)}]{geim2013nature}
\bibinfo{author}{A.~K. Geim}, \bibinfo{author}{I.~V. Grigorieva},
\newblock \bibinfo{title}{{Van der Waals heterostructures}},
\newblock \bibinfo{journal}{Nature} \bibinfo{volume}{499}
  (\bibinfo{year}{2013}) \bibinfo{pages}{419}. \URLprefix
  \url{http://dx.doi.org/10.1038/nature12385}.
  \DOIprefix\doi{10.1038/nature12385}.
%Type = Article
\bibitem[{Novoselov et~al.(2016)Novoselov, Mishchenko, Carvalho, and
  Castro~Neto}]{novoselov2016science}
\bibinfo{author}{K.~S. Novoselov}, \bibinfo{author}{A.~Mishchenko},
  \bibinfo{author}{A.~Carvalho}, \bibinfo{author}{A.~H. Castro~Neto},
\newblock \bibinfo{title}{{2D materials and van der Waals heterostructures}},
\newblock \bibinfo{journal}{Science} \bibinfo{volume}{353}
  (\bibinfo{year}{2016}) \bibinfo{pages}{aac9439}. \URLprefix
  \url{http://science.sciencemag.org/content/353/6298/aac9439}.
  \DOIprefix\doi{10.1126/science.aac9439}.
%Type = Article
\bibitem[{Bordone et~al.(1999)Bordone, Pascoli, Brunetti, Bertoni, Jacoboni,
  and Abramo}]{Bordone1999}
\bibinfo{author}{P.~Bordone}, \bibinfo{author}{M.~Pascoli},
  \bibinfo{author}{R.~Brunetti}, \bibinfo{author}{A.~Bertoni},
  \bibinfo{author}{C.~Jacoboni}, \bibinfo{author}{A.~Abramo},
\newblock \bibinfo{title}{Quantum transport of electrons in open nanostructures
  with the wigner-function formalism},
\newblock \bibinfo{journal}{Phys. Rev. B} \bibinfo{volume}{59}
  (\bibinfo{year}{1999}) \bibinfo{pages}{3060--3069}. \URLprefix
  \url{https://link.aps.org/doi/10.1103/PhysRevB.59.3060}.
  \DOIprefix\doi{10.1103/PhysRevB.59.3060}.
%Type = Article
\bibitem[{Nedjalkov et~al.(2004)Nedjalkov, Kosina, Selberherr, Ringhofer, and
  Ferry}]{Nedjalkov2004}
\bibinfo{author}{M.~Nedjalkov}, \bibinfo{author}{H.~Kosina},
  \bibinfo{author}{S.~Selberherr}, \bibinfo{author}{C.~Ringhofer},
  \bibinfo{author}{D.~K. Ferry},
\newblock \bibinfo{title}{Unified particle approach to wigner-boltzmann
  transport in small semiconductor devices},
\newblock \bibinfo{journal}{Phys. Rev. B} \bibinfo{volume}{70}
  (\bibinfo{year}{2004}) \bibinfo{pages}{115319}. \URLprefix
  \url{https://link.aps.org/doi/10.1103/PhysRevB.70.115319}.
  \DOIprefix\doi{10.1103/PhysRevB.70.115319}.
%Type = Article
\bibitem[{Hershfield and Ambegaokar(1986)}]{Hershfield1986}
\bibinfo{author}{S.~Hershfield}, \bibinfo{author}{V.~Ambegaokar},
\newblock \bibinfo{title}{Transport equation for weakly localized electrons},
\newblock \bibinfo{journal}{Phys. Rev. B} \bibinfo{volume}{34}
  (\bibinfo{year}{1986}) \bibinfo{pages}{2147--2151}. \URLprefix
  \url{https://link.aps.org/doi/10.1103/PhysRevB.34.2147}.
  \DOIprefix\doi{10.1103/PhysRevB.34.2147}.
%Type = Article
\bibitem[{Landauer(1957)}]{landauer1957ibm}
\bibinfo{author}{R.~Landauer},
\newblock \bibinfo{title}{{Spatial Variation of Currents and Fields Due to
  Localized Scatterers in Metallic Conduction}},
\newblock \bibinfo{journal}{IBM J. Res. Dev.} \bibinfo{volume}{1}
  (\bibinfo{year}{1957}) \bibinfo{pages}{223--231}.
  \DOIprefix\doi{doi:10.1147/rd.13.0223}.
%Type = Article
\bibitem[{Landauer(1970)}]{landauer1970pm}
\bibinfo{author}{R.~Landauer},
\newblock \bibinfo{title}{{Electrical resistance of disordered one-dimensional
  lattices}},
\newblock \bibinfo{journal}{Phil. Mag.} \bibinfo{volume}{21}
  (\bibinfo{year}{1970}) \bibinfo{pages}{863--867}.
  \DOIprefix\doi{10.1080/14786437008238472}.
%Type = Article
\bibitem[{B\"uttiker et~al.(1985)B\"uttiker, Imry, Landauer, and
  Pinhas}]{buttiker1985prb}
\bibinfo{author}{M.~B\"uttiker}, \bibinfo{author}{Y.~Imry},
  \bibinfo{author}{R.~Landauer}, \bibinfo{author}{S.~Pinhas},
\newblock \bibinfo{title}{{Generalized many-channel conductance formula with
  application to small rings}},
\newblock \bibinfo{journal}{Phys. Rev. B} \bibinfo{volume}{31}
  (\bibinfo{year}{1985}) \bibinfo{pages}{6207--6215}. \URLprefix
  \url{https://link.aps.org/doi/10.1103/PhysRevB.31.6207}.
  \DOIprefix\doi{10.1103/PhysRevB.31.6207}.
%Type = Article
\bibitem[{Stone and Szafer(1988)}]{StoneIBM1988}
\bibinfo{author}{S.~D. Stone}, \bibinfo{author}{A.~Szafer},
\newblock \bibinfo{title}{{What is Measured when You Measure a Resistance? --
  The Landauer Formula Revisited}},
\newblock \bibinfo{journal}{IBM J. Res. Dev.} \bibinfo{volume}{32}
  (\bibinfo{year}{1988}) \bibinfo{pages}{384--413}.
  \DOIprefix\doi{10.1147/rd.323.0384}.
%Type = Article
\bibitem[{Baranger and Stone(1989)}]{Baranger89}
\bibinfo{author}{H.~U. Baranger}, \bibinfo{author}{A.~D. Stone},
\newblock \bibinfo{title}{Electrical linear-response theory in an arbitrary
  magnetic field: A new fermi-surface formation},
\newblock \bibinfo{journal}{Phys. Rev. B} \bibinfo{volume}{40}
  (\bibinfo{year}{1989}) \bibinfo{pages}{8169--8193}.
  \DOIprefix\doi{10.1103/PhysRevB.40.8169}.
%Type = Article
\bibitem[{Nikoli\ifmmode~\acute{c}\else \'{c}\fi{}(2001)}]{Nikolic2001}
\bibinfo{author}{B.~K. Nikoli\ifmmode~\acute{c}\else \'{c}\fi{}},
\newblock \bibinfo{title}{Deconstructing kubo formula usage: Exact conductance
  of a mesoscopic system from weak to strong disorder limit},
\newblock \bibinfo{journal}{Phys. Rev. B} \bibinfo{volume}{64}
  (\bibinfo{year}{2001}) \bibinfo{pages}{165303}.
  \DOIprefix\doi{10.1103/PhysRevB.64.165303}.
%Type = Book
\bibitem[{Pines and Nozieres(1989)}]{Pines89}
\bibinfo{author}{D.~Pines}, \bibinfo{author}{P.~Nozieres}, \bibinfo{title}{The
  Theory of Quantum Liquids}, \bibinfo{publisher}{Addison-Wesley, California},
  \bibinfo{year}{1989}.
%Type = Book
\bibitem[{Doniach and Sondheimer(1974)}]{Doniach74}
\bibinfo{author}{S.~Doniach}, \bibinfo{author}{E.~H. Sondheimer},
  \bibinfo{title}{Green’s Functions for Solid State Physicists},
  \bibinfo{publisher}{Addison-Wesley}, \bibinfo{year}{1974}.
%Type = Book
\bibitem[{Kubo et~al.(1985)Kubo, Toda, and Hashitsume}]{kubo1985book}
\bibinfo{author}{R.~Kubo}, \bibinfo{author}{M.~Toda},
  \bibinfo{author}{N.~Hashitsume}, \bibinfo{title}{Statistical Physics II:
  Nonequilibrium Statistical Mechanics}, \bibinfo{publisher}{Springer-Verlag
  Berlin Heidelberg}, \bibinfo{year}{1985}.
%Type = Article
\bibitem[{Kubo(1966)}]{kubo1965rpp}
\bibinfo{author}{R.~Kubo},
\newblock \bibinfo{title}{{The fluctuation-dissipation theorem}},
\newblock \bibinfo{journal}{Rep. Prog. Phys.} \bibinfo{volume}{29}
  (\bibinfo{year}{1966}) \bibinfo{pages}{255}. \URLprefix
  \url{http://stacks.iop.org/0034-4885/29/i=1/a=306}.
%Type = Article
\bibitem[{Thouless and Kirkpatrick(1981)}]{thouless1981jpc}
\bibinfo{author}{D.~J. Thouless}, \bibinfo{author}{S.~Kirkpatrick},
\newblock \bibinfo{title}{{Conductivity of the disordered linear chain}},
\newblock \bibinfo{journal}{J. Phys. C: Solid State Phys.} \bibinfo{volume}{14}
  (\bibinfo{year}{1981}) \bibinfo{pages}{235}. \URLprefix
  \url{http://stacks.iop.org/0022-3719/14/i=3/a=007}.
%Type = Article
\bibitem[{Bose et~al.(1993)Bose, Jepsen, and Andersen}]{Bose1993}
\bibinfo{author}{S.~K. Bose}, \bibinfo{author}{O.~Jepsen},
  \bibinfo{author}{O.~K. Andersen},
\newblock \bibinfo{title}{Real-space calculation of the electrical resistivity
  of liquid 3d transition metals using tight-binding linear muffin-tin
  orbitals},
\newblock \bibinfo{journal}{Phys. Rev. B} \bibinfo{volume}{48}
  (\bibinfo{year}{1993}) \bibinfo{pages}{4265--4275}. \URLprefix
  \url{https://link.aps.org/doi/10.1103/PhysRevB.48.4265}.
  \DOIprefix\doi{10.1103/PhysRevB.48.4265}.
%Type = Article
\bibitem[{Mayou(1988)}]{mayou1988epl}
\bibinfo{author}{D.~Mayou},
\newblock \bibinfo{title}{{Calculation of the Conductivity in the
  Short-Mean-Free-Path Regime}},
\newblock \bibinfo{journal}{EPL} \bibinfo{volume}{6} (\bibinfo{year}{1988})
  \bibinfo{pages}{549}. \URLprefix
  \url{http://stacks.iop.org/0295-5075/6/i=6/a=013}.
%Type = Article
\bibitem[{Mayou and Khanna(1995)}]{mayou1995jpi}
\bibinfo{author}{D.~Mayou}, \bibinfo{author}{S.~Khanna},
\newblock \bibinfo{title}{{A Real-Space Approach to Electronic Transport}},
\newblock \bibinfo{journal}{{J. Phys. I}} \bibinfo{volume}{5}
  (\bibinfo{year}{1995}) \bibinfo{pages}{1199--1211}. \URLprefix
  \url{https://hal.archives-ouvertes.fr/jpa-00247129}.
  \DOIprefix\doi{10.1051/jp1:1995191}.
%Type = Article
\bibitem[{Roche and Mayou(1997)}]{roche1997prl}
\bibinfo{author}{S.~Roche}, \bibinfo{author}{D.~Mayou},
\newblock \bibinfo{title}{{Conductivity of Quasiperiodic Systems: A Numerical
  Study}},
\newblock \bibinfo{journal}{Phys. Rev. Lett.} \bibinfo{volume}{79}
  (\bibinfo{year}{1997}) \bibinfo{pages}{2518--2521}. \URLprefix
  \url{https://link.aps.org/doi/10.1103/PhysRevLett.79.2518}.
  \DOIprefix\doi{10.1103/PhysRevLett.79.2518}.
%Type = Article
\bibitem[{Roche(1999)}]{roche1999prb}
\bibinfo{author}{S.~Roche},
\newblock \bibinfo{title}{{Quantum transport by means of $\mathrm{O}(N)$
  real-space methods}},
\newblock \bibinfo{journal}{Phys. Rev. B} \bibinfo{volume}{59}
  (\bibinfo{year}{1999}) \bibinfo{pages}{2284--2291}. \URLprefix
  \url{https://link.aps.org/doi/10.1103/PhysRevB.59.2284}.
  \DOIprefix\doi{10.1103/PhysRevB.59.2284}.
%Type = Article
\bibitem[{Ortmann and Roche(2013)}]{ortmann2013prl}
\bibinfo{author}{F.~Ortmann}, \bibinfo{author}{S.~Roche},
\newblock \bibinfo{title}{{Splitting of the Zero-Energy Landau Level and
  Universal Dissipative Conductivity at Critical Points in Disordered
  Graphene}},
\newblock \bibinfo{journal}{Phys. Rev. Lett.} \bibinfo{volume}{110}
  (\bibinfo{year}{2013}) \bibinfo{pages}{086602}. \URLprefix
  \url{https://link.aps.org/doi/10.1103/PhysRevLett.110.086602}.
  \DOIprefix\doi{10.1103/PhysRevLett.110.086602}.
%Type = Article
\bibitem[{Garc\'{\i}a et~al.(2015)Garc\'{\i}a, Covaci, and
  Rappoport}]{garcia2015prl}
\bibinfo{author}{J.~H. Garc\'{\i}a}, \bibinfo{author}{L.~Covaci},
  \bibinfo{author}{T.~G. Rappoport},
\newblock \bibinfo{title}{{Real-Space Calculation of the Conductivity Tensor
  for Disordered Topological Matter}},
\newblock \bibinfo{journal}{Phys. Rev. Lett.} \bibinfo{volume}{114}
  (\bibinfo{year}{2015}) \bibinfo{pages}{116602}. \URLprefix
  \url{https://link.aps.org/doi/10.1103/PhysRevLett.114.116602}.
  \DOIprefix\doi{10.1103/PhysRevLett.114.116602}.
%Type = Article
\bibitem[{Ortmann et~al.(2015)Ortmann, Leconte, and Roche}]{ortmann2015prb}
\bibinfo{author}{F.~Ortmann}, \bibinfo{author}{N.~Leconte},
  \bibinfo{author}{S.~Roche},
\newblock \bibinfo{title}{{Efficient linear scaling approach for computing the
  Kubo Hall conductivity}},
\newblock \bibinfo{journal}{Phys. Rev. B} \bibinfo{volume}{91}
  (\bibinfo{year}{2015}) \bibinfo{pages}{165117}. \URLprefix
  \url{https://link.aps.org/doi/10.1103/PhysRevB.91.165117}.
  \DOIprefix\doi{10.1103/PhysRevB.91.165117}.
%Type = Article
\bibitem[{Van~Tuan et~al.(2014)Van~Tuan, Ortmann, Soriano, Valenzuela, and
  Roche}]{vantuan2014np}
\bibinfo{author}{D.~Van~Tuan}, \bibinfo{author}{F.~Ortmann},
  \bibinfo{author}{D.~Soriano}, \bibinfo{author}{S.~O. Valenzuela},
  \bibinfo{author}{S.~Roche},
\newblock \bibinfo{title}{{Pseudospin-driven spin relaxation mechanism in
  graphene}},
\newblock \bibinfo{journal}{Nat. Phys.} \bibinfo{volume}{10}
  (\bibinfo{year}{2014}) \bibinfo{pages}{857--863}. \URLprefix
  \url{http://dx.doi.org/10.1038/nphys3083}. \DOIprefix\doi{10.1038/nphys3083}.
%Type = Article
\bibitem[{Cummings et~al.(2017)Cummings, Garc\'{\i}a, Fabian, and
  Roche}]{Cummings2017prl}
\bibinfo{author}{A.~W. Cummings}, \bibinfo{author}{J.~H. Garc\'{\i}a},
  \bibinfo{author}{J.~Fabian}, \bibinfo{author}{S.~Roche},
\newblock \bibinfo{title}{{Giant Spin Lifetime Anisotropy in Graphene Induced
  by Proximity Effects}},
\newblock \bibinfo{journal}{Phys. Rev. Lett.} \bibinfo{volume}{119}
  (\bibinfo{year}{2017}) \bibinfo{pages}{206601}. \URLprefix
  \url{https://link.aps.org/doi/10.1103/PhysRevLett.119.206601}.
  \DOIprefix\doi{10.1103/PhysRevLett.119.206601}.
%Type = Article
\bibitem[{Vierimaa et~al.(2017)Vierimaa, Fan, and Harju}]{vierimaa2017prb}
\bibinfo{author}{V.~Vierimaa}, \bibinfo{author}{Z.~Fan},
  \bibinfo{author}{A.~Harju},
\newblock \bibinfo{title}{{Scattering from spin-polarized charged impurities in
  graphene}},
\newblock \bibinfo{journal}{Phys. Rev. B} \bibinfo{volume}{95}
  (\bibinfo{year}{2017}) \bibinfo{pages}{041401}. \URLprefix
  \url{https://link.aps.org/doi/10.1103/PhysRevB.95.041401}.
  \DOIprefix\doi{10.1103/PhysRevB.95.041401}.
%Type = Article
\bibitem[{Li et~al.(2010)Li, Sevin\ifmmode~\mbox{\c{c}}\else \c{c}\fi{}li,
  Cuniberti, and Roche}]{li2010prb}
\bibinfo{author}{W.~Li}, \bibinfo{author}{H.~Sevin\ifmmode~\mbox{\c{c}}\else
  \c{c}\fi{}li}, \bibinfo{author}{G.~Cuniberti}, \bibinfo{author}{S.~Roche},
\newblock \bibinfo{title}{{Phonon transport in large scale carbon-based
  disordered materials: Implementation of an efficient order-$N$ and real-space
  Kubo methodology}},
\newblock \bibinfo{journal}{Phys. Rev. B} \bibinfo{volume}{82}
  (\bibinfo{year}{2010}) \bibinfo{pages}{041410}. \URLprefix
  \url{https://link.aps.org/doi/10.1103/PhysRevB.82.041410}.
  \DOIprefix\doi{10.1103/PhysRevB.82.041410}.
%Type = Article
\bibitem[{Li et~al.(2011)Li, Sevin\ifmmode~\mbox{\c{c}}\else \c{c}\fi{}li,
  Roche, and Cuniberti}]{li2011prb}
\bibinfo{author}{W.~Li}, \bibinfo{author}{H.~Sevin\ifmmode~\mbox{\c{c}}\else
  \c{c}\fi{}li}, \bibinfo{author}{S.~Roche}, \bibinfo{author}{G.~Cuniberti},
\newblock \bibinfo{title}{{Efficient linear scaling method for computing the
  thermal conductivity of disordered materials}},
\newblock \bibinfo{journal}{Phys. Rev. B} \bibinfo{volume}{83}
  (\bibinfo{year}{2011}) \bibinfo{pages}{155416}. \URLprefix
  \url{https://link.aps.org/doi/10.1103/PhysRevB.83.155416}.
  \DOIprefix\doi{10.1103/PhysRevB.83.155416}.
%Type = Article
\bibitem[{Sevin\ifmmode~\mbox{\c{c}}\else \c{c}\fi{}li
  et~al.(2011)Sevin\ifmmode~\mbox{\c{c}}\else \c{c}\fi{}li, Li, Mingo,
  Cuniberti, and Roche}]{sevincli2011prb}
\bibinfo{author}{H.~Sevin\ifmmode~\mbox{\c{c}}\else \c{c}\fi{}li},
  \bibinfo{author}{W.~Li}, \bibinfo{author}{N.~Mingo},
  \bibinfo{author}{G.~Cuniberti}, \bibinfo{author}{S.~Roche},
\newblock \bibinfo{title}{{Effects of domains in phonon conduction through
  hybrid boron nitride and graphene sheets}},
\newblock \bibinfo{journal}{Phys. Rev. B} \bibinfo{volume}{84}
  (\bibinfo{year}{2011}) \bibinfo{pages}{205444}. \URLprefix
  \url{https://link.aps.org/doi/10.1103/PhysRevB.84.205444}.
  \DOIprefix\doi{10.1103/PhysRevB.84.205444}.
%Type = Article
\bibitem[{Lherbier et~al.(2008)Lherbier, Blase, Niquet, Triozon, and
  Roche}]{lherbier2008prl_chemical}
\bibinfo{author}{A.~Lherbier}, \bibinfo{author}{X.~Blase},
  \bibinfo{author}{Y.-M. Niquet}, \bibinfo{author}{F.~Triozon},
  \bibinfo{author}{S.~Roche},
\newblock \bibinfo{title}{{Charge Transport in Chemically Doped 2D Graphene}},
\newblock \bibinfo{journal}{Phys. Rev. Lett.} \bibinfo{volume}{101}
  (\bibinfo{year}{2008}) \bibinfo{pages}{036808}. \URLprefix
  \url{https://link.aps.org/doi/10.1103/PhysRevLett.101.036808}.
  \DOIprefix\doi{10.1103/PhysRevLett.101.036808}.
%Type = Article
\bibitem[{Wehling et~al.(2010)Wehling, Yuan, Lichtenstein, Geim, and
  Katsnelson}]{WehlingPRL2010}
\bibinfo{author}{T.~O. Wehling}, \bibinfo{author}{S.~Yuan},
  \bibinfo{author}{A.~I. Lichtenstein}, \bibinfo{author}{A.~K. Geim},
  \bibinfo{author}{M.~I. Katsnelson},
\newblock \bibinfo{title}{Resonant scattering by realistic impurities in
  graphene},
\newblock \bibinfo{journal}{Phys. Rev. Lett.} \bibinfo{volume}{105}
  (\bibinfo{year}{2010}) \bibinfo{pages}{056802}. \URLprefix
  \url{https://link.aps.org/doi/10.1103/PhysRevLett.105.056802}.
  \DOIprefix\doi{10.1103/PhysRevLett.105.056802}.
%Type = Article
\bibitem[{Yuan et~al.(2010)Yuan, De~Raedt, and Katsnelson}]{yuan2010prb}
\bibinfo{author}{S.~Yuan}, \bibinfo{author}{H.~De~Raedt},
  \bibinfo{author}{M.~I. Katsnelson},
\newblock \bibinfo{title}{{Modeling electronic structure and transport
  properties of graphene with resonant scattering centers}},
\newblock \bibinfo{journal}{Phys. Rev. B} \bibinfo{volume}{82}
  (\bibinfo{year}{2010}) \bibinfo{pages}{115448}. \URLprefix
  \url{https://link.aps.org/doi/10.1103/PhysRevB.82.115448}.
  \DOIprefix\doi{10.1103/PhysRevB.82.115448}.
%Type = Article
\bibitem[{Radchenko et~al.(2013)Radchenko, Shylau, Zozoulenko, and
  Ferreira}]{radchenko2013prb}
\bibinfo{author}{T.~M. Radchenko}, \bibinfo{author}{A.~A. Shylau},
  \bibinfo{author}{I.~V. Zozoulenko}, \bibinfo{author}{A.~Ferreira},
\newblock \bibinfo{title}{{Effect of charged line defects on conductivity in
  graphene: Numerical Kubo and analytical Boltzmann approaches}},
\newblock \bibinfo{journal}{Phys. Rev. B} \bibinfo{volume}{87}
  (\bibinfo{year}{2013}) \bibinfo{pages}{195448}. \URLprefix
  \url{https://link.aps.org/doi/10.1103/PhysRevB.87.195448}.
  \DOIprefix\doi{10.1103/PhysRevB.87.195448}.
%Type = Article
\bibitem[{Trambly De~Laissardi\`ere and Mayou(2013)}]{laissardiere2013prl}
\bibinfo{author}{G.~Trambly De~Laissardi\`ere}, \bibinfo{author}{D.~Mayou},
\newblock \bibinfo{title}{{Conductivity of Graphene with Resonant and
  Nonresonant Adsorbates}},
\newblock \bibinfo{journal}{Phys. Rev. Lett.} \bibinfo{volume}{111}
  (\bibinfo{year}{2013}) \bibinfo{pages}{146601}. \URLprefix
  \url{https://link.aps.org/doi/10.1103/PhysRevLett.111.146601}.
  \DOIprefix\doi{10.1103/PhysRevLett.111.146601}.
%Type = Article
\bibitem[{Gargiulo et~al.(2014)Gargiulo, Aut\`es, Virk, Barthel, R\"osner,
  Toller, Wehling, and Yazyev}]{GargiuloPRL2014}
\bibinfo{author}{F.~Gargiulo}, \bibinfo{author}{G.~Aut\`es},
  \bibinfo{author}{N.~Virk}, \bibinfo{author}{S.~Barthel},
  \bibinfo{author}{M.~R\"osner}, \bibinfo{author}{L.~R.~M. Toller},
  \bibinfo{author}{T.~O. Wehling}, \bibinfo{author}{O.~V. Yazyev},
\newblock \bibinfo{title}{Electronic transport in graphene with aggregated
  hydrogen adatoms},
\newblock \bibinfo{journal}{Phys. Rev. Lett.} \bibinfo{volume}{113}
  (\bibinfo{year}{2014}) \bibinfo{pages}{246601}.
  \DOIprefix\doi{10.1103/PhysRevLett.113.246601}.
%Type = Article
\bibitem[{Zhao et~al.(2015)Zhao, Yuan, Katsnelson, and De~Raedt}]{zhao2015prb}
\bibinfo{author}{P.-L. Zhao}, \bibinfo{author}{S.~Yuan}, \bibinfo{author}{M.~I.
  Katsnelson}, \bibinfo{author}{H.~De~Raedt},
\newblock \bibinfo{title}{{Fingerprints of disorder source in graphene}},
\newblock \bibinfo{journal}{Phys. Rev. B} \bibinfo{volume}{92}
  (\bibinfo{year}{2015}) \bibinfo{pages}{045437}.
  \DOIprefix\doi{10.1103/PhysRevB.92.045437}.
%Type = Article
\bibitem[{Ferreira and Mucciolo(2015)}]{ferreira2015prl}
\bibinfo{author}{A.~Ferreira}, \bibinfo{author}{E.~R. Mucciolo},
\newblock \bibinfo{title}{{Critical Delocalization of Chiral Zero Energy Modes
  in Graphene}},
\newblock \bibinfo{journal}{Phys. Rev. Lett.} \bibinfo{volume}{115}
  (\bibinfo{year}{2015}) \bibinfo{pages}{106601}. \URLprefix
  \url{https://link.aps.org/doi/10.1103/PhysRevLett.115.106601}.
  \DOIprefix\doi{10.1103/PhysRevLett.115.106601}.
%Type = Article
\bibitem[{Yuan et~al.(2010)Yuan, De~Raedt, and Katsnelson}]{yuan2010prb_b}
\bibinfo{author}{S.~Yuan}, \bibinfo{author}{H.~De~Raedt},
  \bibinfo{author}{M.~I. Katsnelson},
\newblock \bibinfo{title}{{Electronic transport in disordered bilayer and
  trilayer graphene}},
\newblock \bibinfo{journal}{Phys. Rev. B} \bibinfo{volume}{82}
  (\bibinfo{year}{2010}) \bibinfo{pages}{235409}. \URLprefix
  \url{https://link.aps.org/doi/10.1103/PhysRevB.82.235409}.
  \DOIprefix\doi{10.1103/PhysRevB.82.235409}.
%Type = Article
\bibitem[{Missaoui et~al.(2018)Missaoui, Khabthani, Jaidane, Mayou, and
  de~Laissardi{\`{e}}re}]{Missaoui_2018}
\bibinfo{author}{A.~Missaoui}, \bibinfo{author}{J.~J. Khabthani},
  \bibinfo{author}{N.-E. Jaidane}, \bibinfo{author}{D.~Mayou},
  \bibinfo{author}{G.~T. de~Laissardi{\`{e}}re},
\newblock \bibinfo{title}{Mobility gap and quantum transport in a
  functionalized graphene bilayer},
\newblock \bibinfo{journal}{J. Phys. Condens. Matter} \bibinfo{volume}{30}
  (\bibinfo{year}{2018}) \bibinfo{pages}{195701}. \URLprefix
  \url{https://doi.org/10.1088\%2F1361-648x\%2Faaba06}.
  \DOIprefix\doi{10.1088/1361-648x/aaba06}.
%Type = Article
\bibitem[{Ortmann and Roche(2011)}]{ortmann2011prb}
\bibinfo{author}{F.~Ortmann}, \bibinfo{author}{S.~Roche},
\newblock \bibinfo{title}{{Polaron transport in organic crystals: Temperature
  tuning of disorder effects}},
\newblock \bibinfo{journal}{Phys. Rev. B} \bibinfo{volume}{84}
  (\bibinfo{year}{2011}) \bibinfo{pages}{180302}. \URLprefix
  \url{https://link.aps.org/doi/10.1103/PhysRevB.84.180302}.
  \DOIprefix\doi{10.1103/PhysRevB.84.180302}.
%Type = Article
\bibitem[{Ishii et~al.(2015)Ishii, Kobayashi, and Hirose}]{Ishii2015}
\bibinfo{author}{H.~Ishii}, \bibinfo{author}{N.~Kobayashi},
  \bibinfo{author}{K.~Hirose},
\newblock \bibinfo{title}{Wave packet dynamical calculations for charge
  transport of organic semiconductors: Role of molecular vibrations and trap
  potentials},
\newblock \bibinfo{journal}{Mol. Cryst. Liq. Cryst.} \bibinfo{volume}{620}
  (\bibinfo{year}{2015}) \bibinfo{pages}{2--9}.
  \DOIprefix\doi{10.1080/15421406.2015.1094608}.
%Type = Article
\bibitem[{Ishii et~al.(2017)Ishii, Kobayashi, and Hirose}]{Ishii2017}
\bibinfo{author}{H.~Ishii}, \bibinfo{author}{N.~Kobayashi},
  \bibinfo{author}{K.~Hirose},
\newblock \bibinfo{title}{Charge transport calculations by a wave-packet
  dynamical approach using maximally localized wannier functions based on
  density functional theory: Application to high-mobility organic
  semiconductors},
\newblock \bibinfo{journal}{Phys. Rev. B} \bibinfo{volume}{95}
  (\bibinfo{year}{2017}) \bibinfo{pages}{035433}.
  \DOIprefix\doi{10.1103/PhysRevB.95.035433}.
%Type = Article
\bibitem[{Fratini et~al.(2017)Fratini, Ciuchi, Mayou, de~Laissardi{\`e}re, and
  Troisi}]{FratiniNatMat2017}
\bibinfo{author}{S.~Fratini}, \bibinfo{author}{S.~Ciuchi},
  \bibinfo{author}{D.~Mayou}, \bibinfo{author}{G.~T. de~Laissardi{\`e}re},
  \bibinfo{author}{A.~Troisi},
\newblock \bibinfo{title}{{A map of high-mobility molecular semiconductors}},
\newblock \bibinfo{journal}{Nat. Mater.} \bibinfo{volume}{16}
  (\bibinfo{year}{2017}) \bibinfo{pages}{998}. \URLprefix
  \url{https://doi.org/10.1038/nmat4970}, \bibinfo{note}{article}.
%Type = Article
\bibitem[{Ishii et~al.(2018)Ishii, Inoue, Kobayashi, and Hirose}]{Ishii2018}
\bibinfo{author}{H.~Ishii}, \bibinfo{author}{J.-i. Inoue},
  \bibinfo{author}{N.~Kobayashi}, \bibinfo{author}{K.~Hirose},
\newblock \bibinfo{title}{Quantitative mobility evaluation of organic
  semiconductors using quantum dynamics based on density functional theory},
\newblock \bibinfo{journal}{Phys. Rev. B} \bibinfo{volume}{98}
  (\bibinfo{year}{2018}) \bibinfo{pages}{235422}.
  \DOIprefix\doi{10.1103/PhysRevB.98.235422}.
%Type = Article
\bibitem[{Ihnatsenka et~al.(2015)Ihnatsenka, Crispin, and
  Zozoulenko}]{Ihnatsenka2015}
\bibinfo{author}{S.~Ihnatsenka}, \bibinfo{author}{X.~Crispin},
  \bibinfo{author}{I.~V. Zozoulenko},
\newblock \bibinfo{title}{Understanding hopping transport and thermoelectric
  properties of conducting polymers},
\newblock \bibinfo{journal}{Phys. Rev. B} \bibinfo{volume}{92}
  (\bibinfo{year}{2015}) \bibinfo{pages}{035201}.
  \DOIprefix\doi{10.1103/PhysRevB.92.035201}.
%Type = Article
\bibitem[{Adjizian et~al.(2016)Adjizian, Lherbier, M.-M.~Dubois,
  Botello-M{\'e}ndez, and Charlier}]{Adjizian2016}
\bibinfo{author}{J.-J. Adjizian}, \bibinfo{author}{A.~Lherbier},
  \bibinfo{author}{S.~M.-M.~Dubois}, \bibinfo{author}{A.~R.
  Botello-M{\'e}ndez}, \bibinfo{author}{J.-C. Charlier},
\newblock \bibinfo{title}{The electronic and transport properties of
  two-dimensional conjugated polymer networks including disorder},
\newblock \bibinfo{journal}{Nanoscale} \bibinfo{volume}{8}
  (\bibinfo{year}{2016}) \bibinfo{pages}{1642--1651}.
  \DOIprefix\doi{10.1039/C5NR06825H}.
%Type = Article
\bibitem[{Tonnel{\'{e}} et~al.(2019)Tonnel{\'{e}}, Pershin, Gali, Lherbier,
  Charlier, Castet, Muccioli, and Beljonne}]{TonneleJPM2019}
\bibinfo{author}{C.~Tonnel{\'{e}}}, \bibinfo{author}{A.~Pershin},
  \bibinfo{author}{S.~M. Gali}, \bibinfo{author}{A.~Lherbier},
  \bibinfo{author}{J.-C. Charlier}, \bibinfo{author}{F.~Castet},
  \bibinfo{author}{L.~Muccioli}, \bibinfo{author}{D.~Beljonne},
\newblock \bibinfo{title}{Atomistic simulations of charge transport in
  photoswitchable organic-graphene hybrids},
\newblock \bibinfo{journal}{J. Phys. Mater.} \bibinfo{volume}{2}
  (\bibinfo{year}{2019}) \bibinfo{pages}{035001}.
  \DOIprefix\doi{10.1088/2515-7639/ab1314}.
%Type = Article
\bibitem[{Roche et~al.(1997)Roche, Trambly~de Laissardi\`{e}re, and
  Mayou}]{roche1997jmp}
\bibinfo{author}{S.~Roche}, \bibinfo{author}{G.~Trambly~de Laissardi\`{e}re},
  \bibinfo{author}{D.~Mayou},
\newblock \bibinfo{title}{{Electronic transport properties of quasicrystals}},
\newblock \bibinfo{journal}{J. Math. Phys.} \bibinfo{volume}{38}
  (\bibinfo{year}{1997}) \bibinfo{pages}{1794--1822}. \URLprefix
  \url{https://doi.org/10.1063/1.531914}. \DOIprefix\doi{10.1063/1.531914}.
%Type = Article
\bibitem[{Trambly De~Laissardi\`ere and Mayou(2014)}]{DELAISSARDIERE201470}
\bibinfo{author}{G.~Trambly De~Laissardi\`ere}, \bibinfo{author}{D.~Mayou},
\newblock \bibinfo{title}{Anomalous electronic transport in quasicrystals and
  related complex metallic alloys},
\newblock \bibinfo{journal}{Compt. Rendus Phys.} \bibinfo{volume}{15}
  (\bibinfo{year}{2014}) \bibinfo{pages}{70--81}.
  \DOIprefix\doi{10.1016/j.crhy.2013.09.010}.
%Type = Article
\bibitem[{Trambly De~Laissardi\`ere et~al.(2006)Trambly De~Laissardi\`ere,
  Julien, and Mayou}]{TramblyPRL2006}
\bibinfo{author}{G.~Trambly De~Laissardi\`ere}, \bibinfo{author}{J.-P. Julien},
  \bibinfo{author}{D.~Mayou},
\newblock \bibinfo{title}{Quantum transport of slow charge carriers in
  quasicrystals and correlated systems},
\newblock \bibinfo{journal}{Phys. Rev. Lett.} \bibinfo{volume}{97}
  (\bibinfo{year}{2006}) \bibinfo{pages}{026601}. \URLprefix
  \url{https://link.aps.org/doi/10.1103/PhysRevLett.97.026601}.
  \DOIprefix\doi{10.1103/PhysRevLett.97.026601}.
%Type = Article
\bibitem[{Markussen et~al.(2006)Markussen, Rurali, Brandbyge, and
  Jauho}]{markussen2006prb}
\bibinfo{author}{T.~Markussen}, \bibinfo{author}{R.~Rurali},
  \bibinfo{author}{M.~Brandbyge}, \bibinfo{author}{A.-P. Jauho},
\newblock \bibinfo{title}{{Electronic transport through Si nanowires: Role of
  bulk and surface disorder}},
\newblock \bibinfo{journal}{Phys. Rev. B} \bibinfo{volume}{74}
  (\bibinfo{year}{2006}) \bibinfo{pages}{245313}. \URLprefix
  \url{https://link.aps.org/doi/10.1103/PhysRevB.74.245313}.
  \DOIprefix\doi{10.1103/PhysRevB.74.245313}.
%Type = Article
\bibitem[{Persson et~al.(2008)Persson, Lherbier, Niquet, Triozon, and
  Roche}]{persson2008nl}
\bibinfo{author}{M.~P. Persson}, \bibinfo{author}{A.~Lherbier},
  \bibinfo{author}{Y.-M. Niquet}, \bibinfo{author}{F.~Triozon},
  \bibinfo{author}{S.~Roche},
\newblock \bibinfo{title}{{Orientational Dependence of Charge Transport in
  Disordered Silicon Nanowires}},
\newblock \bibinfo{journal}{Nano Lett.} \bibinfo{volume}{8}
  (\bibinfo{year}{2008}) \bibinfo{pages}{4146--4150}. \URLprefix
  \url{https://doi.org/10.1021/nl801128f}. \DOIprefix\doi{10.1021/nl801128f}.
%Type = Article
\bibitem[{Latil et~al.(2004)Latil, Roche, Mayou, and Charlier}]{latil2004prl}
\bibinfo{author}{S.~Latil}, \bibinfo{author}{S.~Roche},
  \bibinfo{author}{D.~Mayou}, \bibinfo{author}{J.-C. Charlier},
\newblock \bibinfo{title}{{Mesoscopic Transport in Chemically Doped Carbon
  Nanotubes}},
\newblock \bibinfo{journal}{Phys. Rev. Lett.} \bibinfo{volume}{92}
  (\bibinfo{year}{2004}) \bibinfo{pages}{256805}. \URLprefix
  \url{https://link.aps.org/doi/10.1103/PhysRevLett.92.256805}.
  \DOIprefix\doi{10.1103/PhysRevLett.92.256805}.
%Type = Article
\bibitem[{Latil et~al.(2005)Latil, Roche, and Charlier}]{LatilNL2005}
\bibinfo{author}{S.~Latil}, \bibinfo{author}{S.~Roche}, \bibinfo{author}{J.-C.
  Charlier},
\newblock \bibinfo{title}{Electronic transport in carbon nanotubes with random
  coverage of physisorbed molecules},
\newblock \bibinfo{journal}{Nano Letters} \bibinfo{volume}{5}
  (\bibinfo{year}{2005}) \bibinfo{pages}{2216--2219}. \URLprefix
  \url{https://doi.org/10.1021/nl0514386}. \DOIprefix\doi{10.1021/nl0514386}.
  \href{http://arxiv.org/abs/https://doi.org/10.1021/nl0514386}{{\tt
  arXiv:https://doi.org/10.1021/nl0514386}}, \bibinfo{note}{pMID: 16277456}.
%Type = Article
\bibitem[{Ishii et~al.(2010)Ishii, Roche, Kobayashi, and Hirose}]{ishii2010prl}
\bibinfo{author}{H.~Ishii}, \bibinfo{author}{S.~Roche},
  \bibinfo{author}{N.~Kobayashi}, \bibinfo{author}{K.~Hirose},
\newblock \bibinfo{title}{{Inelastic Transport in Vibrating Disordered Carbon
  Nanotubes: Scattering Times and Temperature-Dependent Decoherence Effects}},
\newblock \bibinfo{journal}{Phys. Rev. Lett.} \bibinfo{volume}{104}
  (\bibinfo{year}{2010}) \bibinfo{pages}{116801}. \URLprefix
  \url{https://link.aps.org/doi/10.1103/PhysRevLett.104.116801}.
  \DOIprefix\doi{10.1103/PhysRevLett.104.116801}.
%Type = Article
\bibitem[{Soriano et~al.(2012)Soriano, Ortmann, and Roche}]{Soriano2012}
\bibinfo{author}{D.~Soriano}, \bibinfo{author}{F.~Ortmann},
  \bibinfo{author}{S.~Roche},
\newblock \bibinfo{title}{Three-dimensional models of topological insulators:
  Engineering of dirac cones and robustness of the spin texture},
\newblock \bibinfo{journal}{Phys. Rev. Lett.} \bibinfo{volume}{109}
  (\bibinfo{year}{2012}) \bibinfo{pages}{266805}.
  \DOIprefix\doi{10.1103/PhysRevLett.109.266805}.
%Type = Article
\bibitem[{Wehling et~al.(2014)Wehling, Black-Schaffer, and
  Balatsky}]{wehling2014aip}
\bibinfo{author}{T.~Wehling}, \bibinfo{author}{A.~Black-Schaffer},
  \bibinfo{author}{A.~Balatsky},
\newblock \bibinfo{title}{{Dirac materials}},
\newblock \bibinfo{journal}{Adv. Phys.} \bibinfo{volume}{63}
  (\bibinfo{year}{2014}) \bibinfo{pages}{1--76}. \URLprefix
  \url{https://doi.org/10.1080/00018732.2014.927109}.
  \DOIprefix\doi{10.1080/00018732.2014.927109}.
%Type = Article
\bibitem[{Cresti et~al.(2016)Cresti, Nikoli\'{c}, Garc\'{i}a, and
  Roche}]{cresti2016rnc}
\bibinfo{author}{A.~Cresti}, \bibinfo{author}{B.~K. Nikoli\'{c}},
  \bibinfo{author}{J.~H. Garc\'{i}a}, \bibinfo{author}{S.~Roche},
\newblock \bibinfo{title}{{Charge, spin and valley Hall effects in disordered
  graphene}},
\newblock \bibinfo{journal}{Riv. Nuovo Cimento} \bibinfo{volume}{39}
  (\bibinfo{year}{2016}) \bibinfo{pages}{587}. \URLprefix
  \url{https://www.sif.it/riviste/ncr/econtents/2016/039/12/article/0}.
  \DOIprefix\doi{10.1393/ncr/i2016-10130-6}.
%Type = Article
\bibitem[{Kubo(1957)}]{kubo1957jpsj}
\bibinfo{author}{R.~Kubo},
\newblock \bibinfo{title}{{Statistical-Mechanical Theory of Irreversible
  Processes. I. General Theory and Simple Applications to Magnetic and
  Conduction Problems}},
\newblock \bibinfo{journal}{J. Phys. Soc. Jpn.} \bibinfo{volume}{12}
  (\bibinfo{year}{1957}) \bibinfo{pages}{570--586}. \URLprefix
  \url{http://dx.doi.org/10.1143/JPSJ.12.570}.
  \DOIprefix\doi{10.1143/JPSJ.12.570}.
%Type = Article
\bibitem[{Dugaev et~al.(2005)Dugaev, Bruno, Taillefumier, Canals, and
  Lacroix}]{Dugaev2005}
\bibinfo{author}{V.~K. Dugaev}, \bibinfo{author}{P.~Bruno},
  \bibinfo{author}{M.~Taillefumier}, \bibinfo{author}{B.~Canals},
  \bibinfo{author}{C.~Lacroix},
\newblock \bibinfo{title}{{Anomalous Hall effect in a two-dimensional electron
  gas with spin-orbit interaction}},
\newblock \bibinfo{journal}{Phys. Rev. B} \bibinfo{volume}{71}
  (\bibinfo{year}{2005}) \bibinfo{pages}{224423}. \URLprefix
  \url{https://link.aps.org/doi/10.1103/PhysRevB.71.224423}.
  \DOIprefix\doi{10.1103/PhysRevB.71.224423}.
%Type = Article
\bibitem[{Bohr et~al.(2006)Bohr, Schmitteckert, and W{\"{o}}lfle}]{Bohr2006}
\bibinfo{author}{D.~Bohr}, \bibinfo{author}{P.~Schmitteckert},
  \bibinfo{author}{P.~W{\"{o}}lfle},
\newblock \bibinfo{title}{{DMRG evaluation of the Kubo formula - Conductance of
  strongly interacting quantum systems}},
\newblock \bibinfo{journal}{EPL} \bibinfo{volume}{73} (\bibinfo{year}{2006})
  \bibinfo{pages}{246--252}. \URLprefix
  \url{http://stacks.iop.org/0295-5075/73/i=2/a=246?key=crossref.f93a7fee6e29b2cb648f505fe2464d6e}.
  \DOIprefix\doi{10.1209/epl/i2005-10377-6}.
%Type = Article
\bibitem[{Langer(1962)}]{Langer1962}
\bibinfo{author}{J.~S. Langer},
\newblock \bibinfo{title}{{Evaluation of Kubo's formula for the impurity
  resistance of an interacting electron gas}},
\newblock \bibinfo{journal}{Phys. Rev.} \bibinfo{volume}{127}
  (\bibinfo{year}{1962}) \bibinfo{pages}{5--16}. \URLprefix
  \url{https://link.aps.org/doi/10.1103/PhysRev.127.5}.
  \DOIprefix\doi{10.1103/PhysRev.127.5}.
%Type = Book
\bibitem[{Mahan(2000)}]{mahan2000book}
\bibinfo{author}{G.~D. Mahan}, \bibinfo{title}{{Many-Particle Physics}},
  \bibinfo{publisher}{Plenum Press}, \bibinfo{year}{2000}.
%Type = Book
\bibitem[{Di~Ventra(2008)}]{ventra2008book}
\bibinfo{author}{M.~Di~Ventra}, \bibinfo{title}{{Electrical transport in
  nanoscale systems}}, \bibinfo{publisher}{Cambridge University Press},
  \bibinfo{year}{2008}.
%Type = Book
\bibitem[{Rammer(2007)}]{Rammer2007}
\bibinfo{author}{J.~Rammer}, \bibinfo{title}{{Quantum Field Theory of
  Non-Equilibrium States}}, \bibinfo{publisher}{Cambridge University Press},
  \bibinfo{address}{Cambridge}, \bibinfo{year}{2007}. \URLprefix
  \url{http://ebooks.cambridge.org/ref/id/CBO9780511618956}.
  \DOIprefix\doi{10.1017/CBO9780511618956}.
%Type = Book
\bibitem[{Sakurai and Napolitano(2017)}]{sakurai_napolitano_2017}
\bibinfo{author}{J.~J. Sakurai}, \bibinfo{author}{J.~Napolitano},
  \bibinfo{title}{Modern Quantum Mechanics}, \bibinfo{edition}{2} ed.,
  \bibinfo{publisher}{Cambridge University Press}, \bibinfo{year}{2017}.
  \DOIprefix\doi{10.1017/9781108499996}.
%Type = Incollection
\bibitem[{Allen(2006)}]{allen2006book}
\bibinfo{author}{P.~B. Allen},
\newblock \bibinfo{title}{Electron transport},
\newblock in: \bibinfo{editor}{S.~Louie}, \bibinfo{editor}{M.~Cohen} (Eds.),
  \bibinfo{booktitle}{Conceptual Foundations of Materials, Volume 2: A Standard
  Model for Ground- and Excited-State Properties (Contemporary Concepts of
  Condensed Matter Science)}, \bibinfo{publisher}{Elsevier Science},
  \bibinfo{year}{2006}, pp. \bibinfo{pages}{165--218}.
%Type = Article
\bibitem[{Bastin et~al.(1971)Bastin, Lewiner, Betbeder-matibet, and
  Nozieres}]{bastinjphyschem1971}
\bibinfo{author}{A.~Bastin}, \bibinfo{author}{C.~Lewiner},
  \bibinfo{author}{O.~Betbeder-matibet}, \bibinfo{author}{P.~Nozieres},
\newblock \bibinfo{title}{{Quantum oscillations of the hall effect of a fermion
  gas with random impurity scattering}},
\newblock \bibinfo{journal}{J. Phys. Chem. Solids} \bibinfo{volume}{32}
  (\bibinfo{year}{1971}) \bibinfo{pages}{1811--1824}. \URLprefix
  \url{http://linkinghub.elsevier.com/retrieve/pii/S0022369771801476}.
  \DOIprefix\doi{10.1016/S0022-3697(71)80147-6}.
%Type = Article
\bibitem[{Schleede et~al.(2010)Schleede, Schubert, and Fehske}]{Schleede2010}
\bibinfo{author}{J.~Schleede}, \bibinfo{author}{G.~Schubert},
  \bibinfo{author}{H.~Fehske},
\newblock \bibinfo{title}{{Comment on “Anderson transition in disordered
  graphene” by Amini M. et al.}},
\newblock \bibinfo{journal}{EPL} \bibinfo{volume}{90} (\bibinfo{year}{2010})
  \bibinfo{pages}{17002}. \URLprefix
  \url{http://stacks.iop.org/0295-5075/90/i=1/a=17002?key=crossref.7038af6d4b3777c80d01bbe18ff149f0}.
  \DOIprefix\doi{10.1209/0295-5075/90/17002}.
%Type = Article
\bibitem[{Streda(1982)}]{Streda1982PRC}
\bibinfo{author}{P.~Streda},
\newblock \bibinfo{title}{{Theory of quantised Hall conductivity in two
  dimensions}},
\newblock \bibinfo{journal}{J. Phys. C: Solid State Phys.} \bibinfo{volume}{15}
  (\bibinfo{year}{1982}) \bibinfo{pages}{L717--L721}. \URLprefix
  \url{http://stacks.iop.org/0022-3719/15/i=22/a=005?key=crossref.e3d7df72f12035d33249de59a46a32f2}.
  \DOIprefix\doi{10.1088/0022-3719/15/22/005}.
%Type = Article
\bibitem[{Greenwood(1958)}]{greenwood1958pps}
\bibinfo{author}{D.~A. Greenwood},
\newblock \bibinfo{title}{{The Boltzmann Equation in the Theory of Electrical
  Conduction in Metals}},
\newblock \bibinfo{journal}{Proc. Phys. Soc.} \bibinfo{volume}{71}
  (\bibinfo{year}{1958}) \bibinfo{pages}{585}. \URLprefix
  \url{http://stacks.iop.org/0370-1328/71/i=4/a=306}.
%Type = Article
\bibitem[{Chester and Thellung(1959)}]{chester1959pps}
\bibinfo{author}{G.~V. Chester}, \bibinfo{author}{A.~Thellung},
\newblock \bibinfo{title}{{On the Electrical Conductivity of Metals}},
\newblock \bibinfo{journal}{Proc. Phys. Soc.} \bibinfo{volume}{73}
  (\bibinfo{year}{1959}) \bibinfo{pages}{745}. \URLprefix
  \url{http://stacks.iop.org/0370-1328/73/i=5/a=308}.
%Type = Article
\bibitem[{Chester and Thellung(1961)}]{chester1961pps}
\bibinfo{author}{G.~V. Chester}, \bibinfo{author}{A.~Thellung},
\newblock \bibinfo{title}{{The Law of Wiedemann and Franz}},
\newblock \bibinfo{journal}{Proc. Phys. Soc.} \bibinfo{volume}{77}
  (\bibinfo{year}{1961}) \bibinfo{pages}{1005}. \URLprefix
  \url{http://stacks.iop.org/0370-1328/77/i=5/a=309}.
%Type = Article
\bibitem[{Nakajima(1958)}]{Nakajima1958}
\bibinfo{author}{S.~Nakajima},
\newblock \bibinfo{title}{{On Quantum Theory of Transport Phenomena}},
\newblock \bibinfo{journal}{Prog. Theor. Phys.} \bibinfo{volume}{20}
  (\bibinfo{year}{1958}) \bibinfo{pages}{948--959}. \URLprefix
  \url{https://academic.oup.com/ptp/article-lookup/doi/10.1143/PTP.20.948}.
  \DOIprefix\doi{10.1143/PTP.20.948}.
%Type = Incollection
\bibitem[{Beenakker and van Houten(1991)}]{beenakker1991ssp}
\bibinfo{author}{C.~Beenakker}, \bibinfo{author}{H.~van Houten},
\newblock \bibinfo{title}{{Quantum Transport in Semiconductor Nanostructures}},
\newblock in: \bibinfo{editor}{H.~Ehrenreich}, \bibinfo{editor}{D.~Turnbull}
  (Eds.), \bibinfo{booktitle}{Semiconductor Heterostructures and
  Nanostructures}, volume~\bibinfo{volume}{44} of
  \textit{\bibinfo{series}{Solid State Physics}}, \bibinfo{publisher}{Academic
  Press}, \bibinfo{year}{1991}, pp. \bibinfo{pages}{1 -- 228}. \URLprefix
  \url{http://www.sciencedirect.com/science/article/pii/S0081194708600910}.
  \DOIprefix\doi{10.1016/S0081-1947(08)60091-0}.
%Type = Book
\bibitem[{Ashcroft and Mermin(1976)}]{ashcroft1976book}
\bibinfo{author}{N.~Ashcroft}, \bibinfo{author}{N.~Mermin},
  \bibinfo{title}{{Solid State Physics}}, \bibinfo{publisher}{Saunders
  College}, \bibinfo{address}{Philadelphia}, \bibinfo{year}{1976}.
%Type = Article
\bibitem[{Triozon et~al.(2000)Triozon, Roche, and Mayou}]{triozon2000riken}
\bibinfo{author}{F.~Triozon}, \bibinfo{author}{S.~Roche},
  \bibinfo{author}{D.~Mayou},
\newblock \bibinfo{title}{{Wave-packet dynamics by optimized polynomials
  methods}},
\newblock \bibinfo{journal}{RIKEN Rev.} \bibinfo{volume}{29}
  (\bibinfo{year}{2000}) \bibinfo{pages}{73}.
%Type = Article
\bibitem[{Uppstu et~al.(2014)Uppstu, Fan, and Harju}]{uppstu2014prb}
\bibinfo{author}{A.~Uppstu}, \bibinfo{author}{Z.~Fan},
  \bibinfo{author}{A.~Harju},
\newblock \bibinfo{title}{{Obtaining localization properties efficiently using
  the Kubo-Greenwood formalism}},
\newblock \bibinfo{journal}{Phys. Rev. B} \bibinfo{volume}{89}
  (\bibinfo{year}{2014}) \bibinfo{pages}{075420}. \URLprefix
  \url{https://link.aps.org/doi/10.1103/PhysRevB.89.075420}.
  \DOIprefix\doi{10.1103/PhysRevB.89.075420}.
%Type = Article
\bibitem[{Anderson et~al.(1980)Anderson, Thouless, Abrahams, and
  Fisher}]{anderson1980}
\bibinfo{author}{P.~W. Anderson}, \bibinfo{author}{D.~J. Thouless},
  \bibinfo{author}{E.~Abrahams}, \bibinfo{author}{D.~S. Fisher},
\newblock \bibinfo{title}{{New method for a scaling theory of localization}},
\newblock \bibinfo{journal}{Phys. Rev. B} \bibinfo{volume}{22}
  (\bibinfo{year}{1980}) \bibinfo{pages}{3519--3526}. \URLprefix
  \url{https://link.aps.org/doi/10.1103/PhysRevB.22.3519}.
  \DOIprefix\doi{10.1103/PhysRevB.22.3519}.
%Type = Article
\bibitem[{Anderson(1958)}]{anderson1958pr}
\bibinfo{author}{P.~W. Anderson},
\newblock \bibinfo{title}{{Absence of Diffusion in Certain Random Lattices}},
\newblock \bibinfo{journal}{Phys. Rev.} \bibinfo{volume}{109}
  (\bibinfo{year}{1958}) \bibinfo{pages}{1492--1505}. \URLprefix
  \url{https://link.aps.org/doi/10.1103/PhysRev.109.1492}.
  \DOIprefix\doi{10.1103/PhysRev.109.1492}.
%Type = Article
\bibitem[{van Wees et~al.(1988)van Wees, van Houten, Beenakker, Williamson,
  Kouwenhoven, van~der Marel, and Foxon}]{vanWees1988prl}
\bibinfo{author}{B.~J. van Wees}, \bibinfo{author}{H.~van Houten},
  \bibinfo{author}{C.~W.~J. Beenakker}, \bibinfo{author}{J.~G. Williamson},
  \bibinfo{author}{L.~P. Kouwenhoven}, \bibinfo{author}{D.~van~der Marel},
  \bibinfo{author}{C.~T. Foxon},
\newblock \bibinfo{title}{{Quantized conductance of point contacts in a
  two-dimensional electron gas}},
\newblock \bibinfo{journal}{Phys. Rev. Lett.} \bibinfo{volume}{60}
  (\bibinfo{year}{1988}) \bibinfo{pages}{848--850}. \URLprefix
  \url{https://link.aps.org/doi/10.1103/PhysRevLett.60.848}.
  \DOIprefix\doi{10.1103/PhysRevLett.60.848}.
%Type = Article
\bibitem[{Wharam et~al.(1988)Wharam, Thornton, Newbury, Pepper, Ahmed, Frost,
  Hasko, Peacock, Ritchie, and Jones}]{wharam1988jpc}
\bibinfo{author}{D.~A. Wharam}, \bibinfo{author}{T.~J. Thornton},
  \bibinfo{author}{R.~Newbury}, \bibinfo{author}{M.~Pepper},
  \bibinfo{author}{H.~Ahmed}, \bibinfo{author}{J.~E.~F. Frost},
  \bibinfo{author}{D.~G. Hasko}, \bibinfo{author}{D.~C. Peacock},
  \bibinfo{author}{D.~A. Ritchie}, \bibinfo{author}{G.~A.~C. Jones},
\newblock \bibinfo{title}{{One-dimensional transport and the quantisation of
  the ballistic resistance}},
\newblock \bibinfo{journal}{J. Phys. C: Solid State Phys.} \bibinfo{volume}{21}
  (\bibinfo{year}{1988}) \bibinfo{pages}{L209}. \URLprefix
  \url{http://stacks.iop.org/0022-3719/21/i=8/a=002}.
%Type = Article
\bibitem[{Frank et~al.(1998)Frank, Poncharal, Wang, and
  de~Heer}]{frank1998science}
\bibinfo{author}{S.~Frank}, \bibinfo{author}{P.~Poncharal},
  \bibinfo{author}{Z.~L. Wang}, \bibinfo{author}{W.~A. de~Heer},
\newblock \bibinfo{title}{{Carbon Nanotube Quantum Resistors}},
\newblock \bibinfo{journal}{Science} \bibinfo{volume}{280}
  (\bibinfo{year}{1998}) \bibinfo{pages}{1744--1746}.
  \DOIprefix\doi{10.1126/science.280.5370.1744}.
%Type = Article
\bibitem[{Roche and Saito(2001)}]{roche2001prl}
\bibinfo{author}{S.~Roche}, \bibinfo{author}{R.~Saito},
\newblock \bibinfo{title}{{Magnetoresistance of Carbon Nanotubes: From
  Molecular to Mesoscopic Fingerprints}},
\newblock \bibinfo{journal}{Phys. Rev. Lett.} \bibinfo{volume}{87}
  (\bibinfo{year}{2001}) \bibinfo{pages}{246803}. \URLprefix
  \url{https://link.aps.org/doi/10.1103/PhysRevLett.87.246803}.
  \DOIprefix\doi{10.1103/PhysRevLett.87.246803}.
%Type = Article
\bibitem[{Roche et~al.(2001)Roche, Triozon, Rubio, and Mayou}]{roche2001prb}
\bibinfo{author}{S.~Roche}, \bibinfo{author}{F.~Triozon},
  \bibinfo{author}{A.~Rubio}, \bibinfo{author}{D.~Mayou},
\newblock \bibinfo{title}{{Conduction mechanisms and magnetotransport in
  multiwalled carbon nanotubes}},
\newblock \bibinfo{journal}{Phys. Rev. B} \bibinfo{volume}{64}
  (\bibinfo{year}{2001}) \bibinfo{pages}{121401}. \URLprefix
  \url{https://link.aps.org/doi/10.1103/PhysRevB.64.121401}.
  \DOIprefix\doi{10.1103/PhysRevB.64.121401}.
%Type = Article
\bibitem[{Triozon et~al.(2002)Triozon, Vidal, Mosseri, and
  Mayou}]{triozon2002prb}
\bibinfo{author}{F.~Triozon}, \bibinfo{author}{J.~Vidal},
  \bibinfo{author}{R.~Mosseri}, \bibinfo{author}{D.~Mayou},
\newblock \bibinfo{title}{{Quantum dynamics in two- and three-dimensional
  quasiperiodic tilings}},
\newblock \bibinfo{journal}{Phys. Rev. B} \bibinfo{volume}{65}
  (\bibinfo{year}{2002}) \bibinfo{pages}{220202}. \URLprefix
  \url{https://link.aps.org/doi/10.1103/PhysRevB.65.220202}.
  \DOIprefix\doi{10.1103/PhysRevB.65.220202}.
%Type = Article
\bibitem[{Triozon et~al.(2004)Triozon, Roche, Rubio, and
  Mayou}]{triozon2004prb}
\bibinfo{author}{F.~Triozon}, \bibinfo{author}{S.~Roche},
  \bibinfo{author}{A.~Rubio}, \bibinfo{author}{D.~Mayou},
\newblock \bibinfo{title}{{Electrical transport in carbon nanotubes: Role of
  disorder and helical symmetries}},
\newblock \bibinfo{journal}{Phys. Rev. B} \bibinfo{volume}{69}
  (\bibinfo{year}{2004}) \bibinfo{pages}{121410}. \URLprefix
  \url{https://link.aps.org/doi/10.1103/PhysRevB.69.121410}.
  \DOIprefix\doi{10.1103/PhysRevB.69.121410}.
%Type = Incollection
\bibitem[{Skilling(1989)}]{skilling1989}
\bibinfo{author}{J.~Skilling},
\newblock \bibinfo{title}{The eigenvalues of mega-dimensional matrices},
\newblock in: \bibinfo{editor}{J.~Skilling} (Ed.), \bibinfo{booktitle}{Maximum
  Entropy and Bayesian Methods (An International Book Series on The Fundamental
  Theories of Physics: Their Clarification, Development and Application)},
  \bibinfo{publisher}{Springer}, \bibinfo{address}{Dordrecht},
  \bibinfo{year}{1989}, pp. \bibinfo{pages}{455--466}. \URLprefix
  \url{https://link.springer.com/chapter/10.1007/978-94-015-7860-8\_48}.
  \DOIprefix\doi{10.1007/978-94-015-7860-8\_48}.
%Type = Article
\bibitem[{Silver and R\"{o}der(1994)}]{silver1994ijmpc}
\bibinfo{author}{R.~N. Silver}, \bibinfo{author}{H.~R\"{o}der},
\newblock \bibinfo{title}{{Densities of States of Mega-Dimensional Hamiltonian
  Matrices}},
\newblock \bibinfo{journal}{Int. J. Mod. Phys. C} \bibinfo{volume}{05}
  (\bibinfo{year}{1994}) \bibinfo{pages}{735--753}. \URLprefix
  \url{http://www.worldscientific.com/doi/abs/10.1142/S0129183194000842}.
  \DOIprefix\doi{10.1142/S0129183194000842}.
%Type = Article
\bibitem[{Drabold and Sankey(1993)}]{drabold1993prl}
\bibinfo{author}{D.~A. Drabold}, \bibinfo{author}{O.~F. Sankey},
\newblock \bibinfo{title}{{Maximum entropy approach for linear scaling in the
  electronic structure problem}},
\newblock \bibinfo{journal}{Phys. Rev. Lett.} \bibinfo{volume}{70}
  (\bibinfo{year}{1993}) \bibinfo{pages}{3631--3634}. \URLprefix
  \url{https://link.aps.org/doi/10.1103/PhysRevLett.70.3631}.
  \DOIprefix\doi{10.1103/PhysRevLett.70.3631}.
%Type = Article
\bibitem[{Silver et~al.(1996)Silver, Roeder, Voter, and Kress}]{silver1996jcp}
\bibinfo{author}{R.~Silver}, \bibinfo{author}{H.~Roeder},
  \bibinfo{author}{A.~Voter}, \bibinfo{author}{J.~Kress},
\newblock \bibinfo{title}{{Kernel Polynomial Approximations for Densities of
  States and Spectral Functions}},
\newblock \bibinfo{journal}{J. Comp. Phys.} \bibinfo{volume}{124}
  (\bibinfo{year}{1996}) \bibinfo{pages}{115 -- 130}. \URLprefix
  \url{http://www.sciencedirect.com/science/article/pii/S0021999196900480}.
  \DOIprefix\doi{10.1006/jcph.1996.0048}.
%Type = Article
\bibitem[{Iitaka and Ebisuzaki(2004)}]{iitaka2004pre}
\bibinfo{author}{T.~Iitaka}, \bibinfo{author}{T.~Ebisuzaki},
\newblock \bibinfo{title}{{Random phase vector for calculating the trace of a
  large matrix}},
\newblock \bibinfo{journal}{Phys. Rev. E} \bibinfo{volume}{69}
  (\bibinfo{year}{2004}) \bibinfo{pages}{057701}. \URLprefix
  \url{https://link.aps.org/doi/10.1103/PhysRevE.69.057701}.
  \DOIprefix\doi{10.1103/PhysRevE.69.057701}.
%Type = Book
\bibitem[{Boyd(2001)}]{boyd2001book}
\bibinfo{author}{J.~P. Boyd}, \bibinfo{title}{Chebyshev and Fourier Spectral
  Methods: Second Revised Edition (Dover Books on Mathematics)},
  \bibinfo{publisher}{Dover Publications}, \bibinfo{year}{2001}.
%Type = Book
\bibitem[{Gradshteyn and Ryzhik(1975)}]{gradshteyn_book}
\bibinfo{author}{I.~S. Gradshteyn}, \bibinfo{author}{I.~M. Ryzhik},
  \bibinfo{title}{Table of Integrals, Series, and Products, 7th Edition},
  \bibinfo{publisher}{Elsevier, Academic Press}, \bibinfo{year}{1975}.
%Type = Article
\bibitem[{Vijay et~al.(2004)Vijay, Kouri, and Hoffman}]{vijay2004jpca}
\bibinfo{author}{A.~Vijay}, \bibinfo{author}{D.~J. Kouri},
  \bibinfo{author}{D.~K. Hoffman},
\newblock \bibinfo{title}{{Scattering and Bound States: A Lorentzian
  Function-Based Spectral Filter Approach}},
\newblock \bibinfo{journal}{J. Phys. Chem. A} \bibinfo{volume}{108}
  (\bibinfo{year}{2004}) \bibinfo{pages}{8987--9003}. \URLprefix
  \url{https://doi.org/10.1021/jp040356n}. \DOIprefix\doi{10.1021/jp040356n}.
%Type = Article
\bibitem[{Braun and Schmitteckert(2014)}]{braun2014prb}
\bibinfo{author}{A.~Braun}, \bibinfo{author}{P.~Schmitteckert},
\newblock \bibinfo{title}{Numerical evaluation of green's functions based on
  the chebyshev expansion},
\newblock \bibinfo{journal}{Phys. Rev. B} \bibinfo{volume}{90}
  (\bibinfo{year}{2014}) \bibinfo{pages}{165112}. \URLprefix
  \url{https://link.aps.org/doi/10.1103/PhysRevB.90.165112}.
  \DOIprefix\doi{10.1103/PhysRevB.90.165112}.
%Type = Article
\bibitem[{Covaci et~al.(2010)Covaci, Peeters, and Berciu}]{Covaci2010PRL}
\bibinfo{author}{L.~Covaci}, \bibinfo{author}{F.~M. Peeters},
  \bibinfo{author}{M.~Berciu},
\newblock \bibinfo{title}{{Efficient Numerical Approach to Inhomogeneous
  Superconductivity: The Chebyshev-Bogoliubov--de Gennes Method}},
\newblock \bibinfo{journal}{Phys. Rev. Lett.} \bibinfo{volume}{105}
  (\bibinfo{year}{2010}) \bibinfo{pages}{167006}. \URLprefix
  \url{https://link.aps.org/doi/10.1103/PhysRevLett.105.167006}.
  \DOIprefix\doi{10.1103/PhysRevLett.105.167006}.
%Type = Article
\bibitem[{Dagotto(1994)}]{dagotto1994rmp}
\bibinfo{author}{E.~Dagotto},
\newblock \bibinfo{title}{{Correlated electrons in high-temperature
  superconductors}},
\newblock \bibinfo{journal}{Rev. Mod. Phys.} \bibinfo{volume}{66}
  (\bibinfo{year}{1994}) \bibinfo{pages}{763--840}. \URLprefix
  \url{https://link.aps.org/doi/10.1103/RevModPhys.66.763}.
  \DOIprefix\doi{10.1103/RevModPhys.66.763}.
%Type = Article
\bibitem[{Alben et~al.(1975)Alben, Blume, Krakauer, and
  Schwartz}]{alben1975prb}
\bibinfo{author}{R.~Alben}, \bibinfo{author}{M.~Blume},
  \bibinfo{author}{H.~Krakauer}, \bibinfo{author}{L.~Schwartz},
\newblock \bibinfo{title}{{Exact results for a three-dimensional alloy with
  site diagonal disorder: comparison with the coherent potential
  approximation}},
\newblock \bibinfo{journal}{Phys. Rev. B} \bibinfo{volume}{12}
  (\bibinfo{year}{1975}) \bibinfo{pages}{4090--4094}. \URLprefix
  \url{https://link.aps.org/doi/10.1103/PhysRevB.12.4090}.
  \DOIprefix\doi{10.1103/PhysRevB.12.4090}.
%Type = Article
\bibitem[{Feit et~al.(1982)Feit, Fleck, and Steiger}]{feit1982jcp}
\bibinfo{author}{M.~Feit}, \bibinfo{author}{J.~Fleck},
  \bibinfo{author}{A.~Steiger},
\newblock \bibinfo{title}{{Solution of the Schrodinger equation by a spectral
  method}},
\newblock \bibinfo{journal}{J. Comp. Phys.} \bibinfo{volume}{47}
  (\bibinfo{year}{1982}) \bibinfo{pages}{412 -- 433}. \URLprefix
  \url{http://www.sciencedirect.com/science/article/pii/0021999182900912}.
  \DOIprefix\doi{10.1016/0021-9991(82)90091-2}.
%Type = Article
\bibitem[{Hams and De~Raedt(2000)}]{hams2000pre}
\bibinfo{author}{A.~Hams}, \bibinfo{author}{H.~De~Raedt},
\newblock \bibinfo{title}{{Fast algorithm for finding the eigenvalue
  distribution of very large matrices}},
\newblock \bibinfo{journal}{Phys. Rev. E} \bibinfo{volume}{62}
  (\bibinfo{year}{2000}) \bibinfo{pages}{4365--4377}. \URLprefix
  \url{https://link.aps.org/doi/10.1103/PhysRevE.62.4365}.
  \DOIprefix\doi{10.1103/PhysRevE.62.4365}.
%Type = Article
\bibitem[{Wang and Zunger(1994)}]{wang1994prl}
\bibinfo{author}{L.-W. Wang}, \bibinfo{author}{A.~Zunger},
\newblock \bibinfo{title}{{Dielectric Constants of Silicon Quantum Dots}},
\newblock \bibinfo{journal}{Phys. Rev. Lett.} \bibinfo{volume}{73}
  (\bibinfo{year}{1994}) \bibinfo{pages}{1039--1042}. \URLprefix
  \url{https://link.aps.org/doi/10.1103/PhysRevLett.73.1039}.
  \DOIprefix\doi{10.1103/PhysRevLett.73.1039}.
%Type = Article
\bibitem[{Wang(1994)}]{wang1994prb}
\bibinfo{author}{L.-W. Wang},
\newblock \bibinfo{title}{{Calculating the density of states and
  optical-absorption spectra of large quantum systems by the plane-wave moments
  method}},
\newblock \bibinfo{journal}{Phys. Rev. B} \bibinfo{volume}{49}
  (\bibinfo{year}{1994}) \bibinfo{pages}{10154--10158}. \URLprefix
  \url{https://link.aps.org/doi/10.1103/PhysRevB.49.10154}.
  \DOIprefix\doi{10.1103/PhysRevB.49.10154}.
%Type = Article
\bibitem[{Silver and R\"oder(1997)}]{silver1997pre}
\bibinfo{author}{R.~N. Silver}, \bibinfo{author}{H.~R\"oder},
\newblock \bibinfo{title}{{Calculation of densities of states and spectral
  functions by Chebyshev recursion and maximum entropy}},
\newblock \bibinfo{journal}{Phys. Rev. E} \bibinfo{volume}{56}
  (\bibinfo{year}{1997}) \bibinfo{pages}{4822--4829}. \URLprefix
  \url{https://link.aps.org/doi/10.1103/PhysRevE.56.4822}.
  \DOIprefix\doi{10.1103/PhysRevE.56.4822}.
%Type = Article
\bibitem[{Ferreira et~al.(2011)Ferreira, Viana-Gomes, Nilsson, Mucciolo, Peres,
  and Castro~Neto}]{ferreira2011prb}
\bibinfo{author}{A.~Ferreira}, \bibinfo{author}{J.~Viana-Gomes},
  \bibinfo{author}{J.~Nilsson}, \bibinfo{author}{E.~R. Mucciolo},
  \bibinfo{author}{N.~M.~R. Peres}, \bibinfo{author}{A.~H. Castro~Neto},
\newblock \bibinfo{title}{Unified description of the dc conductivity of
  monolayer and bilayer graphene at finite densities based on resonant
  scatterers},
\newblock \bibinfo{journal}{Phys. Rev. B} \bibinfo{volume}{83}
  (\bibinfo{year}{2011}) \bibinfo{pages}{165402}. \URLprefix
  \url{https://link.aps.org/doi/10.1103/PhysRevB.83.165402}.
  \DOIprefix\doi{10.1103/PhysRevB.83.165402}.
%Type = Article
\bibitem[{Fan et~al.(2014)Fan, Uppstu, Siro, and Harju}]{fan2014cpc}
\bibinfo{author}{Z.~Fan}, \bibinfo{author}{A.~Uppstu},
  \bibinfo{author}{T.~Siro}, \bibinfo{author}{A.~Harju},
\newblock \bibinfo{title}{{Efficient linear-scaling quantum transport
  calculations on graphics processing units and applications on electron
  transport in graphene}},
\newblock \bibinfo{journal}{Comput. Phys. Commun.} \bibinfo{volume}{185}
  (\bibinfo{year}{2014}) \bibinfo{pages}{28 -- 39}. \URLprefix
  \url{http://www.sciencedirect.com/science/article/pii/S0010465513002701}.
  \DOIprefix\doi{10.1016/j.cpc.2013.08.009}.
%Type = Article
\bibitem[{Lanczos(1950)}]{lanczos1950jrnbs}
\bibinfo{author}{C.~Lanczos},
\newblock \bibinfo{title}{{An iteration method for the solution of the
  eigenvalue problem of linear differential and integral operators}},
\newblock \bibinfo{journal}{J. Res. Natl. Bur. Stand.} \bibinfo{volume}{45}
  (\bibinfo{year}{1950}) \bibinfo{pages}{255--282}. \URLprefix
  \url{http://dx.doi.org/10.6028/jres.045.026}.
  \DOIprefix\doi{10.6028/jres.045.026}.
%Type = Book
\bibitem[{Cullum and Willoughby(1985)}]{cullum1985}
\bibinfo{author}{J.~K. Cullum}, \bibinfo{author}{R.~A. Willoughby},
  \bibinfo{title}{{Lanczos Algorithms for Large Symmetric Eigenvalue
  Computations}}, \bibinfo{publisher}{Birkh\"{a}user}, \bibinfo{year}{1985}.
%Type = Book
\bibitem[{Saad(2003)}]{saad2003book}
\bibinfo{author}{Y.~Saad}, \bibinfo{title}{Iterative Methods for Sparse Linear
  Systems}, \bibinfo{publisher}{Society for Industrial and Applied
  Mathematics}, \bibinfo{year}{2003}.
%Type = Article
\bibitem[{Benoit(1994)}]{Benoit_1994}
\bibinfo{author}{C.~Benoit},
\newblock \bibinfo{title}{The moments method and damped systems},
\newblock \bibinfo{journal}{Journal of Physics: Condensed Matter}
  \bibinfo{volume}{6} (\bibinfo{year}{1994}) \bibinfo{pages}{3137--3160}.
  \URLprefix \url{https://doi.org/10.1088\%2F0953-8984\%2F6\%2F17\%2F006}.
  \DOIprefix\doi{10.1088/0953-8984/6/17/006}.
%Type = Article
\bibitem[{Benoit et~al.(1995)Benoit, Poussigue, Rousseau, Lakhliai, and
  Chenouni}]{Benoit_1995}
\bibinfo{author}{C.~Benoit}, \bibinfo{author}{G.~Poussigue},
  \bibinfo{author}{V.~Rousseau}, \bibinfo{author}{Z.~Lakhliai},
  \bibinfo{author}{D.~Chenouni},
\newblock \bibinfo{title}{Determination of the green functions for systems with
  large asymmetric matrices by the moments method},
\newblock \bibinfo{journal}{Modelling and Simulation in Materials Science and
  Engineering} \bibinfo{volume}{3} (\bibinfo{year}{1995})
  \bibinfo{pages}{161--185}. \URLprefix
  \url{https://doi.org/10.1088\%2F0965-0393\%2F3\%2F2\%2F003}.
  \DOIprefix\doi{10.1088/0965-0393/3/2/003}.
%Type = Article
\bibitem[{Triozon and Roche(2005)}]{Triozon2005}
\bibinfo{author}{F.~Triozon}, \bibinfo{author}{S.~Roche},
\newblock \bibinfo{title}{Efficient linear scaling method for computing the
  landauer-b{\"u}ttiker conductance},
\newblock \bibinfo{journal}{Eur. Phys. J. B} \bibinfo{volume}{46}
  (\bibinfo{year}{2005}) \bibinfo{pages}{427--431}.
  \DOIprefix\doi{10.1140/epjb/e2005-00260-x}.
%Type = Article
\bibitem[{Schubert et~al.(2010)Schubert, Schleede, Byczuk, Fehske, and
  Vollhardt}]{Schubert2010}
\bibinfo{author}{G.~Schubert}, \bibinfo{author}{J.~Schleede},
  \bibinfo{author}{K.~Byczuk}, \bibinfo{author}{H.~Fehske},
  \bibinfo{author}{D.~Vollhardt},
\newblock \bibinfo{title}{Distribution of the local density of states as a
  criterion for anderson localization: Numerically exact results for various
  lattices in two and three dimensions},
\newblock \bibinfo{journal}{Phys. Rev. B} \bibinfo{volume}{81}
  (\bibinfo{year}{2010}) \bibinfo{pages}{155106}. \URLprefix
  \url{https://link.aps.org/doi/10.1103/PhysRevB.81.155106}.
  \DOIprefix\doi{10.1103/PhysRevB.81.155106}.
%Type = Article
\bibitem[{Joao et~al.(2020)Joao, Andelkovic, Covaci, Rappoport, Lopes, and
  Ferreira}]{Joao2019}
\bibinfo{author}{S.~M. Joao}, \bibinfo{author}{M.~Andelkovic},
  \bibinfo{author}{L.~Covaci}, \bibinfo{author}{T.~G. Rappoport},
  \bibinfo{author}{J.~M. V.~P. Lopes}, \bibinfo{author}{A.~Ferreira},
\newblock \bibinfo{title}{Kite: high-performance accurate modelling of
  electronic structure and response functions of large molecules, disordered
  crystals and heterostructures},
\newblock \bibinfo{journal}{Royal Society Open Science} \bibinfo{volume}{7}
  (\bibinfo{year}{2020}) \bibinfo{pages}{191809}.
  \DOIprefix\doi{10.1098/rsos.191809}.
%Type = Book
\bibitem[{Kadanoff and Baym(1962)}]{kadanoff1962book}
\bibinfo{author}{L.~P. Kadanoff}, \bibinfo{author}{G.~Baym},
  \bibinfo{title}{{Quantum Statistical Mechanics: Green's Function Methods in
  Equilibrium and Nonequilibrium Problems}}, \bibinfo{publisher}{Benjamin},
  \bibinfo{address}{New York}, \bibinfo{year}{1962}.
%Type = Article
\bibitem[{Keldysh(1965)}]{keldysh1965jetp}
\bibinfo{author}{L.~V. Keldysh},
\newblock \bibinfo{title}{{Diagram Technique for Nonequilibrium Processes}},
\newblock \bibinfo{journal}{Sov. Phys. JETP} \bibinfo{volume}{20}
  (\bibinfo{year}{1965}) \bibinfo{pages}{1018}. \URLprefix
  \url{http://www.jetp.ac.ru/cgi-bin/e/index/e/20/4/p1018?a=list}.
%Type = Book
\bibitem[{Haug and Jauho(1996)}]{haug1996book}
\bibinfo{author}{H.~Haug}, \bibinfo{author}{A.-P. Jauho},
  \bibinfo{title}{Quantum Kinetics in Transport and Optics of Semiconductors},
  \bibinfo{publisher}{Springer-Verlag}, \bibinfo{year}{1996}.
%Type = Book
\bibitem[{Ferry and Goodnick(1997)}]{ferry1997book}
\bibinfo{author}{D.~K. Ferry}, \bibinfo{author}{S.~M. Goodnick},
  \bibinfo{title}{{Transport in nanostructures}}, \bibinfo{publisher}{Cambridge
  University Press}, \bibinfo{year}{1997}.
%Type = Article
\bibitem[{Brandbyge et~al.(2002)Brandbyge, Mozos, Ordej\'on, Taylor, and
  Stokbro}]{Brandbyge2002}
\bibinfo{author}{M.~Brandbyge}, \bibinfo{author}{J.-L. Mozos},
  \bibinfo{author}{P.~Ordej\'on}, \bibinfo{author}{J.~Taylor},
  \bibinfo{author}{K.~Stokbro},
\newblock \bibinfo{title}{Density-functional method for nonequilibrium electron
  transport},
\newblock \bibinfo{journal}{Phys. Rev. B} \bibinfo{volume}{65}
  (\bibinfo{year}{2002}) \bibinfo{pages}{165401}. \URLprefix
  \url{https://link.aps.org/doi/10.1103/PhysRevB.65.165401}.
  \DOIprefix\doi{10.1103/PhysRevB.65.165401}.
%Type = Article
\bibitem[{Markussen et~al.(2017)Markussen, Palsgaard, Stradi, Gunst, Brandbyge,
  and Stokbro}]{markussen2017prb}
\bibinfo{author}{T.~Markussen}, \bibinfo{author}{M.~Palsgaard},
  \bibinfo{author}{D.~Stradi}, \bibinfo{author}{T.~Gunst},
  \bibinfo{author}{M.~Brandbyge}, \bibinfo{author}{K.~Stokbro},
\newblock \bibinfo{title}{{Electron-phonon scattering from Green's function
  transport combined with molecular dynamics: Applications to mobility
  predictions}},
\newblock \bibinfo{journal}{Phys. Rev. B} \bibinfo{volume}{95}
  (\bibinfo{year}{2017}) \bibinfo{pages}{245210}. \URLprefix
  \url{https://link.aps.org/doi/10.1103/PhysRevB.95.245210}.
  \DOIprefix\doi{10.1103/PhysRevB.95.245210}.
%Type = Article
\bibitem[{Lopez-Bezanilla et~al.(2009)Lopez-Bezanilla, Triozon, Latil, Blase,
  and Roche}]{Lopez2009}
\bibinfo{author}{A.~Lopez-Bezanilla}, \bibinfo{author}{F.~Triozon},
  \bibinfo{author}{S.~Latil}, \bibinfo{author}{X.~Blase},
  \bibinfo{author}{S.~Roche},
\newblock \bibinfo{title}{{Effect of the Chemical Functionalization on Charge
  Transport in Carbon Nanotubes at the Mesoscopic Scale}},
\newblock \bibinfo{journal}{Nano Lett.} \bibinfo{volume}{9}
  (\bibinfo{year}{2009}) \bibinfo{pages}{940--944}. \URLprefix
  \url{https://doi.org/10.1021/nl802798q}.
%Type = Article
\bibitem[{Lopez-Bezanilla et~al.(2018)Lopez-Bezanilla, Froufe-P\'erez, Roche,
  and S\'aenz}]{Lopez2019}
\bibinfo{author}{A.~Lopez-Bezanilla}, \bibinfo{author}{L.~S. Froufe-P\'erez},
  \bibinfo{author}{S.~Roche}, \bibinfo{author}{J.~J. S\'aenz},
\newblock \bibinfo{title}{Unequivocal signatures of the crossover to anderson
  localization in realistic models of disordered quasi-one-dimensional
  materials},
\newblock \bibinfo{journal}{Phys. Rev. B} \bibinfo{volume}{98}
  (\bibinfo{year}{2018}) \bibinfo{pages}{235423}. \URLprefix
  \url{https://link.aps.org/doi/10.1103/PhysRevB.98.235423}.
  \DOIprefix\doi{10.1103/PhysRevB.98.235423}.
%Type = Article
\bibitem[{Lewenkopf and Mucciolo(2013)}]{lewenkopf2013jce}
\bibinfo{author}{C.~H. Lewenkopf}, \bibinfo{author}{E.~R. Mucciolo},
\newblock \bibinfo{title}{{The recursive Green's function method for
  graphene}},
\newblock \bibinfo{journal}{J. Comput. Electron.} \bibinfo{volume}{12}
  (\bibinfo{year}{2013}) \bibinfo{pages}{203--231}. \URLprefix
  \url{https://doi.org/10.1007/s10825-013-0458-7}.
  \DOIprefix\doi{10.1007/s10825-013-0458-7}.
%Type = Article
\bibitem[{Caroli et~al.(1971)Caroli, Combescot, Nozieres, and
  Saint-James}]{caroli1971jpc}
\bibinfo{author}{C.~Caroli}, \bibinfo{author}{R.~Combescot},
  \bibinfo{author}{P.~Nozieres}, \bibinfo{author}{D.~Saint-James},
\newblock \bibinfo{title}{Direct calculation of the tunneling current},
\newblock \bibinfo{journal}{J. Phys. C: Solid State Phys.} \bibinfo{volume}{4}
  (\bibinfo{year}{1971}) \bibinfo{pages}{916}. \URLprefix
  \url{http://stacks.iop.org/0022-3719/4/i=8/a=018}.
%Type = Article
\bibitem[{Sancho et~al.(1985)Sancho, Sancho, Sancho, and Rubio}]{sancho1985jpf}
\bibinfo{author}{M.~P.~L. Sancho}, \bibinfo{author}{J.~M.~L. Sancho},
  \bibinfo{author}{J.~M.~L. Sancho}, \bibinfo{author}{J.~Rubio},
\newblock \bibinfo{title}{{Highly convergent schemes for the calculation of
  bulk and surface Green functions}},
\newblock \bibinfo{journal}{J. Phys. F} \bibinfo{volume}{15}
  (\bibinfo{year}{1985}) \bibinfo{pages}{851}. \URLprefix
  \url{http://stacks.iop.org/0305-4608/15/i=4/a=009}.
%Type = Article
\bibitem[{Groth et~al.(2014)Groth, Wimmer, Akhmerov, and
  Waintal}]{groth2014njp}
\bibinfo{author}{C.~W. Groth}, \bibinfo{author}{M.~Wimmer},
  \bibinfo{author}{A.~R. Akhmerov}, \bibinfo{author}{X.~Waintal},
\newblock \bibinfo{title}{{Kwant: a software package for quantum transport}},
\newblock \bibinfo{journal}{New J. Phys.} \bibinfo{volume}{16}
  (\bibinfo{year}{2014}) \bibinfo{pages}{063065}. \URLprefix
  \url{http://stacks.iop.org/1367-2630/16/i=6/a=063065}.
%Type = Article
\bibitem[{Gaury et~al.(2014)Gaury, Weston, Santin, Houzet, Groth, and
  Waintal}]{GAURY20141}
\bibinfo{author}{B.~Gaury}, \bibinfo{author}{J.~Weston},
  \bibinfo{author}{M.~Santin}, \bibinfo{author}{M.~Houzet},
  \bibinfo{author}{C.~Groth}, \bibinfo{author}{X.~Waintal},
\newblock \bibinfo{title}{Numerical simulations of time-resolved quantum
  electronics},
\newblock \bibinfo{journal}{Physics Reports} \bibinfo{volume}{534}
  (\bibinfo{year}{2014}) \bibinfo{pages}{1 -- 37}. \URLprefix
  \url{http://www.sciencedirect.com/science/article/pii/S0370157313003451}.
  \DOIprefix\doi{https://doi.org/10.1016/j.physrep.2013.09.001},
  \bibinfo{note}{numerical simulations of time-resolved quantum electronics}.
%Type = Article
\bibitem[{Santos et~al.(2019)Santos, Lima, and Lewenkopf}]{Santos2019}
\bibinfo{author}{T.~P. Santos}, \bibinfo{author}{L.~R. Lima},
  \bibinfo{author}{C.~H. Lewenkopf},
\newblock \bibinfo{title}{An order n numerical method to efficiently calculate
  the transport properties of large systems: An algorithm optimized for sparse
  linear solvers},
\newblock \bibinfo{journal}{Journal of Computational Physics}
  \bibinfo{volume}{394} (\bibinfo{year}{2019}) \bibinfo{pages}{440 -- 455}.
%Type = Article
\bibitem[{Istas et~al.(2018)Istas, Groth, and Waintal}]{Istas2018}
\bibinfo{author}{M.~Istas}, \bibinfo{author}{C.~Groth},
  \bibinfo{author}{X.~Waintal},
\newblock \bibinfo{title}{A general algorithm for computing bound states in
  infinite tight-binding systems},
\newblock \bibinfo{journal}{SciPost Phys.} \bibinfo{volume}{4}
  (\bibinfo{year}{2018}) \bibinfo{pages}{026}.
  \DOIprefix\doi{10.21468/SciPostPhys.4.5.026}.
%Type = Incollection
\bibitem[{Harju et~al.(2013)Harju, Siro, Canova, Hakala, and
  Rantalaiho}]{harju2013lncs}
\bibinfo{author}{A.~Harju}, \bibinfo{author}{T.~Siro}, \bibinfo{author}{F.~F.
  Canova}, \bibinfo{author}{S.~Hakala}, \bibinfo{author}{T.~Rantalaiho},
\newblock \bibinfo{title}{{Computational Physics on Graphics Processing
  Units}},
\newblock in: \bibinfo{editor}{P.~Manninen}, \bibinfo{editor}{P.~\"Oster}
  (Eds.), \bibinfo{booktitle}{Applied Parallel and Scientific Computing, PARA
  2012, Lecture Notes in Computer Science, vol 7782},
  \bibinfo{publisher}{Springer}, \bibinfo{address}{Berlin, Heidelberg},
  \bibinfo{year}{2013}, pp. \bibinfo{pages}{3--26}.
  \DOIprefix\doi{10.1007/978-3-642-36803-5\_1}.
%Type = Article
\bibitem[{Fan et~al.(2018)Fan, Vierimaa, and Harju}]{fan2018cpc}
\bibinfo{author}{Z.~Fan}, \bibinfo{author}{V.~Vierimaa},
  \bibinfo{author}{A.~Harju},
\newblock \bibinfo{title}{{GPUQT: An efficient linear-scaling quantum transport
  code fully implemented on graphics processing units}},
\newblock \bibinfo{journal}{Comput. Phys. Commun.} \bibinfo{volume}{230}
  (\bibinfo{year}{2018}) \bibinfo{pages}{113 -- 120}. \URLprefix
  \url{http://www.sciencedirect.com/science/article/pii/S0010465518301280}.
  \DOIprefix\doi{10.1016/j.cpc.2018.04.013}.
%Type = Book
\bibitem[{Datta(1995)}]{datta1995}
\bibinfo{author}{S.~Datta}, \bibinfo{title}{Electonic transport in mesoscopic
  systems}, \bibinfo{publisher}{Cambridge University Press},
  \bibinfo{year}{1995}.
%Type = Article
\bibitem[{Fan et~al.(2014)Fan, Uppstu, and Harju}]{fan2014prb}
\bibinfo{author}{Z.~Fan}, \bibinfo{author}{A.~Uppstu},
  \bibinfo{author}{A.~Harju},
\newblock \bibinfo{title}{{Anderson localization in two-dimensional graphene
  with short-range disorder: One-parameter scaling and finite-size effects}},
\newblock \bibinfo{journal}{Phys. Rev. B} \bibinfo{volume}{89}
  (\bibinfo{year}{2014}) \bibinfo{pages}{245422}. \URLprefix
  \url{https://link.aps.org/doi/10.1103/PhysRevB.89.245422}.
  \DOIprefix\doi{10.1103/PhysRevB.89.245422}.
%Type = Article
\bibitem[{Weisse(2004)}]{Weisse2004}
\bibinfo{author}{A.~Weisse},
\newblock \bibinfo{title}{{Chebyshev expansion approach to the AC conductivity
  of the Anderson model}},
\newblock \bibinfo{journal}{Eur. Phys. J. B} \bibinfo{volume}{40}
  (\bibinfo{year}{2004}) \bibinfo{pages}{125--128}.
  \DOIprefix\doi{10.1140/epjb/e2004-00250-6}.
%Type = Article
\bibitem[{Yuan et~al.(2011)Yuan, Rold\'an, De~Raedt, and Katsnelson}]{Yuan2011}
\bibinfo{author}{S.~Yuan}, \bibinfo{author}{R.~Rold\'an},
  \bibinfo{author}{H.~De~Raedt}, \bibinfo{author}{M.~Katsnelson},
\newblock \bibinfo{title}{{Optical conductivity of disordered graphene beyond
  the Dirac cone approximation}},
\newblock \bibinfo{journal}{Phys. Rev. B} \bibinfo{volume}{84}
  (\bibinfo{year}{2011}) \bibinfo{pages}{195418}.
  \DOIprefix\doi{10.1103/PhysRevB.84.195418}.
%Type = Article
\bibitem[{Cysne et~al.(2016)Cysne, Rappoport, Ferreira, Viana Parente~Lopes,
  and Peres}]{Cysne2016}
\bibinfo{author}{T.~P. Cysne}, \bibinfo{author}{T.~G. Rappoport},
  \bibinfo{author}{A.~Ferreira}, \bibinfo{author}{J.~M. Viana Parente~Lopes},
  \bibinfo{author}{N.~M.~R. Peres},
\newblock \bibinfo{title}{{Numerical calculation of the Casimir-Polder
  interaction between a graphene sheet with vacancies and an atom}},
\newblock \bibinfo{journal}{Phys. Rev. B} \bibinfo{volume}{94}
  (\bibinfo{year}{2016}) \bibinfo{pages}{235405}.
  \DOIprefix\doi{10.1103/PhysRevB.94.235405}.
%Type = Article
\bibitem[{Wei\ss{}e et~al.(2005)Wei\ss{}e, Schubert, and Fehske}]{Weisse2005}
\bibinfo{author}{A.~Wei\ss{}e}, \bibinfo{author}{G.~Schubert},
  \bibinfo{author}{H.~Fehske},
\newblock \bibinfo{title}{{Optical response of electrons in a random
  potential}},
\newblock \bibinfo{journal}{Physica B Condens. Matter}
  \bibinfo{volume}{359--361} (\bibinfo{year}{2005}) \bibinfo{pages}{786--788}.
  \DOIprefix\doi{10.1016/j.physb.2005.01.227}.
%Type = Article
\bibitem[{Shechtman et~al.(1984)Shechtman, Blech, Gratias, and
  Cahn}]{shechtman1984prl}
\bibinfo{author}{D.~Shechtman}, \bibinfo{author}{I.~Blech},
  \bibinfo{author}{D.~Gratias}, \bibinfo{author}{J.~W. Cahn},
\newblock \bibinfo{title}{{Metallic Phase with Long-Range Orientational Order
  and No Translational Symmetry}},
\newblock \bibinfo{journal}{Phys. Rev. Lett.} \bibinfo{volume}{53}
  (\bibinfo{year}{1984}) \bibinfo{pages}{1951--1953}. \URLprefix
  \url{https://link.aps.org/doi/10.1103/PhysRevLett.53.1951}.
  \DOIprefix\doi{10.1103/PhysRevLett.53.1951}.
%Type = Article
\bibitem[{Roche et~al.(2005)Roche, Jiang, Triozon, and Saito}]{roche2005prl}
\bibinfo{author}{S.~Roche}, \bibinfo{author}{J.~Jiang},
  \bibinfo{author}{F.~Triozon}, \bibinfo{author}{R.~Saito},
\newblock \bibinfo{title}{{Quantum Dephasing in Carbon Nanotubes due to
  Electron-Phonon Coupling}},
\newblock \bibinfo{journal}{Phys. Rev. Lett.} \bibinfo{volume}{95}
  (\bibinfo{year}{2005}) \bibinfo{pages}{076803}. \URLprefix
  \url{https://link.aps.org/doi/10.1103/PhysRevLett.95.076803}.
  \DOIprefix\doi{10.1103/PhysRevLett.95.076803}.
%Type = Article
\bibitem[{Avriller et~al.(2006)Avriller, Latil, Triozon, Blase, and
  Roche}]{avriller2006prb}
\bibinfo{author}{R.~Avriller}, \bibinfo{author}{S.~Latil},
  \bibinfo{author}{F.~Triozon}, \bibinfo{author}{X.~Blase},
  \bibinfo{author}{S.~Roche},
\newblock \bibinfo{title}{{Chemical disorder strength in carbon nanotubes:
  Magnetic tuning of quantum transport regimes}},
\newblock \bibinfo{journal}{Phys. Rev. B} \bibinfo{volume}{74}
  (\bibinfo{year}{2006}) \bibinfo{pages}{121406}. \URLprefix
  \url{https://link.aps.org/doi/10.1103/PhysRevB.74.121406}.
  \DOIprefix\doi{10.1103/PhysRevB.74.121406}.
%Type = Article
\bibitem[{Rozhkov et~al.(2011)Rozhkov, Giavaras, Bliokh, Freilikher, and
  Nori}]{ROZHKOV201177}
\bibinfo{author}{A.~Rozhkov}, \bibinfo{author}{G.~Giavaras},
  \bibinfo{author}{Y.~P. Bliokh}, \bibinfo{author}{V.~Freilikher},
  \bibinfo{author}{F.~Nori},
\newblock \bibinfo{title}{Electronic properties of mesoscopic graphene
  structures: Charge confinement and control of spin and charge transport},
\newblock \bibinfo{journal}{Physics Reports} \bibinfo{volume}{503}
  (\bibinfo{year}{2011}) \bibinfo{pages}{77 -- 114}. \URLprefix
  \url{http://www.sciencedirect.com/science/article/pii/S0370157311000469}.
  \DOIprefix\doi{https://doi.org/10.1016/j.physrep.2011.02.002}.
%Type = Article
\bibitem[{Novoselov et~al.(2004)Novoselov, Geim, Morozov, Jiang, Zhang,
  Dubonos, Grigorieva, and Firsov}]{novoselov2004science}
\bibinfo{author}{K.~S. Novoselov}, \bibinfo{author}{A.~K. Geim},
  \bibinfo{author}{S.~V. Morozov}, \bibinfo{author}{D.~Jiang},
  \bibinfo{author}{Y.~Zhang}, \bibinfo{author}{S.~V. Dubonos},
  \bibinfo{author}{I.~V. Grigorieva}, \bibinfo{author}{A.~A. Firsov},
\newblock \bibinfo{title}{{Electric Field Effect in Atomically Thin Carbon
  Films}},
\newblock \bibinfo{journal}{Science} \bibinfo{volume}{306}
  (\bibinfo{year}{2004}) \bibinfo{pages}{666--669}. \URLprefix
  \url{http://science.sciencemag.org/content/306/5696/666}.
  \DOIprefix\doi{10.1126/science.1102896}.
%Type = Article
\bibitem[{Novoselov et~al.(2005)Novoselov, Jiang, Schedin, Booth, Khotkevich,
  Morozov, and Geim}]{novoselov2005pnas}
\bibinfo{author}{K.~S. Novoselov}, \bibinfo{author}{D.~Jiang},
  \bibinfo{author}{F.~Schedin}, \bibinfo{author}{T.~J. Booth},
  \bibinfo{author}{V.~V. Khotkevich}, \bibinfo{author}{S.~V. Morozov},
  \bibinfo{author}{A.~K. Geim},
\newblock \bibinfo{title}{{Two-dimensional atomic crystals}},
\newblock \bibinfo{journal}{Proc. Natl. Acad. Sci. U.S.A.}
  \bibinfo{volume}{102} (\bibinfo{year}{2005}) \bibinfo{pages}{10451--10453}.
  \URLprefix \url{http://www.pnas.org/content/102/30/10451}.
  \DOIprefix\doi{10.1073/pnas.0502848102}.
%Type = Article
\bibitem[{Peres(2010)}]{peres2010rmp}
\bibinfo{author}{N.~M.~R. Peres},
\newblock \bibinfo{title}{{Colloquium: The transport properties of graphene: An
  introduction}},
\newblock \bibinfo{journal}{Rev. Mod. Phys.} \bibinfo{volume}{82}
  (\bibinfo{year}{2010}) \bibinfo{pages}{2673--2700}. \URLprefix
  \url{https://link.aps.org/doi/10.1103/RevModPhys.82.2673}.
  \DOIprefix\doi{10.1103/RevModPhys.82.2673}.
%Type = Article
\bibitem[{Peres(2009)}]{peres2009jpcm}
\bibinfo{author}{N.~M.~R. Peres},
\newblock \bibinfo{title}{{The transport properties of graphene}},
\newblock \bibinfo{journal}{J. Phys. Condens. Matter} \bibinfo{volume}{21}
  (\bibinfo{year}{2009}) \bibinfo{pages}{323201}. \URLprefix
  \url{http://stacks.iop.org/0953-8984/21/i=32/a=323201}.
%Type = Book
\bibitem[{Torres et~al.(2020)Torres, Roche, and Charlier}]{torres2014book}
\bibinfo{author}{L.~E. F.~F. Torres}, \bibinfo{author}{S.~Roche},
  \bibinfo{author}{J.-C. Charlier}, \bibinfo{title}{Introduction to
  graphene-based nanomaterials}, \bibinfo{publisher}{Cambridge University
  Press}, \bibinfo{year}{2020}.
%Type = Article
\bibitem[{Shon and Ando(1998)}]{shon1998jpsj}
\bibinfo{author}{N.~Shon}, \bibinfo{author}{T.~Ando},
\newblock \bibinfo{title}{{Quantum Transport in Two-Dimensional Graphite
  System}},
\newblock \bibinfo{journal}{J. Phys. Soc. Jpn.} \bibinfo{volume}{67}
  (\bibinfo{year}{1998}) \bibinfo{pages}{2421--2429}. \URLprefix
  \url{https://doi.org/10.1143/JPSJ.67.2421}.
  \DOIprefix\doi{10.1143/JPSJ.67.2421}.
%Type = Article
\bibitem[{Ostrovsky et~al.(2006)Ostrovsky, Gornyi, and
  Mirlin}]{ostrovsky2006prb}
\bibinfo{author}{P.~M. Ostrovsky}, \bibinfo{author}{I.~V. Gornyi},
  \bibinfo{author}{A.~D. Mirlin},
\newblock \bibinfo{title}{Electron transport in disordered graphene},
\newblock \bibinfo{journal}{Phys. Rev. B} \bibinfo{volume}{74}
  (\bibinfo{year}{2006}) \bibinfo{pages}{235443}. \URLprefix
  \url{https://link.aps.org/doi/10.1103/PhysRevB.74.235443}.
  \DOIprefix\doi{10.1103/PhysRevB.74.235443}.
%Type = Article
\bibitem[{Lherbier et~al.(2008)Lherbier, Biel, Niquet, and
  Roche}]{lherbier2008prl}
\bibinfo{author}{A.~Lherbier}, \bibinfo{author}{B.~Biel},
  \bibinfo{author}{Y.-M. Niquet}, \bibinfo{author}{S.~Roche},
\newblock \bibinfo{title}{{Transport Length Scales in Disordered Graphene-Based
  Materials: Strong Localization Regimes and Dimensionality Effects}},
\newblock \bibinfo{journal}{Phys. Rev. Lett.} \bibinfo{volume}{100}
  (\bibinfo{year}{2008}) \bibinfo{pages}{036803}. \URLprefix
  \url{https://link.aps.org/doi/10.1103/PhysRevLett.100.036803}.
  \DOIprefix\doi{10.1103/PhysRevLett.100.036803}.
%Type = Article
\bibitem[{Roche et~al.(2012)Roche, Leconte, Ortmann, Lherbier, Soriano, and
  Charlier}]{roche2012ssc}
\bibinfo{author}{S.~Roche}, \bibinfo{author}{N.~Leconte},
  \bibinfo{author}{F.~Ortmann}, \bibinfo{author}{A.~Lherbier},
  \bibinfo{author}{D.~Soriano}, \bibinfo{author}{J.-C. Charlier},
\newblock \bibinfo{title}{{Quantum transport in disordered graphene: A
  theoretical perspective}},
\newblock \bibinfo{journal}{Solid State Commun.} \bibinfo{volume}{152}
  (\bibinfo{year}{2012}) \bibinfo{pages}{1404 -- 1410}. \URLprefix
  \url{http://www.sciencedirect.com/science/article/pii/S0038109812002360}.
  \DOIprefix\doi{10.1016/j.ssc.2012.04.030}.
%Type = Article
\bibitem[{MacKinnon and Kramer(1981)}]{mackinnon1981prl}
\bibinfo{author}{A.~MacKinnon}, \bibinfo{author}{B.~Kramer},
\newblock \bibinfo{title}{{One-Parameter Scaling of Localization Length and
  Conductance in Disordered Systems}},
\newblock \bibinfo{journal}{Phys. Rev. Lett.} \bibinfo{volume}{47}
  (\bibinfo{year}{1981}) \bibinfo{pages}{1546--1549}. \URLprefix
  \url{https://link.aps.org/doi/10.1103/PhysRevLett.47.1546}.
  \DOIprefix\doi{10.1103/PhysRevLett.47.1546}.
%Type = Article
\bibitem[{Kramer and MacKinnon(1993)}]{kramer1993rpp}
\bibinfo{author}{B.~Kramer}, \bibinfo{author}{A.~MacKinnon},
\newblock \bibinfo{title}{{Localization: theory and experiment}},
\newblock \bibinfo{journal}{Rep. Prog. Phys.} \bibinfo{volume}{56}
  (\bibinfo{year}{1993}) \bibinfo{pages}{1469}. \URLprefix
  \url{http://stacks.iop.org/0034-4885/56/i=12/a=001}.
%Type = Article
\bibitem[{Rycerz et~al.(2007)Rycerz, Tworzydlo, and Beenakker}]{rycerz2007epl}
\bibinfo{author}{A.~Rycerz}, \bibinfo{author}{J.~Tworzydlo},
  \bibinfo{author}{C.~W.~J. Beenakker},
\newblock \bibinfo{title}{Anomalously large conductance fluctuations in weakly
  disordered graphene},
\newblock \bibinfo{journal}{EPL} \bibinfo{volume}{79} (\bibinfo{year}{2007})
  \bibinfo{pages}{57003}. \URLprefix
  \url{http://stacks.iop.org/0295-5075/79/i=5/a=57003}.
%Type = Article
\bibitem[{Radchenko et~al.(2012)Radchenko, Shylau, and
  Zozoulenko}]{radchenko2012prb}
\bibinfo{author}{T.~M. Radchenko}, \bibinfo{author}{A.~A. Shylau},
  \bibinfo{author}{I.~V. Zozoulenko},
\newblock \bibinfo{title}{{Influence of correlated impurities on conductivity
  of graphene sheets: Time-dependent real-space Kubo approach}},
\newblock \bibinfo{journal}{Phys. Rev. B} \bibinfo{volume}{86}
  (\bibinfo{year}{2012}) \bibinfo{pages}{035418}. \URLprefix
  \url{https://link.aps.org/doi/10.1103/PhysRevB.86.035418}.
  \DOIprefix\doi{10.1103/PhysRevB.86.035418}.
%Type = Article
\bibitem[{Zhang et~al.(2009)Zhang, Hu, Bernevig, Wang, Xie, and
  Liu}]{zhang2009prl}
\bibinfo{author}{Y.-Y. Zhang}, \bibinfo{author}{J.~Hu}, \bibinfo{author}{B.~A.
  Bernevig}, \bibinfo{author}{X.~R. Wang}, \bibinfo{author}{X.~C. Xie},
  \bibinfo{author}{W.~M. Liu},
\newblock \bibinfo{title}{{Localization and the Kosterlitz-Thouless Transition
  in Disordered Graphene}},
\newblock \bibinfo{journal}{Phys. Rev. Lett.} \bibinfo{volume}{102}
  (\bibinfo{year}{2009}) \bibinfo{pages}{106401}. \URLprefix
  \url{https://link.aps.org/doi/10.1103/PhysRevLett.102.106401}.
  \DOIprefix\doi{10.1103/PhysRevLett.102.106401}.
%Type = Article
\bibitem[{Ostrovsky et~al.(2007)Ostrovsky, Gornyi, and
  Mirlin}]{ostrovsky2007prl}
\bibinfo{author}{P.~M. Ostrovsky}, \bibinfo{author}{I.~V. Gornyi},
  \bibinfo{author}{A.~D. Mirlin},
\newblock \bibinfo{title}{Quantum criticality and minimal conductivity in
  graphene with long-range disorder},
\newblock \bibinfo{journal}{Phys. Rev. Lett.} \bibinfo{volume}{98}
  (\bibinfo{year}{2007}) \bibinfo{pages}{256801}. \URLprefix
  \url{https://link.aps.org/doi/10.1103/PhysRevLett.98.256801}.
  \DOIprefix\doi{10.1103/PhysRevLett.98.256801}.
%Type = Article
\bibitem[{Bardarson et~al.(2007)Bardarson, Tworzyd\l{}o, Brouwer, and
  Beenakker}]{bardarson2007prl}
\bibinfo{author}{J.~H. Bardarson}, \bibinfo{author}{J.~Tworzyd\l{}o},
  \bibinfo{author}{P.~W. Brouwer}, \bibinfo{author}{C.~W.~J. Beenakker},
\newblock \bibinfo{title}{{One-Parameter Scaling at the Dirac Point in
  Graphene}},
\newblock \bibinfo{journal}{Phys. Rev. Lett.} \bibinfo{volume}{99}
  (\bibinfo{year}{2007}) \bibinfo{pages}{106801}. \URLprefix
  \url{https://link.aps.org/doi/10.1103/PhysRevLett.99.106801}.
  \DOIprefix\doi{10.1103/PhysRevLett.99.106801}.
%Type = Article
\bibitem[{Nomura et~al.(2007)Nomura, Koshino, and Ryu}]{nomura2007prl}
\bibinfo{author}{K.~Nomura}, \bibinfo{author}{M.~Koshino},
  \bibinfo{author}{S.~Ryu},
\newblock \bibinfo{title}{{Topological Delocalization of Two-Dimensional
  Massless Dirac Fermions}},
\newblock \bibinfo{journal}{Phys. Rev. Lett.} \bibinfo{volume}{99}
  (\bibinfo{year}{2007}) \bibinfo{pages}{146806}. \URLprefix
  \url{https://link.aps.org/doi/10.1103/PhysRevLett.99.146806}.
  \DOIprefix\doi{10.1103/PhysRevLett.99.146806}.
%Type = Article
\bibitem[{Chakravarty and Schmid(1986)}]{CHAKRAVARTY1986193}
\bibinfo{author}{S.~Chakravarty}, \bibinfo{author}{A.~Schmid},
\newblock \bibinfo{title}{Weak localization: The quasiclassical theory of
  electrons in a random potential},
\newblock \bibinfo{journal}{Physics Reports} \bibinfo{volume}{140}
  (\bibinfo{year}{1986}) \bibinfo{pages}{193 -- 236}. \URLprefix
  \url{http://www.sciencedirect.com/science/article/pii/037015738690027X}.
  \DOIprefix\doi{https://doi.org/10.1016/0370-1573(86)90027-X}.
%Type = Article
\bibitem[{McCann et~al.(2006)McCann, Kechedzhi, Fal'ko, Suzuura, Ando, and
  Altshuler}]{mccann2006prl}
\bibinfo{author}{E.~McCann}, \bibinfo{author}{K.~Kechedzhi},
  \bibinfo{author}{V.~I. Fal'ko}, \bibinfo{author}{H.~Suzuura},
  \bibinfo{author}{T.~Ando}, \bibinfo{author}{B.~L. Altshuler},
\newblock \bibinfo{title}{{Weak-Localization Magnetoresistance and Valley
  Symmetry in Graphene}},
\newblock \bibinfo{journal}{Phys. Rev. Lett.} \bibinfo{volume}{97}
  (\bibinfo{year}{2006}) \bibinfo{pages}{146805}. \URLprefix
  \url{https://link.aps.org/doi/10.1103/PhysRevLett.97.146805}.
  \DOIprefix\doi{10.1103/PhysRevLett.97.146805}.
%Type = Article
\bibitem[{{Kechedzhi, K.} et~al.(2007){Kechedzhi, K.}, {McCann, E.}, {Fal'ko,
  V. I.}, {Suzuura, H.}, {Ando, T.}, and {Altshuler, B.
  L.}}]{kechedzhi2007epjst}
\bibinfo{author}{{Kechedzhi, K.}}, \bibinfo{author}{{McCann, E.}},
  \bibinfo{author}{{Fal'ko, V. I.}}, \bibinfo{author}{{Suzuura, H.}},
  \bibinfo{author}{{Ando, T.}}, \bibinfo{author}{{Altshuler, B. L.}},
\newblock \bibinfo{title}{Weak localization in monolayer and bilayer graphene},
\newblock \bibinfo{journal}{Eur. Phys. J. Spec. Top.} \bibinfo{volume}{148}
  (\bibinfo{year}{2007}) \bibinfo{pages}{39--54}. \URLprefix
  \url{https://doi.org/10.1140/epjst/e2007-00224-6}.
  \DOIprefix\doi{10.1140/epjst/e2007-00224-6}.
%Type = Article
\bibitem[{Fal'ko et~al.(2007)Fal'ko, Kechedzhi, McCann, Altshuler, Suzuura, and
  Ando}]{falko2007ssc}
\bibinfo{author}{V.~I. Fal'ko}, \bibinfo{author}{K.~Kechedzhi},
  \bibinfo{author}{E.~McCann}, \bibinfo{author}{B.~Altshuler},
  \bibinfo{author}{H.~Suzuura}, \bibinfo{author}{T.~Ando},
\newblock \bibinfo{title}{Weak localization in graphene},
\newblock \bibinfo{journal}{Solid State Commun.} \bibinfo{volume}{143}
  (\bibinfo{year}{2007}) \bibinfo{pages}{33 -- 38}. \URLprefix
  \url{http://www.sciencedirect.com/science/article/pii/S003810980700302X}.
  \DOIprefix\doi{10.1016/j.ssc.2007.03.049}.
%Type = Article
\bibitem[{Lherbier et~al.(2012)Lherbier, Dubois, Declerck, Niquet, Roche, and
  Charlier}]{lherbier2012prb}
\bibinfo{author}{A.~Lherbier}, \bibinfo{author}{S.~M.-M. Dubois},
  \bibinfo{author}{X.~Declerck}, \bibinfo{author}{Y.-M. Niquet},
  \bibinfo{author}{S.~Roche}, \bibinfo{author}{J.-C. Charlier},
\newblock \bibinfo{title}{{Transport properties of graphene containing
  structural defects}},
\newblock \bibinfo{journal}{Phys. Rev. B} \bibinfo{volume}{86}
  (\bibinfo{year}{2012}) \bibinfo{pages}{075402}. \URLprefix
  \url{https://link.aps.org/doi/10.1103/PhysRevB.86.075402}.
  \DOIprefix\doi{10.1103/PhysRevB.86.075402}.
%Type = Article
\bibitem[{Lewenkopf et~al.(2008)Lewenkopf, Mucciolo, and
  Castro~Neto}]{lewenkopf2008prb}
\bibinfo{author}{C.~H. Lewenkopf}, \bibinfo{author}{E.~R. Mucciolo},
  \bibinfo{author}{A.~H. Castro~Neto},
\newblock \bibinfo{title}{{Numerical studies of conductivity and Fano factor in
  disordered graphene}},
\newblock \bibinfo{journal}{Phys. Rev. B} \bibinfo{volume}{77}
  (\bibinfo{year}{2008}) \bibinfo{pages}{081410}. \URLprefix
  \url{https://link.aps.org/doi/10.1103/PhysRevB.77.081410}.
  \DOIprefix\doi{10.1103/PhysRevB.77.081410}.
%Type = Article
\bibitem[{Tikhonenko et~al.(2009)Tikhonenko, Kozikov, Savchenko, and
  Gorbachev}]{tikhonenko2009prl}
\bibinfo{author}{F.~V. Tikhonenko}, \bibinfo{author}{A.~A. Kozikov},
  \bibinfo{author}{A.~K. Savchenko}, \bibinfo{author}{R.~V. Gorbachev},
\newblock \bibinfo{title}{{Transition between Electron Localization and
  Antilocalization in Graphene}},
\newblock \bibinfo{journal}{Phys. Rev. Lett.} \bibinfo{volume}{103}
  (\bibinfo{year}{2009}) \bibinfo{pages}{226801}. \URLprefix
  \url{https://link.aps.org/doi/10.1103/PhysRevLett.103.226801}.
  \DOIprefix\doi{10.1103/PhysRevLett.103.226801}.
%Type = Article
\bibitem[{Ortmann et~al.(2011)Ortmann, Cresti, Montambaux, and
  Roche}]{ortmann2011epl}
\bibinfo{author}{F.~Ortmann}, \bibinfo{author}{A.~Cresti},
  \bibinfo{author}{G.~Montambaux}, \bibinfo{author}{S.~Roche},
\newblock \bibinfo{title}{{Magnetoresistance in disordered graphene: The role
  of pseudospin and dimensionality effects unraveled}},
\newblock \bibinfo{journal}{EPL} \bibinfo{volume}{94} (\bibinfo{year}{2011})
  \bibinfo{pages}{47006}. \URLprefix
  \url{http://stacks.iop.org/0295-5075/94/i=4/a=47006}.
%Type = Article
\bibitem[{Hikami et~al.(1980)Hikami, Larkin, and Nagaoka}]{HikamiLarkinNagaoka}
\bibinfo{author}{S.~Hikami}, \bibinfo{author}{A.~I. Larkin},
  \bibinfo{author}{Y.~Nagaoka},
\newblock \bibinfo{title}{{Spin-Orbit Interaction and Magnetoresistance in the
  Two Dimensional Random System}},
\newblock \bibinfo{journal}{Progress of Theoretical Physics}
  \bibinfo{volume}{63} (\bibinfo{year}{1980}) \bibinfo{pages}{707--710}.
%Type = Article
\bibitem[{Cresti et~al.(2008)Cresti, Nemec, Biel, Niebler, Triozon, Cuniberti,
  and Roche}]{cresti2008nr}
\bibinfo{author}{A.~Cresti}, \bibinfo{author}{N.~Nemec},
  \bibinfo{author}{B.~Biel}, \bibinfo{author}{G.~Niebler},
  \bibinfo{author}{F.~Triozon}, \bibinfo{author}{G.~Cuniberti},
  \bibinfo{author}{S.~Roche},
\newblock \bibinfo{title}{{Charge transport in disordered graphene-based low
  dimensional materials}},
\newblock \bibinfo{journal}{Nano Res.} \bibinfo{volume}{1}
  (\bibinfo{year}{2008}) \bibinfo{pages}{361--394}.
  \DOIprefix\doi{10.1007/s12274-008-8043-2}.
%Type = Article
\bibitem[{Nakaharai et~al.(2013)Nakaharai, Iijima, Ogawa, Suzuki, Li,
  Tsukagoshi, Sato, and Yokoyama}]{nakaharai2013acsnano}
\bibinfo{author}{S.~Nakaharai}, \bibinfo{author}{T.~Iijima},
  \bibinfo{author}{S.~Ogawa}, \bibinfo{author}{S.~Suzuki},
  \bibinfo{author}{S.-L. Li}, \bibinfo{author}{K.~Tsukagoshi},
  \bibinfo{author}{S.~Sato}, \bibinfo{author}{N.~Yokoyama},
\newblock \bibinfo{title}{{Conduction Tuning of Graphene Based on
  Defect-Induced Localization}},
\newblock \bibinfo{journal}{ACS Nano} \bibinfo{volume}{7}
  (\bibinfo{year}{2013}) \bibinfo{pages}{5694--5700}. \URLprefix
  \url{https://doi.org/10.1021/nn401992q}. \DOIprefix\doi{10.1021/nn401992q}.
%Type = Article
\bibitem[{Zhao et~al.(2011)Zhao, He, Rim, Schiros, Kim, Zhou, Guti{\'e}rrez,
  Chockalingam, Arguello, P{\'a}lov{\'a}, Nordlund, Hybertsen, Reichman, Heinz,
  Kim, Pinczuk, Flynn, and Pasupathy}]{zhao2011science}
\bibinfo{author}{L.~Zhao}, \bibinfo{author}{R.~He}, \bibinfo{author}{K.~T.
  Rim}, \bibinfo{author}{T.~Schiros}, \bibinfo{author}{K.~S. Kim},
  \bibinfo{author}{H.~Zhou}, \bibinfo{author}{C.~Guti{\'e}rrez},
  \bibinfo{author}{S.~P. Chockalingam}, \bibinfo{author}{C.~J. Arguello},
  \bibinfo{author}{L.~P{\'a}lov{\'a}}, \bibinfo{author}{D.~Nordlund},
  \bibinfo{author}{M.~S. Hybertsen}, \bibinfo{author}{D.~R. Reichman},
  \bibinfo{author}{T.~F. Heinz}, \bibinfo{author}{P.~Kim},
  \bibinfo{author}{A.~Pinczuk}, \bibinfo{author}{G.~W. Flynn},
  \bibinfo{author}{A.~N. Pasupathy},
\newblock \bibinfo{title}{{Visualizing Individual Nitrogen Dopants in Monolayer
  Graphene}},
\newblock \bibinfo{journal}{Science} \bibinfo{volume}{333}
  (\bibinfo{year}{2011}) \bibinfo{pages}{999--1003}. \URLprefix
  \url{http://science.sciencemag.org/content/333/6045/999}.
  \DOIprefix\doi{10.1126/science.1208759}.
%Type = Article
\bibitem[{Biel et~al.(2009)Biel, Triozon, Blase, and Roche}]{biel2009nl}
\bibinfo{author}{B.~Biel}, \bibinfo{author}{F.~Triozon},
  \bibinfo{author}{X.~Blase}, \bibinfo{author}{S.~Roche},
\newblock \bibinfo{title}{{Effect of the Chemical Functionalization on Charge
  Transport in Carbon Nanotubes at the Mesoscopic Scale}},
\newblock \bibinfo{journal}{Nano Lett.} \bibinfo{volume}{9}
  (\bibinfo{year}{2009}) \bibinfo{pages}{2725--2729}. \URLprefix
  \url{https://doi.org/10.1021/nl901226s}. \DOIprefix\doi{10.1021/nl901226s}.
%Type = Article
\bibitem[{Lherbier et~al.(2013)Lherbier, Botello-M\'endez, and
  Charlier}]{lherbier2013nl}
\bibinfo{author}{A.~Lherbier}, \bibinfo{author}{A.~R. Botello-M\'endez},
  \bibinfo{author}{J.-C. Charlier},
\newblock \bibinfo{title}{{Electronic and Transport Properties of Unbalanced
  Sublattice N-Doping in Graphene}},
\newblock \bibinfo{journal}{Nano Lett.} \bibinfo{volume}{13}
  (\bibinfo{year}{2013}) \bibinfo{pages}{1446--1450}. \URLprefix
  \url{https://doi.org/10.1021/nl304351z}. \DOIprefix\doi{10.1021/nl304351z}.
%Type = Article
\bibitem[{Marconcini et~al.(2012)Marconcini, Cresti, Triozon, Fiori, Biel,
  Niquet, Macucci, and Roche}]{marconcini2012acsnano}
\bibinfo{author}{P.~Marconcini}, \bibinfo{author}{A.~Cresti},
  \bibinfo{author}{F.~Triozon}, \bibinfo{author}{G.~Fiori},
  \bibinfo{author}{B.~Biel}, \bibinfo{author}{Y.-M. Niquet},
  \bibinfo{author}{M.~Macucci}, \bibinfo{author}{S.~Roche},
\newblock \bibinfo{title}{{Atomistic Boron-Doped Graphene Field-Effect
  Transistors: A Route toward Unipolar Characteristics}},
\newblock \bibinfo{journal}{ACS Nano} \bibinfo{volume}{6}
  (\bibinfo{year}{2012}) \bibinfo{pages}{7942--7947}. \URLprefix
  \url{https://doi.org/10.1021/nn3024046}. \DOIprefix\doi{10.1021/nn3024046}.
%Type = Article
\bibitem[{Meyer et~al.(2008)Meyer, Kisielowski, Erni, Rossell, Crommie, and
  Zettl}]{meyer2008nl}
\bibinfo{author}{J.~C. Meyer}, \bibinfo{author}{C.~Kisielowski},
  \bibinfo{author}{R.~Erni}, \bibinfo{author}{M.~D. Rossell},
  \bibinfo{author}{M.~F. Crommie}, \bibinfo{author}{A.~Zettl},
\newblock \bibinfo{title}{{Direct Imaging of Lattice Atoms and Topological
  Defects in Graphene Membranes}},
\newblock \bibinfo{journal}{Nano Lett.} \bibinfo{volume}{8}
  (\bibinfo{year}{2008}) \bibinfo{pages}{3582--3586}. \URLprefix
  \url{https://doi.org/10.1021/nl801386m}. \DOIprefix\doi{10.1021/nl801386m}.
%Type = Article
\bibitem[{Ugeda et~al.(2010)Ugeda, Brihuega, Guinea, and
  G\'omez-Rodr\'{\i}guez}]{ugeda2010prl}
\bibinfo{author}{M.~M. Ugeda}, \bibinfo{author}{I.~Brihuega},
  \bibinfo{author}{F.~Guinea}, \bibinfo{author}{J.~M. G\'omez-Rodr\'{\i}guez},
\newblock \bibinfo{title}{{Missing Atom as a Source of Carbon Magnetism}},
\newblock \bibinfo{journal}{Phys. Rev. Lett.} \bibinfo{volume}{104}
  (\bibinfo{year}{2010}) \bibinfo{pages}{096804}. \URLprefix
  \url{https://link.aps.org/doi/10.1103/PhysRevLett.104.096804}.
  \DOIprefix\doi{10.1103/PhysRevLett.104.096804}.
%Type = Article
\bibitem[{Pereira et~al.(2008)Pereira, Lopes~dos Santos, and
  Castro~Neto}]{pereira2008prb}
\bibinfo{author}{V.~M. Pereira}, \bibinfo{author}{J.~M.~B. Lopes~dos Santos},
  \bibinfo{author}{A.~H. Castro~Neto},
\newblock \bibinfo{title}{{Modeling disorder in graphene}},
\newblock \bibinfo{journal}{Phys. Rev. B} \bibinfo{volume}{77}
  (\bibinfo{year}{2008}) \bibinfo{pages}{115109}. \URLprefix
  \url{https://link.aps.org/doi/10.1103/PhysRevB.77.115109}.
  \DOIprefix\doi{10.1103/PhysRevB.77.115109}.
%Type = Article
\bibitem[{Trambly De~Laissardi\`ere and Mayou(2011)}]{TramblyMPL2011}
\bibinfo{author}{G.~Trambly De~Laissardi\`ere}, \bibinfo{author}{D.~Mayou},
\newblock \bibinfo{title}{Electronic transport in graphene: quantum effects and
  role of local defects},
\newblock \bibinfo{journal}{Modern Physics Letters B} \bibinfo{volume}{25}
  (\bibinfo{year}{2011}) \bibinfo{pages}{1019--1028}. \URLprefix
  \url{https://doi.org/10.1142/S0217984911026747}.
  \DOIprefix\doi{10.1142/S0217984911026747}.
%Type = Article
\bibitem[{Cresti et~al.(2013)Cresti, Ortmann, Louvet, Van~Tuan, and
  Roche}]{cresti2013prl}
\bibinfo{author}{A.~Cresti}, \bibinfo{author}{F.~Ortmann},
  \bibinfo{author}{T.~Louvet}, \bibinfo{author}{D.~Van~Tuan},
  \bibinfo{author}{S.~Roche},
\newblock \bibinfo{title}{{Broken Symmetries, Zero-Energy Modes, and Quantum
  Transport in Disordered Graphene: From Supermetallic to Insulating Regimes}},
\newblock \bibinfo{journal}{Phys. Rev. Lett.} \bibinfo{volume}{110}
  (\bibinfo{year}{2013}) \bibinfo{pages}{196601}. \URLprefix
  \url{https://link.aps.org/doi/10.1103/PhysRevLett.110.196601}.
  \DOIprefix\doi{10.1103/PhysRevLett.110.196601}.
%Type = Article
\bibitem[{Gade and Wegner(1991)}]{gade1991npb}
\bibinfo{author}{R.~Gade}, \bibinfo{author}{F.~Wegner},
\newblock \bibinfo{title}{The n = 0 replica limit of u(n) and u(n)so(n)
  models},
\newblock \bibinfo{journal}{Nucl. Phys. B} \bibinfo{volume}{360}
  (\bibinfo{year}{1991}) \bibinfo{pages}{213 -- 218}. \URLprefix
  \url{http://www.sciencedirect.com/science/article/pii/055032139190401I}.
  \DOIprefix\doi{10.1016/0550-3213(91)90401-I}.
%Type = Article
\bibitem[{Gade(1993)}]{gade1993npb}
\bibinfo{author}{R.~Gade},
\newblock \bibinfo{title}{Anderson localization for sublattice models},
\newblock \bibinfo{journal}{Nucl. Phys. B} \bibinfo{volume}{398}
  (\bibinfo{year}{1993}) \bibinfo{pages}{499 -- 515}. \URLprefix
  \url{http://www.sciencedirect.com/science/article/pii/055032139390601K}.
  \DOIprefix\doi{10.1016/0550-3213(93)90601-K}.
%Type = Article
\bibitem[{Ostrovsky et~al.(2010)Ostrovsky, Titov, Bera, Gornyi, and
  Mirlin}]{ostrovsky2010prl}
\bibinfo{author}{P.~M. Ostrovsky}, \bibinfo{author}{M.~Titov},
  \bibinfo{author}{S.~Bera}, \bibinfo{author}{I.~V. Gornyi},
  \bibinfo{author}{A.~D. Mirlin},
\newblock \bibinfo{title}{{Diffusion and Criticality in Undoped Graphene with
  Resonant Scatterers}},
\newblock \bibinfo{journal}{Phys. Rev. Lett.} \bibinfo{volume}{105}
  (\bibinfo{year}{2010}) \bibinfo{pages}{266803}. \URLprefix
  \url{https://link.aps.org/doi/10.1103/PhysRevLett.105.266803}.
  \DOIprefix\doi{10.1103/PhysRevLett.105.266803}.
%Type = Article
\bibitem[{Ostrovsky et~al.(2014)Ostrovsky, Protopopov, K\"onig, Gornyi, Mirlin,
  and Skvortsov}]{Ostrovsky2014}
\bibinfo{author}{P.~M. Ostrovsky}, \bibinfo{author}{I.~V. Protopopov},
  \bibinfo{author}{E.~J. K\"onig}, \bibinfo{author}{I.~V. Gornyi},
  \bibinfo{author}{A.~D. Mirlin}, \bibinfo{author}{M.~A. Skvortsov},
\newblock \bibinfo{title}{Density of states in a two-dimensional chiral metal
  with vacancies},
\newblock \bibinfo{journal}{Phys. Rev. Lett.} \bibinfo{volume}{113}
  (\bibinfo{year}{2014}) \bibinfo{pages}{186803}. \URLprefix
  \url{https://link.aps.org/doi/10.1103/PhysRevLett.113.186803}.
  \DOIprefix\doi{10.1103/PhysRevLett.113.186803}.
%Type = Article
\bibitem[{Lherbier et~al.(2011)Lherbier, Dubois, Declerck, Roche, Niquet, and
  Charlier}]{LherbierPRL2011}
\bibinfo{author}{A.~Lherbier}, \bibinfo{author}{S.~M.-M. Dubois},
  \bibinfo{author}{X.~Declerck}, \bibinfo{author}{S.~Roche},
  \bibinfo{author}{Y.-M. Niquet}, \bibinfo{author}{J.-C. Charlier},
\newblock \bibinfo{title}{Two-dimensional graphene with structural defects:
  Elastic mean free path, minimum conductivity, and anderson transition},
\newblock \bibinfo{journal}{Phys. Rev. Lett.} \bibinfo{volume}{106}
  (\bibinfo{year}{2011}) \bibinfo{pages}{046803}. \URLprefix
  \url{https://link.aps.org/doi/10.1103/PhysRevLett.106.046803}.
%Type = Article
\bibitem[{Pedersen et~al.(2008)Pedersen, Flindt, Pedersen, Mortensen, Jauho,
  and Pedersen}]{pedersen2008prl}
\bibinfo{author}{T.~G. Pedersen}, \bibinfo{author}{C.~Flindt},
  \bibinfo{author}{J.~Pedersen}, \bibinfo{author}{N.~A. Mortensen},
  \bibinfo{author}{A.-P. Jauho}, \bibinfo{author}{K.~Pedersen},
\newblock \bibinfo{title}{{Graphene Antidot Lattices: Designed Defects and Spin
  Qubits}},
\newblock \bibinfo{journal}{Phys. Rev. Lett.} \bibinfo{volume}{100}
  (\bibinfo{year}{2008}) \bibinfo{pages}{136804}. \URLprefix
  \url{https://link.aps.org/doi/10.1103/PhysRevLett.100.136804}.
  \DOIprefix\doi{10.1103/PhysRevLett.100.136804}.
%Type = Article
\bibitem[{Bai et~al.(2010)Bai, Zhong, Jiang, Huang, and Duan}]{bai2010nnt}
\bibinfo{author}{J.~Bai}, \bibinfo{author}{X.~Zhong},
  \bibinfo{author}{S.~Jiang}, \bibinfo{author}{Y.~Huang},
  \bibinfo{author}{X.~Duan},
\newblock \bibinfo{title}{{Graphene nanomesh}},
\newblock \bibinfo{journal}{Nat. Nanotechnol.} \bibinfo{volume}{5}
  (\bibinfo{year}{2010}) \bibinfo{pages}{190}. \URLprefix
  \url{http://dx.doi.org/10.1038/nnano.2010.8}.
  \DOIprefix\doi{10.1038/nnano.2010.8}.
%Type = Article
\bibitem[{Zhang et~al.(2013)Zhang, Zhang, and Zhou}]{2013zhang_cvdreview}
\bibinfo{author}{Y.~Zhang}, \bibinfo{author}{L.~Zhang},
  \bibinfo{author}{C.~Zhou},
\newblock \bibinfo{title}{{Review of Chemical Vapor Deposition of Graphene and
  Related Applications}},
\newblock \bibinfo{journal}{Acc. Chem. Res.} \bibinfo{volume}{46}
  (\bibinfo{year}{2013}) \bibinfo{pages}{2329--2339}. \URLprefix
  \url{https://doi.org/10.1021/ar300203n}. \DOIprefix\doi{10.1021/ar300203n}.
%Type = Article
\bibitem[{Arjmandi-Tash et~al.(2018)Arjmandi-Tash, Kalita, Han, Othmen, Nayak,
  Berne, Landers, Watanabe, Taniguchi, Marty, Coraux, Bendiab, and
  Bouchiat}]{Arjmandi_Tash_2018}
\bibinfo{author}{H.~Arjmandi-Tash}, \bibinfo{author}{D.~Kalita},
  \bibinfo{author}{Z.~Han}, \bibinfo{author}{R.~Othmen},
  \bibinfo{author}{G.~Nayak}, \bibinfo{author}{C.~Berne},
  \bibinfo{author}{J.~Landers}, \bibinfo{author}{K.~Watanabe},
  \bibinfo{author}{T.~Taniguchi}, \bibinfo{author}{L.~Marty},
  \bibinfo{author}{J.~Coraux}, \bibinfo{author}{N.~Bendiab},
  \bibinfo{author}{V.~Bouchiat},
\newblock \bibinfo{title}{Large scale graphene/h-{BN} heterostructures obtained
  by direct {CVD} growth of graphene using high-yield proximity-catalytic
  process},
\newblock \bibinfo{journal}{J. Phys. Mater.} \bibinfo{volume}{1}
  (\bibinfo{year}{2018}) \bibinfo{pages}{015003}. \URLprefix
  \url{https://doi.org/10.1088\%2F2515-7639\%2Faac66e}.
  \DOIprefix\doi{10.1088/2515-7639/aac66e}.
%Type = Article
\bibitem[{Isacsson et~al.(2017)Isacsson, Cummings, Colombo, Colombo, Kinaret,
  and Roche}]{Isacsson2017_2dmater}
\bibinfo{author}{A.~Isacsson}, \bibinfo{author}{A.~W. Cummings},
  \bibinfo{author}{L.~Colombo}, \bibinfo{author}{L.~Colombo},
  \bibinfo{author}{J.~M. Kinaret}, \bibinfo{author}{S.~Roche},
\newblock \bibinfo{title}{{Scaling properties of polycrystalline graphene: a
  review}},
\newblock \bibinfo{journal}{2D Mater.} \bibinfo{volume}{4}
  (\bibinfo{year}{2017}) \bibinfo{pages}{012002}. \URLprefix
  \url{http://stacks.iop.org/2053-1583/4/i=1/a=012002}.
%Type = Article
\bibitem[{Van~Tuan et~al.(2013)Van~Tuan, Kotakoski, Louvet, Ortmann, Meyer, and
  Roche}]{vantuan2013nl}
\bibinfo{author}{D.~Van~Tuan}, \bibinfo{author}{J.~Kotakoski},
  \bibinfo{author}{T.~Louvet}, \bibinfo{author}{F.~Ortmann},
  \bibinfo{author}{J.~C. Meyer}, \bibinfo{author}{S.~Roche},
\newblock \bibinfo{title}{{Scaling Properties of Charge Transport in
  Polycrystalline Graphene}},
\newblock \bibinfo{journal}{Nano Lett.} \bibinfo{volume}{13}
  (\bibinfo{year}{2013}) \bibinfo{pages}{1730--1735}. \URLprefix
  \url{https://doi.org/10.1021/nl400321r}. \DOIprefix\doi{10.1021/nl400321r}.
%Type = Article
\bibitem[{Cummings et~al.(2014{\natexlab{a}})Cummings, Duong, Nguyen, Van~Tuan,
  Kotakoski, Barrios~Vargas, Lee, and Roche}]{cummings2014am}
\bibinfo{author}{A.~W. Cummings}, \bibinfo{author}{D.~L. Duong},
  \bibinfo{author}{V.~L. Nguyen}, \bibinfo{author}{D.~Van~Tuan},
  \bibinfo{author}{J.~Kotakoski}, \bibinfo{author}{J.~E. Barrios~Vargas},
  \bibinfo{author}{Y.~H. Lee}, \bibinfo{author}{S.~Roche},
\newblock \bibinfo{title}{{Charge Transport in Polycrystalline Graphene:
  Challenges and Opportunities}},
\newblock \bibinfo{journal}{Adv. Mater.} \bibinfo{volume}{26}
  (\bibinfo{year}{2014}{\natexlab{a}}) \bibinfo{pages}{5079--5094}. \URLprefix
  \url{http://dx.doi.org/10.1002/adma.201401389}.
  \DOIprefix\doi{10.1002/adma.201401389}.
%Type = Article
\bibitem[{Cummings et~al.(2014{\natexlab{b}})Cummings, Cresti, and
  Roche}]{cummings2014prb}
\bibinfo{author}{A.~W. Cummings}, \bibinfo{author}{A.~Cresti},
  \bibinfo{author}{S.~Roche},
\newblock \bibinfo{title}{{Quantum Hall effect in polycrystalline graphene: The
  role of grain boundaries}},
\newblock \bibinfo{journal}{Phys. Rev. B} \bibinfo{volume}{90}
  (\bibinfo{year}{2014}{\natexlab{b}}) \bibinfo{pages}{161401}. \URLprefix
  \url{https://link.aps.org/doi/10.1103/PhysRevB.90.161401}.
  \DOIprefix\doi{10.1103/PhysRevB.90.161401}.
%Type = Article
\bibitem[{Seifert et~al.(2015)Seifert, Vargas, Bobinger, Sachsenhauser,
  Cummings, Roche, and Garrido}]{seifert2015twodm}
\bibinfo{author}{M.~Seifert}, \bibinfo{author}{J.~E.~B. Vargas},
  \bibinfo{author}{M.~Bobinger}, \bibinfo{author}{M.~Sachsenhauser},
  \bibinfo{author}{A.~W. Cummings}, \bibinfo{author}{S.~Roche},
  \bibinfo{author}{J.~A. Garrido},
\newblock \bibinfo{title}{{Role of grain boundaries in tailoring electronic
  properties of polycrystalline graphene by chemical functionalization}},
\newblock \bibinfo{journal}{2D Mater.} \bibinfo{volume}{2}
  (\bibinfo{year}{2015}) \bibinfo{pages}{024008}. \URLprefix
  \url{http://stacks.iop.org/2053-1583/2/i=2/a=024008}.
%Type = Article
\bibitem[{Barrios-Vargas et~al.(2017)Barrios-Vargas, Mortazavi, Cummings,
  Martinez-Gordillo, Pruneda, Colombo, Rabczuk, and
  Roche}]{barrios-vargas2017nl}
\bibinfo{author}{J.~E. Barrios-Vargas}, \bibinfo{author}{B.~Mortazavi},
  \bibinfo{author}{A.~W. Cummings}, \bibinfo{author}{R.~Martinez-Gordillo},
  \bibinfo{author}{M.~Pruneda}, \bibinfo{author}{L.~Colombo},
  \bibinfo{author}{T.~Rabczuk}, \bibinfo{author}{S.~Roche},
\newblock \bibinfo{title}{{Electrical and Thermal Transport in Coplanar
  Polycrystalline Graphene-hBN Heterostructures}},
\newblock \bibinfo{journal}{Nano Lett.} \bibinfo{volume}{17}
  (\bibinfo{year}{2017}) \bibinfo{pages}{1660--1664}. \URLprefix
  \url{http://dx.doi.org/10.1021/acs.nanolett.6b04936}.
  \DOIprefix\doi{10.1021/acs.nanolett.6b04936}.
%Type = Article
\bibitem[{Power and Jauho(2014)}]{power2014prb}
\bibinfo{author}{S.~R. Power}, \bibinfo{author}{A.-P. Jauho},
\newblock \bibinfo{title}{{Electronic transport in disordered graphene antidot
  lattice devices}},
\newblock \bibinfo{journal}{Phys. Rev. B} \bibinfo{volume}{90}
  (\bibinfo{year}{2014}) \bibinfo{pages}{115408}. \URLprefix
  \url{https://link.aps.org/doi/10.1103/PhysRevB.90.115408}.
  \DOIprefix\doi{10.1103/PhysRevB.90.115408}.
%Type = Article
\bibitem[{Fan et~al.(2015)Fan, Uppstu, and Harju}]{fan2015prb}
\bibinfo{author}{Z.~Fan}, \bibinfo{author}{A.~Uppstu},
  \bibinfo{author}{A.~Harju},
\newblock \bibinfo{title}{{Electronic and transport properties in geometrically
  disordered graphene antidot lattices}},
\newblock \bibinfo{journal}{Phys. Rev. B} \bibinfo{volume}{91}
  (\bibinfo{year}{2015}) \bibinfo{pages}{125434}. \URLprefix
  \url{https://link.aps.org/doi/10.1103/PhysRevB.91.125434}.
  \DOIprefix\doi{10.1103/PhysRevB.91.125434}.
%Type = Article
\bibitem[{Eroms and Weiss(2009)}]{eroms2009njp}
\bibinfo{author}{J.~Eroms}, \bibinfo{author}{D.~Weiss},
\newblock \bibinfo{title}{{Weak localization and transport gap in graphene
  antidot lattices}},
\newblock \bibinfo{journal}{New J. Phys.} \bibinfo{volume}{11}
  (\bibinfo{year}{2009}) \bibinfo{pages}{095021}. \URLprefix
  \url{http://stacks.iop.org/1367-2630/11/i=9/a=095021}.
%Type = Article
\bibitem[{Giesbers et~al.(2012)Giesbers, Peters, Burghard, and
  Kern}]{giesbers2012prb}
\bibinfo{author}{A.~J.~M. Giesbers}, \bibinfo{author}{E.~C. Peters},
  \bibinfo{author}{M.~Burghard}, \bibinfo{author}{K.~Kern},
\newblock \bibinfo{title}{{Charge transport gap in graphene antidot lattices}},
\newblock \bibinfo{journal}{Phys. Rev. B} \bibinfo{volume}{86}
  (\bibinfo{year}{2012}) \bibinfo{pages}{045445}. \URLprefix
  \url{https://link.aps.org/doi/10.1103/PhysRevB.86.045445}.
  \DOIprefix\doi{10.1103/PhysRevB.86.045445}.
%Type = Article
\bibitem[{Zhang et~al.(2013)Zhang, Lu, Shi, Wang, Zhang, Sun, Zheng, Chen,
  Wang, Lin, and Sheng}]{zhang2013prl}
\bibinfo{author}{H.~Zhang}, \bibinfo{author}{J.~Lu}, \bibinfo{author}{W.~Shi},
  \bibinfo{author}{Z.~Wang}, \bibinfo{author}{T.~Zhang},
  \bibinfo{author}{M.~Sun}, \bibinfo{author}{Y.~Zheng},
  \bibinfo{author}{Q.~Chen}, \bibinfo{author}{N.~Wang}, \bibinfo{author}{J.-J.
  Lin}, \bibinfo{author}{P.~Sheng},
\newblock \bibinfo{title}{{Large-scale Mesoscopic Transport in Nanostructured
  Graphene}},
\newblock \bibinfo{journal}{Phys. Rev. Lett.} \bibinfo{volume}{110}
  (\bibinfo{year}{2013}) \bibinfo{pages}{066805}. \URLprefix
  \url{https://link.aps.org/doi/10.1103/PhysRevLett.110.066805}.
  \DOIprefix\doi{10.1103/PhysRevLett.110.066805}.
%Type = Article
\bibitem[{Pedersen et~al.(2014)Pedersen, Cummings, and Roche}]{pedersen2014prb}
\bibinfo{author}{J.~G. Pedersen}, \bibinfo{author}{A.~W. Cummings},
  \bibinfo{author}{S.~Roche},
\newblock \bibinfo{title}{{Anisotropic behavior of quantum transport in
  graphene superlattices: Coexistence of ballistic conduction with Anderson
  insulating regime}},
\newblock \bibinfo{journal}{Phys. Rev. B} \bibinfo{volume}{89}
  (\bibinfo{year}{2014}) \bibinfo{pages}{165401}.
  \DOIprefix\doi{10.1103/PhysRevB.89.165401}.
%Type = Article
\bibitem[{Joo et~al.(2017)Joo, Lee, Jang, Kang, Kwon, Chung, Lee, Kim, Kim,
  Yang, Kim, Choi, Whang, and Hwang}]{Jooe1601821}
\bibinfo{author}{W.-J. Joo}, \bibinfo{author}{J.-H. Lee},
  \bibinfo{author}{Y.~Jang}, \bibinfo{author}{S.-G. Kang},
  \bibinfo{author}{Y.-N. Kwon}, \bibinfo{author}{J.~Chung},
  \bibinfo{author}{S.~Lee}, \bibinfo{author}{C.~Kim}, \bibinfo{author}{T.-H.
  Kim}, \bibinfo{author}{C.-W. Yang}, \bibinfo{author}{U.~J. Kim},
  \bibinfo{author}{B.~L. Choi}, \bibinfo{author}{D.~Whang},
  \bibinfo{author}{S.-W. Hwang},
\newblock \bibinfo{title}{Realization of continuous zachariasen carbon
  monolayer},
\newblock \bibinfo{journal}{Science Advances} \bibinfo{volume}{3}
  (\bibinfo{year}{2017}). \URLprefix
  \url{https://advances.sciencemag.org/content/3/2/e1601821}.
  \DOIprefix\doi{10.1126/sciadv.1601821}.
  \href{http://arxiv.org/abs/https://advances.sciencemag.org/content/3/2/e1601821.full.pdf}{{\tt
  arXiv:https://advances.sciencemag.org/content/3/2/e1601821.full.pdf}}.
%Type = Article
\bibitem[{Toh et~al.(2020)Toh, Zhang, Lin, Mayorov, Wang, Orofeo, Ferry,
  Andersen, Kakenov, Guo, Abidi, Sims, Suenaga, Pantelides, and
  {\"O}zyilmaz}]{Toh2020}
\bibinfo{author}{C.-T. Toh}, \bibinfo{author}{H.~Zhang},
  \bibinfo{author}{J.~Lin}, \bibinfo{author}{A.~S. Mayorov},
  \bibinfo{author}{Y.-P. Wang}, \bibinfo{author}{C.~M. Orofeo},
  \bibinfo{author}{D.~B. Ferry}, \bibinfo{author}{H.~Andersen},
  \bibinfo{author}{N.~Kakenov}, \bibinfo{author}{Z.~Guo},
  \bibinfo{author}{I.~H. Abidi}, \bibinfo{author}{H.~Sims},
  \bibinfo{author}{K.~Suenaga}, \bibinfo{author}{S.~T. Pantelides},
  \bibinfo{author}{B.~{\"O}zyilmaz},
\newblock \bibinfo{title}{Synthesis and properties of free-standing monolayer
  amorphous carbon},
\newblock \bibinfo{journal}{Nature} \bibinfo{volume}{577}
  (\bibinfo{year}{2020}) \bibinfo{pages}{199--203}. \URLprefix
  \url{https://doi.org/10.1038/s41586-019-1871-2}.
  \DOIprefix\doi{10.1038/s41586-019-1871-2}.
%Type = Article
\bibitem[{Van~Tuan et~al.(2012)Van~Tuan, Kumar, Roche, Ortmann, Thorpe, and
  Ordejon}]{VanTuanPRB2012}
\bibinfo{author}{D.~Van~Tuan}, \bibinfo{author}{A.~Kumar},
  \bibinfo{author}{S.~Roche}, \bibinfo{author}{F.~Ortmann},
  \bibinfo{author}{M.~F. Thorpe}, \bibinfo{author}{P.~Ordejon},
\newblock \bibinfo{title}{Insulating behavior of an amorphous graphene
  membrane},
\newblock \bibinfo{journal}{Phys. Rev. B} \bibinfo{volume}{86}
  (\bibinfo{year}{2012}) \bibinfo{pages}{121408}. \URLprefix
  \url{https://link.aps.org/doi/10.1103/PhysRevB.86.121408}.
  \DOIprefix\doi{10.1103/PhysRevB.86.121408}.
%Type = Article
\bibitem[{Lherbier et~al.(2013)Lherbier, Roche, Restrepo, Niquet, Delcorte, and
  Charlier}]{LherbierNR2013}
\bibinfo{author}{A.~Lherbier}, \bibinfo{author}{S.~Roche},
  \bibinfo{author}{O.~A. Restrepo}, \bibinfo{author}{Y.-M. Niquet},
  \bibinfo{author}{A.~Delcorte}, \bibinfo{author}{J.-C. Charlier},
\newblock \bibinfo{title}{Highly defective graphene: A key prototype of
  two-dimensional anderson insulators},
\newblock \bibinfo{journal}{Nano Research} \bibinfo{volume}{6}
  (\bibinfo{year}{2013}) \bibinfo{pages}{326--334}. \URLprefix
  \url{https://doi.org/10.1007/s12274-013-0309-7}.
  \DOIprefix\doi{10.1007/s12274-013-0309-7}.
%Type = Article
\bibitem[{Bose et~al.(1994)Bose, Jepsen, and Andersen}]{Bose_1994}
\bibinfo{author}{S.~K. Bose}, \bibinfo{author}{O.~Jepsen},
  \bibinfo{author}{O.~K. Andersen},
\newblock \bibinfo{title}{{An electronic structure and resistivity calculation
  for liquid La}},
\newblock \bibinfo{journal}{J. Phys. Condens. Matter} \bibinfo{volume}{6}
  (\bibinfo{year}{1994}) \bibinfo{pages}{2145--2158}.
  \DOIprefix\doi{10.1088/0953-8984/6/11/004}.
%Type = Article
\bibitem[{Bose(1998)}]{Bose1998}
\bibinfo{author}{S.~K. Bose},
\newblock \bibinfo{title}{Electronic structure and related properties of
  metallic glasses: Linear muffin-tin orbital approach},
\newblock \bibinfo{journal}{Metall. Mater. Trans. A} \bibinfo{volume}{29}
  (\bibinfo{year}{1998}) \bibinfo{pages}{1853}. \URLprefix
  \url{https://doi.org/10.1007/s11661-998-0010-8}.
  \DOIprefix\doi{10.1007/s11661-998-0010-8}.
%Type = Article
\bibitem[{Bose(1999)}]{Bose_1999}
\bibinfo{author}{S.~K. Bose},
\newblock \bibinfo{title}{Electronic structure of liquid mercury},
\newblock \bibinfo{journal}{J. Phys. Condens. Matter} \bibinfo{volume}{11}
  (\bibinfo{year}{1999}) \bibinfo{pages}{4597--4615}. \URLprefix
  \url{https://doi.org/10.1088\%2F0953-8984\%2F11\%2F24\%2F303}.
  \DOIprefix\doi{10.1088/0953-8984/11/24/303}.
%Type = Article
\bibitem[{Pixley et~al.(2015)Pixley, Goswami, and Das~Sarma}]{PixleyPRL2015}
\bibinfo{author}{J.~H. Pixley}, \bibinfo{author}{P.~Goswami},
  \bibinfo{author}{S.~Das~Sarma},
\newblock \bibinfo{title}{Anderson localization and the quantum phase diagram
  of three dimensional disordered dirac semimetals},
\newblock \bibinfo{journal}{Phys. Rev. Lett.} \bibinfo{volume}{115}
  (\bibinfo{year}{2015}) \bibinfo{pages}{076601}.
  \DOIprefix\doi{10.1103/PhysRevLett.115.076601}.
%Type = Article
\bibitem[{Pixley et~al.(2016)Pixley, Goswami, and Das~Sarma}]{DasSarma20162}
\bibinfo{author}{J.~H. Pixley}, \bibinfo{author}{P.~Goswami},
  \bibinfo{author}{S.~Das~Sarma},
\newblock \bibinfo{title}{Disorder-driven itinerant quantum criticality of
  three-dimensional massless dirac fermions},
\newblock \bibinfo{journal}{Phys. Rev. B} \bibinfo{volume}{93}
  (\bibinfo{year}{2016}) \bibinfo{pages}{085103}. \URLprefix
  \url{https://link.aps.org/doi/10.1103/PhysRevB.93.085103}.
  \DOIprefix\doi{10.1103/PhysRevB.93.085103}.
%Type = Article
\bibitem[{Pixley et~al.(2017)Pixley, Chou, Goswami, Huse, Nandkishore,
  Radzihovsky, and Das~Sarma}]{DasSarma2017}
\bibinfo{author}{J.~H. Pixley}, \bibinfo{author}{Y.-Z. Chou},
  \bibinfo{author}{P.~Goswami}, \bibinfo{author}{D.~A. Huse},
  \bibinfo{author}{R.~Nandkishore}, \bibinfo{author}{L.~Radzihovsky},
  \bibinfo{author}{S.~Das~Sarma},
\newblock \bibinfo{title}{Single-particle excitations in disordered weyl
  fluids},
\newblock \bibinfo{journal}{Phys. Rev. B} \bibinfo{volume}{95}
  (\bibinfo{year}{2017}) \bibinfo{pages}{235101}. \URLprefix
  \url{https://link.aps.org/doi/10.1103/PhysRevB.95.235101}.
  \DOIprefix\doi{10.1103/PhysRevB.95.235101}.
%Type = Article
\bibitem[{Wilson et~al.(2017)Wilson, Pixley, Goswami, and
  Das~Sarma}]{DasSarma20172}
\bibinfo{author}{J.~H. Wilson}, \bibinfo{author}{J.~H. Pixley},
  \bibinfo{author}{P.~Goswami}, \bibinfo{author}{S.~Das~Sarma},
\newblock \bibinfo{title}{Quantum phases of disordered three-dimensional
  majorana-weyl fermions},
\newblock \bibinfo{journal}{Phys. Rev. B} \bibinfo{volume}{95}
  (\bibinfo{year}{2017}) \bibinfo{pages}{155122}. \URLprefix
  \url{https://link.aps.org/doi/10.1103/PhysRevB.95.155122}.
  \DOIprefix\doi{10.1103/PhysRevB.95.155122}.
%Type = Article
\bibitem[{Pixley et~al.(2018)Pixley, Wilson, Huse, and
  Gopalakrishnan}]{DasSarma2018}
\bibinfo{author}{J.~H. Pixley}, \bibinfo{author}{J.~H. Wilson},
  \bibinfo{author}{D.~A. Huse}, \bibinfo{author}{S.~Gopalakrishnan},
\newblock \bibinfo{title}{Weyl semimetal to metal phase transitions driven by
  quasiperiodic potentials},
\newblock \bibinfo{journal}{Phys. Rev. Lett.} \bibinfo{volume}{120}
  (\bibinfo{year}{2018}) \bibinfo{pages}{207604}. \URLprefix
  \url{https://link.aps.org/doi/10.1103/PhysRevLett.120.207604}.
  \DOIprefix\doi{10.1103/PhysRevLett.120.207604}.
%Type = Article
\bibitem[{Giustino(2017)}]{giustino2017rmp}
\bibinfo{author}{F.~Giustino},
\newblock \bibinfo{title}{{Electron-phonon interactions from first
  principles}},
\newblock \bibinfo{journal}{Rev. Mod. Phys.} \bibinfo{volume}{89}
  (\bibinfo{year}{2017}) \bibinfo{pages}{015003}. \URLprefix
  \url{https://link.aps.org/doi/10.1103/RevModPhys.89.015003}.
  \DOIprefix\doi{10.1103/RevModPhys.89.015003}.
%Type = Article
\bibitem[{Bardeen et~al.(1957)Bardeen, Cooper, and Schrieffer}]{bardeen1958pr}
\bibinfo{author}{J.~Bardeen}, \bibinfo{author}{L.~N. Cooper},
  \bibinfo{author}{J.~R. Schrieffer},
\newblock \bibinfo{title}{{Theory of Superconductivity}},
\newblock \bibinfo{journal}{Phys. Rev.} \bibinfo{volume}{108}
  (\bibinfo{year}{1957}) \bibinfo{pages}{1175--1204}. \URLprefix
  \url{https://link.aps.org/doi/10.1103/PhysRev.108.1175}.
  \DOIprefix\doi{10.1103/PhysRev.108.1175}.
%Type = Article
\bibitem[{Frederiksen et~al.(2007)Frederiksen, Paulsson, Brandbyge, and
  Jauho}]{frederiksen2007prb}
\bibinfo{author}{T.~Frederiksen}, \bibinfo{author}{M.~Paulsson},
  \bibinfo{author}{M.~Brandbyge}, \bibinfo{author}{A.-P. Jauho},
\newblock \bibinfo{title}{{Inelastic transport theory from first principles:
  Methodology and application to nanoscale devices}},
\newblock \bibinfo{journal}{Phys. Rev. B} \bibinfo{volume}{75}
  (\bibinfo{year}{2007}) \bibinfo{pages}{205413}. \URLprefix
  \url{https://link.aps.org/doi/10.1103/PhysRevB.75.205413}.
  \DOIprefix\doi{10.1103/PhysRevB.75.205413}.
%Type = Article
\bibitem[{Luisier and Klimeck(2009)}]{luisier2009prb}
\bibinfo{author}{M.~Luisier}, \bibinfo{author}{G.~Klimeck},
\newblock \bibinfo{title}{{Atomistic full-band simulations of silicon nanowire
  transistors: Effects of electron-phonon scattering}},
\newblock \bibinfo{journal}{Phys. Rev. B} \bibinfo{volume}{80}
  (\bibinfo{year}{2009}) \bibinfo{pages}{155430}. \URLprefix
  \url{https://link.aps.org/doi/10.1103/PhysRevB.80.155430}.
  \DOIprefix\doi{10.1103/PhysRevB.80.155430}.
%Type = Article
\bibitem[{Rhyner and Luisier(2014)}]{rhyner2014prb}
\bibinfo{author}{R.~Rhyner}, \bibinfo{author}{M.~Luisier},
\newblock \bibinfo{title}{{Atomistic modeling of coupled electron-phonon
  transport in nanowire transistors}},
\newblock \bibinfo{journal}{Phys. Rev. B} \bibinfo{volume}{89}
  (\bibinfo{year}{2014}) \bibinfo{pages}{235311}.
  \DOIprefix\doi{10.1103/PhysRevB.89.235311}.
%Type = Book
\bibitem[{Harrison(1989)}]{harrison1989book}
\bibinfo{author}{W.~A. Harrison}, \bibinfo{title}{{Electronic Structure and the
  Properties of Solids: The Physics of the Chemical Bond}},
  \bibinfo{publisher}{Dover Publications}, \bibinfo{year}{1989}.
%Type = Article
\bibitem[{Porezag et~al.(1995)Porezag, Frauenheim, K\"ohler, Seifert, and
  Kaschner}]{porezag1995prb}
\bibinfo{author}{D.~Porezag}, \bibinfo{author}{T.~Frauenheim},
  \bibinfo{author}{T.~K\"ohler}, \bibinfo{author}{G.~Seifert},
  \bibinfo{author}{R.~Kaschner},
\newblock \bibinfo{title}{{Construction of tight-binding-like potentials on the
  basis of density-functional theory: Application to carbon}},
\newblock \bibinfo{journal}{Phys. Rev. B} \bibinfo{volume}{51}
  (\bibinfo{year}{1995}) \bibinfo{pages}{12947--12957}. \URLprefix
  \url{https://link.aps.org/doi/10.1103/PhysRevB.51.12947}.
  \DOIprefix\doi{10.1103/PhysRevB.51.12947}.
%Type = Article
\bibitem[{Roche et~al.(2005)Roche, Jiang, Triozon, and Saito}]{roche2005prb}
\bibinfo{author}{S.~Roche}, \bibinfo{author}{J.~Jiang},
  \bibinfo{author}{F.~Triozon}, \bibinfo{author}{R.~Saito},
\newblock \bibinfo{title}{{Conductance and coherence lengths in disordered
  carbon nanotubes: Role of lattice defects and phonon vibrations}},
\newblock \bibinfo{journal}{Phys. Rev. B} \bibinfo{volume}{72}
  (\bibinfo{year}{2005}) \bibinfo{pages}{113410}. \URLprefix
  \url{https://link.aps.org/doi/10.1103/PhysRevB.72.113410}.
  \DOIprefix\doi{10.1103/PhysRevB.72.113410}.
%Type = Article
\bibitem[{Troisi and Orlandi(2006)}]{troisi2006prl}
\bibinfo{author}{A.~Troisi}, \bibinfo{author}{G.~Orlandi},
\newblock \bibinfo{title}{{Charge-Transport Regime of Crystalline Organic
  Semiconductors: Diffusion Limited by Thermal Off-Diagonal Electronic
  Disorder}},
\newblock \bibinfo{journal}{Phys. Rev. Lett.} \bibinfo{volume}{96}
  (\bibinfo{year}{2006}) \bibinfo{pages}{086601}. \URLprefix
  \url{https://link.aps.org/doi/10.1103/PhysRevLett.96.086601}.
  \DOIprefix\doi{10.1103/PhysRevLett.96.086601}.
%Type = Article
\bibitem[{Ishii et~al.(2009)Ishii, Triozon, Kobayashi, Hirose, and
  Roche}]{ishii2009crp}
\bibinfo{author}{H.~Ishii}, \bibinfo{author}{F.~Triozon},
  \bibinfo{author}{N.~Kobayashi}, \bibinfo{author}{K.~Hirose},
  \bibinfo{author}{S.~Roche},
\newblock \bibinfo{title}{{Charge transport in carbon nanotubes based
  materials: a Kubo-Greenwood computational approach}},
\newblock \bibinfo{journal}{Compt. Rendus Phys.} \bibinfo{volume}{10}
  (\bibinfo{year}{2009}) \bibinfo{pages}{283 -- 296}. \URLprefix
  \url{http://www.sciencedirect.com/science/article/pii/S1631070509000711}.
  \DOIprefix\doi{10.1016/j.crhy.2009.04.003}.
%Type = Article
\bibitem[{Ishii et~al.(2010)Ishii, Kobayashi, and Hirose}]{ishii2010prb}
\bibinfo{author}{H.~Ishii}, \bibinfo{author}{N.~Kobayashi},
  \bibinfo{author}{K.~Hirose},
\newblock \bibinfo{title}{{Order-$N$ electron transport calculations from
  ballistic to diffusive regimes by a time-dependent wave-packet diffusion
  method: Application to transport properties of carbon nanotubes}},
\newblock \bibinfo{journal}{Phys. Rev. B} \bibinfo{volume}{82}
  (\bibinfo{year}{2010}) \bibinfo{pages}{085435}. \URLprefix
  \url{https://link.aps.org/doi/10.1103/PhysRevB.82.085435}.
  \DOIprefix\doi{10.1103/PhysRevB.82.085435}.
%Type = Article
\bibitem[{Roche et~al.(2007)Roche, Jiang, Torres, and Saito}]{roche2007jpcm}
\bibinfo{author}{S.~Roche}, \bibinfo{author}{J.~Jiang}, \bibinfo{author}{L.~E.
  F.~F. Torres}, \bibinfo{author}{R.~Saito},
\newblock \bibinfo{title}{{Charge transport in carbon nanotubes: quantum
  effects of electron-phonon coupling}},
\newblock \bibinfo{journal}{J. Phys. Condens. Matter} \bibinfo{volume}{19}
  (\bibinfo{year}{2007}) \bibinfo{pages}{183203}. \URLprefix
  \url{http://stacks.iop.org/0953-8984/19/i=18/a=183203}.
%Type = Article
\bibitem[{Fan et~al.(2017)Fan, Uppstu, and Harju}]{fan2017tdm}
\bibinfo{author}{Z.~Fan}, \bibinfo{author}{A.~Uppstu},
  \bibinfo{author}{A.~Harju},
\newblock \bibinfo{title}{{Dominant source of disorder in graphene: charged
  impurities or ripples?}},
\newblock \bibinfo{journal}{2D Mater.} \bibinfo{volume}{4}
  (\bibinfo{year}{2017}) \bibinfo{pages}{025004}. \URLprefix
  \url{http://stacks.iop.org/2053-1583/4/i=2/a=025004}.
%Type = Article
\bibitem[{Li and Das~Sarma(2013)}]{li2013prb}
\bibinfo{author}{Q.~Li}, \bibinfo{author}{S.~Das~Sarma},
\newblock \bibinfo{title}{{Finite temperature inelastic mean free path and
  quasiparticle lifetime in graphene}},
\newblock \bibinfo{journal}{Phys. Rev. B} \bibinfo{volume}{87}
  (\bibinfo{year}{2013}) \bibinfo{pages}{085406}. \URLprefix
  \url{https://link.aps.org/doi/10.1103/PhysRevB.87.085406}.
  \DOIprefix\doi{10.1103/PhysRevB.87.085406}.
%Type = Article
\bibitem[{Liu et~al.(2015)Liu, Yuan, Wesselink, Starikov, van Schilfgaarde, and
  Kelly}]{liu2015prb}
\bibinfo{author}{Y.~Liu}, \bibinfo{author}{Z.~Yuan}, \bibinfo{author}{R.~J.~H.
  Wesselink}, \bibinfo{author}{A.~A. Starikov}, \bibinfo{author}{M.~van
  Schilfgaarde}, \bibinfo{author}{P.~J. Kelly},
\newblock \bibinfo{title}{{Direct method for calculating temperature-dependent
  transport properties}},
\newblock \bibinfo{journal}{Phys. Rev. B} \bibinfo{volume}{91}
  (\bibinfo{year}{2015}) \bibinfo{pages}{220405}.
  \DOIprefix\doi{10.1103/PhysRevB.91.220405}.
%Type = Article
\bibitem[{Gunst et~al.(2017)Gunst, Markussen, Palsgaard, Stokbro, and
  Brandbyge}]{gunst2017prb}
\bibinfo{author}{T.~Gunst}, \bibinfo{author}{T.~Markussen},
  \bibinfo{author}{M.~L.~N. Palsgaard}, \bibinfo{author}{K.~Stokbro},
  \bibinfo{author}{M.~Brandbyge},
\newblock \bibinfo{title}{{First-principles electron transport with phonon
  coupling: Large scale at low cost}},
\newblock \bibinfo{journal}{Phys. Rev. B} \bibinfo{volume}{96}
  (\bibinfo{year}{2017}) \bibinfo{pages}{161404}. \URLprefix
  \url{https://link.aps.org/doi/10.1103/PhysRevB.96.161404}.
  \DOIprefix\doi{10.1103/PhysRevB.96.161404}.
%Type = Article
\bibitem[{Park et~al.(2004)Park, Rosenblatt, Yaish, Sazonova, Ustunel, Braig,
  Arias, Brouwer, and McEuen}]{park2004nl}
\bibinfo{author}{J.-Y. Park}, \bibinfo{author}{S.~Rosenblatt},
  \bibinfo{author}{Y.~Yaish}, \bibinfo{author}{V.~Sazonova},
  \bibinfo{author}{H.~Ustunel}, \bibinfo{author}{S.~Braig},
  \bibinfo{author}{T.~A. Arias}, \bibinfo{author}{P.~W. Brouwer},
  \bibinfo{author}{P.~L. McEuen},
\newblock \bibinfo{title}{{Electron-Phonon Scattering in Metallic Single-Walled
  Carbon Nanotubes}},
\newblock \bibinfo{journal}{Nano Lett.} \bibinfo{volume}{4}
  (\bibinfo{year}{2004}) \bibinfo{pages}{517--520}. \URLprefix
  \url{https://doi.org/10.1021/nl035258c}. \DOIprefix\doi{10.1021/nl035258c}.
%Type = Article
\bibitem[{Gershenson et~al.(2006)Gershenson, Podzorov, and
  Morpurgo}]{PodzorovRMP2006}
\bibinfo{author}{M.~E. Gershenson}, \bibinfo{author}{V.~Podzorov},
  \bibinfo{author}{A.~F. Morpurgo},
\newblock \bibinfo{title}{{Colloquium: Electronic transport in single-crystal
  organic transistors}},
\newblock \bibinfo{journal}{Rev. Mod. Phys.} \bibinfo{volume}{78}
  (\bibinfo{year}{2006}) \bibinfo{pages}{973--989}. \URLprefix
  \url{https://link.aps.org/doi/10.1103/RevModPhys.78.973}.
  \DOIprefix\doi{10.1103/RevModPhys.78.973}.
%Type = Article
\bibitem[{Ortmann et~al.(2009)Ortmann, Bechstedt, and
  Hannewald}]{ortmann2009prb}
\bibinfo{author}{F.~Ortmann}, \bibinfo{author}{F.~Bechstedt},
  \bibinfo{author}{K.~Hannewald},
\newblock \bibinfo{title}{{Theory of charge transport in organic crystals:
  Beyond Holstein's small-polaron model}},
\newblock \bibinfo{journal}{Phys. Rev. B} \bibinfo{volume}{79}
  (\bibinfo{year}{2009}) \bibinfo{pages}{235206}. \URLprefix
  \url{https://link.aps.org/doi/10.1103/PhysRevB.79.235206}.
  \DOIprefix\doi{10.1103/PhysRevB.79.235206}.
%Type = Article
\bibitem[{Ciuchi et~al.(2011)Ciuchi, Fratini, and Mayou}]{ciuchi2011prb}
\bibinfo{author}{S.~Ciuchi}, \bibinfo{author}{S.~Fratini},
  \bibinfo{author}{D.~Mayou},
\newblock \bibinfo{title}{{Transient localization in crystalline organic
  semiconductors}},
\newblock \bibinfo{journal}{Phys. Rev. B} \bibinfo{volume}{83}
  (\bibinfo{year}{2011}) \bibinfo{pages}{081202}. \URLprefix
  \url{https://link.aps.org/doi/10.1103/PhysRevB.83.081202}.
  \DOIprefix\doi{10.1103/PhysRevB.83.081202}.
%Type = Article
\bibitem[{Ishii et~al.(2012)Ishii, Honma, Kobayashi, and Hirose}]{ishii2012prb}
\bibinfo{author}{H.~Ishii}, \bibinfo{author}{K.~Honma},
  \bibinfo{author}{N.~Kobayashi}, \bibinfo{author}{K.~Hirose},
\newblock \bibinfo{title}{{Wave-packet approach to transport properties of
  carrier coupled with intermolecular and intramolecular vibrations of organic
  semiconductors}},
\newblock \bibinfo{journal}{Phys. Rev. B} \bibinfo{volume}{85}
  (\bibinfo{year}{2012}) \bibinfo{pages}{245206}. \URLprefix
  \url{https://link.aps.org/doi/10.1103/PhysRevB.85.245206}.
  \DOIprefix\doi{10.1103/PhysRevB.85.245206}.
%Type = Article
\bibitem[{Holstein(1959)}]{Holstein1959}
\bibinfo{author}{T.~Holstein},
\newblock \bibinfo{title}{{Studies of polaron motion: Part II. The "small"
  polaron}},
\newblock \bibinfo{journal}{Ann. Phys.} \bibinfo{volume}{8}
  (\bibinfo{year}{1959}) \bibinfo{pages}{343--389}. \URLprefix
  \url{http://www.sciencedirect.com/science/article/pii/000349165990003X}.
%Type = Article
\bibitem[{Hannewald and Bobbert(2004)}]{HannewaldBobbertPRB2004}
\bibinfo{author}{K.~Hannewald}, \bibinfo{author}{P.~A. Bobbert},
\newblock \bibinfo{title}{{Anisotropy effects in phonon-assisted charge-carrier
  transport in organic molecular crystals}},
\newblock \bibinfo{journal}{Phys. Rev. B} \bibinfo{volume}{69}
  (\bibinfo{year}{2004}) \bibinfo{pages}{075212}. \URLprefix
  \url{https://link.aps.org/doi/10.1103/PhysRevB.69.075212}.
  \DOIprefix\doi{10.1103/PhysRevB.69.075212}.
%Type = Article
\bibitem[{Cheng and Silbey(2008)}]{Silbey2008}
\bibinfo{author}{Y.-C. Cheng}, \bibinfo{author}{R.~J. Silbey},
\newblock \bibinfo{title}{{A unified theory for charge-carrier transport in
  organic crystals}},
\newblock \bibinfo{journal}{J. Chem. Phys.} \bibinfo{volume}{128}
  (\bibinfo{year}{2008}) \bibinfo{pages}{114713}. \URLprefix
  \url{https://doi.org/10.1063/1.2894840}. \DOIprefix\doi{10.1063/1.2894840}.
%Type = Article
\bibitem[{Sundar et~al.(2004)Sundar, Zaumseil, Podzorov, Menard, Willett,
  Someya, Gershenson, and Rogers}]{SundarScience2004}
\bibinfo{author}{V.~C. Sundar}, \bibinfo{author}{J.~Zaumseil},
  \bibinfo{author}{V.~Podzorov}, \bibinfo{author}{E.~Menard},
  \bibinfo{author}{R.~L. Willett}, \bibinfo{author}{T.~Someya},
  \bibinfo{author}{M.~E. Gershenson}, \bibinfo{author}{J.~A. Rogers},
\newblock \bibinfo{title}{{Elastomeric Transistor Stamps: Reversible Probing of
  Charge Transport in Organic Crystals}},
\newblock \bibinfo{journal}{Science} \bibinfo{volume}{303}
  (\bibinfo{year}{2004}) \bibinfo{pages}{1644--1646}. \URLprefix
  \url{http://science.sciencemag.org/content/303/5664/1644}.
  \DOIprefix\doi{10.1126/science.1094196}.
%Type = Article
\bibitem[{Podzorov et~al.(2004)Podzorov, Menard, Borissov, Kiryukhin, Rogers,
  and Gershenson}]{PodzorovPRL2004}
\bibinfo{author}{V.~Podzorov}, \bibinfo{author}{E.~Menard},
  \bibinfo{author}{A.~Borissov}, \bibinfo{author}{V.~Kiryukhin},
  \bibinfo{author}{J.~A. Rogers}, \bibinfo{author}{M.~E. Gershenson},
\newblock \bibinfo{title}{{Intrinsic Charge Transport on the Surface of Organic
  Semiconductors}},
\newblock \bibinfo{journal}{Phys. Rev. Lett.} \bibinfo{volume}{93}
  (\bibinfo{year}{2004}) \bibinfo{pages}{086602}. \URLprefix
  \url{https://link.aps.org/doi/10.1103/PhysRevLett.93.086602}.
  \DOIprefix\doi{10.1103/PhysRevLett.93.086602}.
%Type = Article
\bibitem[{Troisi(2007)}]{TroisiAdvMat2007}
\bibinfo{author}{A.~Troisi},
\newblock \bibinfo{title}{{Prediction of the Absolute Charge Mobility of
  Molecular Semiconductors: the Case of Rubrene}},
\newblock \bibinfo{journal}{Adv. Mater.} \bibinfo{volume}{19}
  (\bibinfo{year}{2007}) \bibinfo{pages}{2000--2004}. \URLprefix
  \url{https://doi.org/10.1002/adma.200700550}.
  \DOIprefix\doi{10.1002/adma.200700550}.
%Type = Article
\bibitem[{Machida et~al.(2010)Machida, Nakayama, Duhm, Xin, Funakoshi, Ogawa,
  Kera, Ueno, and Ishii}]{KeraPRL2010}
\bibinfo{author}{S.-i. Machida}, \bibinfo{author}{Y.~Nakayama},
  \bibinfo{author}{S.~Duhm}, \bibinfo{author}{Q.~Xin},
  \bibinfo{author}{A.~Funakoshi}, \bibinfo{author}{N.~Ogawa},
  \bibinfo{author}{S.~Kera}, \bibinfo{author}{N.~Ueno},
  \bibinfo{author}{H.~Ishii},
\newblock \bibinfo{title}{{Highest-Occupied-Molecular-Orbital Band Dispersion
  of Rubrene Single Crystals as Observed by Angle-Resolved Ultraviolet
  Photoelectron Spectroscopy}},
\newblock \bibinfo{journal}{Phys. Rev. Lett.} \bibinfo{volume}{104}
  (\bibinfo{year}{2010}) \bibinfo{pages}{156401}. \URLprefix
  \url{https://link.aps.org/doi/10.1103/PhysRevLett.104.156401}.
  \DOIprefix\doi{10.1103/PhysRevLett.104.156401}.
%Type = Article
\bibitem[{Girlando et~al.(2010)Girlando, Grisanti, Masino, Bilotti, Brillante,
  Della~Valle, and Venuti}]{MasinoPRB2010}
\bibinfo{author}{A.~Girlando}, \bibinfo{author}{L.~Grisanti},
  \bibinfo{author}{M.~Masino}, \bibinfo{author}{I.~Bilotti},
  \bibinfo{author}{A.~Brillante}, \bibinfo{author}{R.~G. Della~Valle},
  \bibinfo{author}{E.~Venuti},
\newblock \bibinfo{title}{{Peierls and Holstein carrier-phonon coupling in
  crystalline rubrene}},
\newblock \bibinfo{journal}{Phys. Rev. B} \bibinfo{volume}{82}
  (\bibinfo{year}{2010}) \bibinfo{pages}{035208}. \URLprefix
  \url{https://link.aps.org/doi/10.1103/PhysRevB.82.035208}.
  \DOIprefix\doi{10.1103/PhysRevB.82.035208}.
%Type = Article
\bibitem[{Ordej\'on et~al.(2017)Ordej\'on, Boskovic, Panhans, and
  Ortmann}]{OrdejonPRB2017}
\bibinfo{author}{P.~Ordej\'on}, \bibinfo{author}{D.~Boskovic},
  \bibinfo{author}{M.~Panhans}, \bibinfo{author}{F.~Ortmann},
\newblock \bibinfo{title}{{Ab initio study of electron-phonon coupling in
  rubrene}},
\newblock \bibinfo{journal}{Phys. Rev. B} \bibinfo{volume}{96}
  (\bibinfo{year}{2017}) \bibinfo{pages}{035202}.
  \DOIprefix\doi{10.1103/PhysRevB.96.035202}.
%Type = Article
\bibitem[{Panhans et~al.(2020)Panhans, Hutsch, Benduhn, Schellhammer, Nikolis,
  Vangerven, Vandewal, and Ortmann}]{Panhans2020NatCommun}
\bibinfo{author}{M.~Panhans}, \bibinfo{author}{S.~Hutsch},
  \bibinfo{author}{J.~Benduhn}, \bibinfo{author}{K.~S. Schellhammer},
  \bibinfo{author}{V.~C. Nikolis}, \bibinfo{author}{T.~Vangerven},
  \bibinfo{author}{K.~Vandewal}, \bibinfo{author}{F.~Ortmann},
\newblock \bibinfo{title}{Molecular vibrations reduce the maximum achievable
  photovoltage in organic solar cells},
\newblock \bibinfo{journal}{Nature Communications} \bibinfo{volume}{11}
  (\bibinfo{year}{2020}) \bibinfo{pages}{1488}. \URLprefix
  \url{https://doi.org/10.1038/s41467-020-15215-x}.
  \DOIprefix\doi{10.1038/s41467-020-15215-x}.
%Type = Article
\bibitem[{Lacroix et~al.(2020)Lacroix, de~Laissardi\`ere, Qu\'emerais, Julien,
  and Mayou}]{Mayou2020PRL}
\bibinfo{author}{A.~Lacroix}, \bibinfo{author}{G.~T. de~Laissardi\`ere},
  \bibinfo{author}{P.~Qu\'emerais}, \bibinfo{author}{J.-P. Julien},
  \bibinfo{author}{D.~Mayou},
\newblock \bibinfo{title}{Modeling of electronic mobilities in halide
  perovskites: Adiabatic quantum localization scenario},
\newblock \bibinfo{journal}{Phys. Rev. Lett.} \bibinfo{volume}{124}
  (\bibinfo{year}{2020}) \bibinfo{pages}{196601}. \URLprefix
  \url{https://link.aps.org/doi/10.1103/PhysRevLett.124.196601}.
  \DOIprefix\doi{10.1103/PhysRevLett.124.196601}.
%Type = Article
\bibitem[{Thouless et~al.(1982)Thouless, Kohmoto, Nightingale, and den
  Nijs}]{ThoulessPRL1982}
\bibinfo{author}{D.~J. Thouless}, \bibinfo{author}{M.~Kohmoto},
  \bibinfo{author}{M.~P. Nightingale}, \bibinfo{author}{M.~den Nijs},
\newblock \bibinfo{title}{{Quantized Hall Conductance in a Two-Dimensional
  Periodic Potential}},
\newblock \bibinfo{journal}{Phys. Rev. Lett.} \bibinfo{volume}{49}
  (\bibinfo{year}{1982}) \bibinfo{pages}{405--408}. \URLprefix
  \url{https://link.aps.org/doi/10.1103/PhysRevLett.49.405}.
  \DOIprefix\doi{10.1103/PhysRevLett.49.405}.
%Type = Article
\bibitem[{Cr\'epieux and Bruno(2001)}]{crepieux2001prb}
\bibinfo{author}{A.~Cr\'epieux}, \bibinfo{author}{P.~Bruno},
\newblock \bibinfo{title}{{Theory of the anomalous Hall effect from the Kubo
  formula and the Dirac equation}},
\newblock \bibinfo{journal}{Phys. Rev. B} \bibinfo{volume}{64}
  (\bibinfo{year}{2001}) \bibinfo{pages}{014416}. \URLprefix
  \url{https://link.aps.org/doi/10.1103/PhysRevB.64.014416}.
  \DOIprefix\doi{10.1103/PhysRevB.64.014416}.
%Type = Article
\bibitem[{Bonbien(????)}]{Bonbien202Arxvi}
\bibinfo{author}{V.~Bonbien},
\newblock \bibinfo{title}{{A new decomposition of the Kubo-Bastin formula}}
  (????). \href{http://arxiv.org/abs/arXiv:2005.04678v1}{{\tt
  arXiv:arXiv:2005.04678v1}}.
%Type = Article
\bibitem[{Kane and Mele(2005)}]{KaneMele2005}
\bibinfo{author}{C.~L. Kane}, \bibinfo{author}{E.~J. Mele},
\newblock \bibinfo{title}{{Quantum Spin Hall Effect in Graphene}},
\newblock \bibinfo{journal}{Phys. Rev. Lett.} \bibinfo{volume}{95}
  (\bibinfo{year}{2005}) \bibinfo{pages}{226801}. \URLprefix
  \url{https://link.aps.org/doi/10.1103/PhysRevLett.95.226801}.
  \DOIprefix\doi{10.1103/PhysRevLett.95.226801}.
%Type = Article
\bibitem[{Aoki and Ando(1981)}]{aokisscomm1981}
\bibinfo{author}{H.~Aoki}, \bibinfo{author}{T.~Ando},
\newblock \bibinfo{title}{{Effect of localization on the hall conductivity in
  the two-dimensional system in strong magnetic fields}},
\newblock \bibinfo{journal}{Solid State Commun.} \bibinfo{volume}{38}
  (\bibinfo{year}{1981}) \bibinfo{pages}{1079 -- 1082}. \URLprefix
  \url{http://www.sciencedirect.com/science/article/pii/0038109881900211}.
  \DOIprefix\doi{10.1016/0038-1098(81)90021-1}.
%Type = Article
\bibitem[{Aoki(1985)}]{aoki1985prl}
\bibinfo{author}{H.~Aoki},
\newblock \bibinfo{title}{{Aharonov-Bohm Effect for the Quantum Hall
  Conductivity on a Disordered Lattice}},
\newblock \bibinfo{journal}{Phys. Rev. Lett.} \bibinfo{volume}{55}
  (\bibinfo{year}{1985}) \bibinfo{pages}{1136--1139}. \URLprefix
  \url{https://link.aps.org/doi/10.1103/PhysRevLett.55.1136}.
  \DOIprefix\doi{10.1103/PhysRevLett.55.1136}.
%Type = Article
\bibitem[{Luttinger(1951)}]{Luttinger1951PRL}
\bibinfo{author}{J.~M. Luttinger},
\newblock \bibinfo{title}{{The Effect of a Magnetic Field on Electrons in a
  Periodic Potential}},
\newblock \bibinfo{journal}{Phys. Rev.} \bibinfo{volume}{84}
  (\bibinfo{year}{1951}) \bibinfo{pages}{814--817}. \URLprefix
  \url{https://link.aps.org/doi/10.1103/PhysRev.84.814}.
  \DOIprefix\doi{10.1103/PhysRev.84.814}.
%Type = Article
\bibitem[{Novoselov et~al.(2005)Novoselov, Geim, Morozov, Jiang, Katsnelson,
  Grigorieva, Dubonos, and Firsov}]{Novoselov2005Nature}
\bibinfo{author}{K.~S. Novoselov}, \bibinfo{author}{A.~K. Geim},
  \bibinfo{author}{S.~V. Morozov}, \bibinfo{author}{D.~Jiang},
  \bibinfo{author}{M.~I. Katsnelson}, \bibinfo{author}{I.~V. Grigorieva},
  \bibinfo{author}{S.~V. Dubonos}, \bibinfo{author}{A.~A. Firsov},
\newblock \bibinfo{title}{{Two-dimensional gas of massless Dirac fermions in
  graphene}},
\newblock \bibinfo{journal}{Nature} \bibinfo{volume}{438}
  (\bibinfo{year}{2005}) \bibinfo{pages}{197--200}. \URLprefix
  \url{http://dx.doi.org/10.1038/nature04233}.
%Type = Article
\bibitem[{Zhang et~al.(2005)Zhang, Tan, Stormer, and Kim}]{Zhang2005}
\bibinfo{author}{Y.~Zhang}, \bibinfo{author}{Y.-W. Tan}, \bibinfo{author}{H.~L.
  Stormer}, \bibinfo{author}{P.~Kim},
\newblock \bibinfo{title}{{Experimental observation of the quantum Hall effect
  and {B}erry's phase in graphene}},
\newblock \bibinfo{journal}{Nature} \bibinfo{volume}{438}
  (\bibinfo{year}{2005}) \bibinfo{pages}{201--204}. \URLprefix
  \url{http://dx.doi.org/10.1038/nature04235}.
%Type = Article
\bibitem[{Sinitsyn et~al.(2006)Sinitsyn, Hill, Min, Sinova, and
  MacDonald}]{Sinitsyn2006}
\bibinfo{author}{N.~A. Sinitsyn}, \bibinfo{author}{J.~E. Hill},
  \bibinfo{author}{H.~Min}, \bibinfo{author}{J.~Sinova}, \bibinfo{author}{A.~H.
  MacDonald},
\newblock \bibinfo{title}{{Charge and spin hall conductivity in metallic
  graphene}},
\newblock \bibinfo{journal}{Phys. Rev. Lett.} \bibinfo{volume}{97}
  (\bibinfo{year}{2006}) \bibinfo{pages}{1--4}.
  \DOIprefix\doi{10.1103/PhysRevLett.97.106804}.
%Type = Article
\bibitem[{Levy et~al.(2010)Levy, Burke, Meaker, Panlasigui, Zettl, Guinea,
  Neto, and Crommie}]{Levy2010}
\bibinfo{author}{N.~Levy}, \bibinfo{author}{S.~A. Burke},
  \bibinfo{author}{K.~L. Meaker}, \bibinfo{author}{M.~Panlasigui},
  \bibinfo{author}{A.~Zettl}, \bibinfo{author}{F.~Guinea},
  \bibinfo{author}{A.~H.~C. Neto}, \bibinfo{author}{M.~F. Crommie},
\newblock \bibinfo{title}{{Strain-Induced Pseudo-Magnetic Fields Greater Than
  300 Tesla in Graphene Nanobubbles}},
\newblock \bibinfo{journal}{Science} \bibinfo{volume}{329}
  (\bibinfo{year}{2010}) \bibinfo{pages}{544--547}. \URLprefix
  \url{http://www.sciencemag.org/cgi/doi/10.1126/science.1191700}.
  \DOIprefix\doi{10.1126/science.1191700}.
%Type = Article
\bibitem[{Settnes et~al.(2017)Settnes, Garc\'{\i}a, and Roche}]{settnes2017tdm}
\bibinfo{author}{M.~Settnes}, \bibinfo{author}{J.~H. Garc\'{\i}a},
  \bibinfo{author}{S.~Roche},
\newblock \bibinfo{title}{{Valley-polarized quantum transport generated by
  gauge fields in graphene}},
\newblock \bibinfo{journal}{2D Mater.} \bibinfo{volume}{4}
  (\bibinfo{year}{2017}) \bibinfo{pages}{031006}. \URLprefix
  \url{http://stacks.iop.org/2053-1583/4/i=3/a=031006}.
%Type = Article
\bibitem[{Guinea et~al.(2010)Guinea, Katsnelson, and Geim}]{Guinea2010}
\bibinfo{author}{F.~Guinea}, \bibinfo{author}{M.~I. Katsnelson},
  \bibinfo{author}{A.~K. Geim},
\newblock \bibinfo{title}{{Energy gaps and a zero-field quantum Hall effect in
  graphene by strain engineering}},
\newblock \bibinfo{journal}{Nat. Phys.} \bibinfo{volume}{6}
  (\bibinfo{year}{2010}) \bibinfo{pages}{30--33}. \URLprefix
  \url{http://www.nature.com/articles/nphys1420}.
  \DOIprefix\doi{10.1038/nphys1420}.
%Type = Article
\bibitem[{Vozmediano et~al.(2010)Vozmediano, Katsnelson, and
  Guinea}]{Vozmediano2010}
\bibinfo{author}{M.~A. Vozmediano}, \bibinfo{author}{M.~I. Katsnelson},
  \bibinfo{author}{F.~Guinea},
\newblock \bibinfo{title}{{Gauge fields in graphene}},
\newblock \bibinfo{journal}{Phys. Rep.} \bibinfo{volume}{496}
  (\bibinfo{year}{2010}) \bibinfo{pages}{109--148}. \URLprefix
  \url{http://linkinghub.elsevier.com/retrieve/pii/S0370157310001729}.
  \DOIprefix\doi{10.1016/j.physrep.2010.07.003}.
%Type = Article
\bibitem[{Fujita et~al.(2011)Fujita, Jalil, Tan, and Murakami}]{Fujita2011}
\bibinfo{author}{T.~Fujita}, \bibinfo{author}{M.~B.~A. Jalil},
  \bibinfo{author}{S.~G. Tan}, \bibinfo{author}{S.~Murakami},
\newblock \bibinfo{title}{{Gauge fields in spintronics}},
\newblock \bibinfo{journal}{J. Appl. Phys.} \bibinfo{volume}{110}
  (\bibinfo{year}{2011}) \bibinfo{pages}{121301}. \URLprefix
  \url{http://aip.scitation.org/doi/10.1063/1.3665219}.
  \DOIprefix\doi{10.1063/1.3665219}.
%Type = Article
\bibitem[{Zhu et~al.(2015)Zhu, Stroscio, and Li}]{Zhu2015}
\bibinfo{author}{S.~Zhu}, \bibinfo{author}{J.~A. Stroscio},
  \bibinfo{author}{T.~Li},
\newblock \bibinfo{title}{{Programmable Extreme Pseudomagnetic Fields in
  Graphene by a Uniaxial Stretch}},
\newblock \bibinfo{journal}{Phys. Rev. Lett.} \bibinfo{volume}{115}
  (\bibinfo{year}{2015}) \bibinfo{pages}{245501}.
  \DOIprefix\doi{10.1103/PhysRevLett.115.245501}.
%Type = Article
\bibitem[{\ifmmode \check{Z}\else \v{Z}\fi{}uti\ifmmode~\acute{c}\else
  \'{c}\fi{} et~al.(2004)\ifmmode \check{Z}\else
  \v{Z}\fi{}uti\ifmmode~\acute{c}\else \'{c}\fi{}, Fabian, and
  Das~Sarma}]{ZuticRMP2004}
\bibinfo{author}{I.~\ifmmode \check{Z}\else
  \v{Z}\fi{}uti\ifmmode~\acute{c}\else \'{c}\fi{}},
  \bibinfo{author}{J.~Fabian}, \bibinfo{author}{S.~Das~Sarma},
\newblock \bibinfo{title}{Spintronics: Fundamentals and applications},
\newblock \bibinfo{journal}{Rev. Mod. Phys.} \bibinfo{volume}{76}
  (\bibinfo{year}{2004}) \bibinfo{pages}{323--410}.
  \DOIprefix\doi{10.1103/RevModPhys.76.323}.
%Type = Article
\bibitem[{Fabian et~al.(2007)Fabian, Matos-Abiague, Ertler, Stano, and
  Zutic}]{Fabian2007}
\bibinfo{author}{F.~Fabian}, \bibinfo{author}{A.~Matos-Abiague},
  \bibinfo{author}{C.~Ertler}, \bibinfo{author}{P.~Stano},
  \bibinfo{author}{I.~Zutic},
\newblock \bibinfo{title}{Semiconductor spintronics},
\newblock \bibinfo{journal}{Acta Phys. Slovaca} \bibinfo{volume}{57}
  (\bibinfo{year}{2007}) \bibinfo{pages}{565}. \URLprefix
  \url{http://www.physics.sk/aps/pub.php?y=2007\&pub=aps-07-04}.
%Type = Article
\bibitem[{Elliott(1954)}]{Elliott1954}
\bibinfo{author}{R.~J. Elliott},
\newblock \bibinfo{title}{{Theory of the effect of spin-orbit coupling on
  magnetic resonance in some semiconductors}},
\newblock \bibinfo{journal}{Phys. Rev.} \bibinfo{volume}{96}
  (\bibinfo{year}{1954}) \bibinfo{pages}{266--279}. \URLprefix
  \url{https://journals.aps.org/pr/pdf/10.1103/PhysRev.96.266}.
  \DOIprefix\doi{10.1103/PhysRev.96.266}.
%Type = Incollection
\bibitem[{Yafet(1963)}]{Yafet1963}
\bibinfo{author}{Y.~Yafet},
\newblock \bibinfo{title}{{g Factors and Spin-Lattice Relaxation of Conduction
  Electrons}},
\newblock in: \bibinfo{editor}{F.~Seitz}, \bibinfo{editor}{D.~Turnbull} (Eds.),
  \bibinfo{booktitle}{Solid State Physics}, volume~\bibinfo{volume}{14} of
  \textit{\bibinfo{series}{Solid State Physics}}, \bibinfo{publisher}{Academic
  Press}, \bibinfo{year}{1963}, pp. \bibinfo{pages}{1 -- 98}. \URLprefix
  \url{http://www.sciencedirect.com/science/article/pii/S0081194708602593}.
%Type = Article
\bibitem[{D'yakonov and Perel'(1971)}]{dyakonov1971spin}
\bibinfo{author}{M.~I. D'yakonov}, \bibinfo{author}{V.~I. Perel'},
\newblock \bibinfo{title}{{Spin Orientation of Electrons Associated with the
  Interband Absorption of Light in Semiconductors}},
\newblock \bibinfo{journal}{Zh. Eksp. Teor. Fiz.} \bibinfo{volume}{60}
  (\bibinfo{year}{1971}) \bibinfo{pages}{1954--1965}. \URLprefix
  \url{http://www.jetp.ac.ru/cgi-bin/e/index/e/33/5/p1053?a=list}.
%Type = Article
\bibitem[{Gridnev(2001)}]{Gridnev2001}
\bibinfo{author}{V.~N. Gridnev},
\newblock \bibinfo{title}{{Theory of Faraday rotation beats in quantum wells
  with large spin splitting}},
\newblock \bibinfo{journal}{JETP Lett.} \bibinfo{volume}{74}
  (\bibinfo{year}{2001}) \bibinfo{pages}{380--383}. \URLprefix
  \url{https://doi.org/10.1134/1.1427126}.
%Type = Article
\bibitem[{Cummings and Roche(2016)}]{cummings2016prl}
\bibinfo{author}{A.~W. Cummings}, \bibinfo{author}{S.~Roche},
\newblock \bibinfo{title}{{Effects of Dephasing on Spin Lifetime in Ballistic
  Spin-Orbit Materials}},
\newblock \bibinfo{journal}{Phys. Rev. Lett.} \bibinfo{volume}{116}
  (\bibinfo{year}{2016}) \bibinfo{pages}{086602}. \URLprefix
  \url{https://journals.aps.org/prl/pdf/10.1103/PhysRevLett.116.086602
  https://link.aps.org/doi/10.1103/PhysRevLett.116.086602}.
  \DOIprefix\doi{10.1103/PhysRevLett.116.086602}.
%Type = Article
\bibitem[{Van~Tuan et~al.(2016)Van~Tuan, Ortmann, Cummings, Soriano, and
  Roche}]{vantuan2016scirep}
\bibinfo{author}{D.~Van~Tuan}, \bibinfo{author}{F.~Ortmann},
  \bibinfo{author}{A.~W. Cummings}, \bibinfo{author}{D.~Soriano},
  \bibinfo{author}{S.~Roche},
\newblock \bibinfo{title}{{Spin dynamics and relaxation in graphene dictated by
  electron-hole puddles}},
\newblock \bibinfo{journal}{Sci. Rep.} \bibinfo{volume}{6}
  (\bibinfo{year}{2016}) \bibinfo{pages}{21046}. \URLprefix
  \url{http://dx.doi.org/10.1038/srep21046}.
%Type = Article
\bibitem[{Van~Tuan and Roche(2016)}]{vantuan2016prl}
\bibinfo{author}{D.~Van~Tuan}, \bibinfo{author}{S.~Roche},
\newblock \bibinfo{title}{{Spin Manipulation in Graphene by Chemically Induced
  Pseudospin Polarization}},
\newblock \bibinfo{journal}{Phys. Rev. Lett.} \bibinfo{volume}{116}
  (\bibinfo{year}{2016}) \bibinfo{pages}{106601}. \URLprefix
  \url{https://link.aps.org/doi/10.1103/PhysRevLett.116.106601}.
  \DOIprefix\doi{10.1103/PhysRevLett.116.106601}.
%Type = Article
\bibitem[{Vila et~al.(2020)Vila, Garcia, Cummings, Power, Groth, Waintal, and
  Roche}]{Vila2020}
\bibinfo{author}{M.~Vila}, \bibinfo{author}{J.~H. Garcia},
  \bibinfo{author}{A.~W. Cummings}, \bibinfo{author}{S.~R. Power},
  \bibinfo{author}{C.~W. Groth}, \bibinfo{author}{X.~Waintal},
  \bibinfo{author}{S.~Roche},
\newblock \bibinfo{title}{Nonlocal spin dynamics in the crossover from
  diffusive to ballistic transport},
\newblock \bibinfo{journal}{Phys. Rev. Lett.} \bibinfo{volume}{124}
  (\bibinfo{year}{2020}) \bibinfo{pages}{196602}. \URLprefix
  \url{https://link.aps.org/doi/10.1103/PhysRevLett.124.196602}.
  \DOIprefix\doi{10.1103/PhysRevLett.124.196602}.
%Type = Article
\bibitem[{Raes et~al.(2016)Raes, Scheerder, Costache, Bonell, Sierra, Cuppens,
  Van~de Vondel, and Valenzuela}]{Raes2016ncomm}
\bibinfo{author}{B.~Raes}, \bibinfo{author}{J.~E. Scheerder},
  \bibinfo{author}{M.~V. Costache}, \bibinfo{author}{F.~Bonell},
  \bibinfo{author}{J.~F. Sierra}, \bibinfo{author}{J.~Cuppens},
  \bibinfo{author}{J.~Van~de Vondel}, \bibinfo{author}{S.~O. Valenzuela},
\newblock \bibinfo{title}{{Determination of the spin-lifetime anisotropy in
  graphene using oblique spin precession}},
\newblock \bibinfo{journal}{Nat. Commun.} \bibinfo{volume}{7}
  (\bibinfo{year}{2016}) \bibinfo{pages}{11444}. \URLprefix
  \url{http://dx.doi.org/10.1038/ncomms11444}.
%Type = Article
\bibitem[{Offidani and Ferreira(2018)}]{Offidanispin2018}
\bibinfo{author}{M.~Offidani}, \bibinfo{author}{A.~Ferreira},
\newblock \bibinfo{title}{Microscopic theory of spin relaxation anisotropy in
  graphene with proximity-induced spin-orbit coupling},
\newblock \bibinfo{journal}{Phys. Rev. B} \bibinfo{volume}{98}
  (\bibinfo{year}{2018}) \bibinfo{pages}{245408}.
  \DOIprefix\doi{10.1103/PhysRevB.98.245408}.
%Type = Article
\bibitem[{Ben{\'i}tez et~al.(2018)Ben{\'i}tez, Sierra, Savero~Torres, Arrighi,
  Bonell, Costache, and Valenzuela}]{Benitez2018np}
\bibinfo{author}{L.~A. Ben{\'i}tez}, \bibinfo{author}{J.~F. Sierra},
  \bibinfo{author}{W.~Savero~Torres}, \bibinfo{author}{A.~Arrighi},
  \bibinfo{author}{F.~Bonell}, \bibinfo{author}{M.~V. Costache},
  \bibinfo{author}{S.~O. Valenzuela},
\newblock \bibinfo{title}{{Strongly anisotropic spin relaxation in
  graphene-transition metal dichalcogenide heterostructures at room
  temperature}},
\newblock \bibinfo{journal}{Nat. Phys.} \bibinfo{volume}{14}
  (\bibinfo{year}{2018}) \bibinfo{pages}{303--308}. \URLprefix
  \url{https://doi.org/10.1038/s41567-017-0019-2}.
  \DOIprefix\doi{10.1038/s41567-017-0019-2}.
%Type = Article
\bibitem[{Ghiasi et~al.(2017)Ghiasi, Ingla-Ayn{\'e}s, Kaverzin, and van
  Wees}]{ghiasi2017nl}
\bibinfo{author}{T.~S. Ghiasi}, \bibinfo{author}{J.~Ingla-Ayn{\'e}s},
  \bibinfo{author}{A.~A. Kaverzin}, \bibinfo{author}{B.~J. van Wees},
\newblock \bibinfo{title}{{Large Proximity-Induced Spin Lifetime Anisotropy in
  Transition-Metal Dichalcogenide/Graphene Heterostructures}},
\newblock \bibinfo{journal}{Nano Lett.} \bibinfo{volume}{17}
  (\bibinfo{year}{2017}) \bibinfo{pages}{7528--7532}. \URLprefix
  \url{https://doi.org/10.1021/acs.nanolett.7b03460}.
  \DOIprefix\doi{10.1021/acs.nanolett.7b03460}.
%Type = Article
\bibitem[{Cresti et~al.(2014)Cresti, Van~Tuan, Soriano, Cummings, and
  Roche}]{cresti2014prl}
\bibinfo{author}{A.~Cresti}, \bibinfo{author}{D.~Van~Tuan},
  \bibinfo{author}{D.~Soriano}, \bibinfo{author}{A.~W. Cummings},
  \bibinfo{author}{S.~Roche},
\newblock \bibinfo{title}{{Multiple Quantum Phases in Graphene with Enhanced
  Spin-Orbit Coupling: From the Quantum Spin Hall Regime to the Spin Hall
  Effect and a Robust Metallic State}},
\newblock \bibinfo{journal}{Phys. Rev. Lett.} \bibinfo{volume}{113}
  (\bibinfo{year}{2014}) \bibinfo{pages}{246603}. \URLprefix
  \url{https://link.aps.org/doi/10.1103/PhysRevLett.113.246603}.
  \DOIprefix\doi{10.1103/PhysRevLett.113.246603}.
%Type = Article
\bibitem[{Soriano et~al.(2015)Soriano, Tuan, Dubois, Gmitra, Cummings, Kochan,
  Ortmann, Charlier, Fabian, and Roche}]{soriano2015twodm}
\bibinfo{author}{D.~Soriano}, \bibinfo{author}{D.~V. Tuan},
  \bibinfo{author}{S.~M.-M. Dubois}, \bibinfo{author}{M.~Gmitra},
  \bibinfo{author}{A.~W. Cummings}, \bibinfo{author}{D.~Kochan},
  \bibinfo{author}{F.~Ortmann}, \bibinfo{author}{J.-C. Charlier},
  \bibinfo{author}{J.~Fabian}, \bibinfo{author}{S.~Roche},
\newblock \bibinfo{title}{{Spin transport in hydrogenated graphene}},
\newblock \bibinfo{journal}{2D Mater.} \bibinfo{volume}{2}
  (\bibinfo{year}{2015}) \bibinfo{pages}{022002}. \URLprefix
  \url{http://stacks.iop.org/2053-1583/2/i=2/a=022002}.
%Type = Article
\bibitem[{Hirsch(1999)}]{Hirsch1999}
\bibinfo{author}{J.~E. Hirsch},
\newblock \bibinfo{title}{Spin hall effect},
\newblock \bibinfo{journal}{Phys. Rev. Lett.} \bibinfo{volume}{83}
  (\bibinfo{year}{1999}) \bibinfo{pages}{1834--1837}.
  \DOIprefix\doi{10.1103/PhysRevLett.83.1834}.
%Type = Article
\bibitem[{Milletar{\`{i}} et~al.(2017)Milletar{\`{i}}, Offidani, Ferreira, and
  Raimondi}]{PRLMilletari2017}
\bibinfo{author}{M.~Milletar{\`{i}}}, \bibinfo{author}{M.~Offidani},
  \bibinfo{author}{A.~Ferreira}, \bibinfo{author}{R.~Raimondi},
\newblock \bibinfo{title}{{Covariant conservation laws and spin Hall effect in
  the Dirac-Rashba model}},
\newblock \bibinfo{journal}{Phys. Rev. Lett.} \bibinfo{volume}{119}
  (\bibinfo{year}{2017}) \bibinfo{pages}{246801}.
  \DOIprefix\doi{10.1103/PhysRevLett.119.246801}.
%Type = Article
\bibitem[{Sinova et~al.(2015)Sinova, Valenzuela, Wunderlich, Back, and
  Jungwirth}]{Sinova2015}
\bibinfo{author}{J.~Sinova}, \bibinfo{author}{S.~O. Valenzuela},
  \bibinfo{author}{J.~Wunderlich}, \bibinfo{author}{C.~H. Back},
  \bibinfo{author}{T.~Jungwirth},
\newblock \bibinfo{title}{Spin hall effects},
\newblock \bibinfo{journal}{Rev. Mod. Phys.} \bibinfo{volume}{87}
  (\bibinfo{year}{2015}) \bibinfo{pages}{1213--1260}.
  \DOIprefix\doi{10.1103/RevModPhys.87.1213}.
%Type = Article
\bibitem[{Mishchenko et~al.(2004)Mishchenko, Shytov, and
  Halperin}]{PRLMishchenko2004}
\bibinfo{author}{E.~G. Mishchenko}, \bibinfo{author}{A.~V. Shytov},
  \bibinfo{author}{B.~I. Halperin},
\newblock \bibinfo{title}{{Spin current and polarization in impure
  two-dimensional electron systems with spin-orbit coupling}},
\newblock \bibinfo{journal}{Phys. Rev. Lett.} \bibinfo{volume}{93}
  (\bibinfo{year}{2004}) \bibinfo{pages}{226602}.
  \DOIprefix\doi{10.1103/PhysRevLett.93.226602}.
%Type = Article
\bibitem[{Inoue et~al.(2004)Inoue, Bauer, and Molenkamp}]{PRBInoue2004}
\bibinfo{author}{J.~I. Inoue}, \bibinfo{author}{G.~E.~W. Bauer},
  \bibinfo{author}{L.~W. Molenkamp},
\newblock \bibinfo{title}{{Suppression of the persistent spin Hall current by
  defect scattering}},
\newblock \bibinfo{journal}{Phys. Rev. B} \bibinfo{volume}{70}
  (\bibinfo{year}{2004}) \bibinfo{pages}{041303}. \URLprefix
  \url{https://link.aps.org/doi/10.1103/PhysRevB.70.041303}.
  \DOIprefix\doi{10.1103/PhysRevB.70.041303}.
%Type = Article
\bibitem[{van~den Berg et~al.(2011)van~den Berg, Raymond, and Verga}]{Berg2011}
\bibinfo{author}{T.~L. van~den Berg}, \bibinfo{author}{L.~Raymond},
  \bibinfo{author}{A.~Verga},
\newblock \bibinfo{title}{Dynamical spin hall conductivity in a magnetic
  disordered system},
\newblock \bibinfo{journal}{Phys. Rev. B} \bibinfo{volume}{84}
  (\bibinfo{year}{2011}) \bibinfo{pages}{245210}.
  \DOIprefix\doi{10.1103/PhysRevB.84.245210}.
%Type = Article
\bibitem[{Van~Tuan et~al.(2016)Van~Tuan, Marmolejo-Tejada, Waintal,
  Nikoli\ifmmode~\acute{c}\else \'{c}\fi{}, Valenzuela, and
  Roche}]{DinhPRL2016c}
\bibinfo{author}{D.~Van~Tuan}, \bibinfo{author}{J.~M. Marmolejo-Tejada},
  \bibinfo{author}{X.~Waintal}, \bibinfo{author}{B.~K.
  Nikoli\ifmmode~\acute{c}\else \'{c}\fi{}}, \bibinfo{author}{S.~O.
  Valenzuela}, \bibinfo{author}{S.~Roche},
\newblock \bibinfo{title}{Spin hall effect and origins of nonlocal resistance
  in adatom-decorated graphene},
\newblock \bibinfo{journal}{Phys. Rev. Lett.} \bibinfo{volume}{117}
  (\bibinfo{year}{2016}) \bibinfo{pages}{176602}.
  \DOIprefix\doi{10.1103/PhysRevLett.117.176602}.
%Type = Article
\bibitem[{Garc\'{\i}a and Rappoport(2016)}]{garcia2016tdmat}
\bibinfo{author}{J.~H. Garc\'{\i}a}, \bibinfo{author}{T.~G. Rappoport},
\newblock \bibinfo{title}{{Kubo-Bastin approach for the spin Hall conductivity
  of decorated graphene}},
\newblock \bibinfo{journal}{2D Mater.} \bibinfo{volume}{3}
  (\bibinfo{year}{2016}) \bibinfo{pages}{024007}. \URLprefix
  \url{http://stacks.iop.org/2053-1583/3/i=2/a=024007?key=crossref.23a8bb0514d5e230cffb8f5a4d3d1e38}.
  \DOIprefix\doi{10.1088/2053-1583/3/2/024007}.
%Type = Article
\bibitem[{Savero~Torres et~al.(2020)Savero~Torres, Sierra, Benítez, Bonell,
  García, Roche, and Valenzuela}]{savero2020}
\bibinfo{author}{W.~Savero~Torres}, \bibinfo{author}{J.~Sierra},
  \bibinfo{author}{L.~Benítez}, \bibinfo{author}{F.~Bonell},
  \bibinfo{author}{J.~García}, \bibinfo{author}{S.~Roche},
  \bibinfo{author}{S.~Valenzuela},
\newblock \bibinfo{title}{Magnetism, spin dynamics, and quantum transport in
  two-dimensional systems},
\newblock \bibinfo{journal}{MRS Bulletin} \bibinfo{volume}{45}
  (\bibinfo{year}{2020}) \bibinfo{pages}{357--365}.
  \DOIprefix\doi{10.1557/mrs.2020.121}.
%Type = Article
\bibitem[{Garc\'{\i}a et~al.(2017)Garc\'{\i}a, Cummings, and
  Roche}]{garcia2017nanolet}
\bibinfo{author}{J.~H. Garc\'{\i}a}, \bibinfo{author}{A.~W. Cummings},
  \bibinfo{author}{S.~Roche},
\newblock \bibinfo{title}{{Spin hall effect and weak antilocalization in
  graphene/transition metal dichalcogenide heterostructures}},
\newblock \bibinfo{journal}{Nano Lett.} \bibinfo{volume}{17}
  (\bibinfo{year}{2017}) \bibinfo{pages}{5078--5083}.
  \DOIprefix\doi{10.1021/acs.nanolett.7b02364}.
%Type = Article
\bibitem[{Garc\'{\i}a et~al.(2018)Garc\'{\i}a, Vila, Cummings, and
  Roche}]{garcia2018csr}
\bibinfo{author}{J.~H. Garc\'{\i}a}, \bibinfo{author}{M.~Vila},
  \bibinfo{author}{A.~W. Cummings}, \bibinfo{author}{S.~Roche},
\newblock \bibinfo{title}{{Spin transport in graphene/transition metal
  dichalcogenide heterostructures}},
\newblock \bibinfo{journal}{Chem. Soc. Rev.} \bibinfo{volume}{47}
  (\bibinfo{year}{2018}) \bibinfo{pages}{3359--3379}.
  \DOIprefix\doi{10.1039/C7CS00864C}.
%Type = Article
\bibitem[{Safeer et~al.(2019)Safeer, Ingla-Ayn{\'e}s, Herling, Garcia, Vila,
  Ontoso, Calvo, Roche, Hueso, and Casanova}]{Safeer2019}
\bibinfo{author}{C.~K. Safeer}, \bibinfo{author}{J.~Ingla-Ayn{\'e}s},
  \bibinfo{author}{F.~Herling}, \bibinfo{author}{J.~H. Garcia},
  \bibinfo{author}{M.~Vila}, \bibinfo{author}{N.~Ontoso},
  \bibinfo{author}{M.~R. Calvo}, \bibinfo{author}{S.~Roche},
  \bibinfo{author}{L.~E. Hueso}, \bibinfo{author}{F.~Casanova},
\newblock \bibinfo{title}{Room-temperature spin hall effect in graphene/mos2
  van der waals heterostructures},
\newblock \bibinfo{journal}{Nano Lett.} \bibinfo{volume}{19}
  (\bibinfo{year}{2019}) \bibinfo{pages}{1074--1082}.
  \DOIprefix\doi{10.1021/acs.nanolett.8b04368}.
%Type = Article
\bibitem[{Ghiasi et~al.(2019)Ghiasi, Kaverzin, Blah, and van Wees}]{Ghiasi2019}
\bibinfo{author}{T.~S. Ghiasi}, \bibinfo{author}{A.~A. Kaverzin},
  \bibinfo{author}{P.~J. Blah}, \bibinfo{author}{B.~J. van Wees},
\newblock \bibinfo{title}{Charge-to-spin conversion by the rashba--edelstein
  effect in two-dimensional van der waals heterostructures up to room
  temperature},
\newblock \bibinfo{journal}{Nano Letters} \bibinfo{volume}{19}
  (\bibinfo{year}{2019}) \bibinfo{pages}{5959--5966}. \URLprefix
  \url{https://doi.org/10.1021/acs.nanolett.9b01611}.
  \DOIprefix\doi{10.1021/acs.nanolett.9b01611}.
%Type = Article
\bibitem[{Ben{\'i}tez et~al.(2020)Ben{\'i}tez, Savero~Torres, Sierra,
  Timmermans, Garcia, Roche, Costache, and Valenzuela}]{Benitez2020}
\bibinfo{author}{L.~A. Ben{\'i}tez}, \bibinfo{author}{W.~Savero~Torres},
  \bibinfo{author}{J.~F. Sierra}, \bibinfo{author}{M.~Timmermans},
  \bibinfo{author}{J.~H. Garcia}, \bibinfo{author}{S.~Roche},
  \bibinfo{author}{M.~V. Costache}, \bibinfo{author}{S.~O. Valenzuela},
\newblock \bibinfo{title}{Tunable room-temperature spin galvanic and spin hall
  effects in van der waals heterostructures},
\newblock \bibinfo{journal}{Nature Materials} \bibinfo{volume}{19}
  (\bibinfo{year}{2020}) \bibinfo{pages}{170--175}. \URLprefix
  \url{https://doi.org/10.1038/s41563-019-0575-1}.
  \DOIprefix\doi{10.1038/s41563-019-0575-1}.
%Type = Article
\bibitem[{Safeer et~al.(2020)Safeer, Ingla-Aynes, Ontoso, Herling, Yan, Hueso,
  and Casanova}]{Safeer2020}
\bibinfo{author}{C.~K. Safeer}, \bibinfo{author}{J.~Ingla-Aynes},
  \bibinfo{author}{N.~Ontoso}, \bibinfo{author}{F.~Herling},
  \bibinfo{author}{W.~Yan}, \bibinfo{author}{L.~E. Hueso},
  \bibinfo{author}{F.~Casanova},
\newblock \bibinfo{title}{Spin hall effect in bilayer graphene combined with an
  insulator up to room temperature},
\newblock \bibinfo{journal}{Nano Letters}  (\bibinfo{year}{2020}). \URLprefix
  \url{https://doi.org/10.1021/acs.nanolett.0c01428}.
  \DOIprefix\doi{10.1021/acs.nanolett.0c01428}.
%Type = Article
\bibitem[{Gregersen et~al.(2018)Gregersen, Garc\'{\i}a, Jauho, Roche, and
  Power}]{Gregersen2018}
\bibinfo{author}{S.~S. Gregersen}, \bibinfo{author}{J.~H. Garc\'{\i}a},
  \bibinfo{author}{A.-P. Jauho}, \bibinfo{author}{S.~Roche},
  \bibinfo{author}{S.~R. Power},
\newblock \bibinfo{title}{Charge and spin transport anisotropy in nanopatterned
  graphene},
\newblock \bibinfo{journal}{J. Phys. Mater.} \bibinfo{volume}{1}
  (\bibinfo{year}{2018}) \bibinfo{pages}{015005}.
  \DOIprefix\doi{10.1088/2515-7639/aadca3}.
%Type = Article
\bibitem[{Song et~al.(2015)Song, Samutpraphoot, and Levitov}]{Song10879}
\bibinfo{author}{J.~C.~W. Song}, \bibinfo{author}{P.~Samutpraphoot},
  \bibinfo{author}{L.~S. Levitov},
\newblock \bibinfo{title}{Topological bloch bands in graphene superlattices},
\newblock \bibinfo{journal}{Proc. Natl. Acad. Sci. U.S.A.}
  \bibinfo{volume}{112} (\bibinfo{year}{2015}) \bibinfo{pages}{10879--10883}.
  \DOIprefix\doi{10.1073/pnas.1424760112}.
%Type = Article
\bibitem[{Beconcini et~al.(2016)Beconcini, Taddei, and Polini}]{Beconcini2016}
\bibinfo{author}{M.~Beconcini}, \bibinfo{author}{F.~Taddei},
  \bibinfo{author}{M.~Polini},
\newblock \bibinfo{title}{Nonlocal topological valley transport at large valley
  hall angles},
\newblock \bibinfo{journal}{Phys. Rev. B} \bibinfo{volume}{94}
  (\bibinfo{year}{2016}) \bibinfo{pages}{121408}.
  \DOIprefix\doi{10.1103/PhysRevB.94.121408}.
%Type = Article
\bibitem[{Gorbachev et~al.(2014)Gorbachev, Song, Yu, Kretinin, Withers, Cao,
  Mishchenko, Grigorieva, Novoselov, Levitov, and Geim}]{Gorbachev448}
\bibinfo{author}{R.~V. Gorbachev}, \bibinfo{author}{J.~C.~W. Song},
  \bibinfo{author}{G.~L. Yu}, \bibinfo{author}{A.~V. Kretinin},
  \bibinfo{author}{F.~Withers}, \bibinfo{author}{Y.~Cao},
  \bibinfo{author}{A.~Mishchenko}, \bibinfo{author}{I.~V. Grigorieva},
  \bibinfo{author}{K.~S. Novoselov}, \bibinfo{author}{L.~S. Levitov},
  \bibinfo{author}{A.~K. Geim},
\newblock \bibinfo{title}{Detecting topological currents in graphene
  superlattices},
\newblock \bibinfo{journal}{Science} \bibinfo{volume}{346}
  (\bibinfo{year}{2014}) \bibinfo{pages}{448--451}.
  \DOIprefix\doi{10.1126/science.1254966}.
%Type = Article
\bibitem[{Marmolejo-Tejada et~al.(2018)Marmolejo-Tejada, Garc\'{\i}a,
  Petrovi\'{c}, Chang, Sheng, Cresti, Plechac, Roche, and
  Nikoli\'{c}}]{Marmolejo2018}
\bibinfo{author}{J.~M. Marmolejo-Tejada}, \bibinfo{author}{J.~H. Garc\'{\i}a},
  \bibinfo{author}{M.~D. Petrovi\'{c}}, \bibinfo{author}{P.-H. Chang},
  \bibinfo{author}{X.-L. Sheng}, \bibinfo{author}{A.~Cresti},
  \bibinfo{author}{P.~Plechac}, \bibinfo{author}{S.~Roche},
  \bibinfo{author}{B.~K. Nikoli\'{c}},
\newblock \bibinfo{title}{Deciphering the origin of nonlocal resistance in
  multiterminal graphene on hexagonal-boron-nitride with ab initio quantum
  transport: Fermi surface edge currents rather than fermi sea topological
  valley currents},
\newblock \bibinfo{journal}{J. Phys. Mater.} \bibinfo{volume}{1}
  (\bibinfo{year}{2018}) \bibinfo{pages}{015006}.
  \DOIprefix\doi{10.1088/2515-7639/aad585}.
%Type = Article
\bibitem[{Haldane(1988)}]{Haldane1988}
\bibinfo{author}{F.~D.~M. Haldane},
\newblock \bibinfo{title}{Model for a quantum hall effect without landau
  levels: Condensed-matter realization of the "parity anomaly"},
\newblock \bibinfo{journal}{Phys. Rev. Lett.} \bibinfo{volume}{61}
  (\bibinfo{year}{1988}) \bibinfo{pages}{2015--2018}. \URLprefix
  \url{https://link.aps.org/doi/10.1103/PhysRevLett.61.2015}.
  \DOIprefix\doi{10.1103/PhysRevLett.61.2015}.
%Type = Article
\bibitem[{Hasan and Kane(2010)}]{Hasan2010RMP}
\bibinfo{author}{M.~Z. Hasan}, \bibinfo{author}{C.~L. Kane},
\newblock \bibinfo{title}{Colloquium: Topological insulators},
\newblock \bibinfo{journal}{Rev. Mod. Phys.} \bibinfo{volume}{82}
  (\bibinfo{year}{2010}) \bibinfo{pages}{3045--3067}. \URLprefix
  \url{https://link.aps.org/doi/10.1103/RevModPhys.82.3045}.
  \DOIprefix\doi{10.1103/RevModPhys.82.3045}.
%Type = Inbook
\bibitem[{Ortmann et~al.(2015)Ortmann, Roche, and Valenzuela}]{ORTMANNti}
\bibinfo{author}{F.~Ortmann}, \bibinfo{author}{S.~Roche},
  \bibinfo{author}{S.~O. Valenzuela}, \bibinfo{title}{Quantum Spin Hall Effect
  and Topological Insulators}, \bibinfo{publisher}{John Wiley and Sons, Ltd},
  \bibinfo{year}{2015}, pp. \bibinfo{pages}{1--407}. \URLprefix
  \url{https://onlinelibrary.wiley.com/doi/abs/10.1002/9783527681594.ch1}.
  \DOIprefix\doi{10.1002/9783527681594.ch1}.
%Type = Article
\bibitem[{Ren et~al.(2016)Ren, Qiao, and Niu}]{Ren_2016}
\bibinfo{author}{Y.~Ren}, \bibinfo{author}{Z.~Qiao}, \bibinfo{author}{Q.~Niu},
\newblock \bibinfo{title}{Topological phases in two-dimensional materials: a
  review},
\newblock \bibinfo{journal}{Reports on Progress in Physics}
  \bibinfo{volume}{79} (\bibinfo{year}{2016}) \bibinfo{pages}{066501}.
  \URLprefix \url{https://doi.org/10.1088\%2F0034-4885\%2F79\%2F6\%2F066501}.
  \DOIprefix\doi{10.1088/0034-4885/79/6/066501}.
%Type = Article
\bibitem[{Olsen et~al.(2019)Olsen, Andersen, Okugawa, Torelli, Deilmann, and
  Thygesen}]{Olsen2019}
\bibinfo{author}{T.~Olsen}, \bibinfo{author}{E.~Andersen},
  \bibinfo{author}{T.~Okugawa}, \bibinfo{author}{D.~Torelli},
  \bibinfo{author}{T.~Deilmann}, \bibinfo{author}{K.~S. Thygesen},
\newblock \bibinfo{title}{Discovering two-dimensional topological insulators
  from high-throughput computations},
\newblock \bibinfo{journal}{Phys. Rev. Materials} \bibinfo{volume}{3}
  (\bibinfo{year}{2019}) \bibinfo{pages}{024005}. \URLprefix
  \url{https://link.aps.org/doi/10.1103/PhysRevMaterials.3.024005}.
  \DOIprefix\doi{10.1103/PhysRevMaterials.3.024005}.
%Type = Article
\bibitem[{Weeks et~al.(2011)Weeks, Hu, Alicea, Franz, and Wu}]{Weeks2011}
\bibinfo{author}{C.~Weeks}, \bibinfo{author}{J.~Hu},
  \bibinfo{author}{J.~Alicea}, \bibinfo{author}{M.~Franz},
  \bibinfo{author}{R.~Wu},
\newblock \bibinfo{title}{Engineering a robust quantum spin hall state in
  graphene via adatom deposition},
\newblock \bibinfo{journal}{Phys. Rev. X} \bibinfo{volume}{1}
  (\bibinfo{year}{2011}) \bibinfo{pages}{021001}. \URLprefix
  \url{https://link.aps.org/doi/10.1103/PhysRevX.1.021001}.
  \DOIprefix\doi{10.1103/PhysRevX.1.021001}.
%Type = Article
\bibitem[{Jiang et~al.(2012)Jiang, Qiao, Liu, Shi, and Niu}]{JiangPRL2012}
\bibinfo{author}{H.~Jiang}, \bibinfo{author}{Z.~Qiao},
  \bibinfo{author}{H.~Liu}, \bibinfo{author}{J.~Shi}, \bibinfo{author}{Q.~Niu},
\newblock \bibinfo{title}{Stabilizing topological phases in graphene via random
  adsorption},
\newblock \bibinfo{journal}{Phys. Rev. Lett.} \bibinfo{volume}{109}
  (\bibinfo{year}{2012}) \bibinfo{pages}{116803}. \URLprefix
  \url{https://link.aps.org/doi/10.1103/PhysRevLett.109.116803}.
  \DOIprefix\doi{10.1103/PhysRevLett.109.116803}.
%Type = Article
\bibitem[{Liu et~al.(2015)Liu, Zhu, and Zheng}]{Liu2015QSHE}
\bibinfo{author}{Z.~Liu}, \bibinfo{author}{M.~Zhu}, \bibinfo{author}{Y.~Zheng},
\newblock \bibinfo{title}{Quantum transport properties of graphene in the
  presence of randomly distributed spin-orbit coupling impurities},
\newblock \bibinfo{journal}{Phys. Rev. B} \bibinfo{volume}{92}
  (\bibinfo{year}{2015}) \bibinfo{pages}{245438}. \URLprefix
  \url{https://link.aps.org/doi/10.1103/PhysRevB.92.245438}.
  \DOIprefix\doi{10.1103/PhysRevB.92.245438}.
%Type = Article
\bibitem[{Santos et~al.(2018)Santos, Bahamon, Muniz, McKenna, Castro, Lischner,
  and Ferreira}]{Santos2018}
\bibinfo{author}{F.~J.~d. Santos}, \bibinfo{author}{D.~A. Bahamon},
  \bibinfo{author}{R.~B. Muniz}, \bibinfo{author}{K.~McKenna},
  \bibinfo{author}{E.~V. Castro}, \bibinfo{author}{J.~Lischner},
  \bibinfo{author}{A.~Ferreira},
\newblock \bibinfo{title}{Impact of complex adatom-induced interactions on
  quantum spin hall phases},
\newblock \bibinfo{journal}{Phys. Rev. B} \bibinfo{volume}{98}
  (\bibinfo{year}{2018}) \bibinfo{pages}{081407}. \URLprefix
  \url{https://link.aps.org/doi/10.1103/PhysRevB.98.081407}.
  \DOIprefix\doi{10.1103/PhysRevB.98.081407}.
%Type = Article
\bibitem[{Groth et~al.(2009)Groth, Wimmer, Akhmerov, Tworzyd\l{}o, and
  Beenakker}]{GrothPRL2009}
\bibinfo{author}{C.~W. Groth}, \bibinfo{author}{M.~Wimmer},
  \bibinfo{author}{A.~R. Akhmerov}, \bibinfo{author}{J.~Tworzyd\l{}o},
  \bibinfo{author}{C.~W.~J. Beenakker},
\newblock \bibinfo{title}{Theory of the topological anderson insulator},
\newblock \bibinfo{journal}{Phys. Rev. Lett.} \bibinfo{volume}{103}
  (\bibinfo{year}{2009}) \bibinfo{pages}{196805}.
  \DOIprefix\doi{10.1103/PhysRevLett.103.196805}.
%Type = Article
\bibitem[{Li et~al.(2009)Li, Chu, Jain, and Shen}]{JianPRL2009}
\bibinfo{author}{J.~Li}, \bibinfo{author}{R.-L. Chu}, \bibinfo{author}{J.~K.
  Jain}, \bibinfo{author}{S.-Q. Shen},
\newblock \bibinfo{title}{Topological anderson insulator},
\newblock \bibinfo{journal}{Phys. Rev. Lett.} \bibinfo{volume}{102}
  (\bibinfo{year}{2009}) \bibinfo{pages}{136806}. \URLprefix
  \url{https://link.aps.org/doi/10.1103/PhysRevLett.102.136806}.
  \DOIprefix\doi{10.1103/PhysRevLett.102.136806}.
%Type = Article
\bibitem[{Zhang et~al.(2012)Zhang, Chu, Zhang, and Shen}]{ZhangPRB2012}
\bibinfo{author}{Y.-Y. Zhang}, \bibinfo{author}{R.-L. Chu},
  \bibinfo{author}{F.-C. Zhang}, \bibinfo{author}{S.-Q. Shen},
\newblock \bibinfo{title}{Localization and mobility gap in the topological
  anderson insulator},
\newblock \bibinfo{journal}{Phys. Rev. B} \bibinfo{volume}{85}
  (\bibinfo{year}{2012}) \bibinfo{pages}{035107}.
  \DOIprefix\doi{10.1103/PhysRevB.85.035107}.
%Type = Article
\bibitem[{Young et~al.(2012)Young, Zaheer, Teo, Kane, Mele, and
  Rappe}]{YoungPRL2012}
\bibinfo{author}{S.~M. Young}, \bibinfo{author}{S.~Zaheer},
  \bibinfo{author}{J.~C.~Y. Teo}, \bibinfo{author}{C.~L. Kane},
  \bibinfo{author}{E.~J. Mele}, \bibinfo{author}{A.~M. Rappe},
\newblock \bibinfo{title}{Dirac semimetal in three dimensions},
\newblock \bibinfo{journal}{Phys. Rev. Lett.} \bibinfo{volume}{108}
  (\bibinfo{year}{2012}) \bibinfo{pages}{140405}.
  \DOIprefix\doi{10.1103/PhysRevLett.108.140405}.
%Type = Article
\bibitem[{Kobayashi et~al.(2014)Kobayashi, Ohtsuki, Imura, and
  Herbut}]{KobayashiPRL2014}
\bibinfo{author}{K.~Kobayashi}, \bibinfo{author}{T.~Ohtsuki},
  \bibinfo{author}{K.-I. Imura}, \bibinfo{author}{I.~F. Herbut},
\newblock \bibinfo{title}{Density of states scaling at the semimetal to metal
  transition in three dimensional topological insulators},
\newblock \bibinfo{journal}{Phys. Rev. Lett.} \bibinfo{volume}{112}
  (\bibinfo{year}{2014}) \bibinfo{pages}{016402}.
  \DOIprefix\doi{10.1103/PhysRevLett.112.016402}.
%Type = Article
\bibitem[{Louvet et~al.(2018)Louvet, Houzet, and Carpentier}]{Louvet_2018}
\bibinfo{author}{T.~Louvet}, \bibinfo{author}{M.~Houzet},
  \bibinfo{author}{D.~Carpentier},
\newblock \bibinfo{title}{Signature of the chiral anomaly in ballistic weyl
  junctions},
\newblock \bibinfo{journal}{Journal of Physics: Materials} \bibinfo{volume}{1}
  (\bibinfo{year}{2018}) \bibinfo{pages}{015008}. \URLprefix
  \url{https://doi.org/10.1088\%2F2515-7639\%2Faadd61}.
%Type = Article
\bibitem[{Hasan and Kane(2010)}]{HasanRMP2010}
\bibinfo{author}{M.~Z. Hasan}, \bibinfo{author}{C.~L. Kane},
\newblock \bibinfo{title}{Colloquium: Topological insulators},
\newblock \bibinfo{journal}{Rev. Mod. Phys.} \bibinfo{volume}{82}
  (\bibinfo{year}{2010}) \bibinfo{pages}{3045--3067}.
  \DOIprefix\doi{10.1103/RevModPhys.82.3045}.
%Type = Article
\bibitem[{Fu et~al.(2007)Fu, Kane, and Mele}]{FuPRL2007}
\bibinfo{author}{L.~Fu}, \bibinfo{author}{C.~L. Kane}, \bibinfo{author}{E.~J.
  Mele},
\newblock \bibinfo{title}{Topological insulators in three dimensions},
\newblock \bibinfo{journal}{Phys. Rev. Lett.} \bibinfo{volume}{98}
  (\bibinfo{year}{2007}) \bibinfo{pages}{106803}.
  \DOIprefix\doi{10.1103/PhysRevLett.98.106803}.
%Type = Article
\bibitem[{Liao et~al.(2015)Liao, Ou, Feng, Yang, Lin, Yang, Wu, He, Ma, Xue,
  and Li}]{Liao2015}
\bibinfo{author}{J.~Liao}, \bibinfo{author}{Y.~Ou}, \bibinfo{author}{X.~Feng},
  \bibinfo{author}{S.~Yang}, \bibinfo{author}{C.~Lin},
  \bibinfo{author}{W.~Yang}, \bibinfo{author}{K.~Wu}, \bibinfo{author}{K.~He},
  \bibinfo{author}{X.~Ma}, \bibinfo{author}{Q.-K. Xue},
  \bibinfo{author}{Y.~Li},
\newblock \bibinfo{title}{Observation of anderson localization in ultrathin
  films of three-dimensional topological insulators},
\newblock \bibinfo{journal}{Phys. Rev. Lett.} \bibinfo{volume}{114}
  (\bibinfo{year}{2015}) \bibinfo{pages}{216601}.
  \DOIprefix\doi{10.1103/PhysRevLett.114.216601}.
%Type = Article
\bibitem[{Chiu et~al.(2016)Chiu, Teo, Schnyder, and Ryu}]{Chiu2016}
\bibinfo{author}{C.-K. Chiu}, \bibinfo{author}{J.~C.~Y. Teo},
  \bibinfo{author}{A.~P. Schnyder}, \bibinfo{author}{S.~Ryu},
\newblock \bibinfo{title}{Classification of topological quantum matter with
  symmetries},
\newblock \bibinfo{journal}{Rev. Mod. Phys.} \bibinfo{volume}{88}
  (\bibinfo{year}{2016}) \bibinfo{pages}{035005}.
  \DOIprefix\doi{10.1103/RevModPhys.88.035005}.
%Type = Article
\bibitem[{Araki et~al.(2019)Araki, Mizoguchi, and Hatsugai}]{Araki2019}
\bibinfo{author}{H.~Araki}, \bibinfo{author}{T.~Mizoguchi},
  \bibinfo{author}{Y.~Hatsugai},
\newblock \bibinfo{title}{Phase diagram of a disordered higher-order
  topological insulator: A machine learning study},
\newblock \bibinfo{journal}{Phys. Rev. B} \bibinfo{volume}{99}
  (\bibinfo{year}{2019}) \bibinfo{pages}{085406}.
  \DOIprefix\doi{10.1103/PhysRevB.99.085406}.
%Type = Article
\bibitem[{Schleder et~al.(2019)Schleder, Padilha, Acosta, Costa, and
  Fazzio}]{Schleder2019}
\bibinfo{author}{G.~R. Schleder}, \bibinfo{author}{A.~C.~M. Padilha},
  \bibinfo{author}{C.~M. Acosta}, \bibinfo{author}{M.~Costa},
  \bibinfo{author}{A.~Fazzio},
\newblock \bibinfo{title}{From {DFT} to machine learning: recent approaches to
  materials science{\textendash}a review},
\newblock \bibinfo{journal}{J. Phys. Mater.} \bibinfo{volume}{2}
  (\bibinfo{year}{2019}) \bibinfo{pages}{032001}.
  \DOIprefix\doi{10.1088/2515-7639/ab084b}.

\end{thebibliography}
